\documentclass[preprint,floatfix] {revtex4} 
\usepackage{graphicx}
\usepackage{subfigure}
\usepackage{xcolor}
\usepackage{enumerate}

\usepackage{color, soul}
\definecolor{darkblue}{rgb}{0,0,0.5}
\setulcolor{darkblue}

\begin{document}

\title{Quantum confinement in an asymmetric double-well potential through energy analysis and information entropic measure}

\author{Neetik Mukherjee}
\altaffiliation{Email: neetik.mukherjee@iiserkol.ac.in.}

\author{Amlan K.~Roy}
\altaffiliation{Corresponding author. Email: akroy@iiserkol.ac.in.}
\affiliation{Department of Chemical Sciences\\
Indian Institute of Science Education and Research (IISER) Kolkata, 
Mohanpur-741246, Nadia, West Bengal, India}

\begin{abstract}
Localization of a particle in the wells of an asymmetric double-well (DW) potential is investigated here. 
Information entropy-based uncertainty measures, such as Shannon entropy, Fisher information, Onicescu energy, etc., and phase-space area,
are utilized to explain the contrasting effect of localization-delocalization and role of asymmetric term in such two-well 
potentials. In asymmetric situation, two wells behaves like two different potentials. A general rule has been proposed 
for arrangement of quasi-degenerate pairs, in terms of asymmetry parameter. Further, it enables   
to describe the distribution of particle in either of the deeper or shallow wells in various energy states. One finds 
that, all states eventually get localized to the deeper well, provided the asymmetry parameter attains certain 
threshold value. This generalization produces symmetric DW as a natural consequence of asymmetric DW. 
Eigenfunctions, eigenvalues are obtained by means of a simple, accurate variation-induced exact diagonalization 
method. In brief, information measures and phase-space analysis can provide valuable insight toward the understanding of such potentials.  

\vspace{5mm}

\vspace{5mm}

\end{abstract}
\maketitle

\section{Introduction}
Uncertainty in position space is a well known measure of spatial delocalization 
\cite{razavy2003quantum,vidal2000interaction} of a quantum particle. This relation provides a lower bound on 
information content, but does not quantify the full information content of a particle in a given quantum state.
The equation, 
\begin{equation}\label{uncer}
\triangle x \triangle p \geq \frac{1}{2}, 
\end{equation}
implies that momentum and position of a particle in a system cannot be measured simultaneously. If position is measured 
with complete precision, momentum will be infinitely uncertain and vice-versa. This has been used successfully 
in studying uncertainty-like relationship in numerous quantum systems, such as  
spherically symmetric potentials including radial position-momentum uncertainties in Klein-Gordon H-like atoms 
\citep{tsapline1970expectation,majernik1997entropic,grypeos2004hvt,kuo2005uncertainties,qiang2006radial,qiang2008radial}. 

It is a well known fact that, information entropy (IE)-based uncertainty relations may offer more efficient and stronger 
bounds \citep{patil2007characteristic} than that provided by conventional uncertainty relation, Eq.~(1). In case of IE, 
measurements are carried out in conjugate position and momentum space. Our curiosity and thrust of using concept 
of IE thrives with development of various entropic uncertainty relations 
\citep{bialynicki1975uncertainty, aydiner2008quantum}. Lately, this mixed entropic uncertainty relation has been 
extensively used to investigate a variety of confined quantum systems ranging from confined harmonic oscillator 
\citep{laguna2014quantum} to confined atom \citep{patil2007characteristic, romera2005fisher, toranzo2014pauli}. Day by 
day, interest in the topic continues to expand as evidenced by a substantial amount of existing literature. In past 
few years, Shannon entropy has been employed on a vast number of physically and chemically important potentials, 
such as, P\"oschl-Teller-like \cite{sun2013quantum}, Rosen-Morse \cite{sun2013quantum1}, squared tangent well 
\cite{dong2014quantum}, position-dependent mass Schr\"odinger equation \cite{yanez2014quantum,chinphysb}, hyperbolic 
\cite{valencia2015quantum}, infinite circular well \cite{song2015shannon}, hyperbolic double-well (DW) potential
\cite{sun2015shannon}, etc.  

A DW potential has relevance in many important physical, chemical phenomena where the potential can be 
modeled as a two-state system \cite{verguilla1993tunneling,childsymm}, wherein two wells may be identified as two different states 
of a quantum particle. Thus, significant attention was paid for their understanding, starting from the dawn of quantum 
mechanics till today (see, for example, \cite{griffiths1995introduction,merzbacher98,jelic2012}). In a previous work 
\citep{doublewell}, several information measures like Fisher information (I) \citep{frieden2004science, 
romera2005fisher}, Shannon entropy (S) \citep{shannon1951prediction}, Onicescu energy (E) \citep{agop2014implications} 
and Onicescu-Shannon information entropy (OS) \citep{LpezRuiz1995321, chatzisavvas2005information, panos2007comparison} 
were invoked to understand the competing effect of localization-delocalization in a symmetric DW potential, represented 
by, $V(x)=\alpha x^4-\beta x^2 +h,$ where $\alpha, \beta$ are two potential parameters and $h=\frac{\beta^2}{4\alpha}$
controls barrier height. It was found that, while traditional uncertainty relation, Eq.~(1), and Fisher 
information measures can qualitatively indicate quasi-degeneracy \citep{quasi-degeneracy}, they are generally inadequate 
to interpret the role of competing effects which causes quasi-degeneracy in such a system. However, it turns out that, 
measures such as $S, I, E, OS$ in the conjugate spaces bear distinct signatures of these contrasting processes. For 
example, there appears characteristic extrema in $S, E, OS$ values corresponding to the balancing of these opposing 
effects; initially delocalization predominates but finally localization overcoming it. Furthermore, a phase-space 
area analysis explained the tunneling in such a potential. In presence of asymmetry, such interplay is expected to lead 
to more interesting and complicated results. A key to our understanding would be to determine how quasi-degeneracy is 
affected in such a scenario. 

In case of a symmetric DW, aforesaid competing effects lead to a number of quasi-degenerate pair states. An increase in 
positive term (strength) reduces spacing between classical turning points but at the same time, reduces barrier area 
as well as barrier height (refer to Eq.~(2) later). Conversely, negative term increases spacing between classical 
turning points but also increases barrier area and barrier height. Introduction of an asymmetric term   
makes the system more anomalous and interesting. Because it gives rise to two asymmetric wells in the system. From a 
classical mechanical point of view, a particle will always reside in deeper well, but in quantum mechanics the 
situation is not that straightforward. Therefore, a careful analysis is desirable to understand the quantum effects 
in such system. In this context, our primary motivation is to study energy and distribution of a particle in 
an asymmetric DW potential; then to use information-based uncertainty measures to broaden our knowledge on the 
overall situation of system. For our purposes, such a potential can be written conveniently in following form,
\begin{equation}\label{dw}
V(x) = \alpha x^{2p}-\beta x^{2q} +\gamma x^{r},
\end{equation}
where $p, q$ are two positive integers such that $p \! > \! q$, and $r$ is an odd integer with $q \! > \! r$. In what 
follows, we concentrate on the specific case of $p \! = \! 2,q \! = \! r \! = \! 1$, unless mentioned otherwise. This 
choice corresponds to the simplest 
possible asymmetric DW case within a polynomial potential family in Eq.~(2). Note that one requires a quartic, 
quadratic and linear term. \emph{Exact} analytical solution of our asymmetric DW potential has not been reported 
as yet. We are aware of the lone \emph{exactly solvable} work of \citep{selg2000exactly} for a special case of asymmetric
DW potential, constructed from several smoothly joined Morse-type components. 

Application of asymmetric DW potential ranges from hydrogen bonded solid \citep{somorzai}, anomalous optical lattice 
vibrations in HgTe \cite{kozyrev2010}, electron in a double quantum dot \citep{burkard}, proton transfer in 
DNA \cite{RevModPhys.35.724} to model brain micro-tubules \cite{faber2006information}, as well as internal rotation 
\cite{hammons1979electron}, etc., to name a few. It has relevance also in physics of solid-state devices 
\cite{holmes1980addendum}, solar cells and electron tunneling microscopes \cite{holmes1980addendum}. In last two decades, 
quantum tunneling in asymmetric DW potential was studied vigorously 
\citep{tunnel1,tunnel2,tunnel3,tunnel4,tunnel5,vybornyi2014tunnel,song2008tunneling}. Tunneling is also used to locate 
diabolic points of magnetic molecule Fe$_8$, where the bottoms of wells can be moved around by applying magnetic fields 
\citep{garg2001quenched}. Some other example applications are: Bose-Einstein condensation
in trapped potentials \citep{levy2007ac,anker2005nonlinear,albiez2005direct,shin2005optical,hall2007condensate},
quantum superconducting circuit based on Josephson junction \citep{joseph1,joseph2,joseph3,joseph4,joseph5,joseph6}, 
quantum computing devices \citep{ladd}, etc. 

In this endeavor, our objective is threefold. First, to investigate distribution of energy as a function of 
asymmetric term $\gamma$, followed by an attempt to frame a general rule for appearance of accidental quasi-degeneracy 
in such systems. Secondly, we try to establish localization of particle in one of the wells from a consideration 
of probability density as a function of $\gamma$; in this case also some simple general guidelines have been provided.  
As expected, distribution of particle is not uniform in two sides of potential. Thus there is a possibility that 
these two wells behave as two different potentials. This is examined by calculating number of effective nodes 
corresponding to a given energy state. Thirdly, variations in $S,I, E, OS$ as function of $\beta$, $\gamma$ 
parameters in both position, momentum space are followed. Results from above mentioned net information measures 
are then correlated with observations on energy and probability distribution. If $\gamma=0$ is chosen, 
Eq.~(2) leads to a symmetric DW. Thus, attempt is made to connect symmetric DW with asymmetric DW. Additionally, 
these IEs are employed to interpret confinement and trapping \cite{karski2009quantum} of particle inside one of 
the potential wells along the lines of our previous work in a symmetric DW \cite{doublewell}.  

\begin{equation}
A_{n} =  \int \sqrt{V(x)-E_{n}} \ dx
\end{equation}

Further, in order to get a semi-classical viewpoint for the distribution of particle in an asymmetric DW,
we offer phase-space area ($A_n$), calculated through Eq.~(3). Recently Wigner probability distribution has been 
reported for 
quartic DW potential \cite{wigner}. In the current work, our focus lies mostly to understand the effect of parameters 
$\gamma$, $\beta$ on phase-space area, and to correlate them with quantum mechanical results.  

\section{Methodology}
Without any loss of generality, the DW potential is simply written as, 
\begin{eqnarray}
V(x) = \alpha x^{4}-\beta x^{2} + \gamma x + V_0,\\
V_{1}(x) = \alpha x^{4}-\beta x^{2} - \gamma x + V_0.
\end{eqnarray}
Here $V_0$ refers to the minimum value of potential, added to make total energy positive. 
Clearly, $V$ and $V_{1}$ constitute a mirror-image pair; substitution of $x$ by $-x$ converts one to other. Eigenspectra 
of these potentials are very similar; energies possessed by all states are identical for both. Moreover, their wave 
functions are also mirror images to each other. Thus, it suffices to study the behavior of any one of them; other one  
automatically follows. This leads us to choose $V(x)$ as our model asymmetric DW potential, which is explored in remaining
part of this communication. Introduction of a linear term in a symmetric DW brings asymmetry in the system. With an 
increase in quartic parameter $\alpha$, separation between classical turning points and barrier strength decrease.
Thus, in one hand, it increases localization and on other hand, decreases particle's confinement within the well. An 
increase in $\beta$ too brings competing effects into the picture. It causes an increase in separation of classical 
turning points, implying added delocalization of particle, whereas at the same time, barrier area and barrier height 
increase, which apparently promote localization into one of the wells. Asymmetric term shifts the maximum 
of potential from zero to either right or left sides depending on whether $\gamma$ is positive or negative. In general, 
there exists a deeper (I) and a shallow (II) well. Figure~(1) illustrates two typical scenario at two different 
$\gamma$. It also displays that maximum of $V$ shifts towards right as $\gamma$ increases. 

\begin{figure}                         
\begin{minipage}[c]{0.4\textwidth}\centering
\includegraphics[scale=0.75]{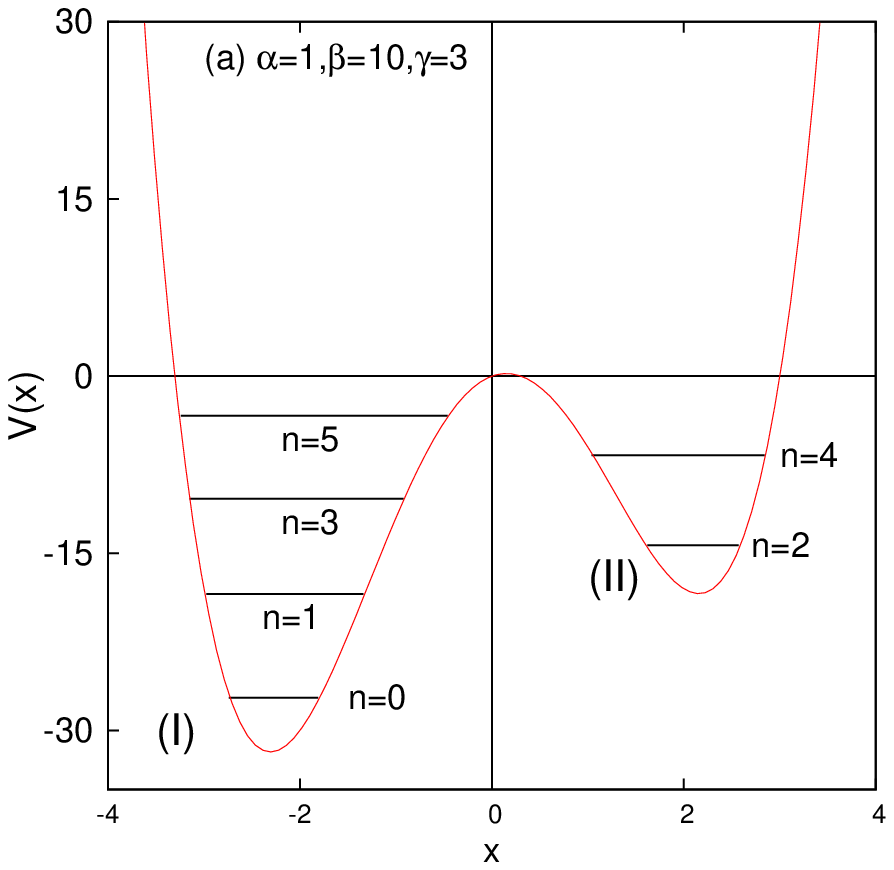}
\end{minipage}%
\hspace{0.1in}
\begin{minipage}[c]{0.5\textwidth}\centering
\includegraphics[scale=0.75]{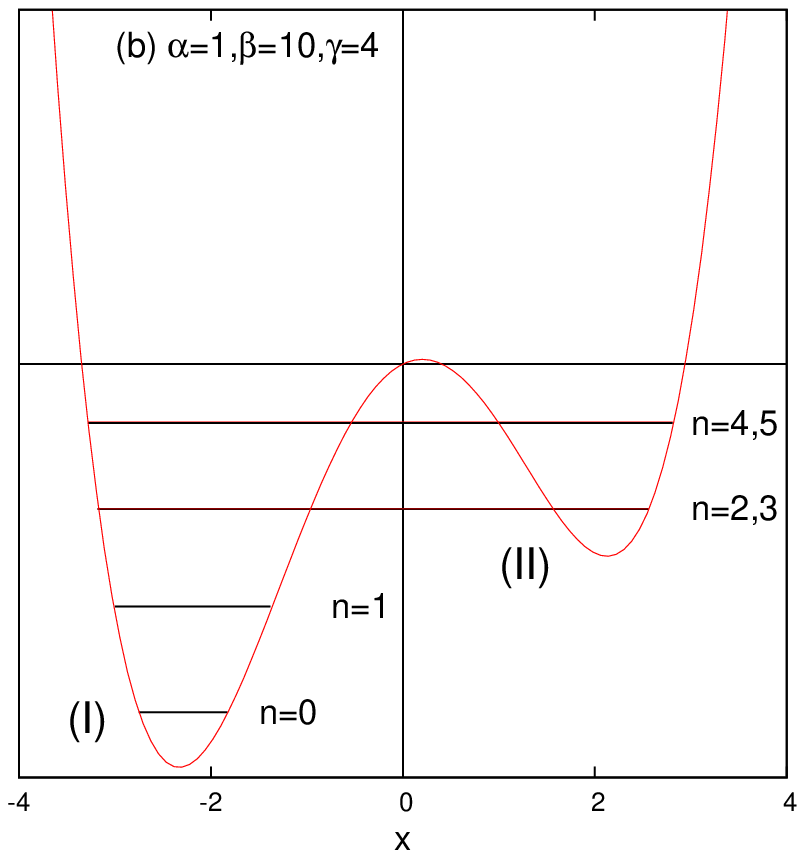}
\end{minipage}%
\caption{Schematic representation of a DW potential, in Eq.~(4), plotted at two different parameter sets: (a) 
$\alpha \! = \! 1, \beta \! = \! 10, \gamma \! = \! 3$ (b) $\alpha \! = \! 1, \beta \! = \! 10, \gamma \! = \! 4$. (I), 
(II) signify larger, smaller wells respectively.}
\end{figure}

It may be noted that, most prominent $\beta$, $\gamma$-variation is likely to be observed for lowest states; hence
present study mainly focuses on first four energy states. However, in some occasions, we also analyze these changes
for higher excited states as well; mostly to verify if the qualitative trends present in lower states still persist in 
other states. In this connection, it is worthwhile to study how the successive appearance of nodes in wave function
impacts behavior of IE. Note that, last term shifts minimum at zero to make energy value positive. 
 
The Hamiltonian in position space is:
\begin{equation}
\hat{H_{x}}=-\frac{d^{2}}{dx^{2}}+\alpha x^{4}-\beta x^{2} + \gamma x,
\end{equation}
whereas in momentum space, this reads, 
\begin{equation}
\hat{H_{p}}=p^{2}+\alpha\frac{d^{4}}{dp^{4}}+\beta\frac{d^{2}}{dp^{2}}-i\gamma\frac{d}{dp}.
\end{equation}
For sake of convenience, we choose $2m \! = \! 1$ and $\hbar=1$.

\subsection{Variation-induced exact diagonalization method}
Eigenvalues and eigenvectors needed for our calculation are generated using a diagonalization method with Hermite basis of a
quantum harmonic oscillator (QHO). It \cite{griffiths1995introduction} is simple, easy to implement numerically; yet produces 
accurate, reliable solutions without requiring much complication. Details of methodology as well its feasibility, 
performance and relevance in our present context have been amply demonstrated earlier \citep{doublewell}. Thus these
are not repeated here; instead only basic equations are given. \emph{Near-exact} solutions (for 
a number of parameter sets) were offered for symmetric DW, using a modest number of basis functions. 
These were comparable to some of the best results available in literature. Essentially, the full Hamiltonian 
is represented in terms of raising and lowering operators of a conventional QHO, containing a single non-linear parameter 
$\sigma$ (see Eq.~(16) of \cite{doublewell}). Dimension of our Hamiltonian matrix is set to $N=100$ (see Table~I later) to 
guarantee convergence of energies and wave functions. After some straightforward algebra, analytical expressions of matrix 
elements in above number-operator basis, are easily obtained as ($l,m$ denote state index), 
\begin{equation}
h_{lm} = \langle l|\hat{\mbox{H}}|m\rangle.
\end{equation}  

In position space, the non-zero matrix elements can be written as follows,  
\begin{eqnarray}
h_{lm} & = & \frac{\alpha}{16\sigma^2}\sqrt{l(l-1)(l-2)(l-3)}, \hspace{2.05in} \mathrm{if} \  l-m=4  \\
        & = & \frac{1}{8\sigma^2}[\alpha(2l-1)-2(\beta+4\sigma^2) \sigma]\sqrt{l(l-1)}, \hspace{1.3in}  
\mathrm{if} \ l-m=2  \nonumber \\
        & =& \gamma \sqrt{\frac{l}{4\sigma}}, \hspace{3.65in} \mathrm{if} \ l-m=1  \nonumber  \\
        &= & \frac{3\alpha}{16\sigma^2}(2l^2+2l+1)-(\beta+4\sigma^2) \frac{(2l+1)}{4\sigma}+2 \sigma (2l+1),  
\hspace{0.35in} \ \mathrm{if} \ l-m=0   \nonumber \\
        & = & \gamma \sqrt{\frac{(l+1)}{4\sigma}}, \hspace{3.35in}  \mathrm{if} \ l-m=-1 \nonumber \\
        & = & \frac{1}{8\sigma^2}[\alpha(2l+3)-2(\beta+4\sigma^2) \sigma]\sqrt{(l+1)(l+2)},  \hspace{0.95in}  
\mathrm{if} \ l-m=-2 \nonumber \\
        & = & \frac{\alpha}{16\sigma^2}\sqrt{(l+1)(l+2)(l+3)(l+4)}.  \hspace{1.78in} \mathrm{if} \ l-m=-4  \nonumber 
\end{eqnarray}
On the other hand, in momentum space, the non-zero elements are obtained as, 
\begin{eqnarray}
g_{lm} & = &  h_{lm}, \hspace{2.2in} \mathrm{if} \  (l-m)=4,-4 \\ 
       & = & -h_{lm}, \hspace{2.05in} \mathrm{if} \  (l-m)=2,-2 \nonumber \\ 
       & = &-ih_{lm}, \hspace{2in} \mathrm{if} \  (l-m)=1  \nonumber    \\
       & = & ih_{lm}. \hspace{2.12in} \mathrm{if} \  (l-m)=-1  \nonumber
\end{eqnarray}

Diagonalization of the \emph{symmetric} matrix, {\bf h} was accomplished efficiently by MATHEMATICA, leading to accurate 
energy eigenvalues and corresponding eigenvectors. We adopt a Manifold-Energy minimization approach due to 
\cite{hendekovic1983reply,pathak1994nonlinear}, where instead of minimizing a particular energy state, one minimizes 
trace of the matrix, which is given below, 
\begin{equation}
Tr[h]=\sum_{l}h_{ll} = \sum_{l}\left[\frac{3\alpha}{16\sigma^{2}}(2l^{2}+2l+1)-
\frac{(\beta + 4\sigma^2)(2l+1)}{4\sigma}+2\sigma(2l+1)\right]. 
\end{equation}
with respect to $\sigma$. This leads to a cubic equation in $\sigma$ having a single real root. Finally the process is
completed by minimizing the desired matrix in Eq.~(8), for above value of $\sigma$.

In order to validate and assess the performance of above approach, sample calculations are done on energy spectrum 
of an asymmetric DW potential having following form,
\begin{equation}
V_{2}(x)=0.01(x^{4}-0.75x^{3}-0.25x^{2}). 
\end{equation}
Table~I shows the convergence of energy values for first four states. These are compared with the very 
accurate results \citep{taseli}, obtained by using trigonometric basis within a Rayleigh-Ritz variational framework.
With an increase in basis, energies steadily improve and proceeds towards converged values. In principle, it is
possible to better these results further by employing larger basis. However, that is aside from the purpose of our
present work. It is sufficient for us to have $N=100$; all analysis henceforth is done using this basis. 

\begingroup      
\squeezetable
\begin{table}
\caption{Convergence of eigenvalues for ground and first three excited states of asymmetric DW potential in Eq.~(12). 
Accurate reference energies are quoted from \citep{taseli}.}
\centering
\begin{ruledtabular}
\begin{tabular}{cllll}
$N$ &  E$_0^{\S}$  &  E$_1^\dag$   &  E$_2^\ddag$    &   E$_3^\P$   \\
\hline
25  &  0.2204969              & 0.7990763               &  1.579426   &  2.475227      \\
50  &  0.22049693355          & 0.79907615613           &  1.579425872716    &  2.475227126        \\
75  &  0.22049693355138       & 0.79907615613404        &  1.579425872715042  &  2.47522712629      \\
100 &  0.22049693355138318  &  0.799076156134041042     &  1.5794258727150421868     &  2.47522712627695    \\
\end{tabular}
\end{ruledtabular}
\begin{tabbing}
$^{\S}$Reference value is: 0.22049693355138318. \hspace{25pt} \= 
$^\dag$Reference value is: 0.799076156134041042.  \\
$^\ddag$Reference value is: 1.5794258727150421868. \hspace{15pt} \=
$^\P$Reference value is: 2.47522712627695799794.
\end{tabbing}
\end{table} 
\endgroup

\section{Uncertainty-like information measures}
IEs are properly weighted measure of quantum probability distribution function, $\rho(x) = |\psi(x)|^2$. In position 
space, whereas uncertainty contains up to \emph{second} moments of $\rho(x)$, IE contains contributions from \emph{all} 
moments that are relevant and present in quantum probability distribution \cite{diosi2011short}. Thus intuitively, IE 
gives a better and more complete description of all competitive moments present in wave function. These uncertainty 
measures are significantly useful in predicting localization of particle in position space. An analogous description 
exists in momentum space as well. Like conventional uncertainty product, some of the information measures also satisfy 
certain lower bounds. Further, it is well known that total Shannon entropy remains conserved under uniform scaling of 
particle coordinates \cite{gadre1985some}.

\subsection{Conventional uncertainty relation}
Conventional uncertainty is the standard deviation in a measurement of an observable, say position or momentum. They are 
expressed in the form ($\hbar=1$),  
\begin{equation}
(\triangle x)^2   =  \langle\left(x -\langle x \rangle\right)^2 \rangle, \ \ \ \ \ 
(\triangle p)^2 = \langle\left(p -\langle p \rangle\right)^2 \rangle, \ \ \ \ \ \ \
\triangle x \triangle p \geq \frac{1}{2}.
\end{equation}

\subsection{Shannon entropy ($S$)}
Shannon entropy \citep{shannon1951prediction} is the expectation value of logarithmic probability density function. 
The expressions is position and momentum space are given by, 
\begin{equation}
S_{x} =  -\int \rho(x) \mbox{ln}\left[\rho(x)\right] dx,   \ \ \ \ \ \ \ \ \
S_{p} =  -\int \rho(p) \mbox{ln}\left[\rho(p)\right] dp. 
\end{equation}
The total Shannon entropy ($S$), defined as below, obeys the following bound, 
\begin{equation}
S=S_{x}+S_{p} \geq (1+ \mbox{ln}\pi).  
\end{equation}

\subsection{Fisher information ($I$)}
Fisher information is a gradient functional of density \citep{cover2012elements}. In position and momentum space, they read, 
\begin{equation}
I_{x} =  \int \left[\frac{|\nabla\rho(x)|^2}{\rho(x)}\right] dx,  \ \ \ \ \ \ \ \ \
I_{p} =  \int \left[\frac{|\nabla\rho(p)|^2}{\rho(p)}\right] dp. 
\end{equation}
The net Fisher information ($I$), as obtained below, satisfies the bound,  
\begin{equation}
I=I_{x}I_{p} \geq 4.
\end{equation}

\subsection{Onicescu energy ($E$)} 
Onicescu energy \citep{agop2014implications, alipour2012onicescu} is a quadratic functional of density. In position 
and momentum space, it is expressed as,   
\begin{equation}
E_{x} =  \int \left[|\rho(x)|^2\right] dx,  \ \ \ \ \ \ \ \ \ \ \
E_{p} =  \int \left[|\rho(p)|^2\right] dp. 
\end{equation}
The total quantity, $E$ is then, defined as, 
\begin{equation}
E=E_{x}E_{p} \geq \frac{1}{2\pi}
\end{equation}

\subsection{Onicescu-Shannon information ($OS$)}
Onicescu-Shannon information measure \citep{LpezRuiz1995321, chatzisavvas2005information, panos2007comparison}
is derived from Shannon entropy and Onicescu energy. In position and momentum space, it assumes the form, 
\begin{equation}
OS_{x} = \exp\left[\frac{2}{3}{S_{x}}\right]E_{x},  \ \ \ \ \ \ \ \ \ 
OS_{p} =  \exp\left[\frac{2}{3}{S_{p}}\right]E_{p}. 
\end{equation}
The total measure, gives below, satisfies the bound as, 
\begin{equation}
OS=\exp\left[\frac{2}{3}{S}\right]E \geq
\left[\frac{1}{2}\left({\frac{1}{\pi}}\right)^{\frac{1}{3}}\exp\left[{\frac{2}{3}}\right]. 
\right]
\end{equation}

\begin{figure}             
\centering
\begin{minipage}[c]{0.30\textwidth}\centering
\includegraphics[scale=0.48]{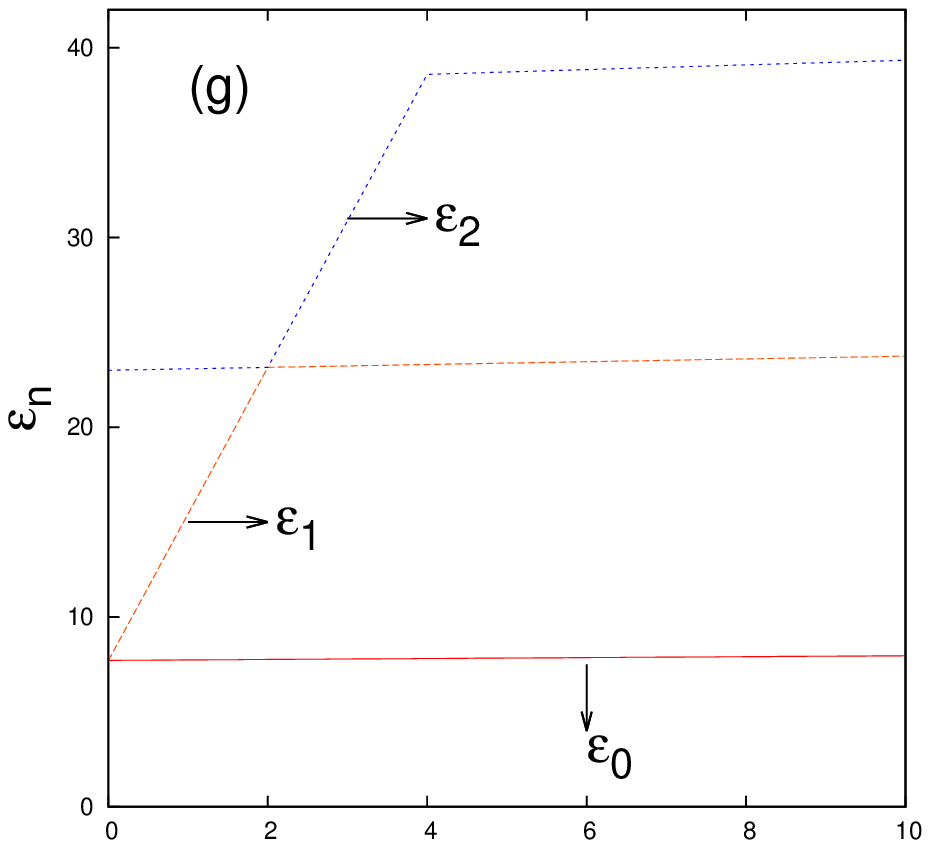}
\end{minipage}\hspace{0.05in}
\begin{minipage}[c]{0.30\textwidth}\centering
\includegraphics[scale=0.48]{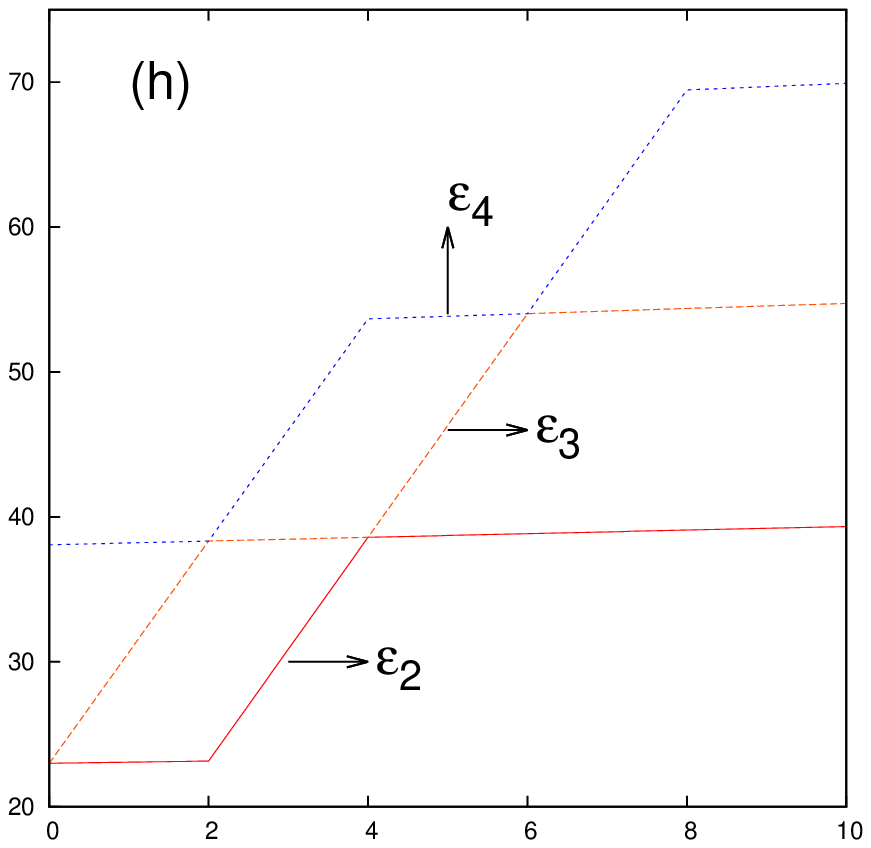}
\end{minipage}\hspace{0.05in}
\begin{minipage}[c]{0.30\textwidth}\centering
\includegraphics[scale=0.48]{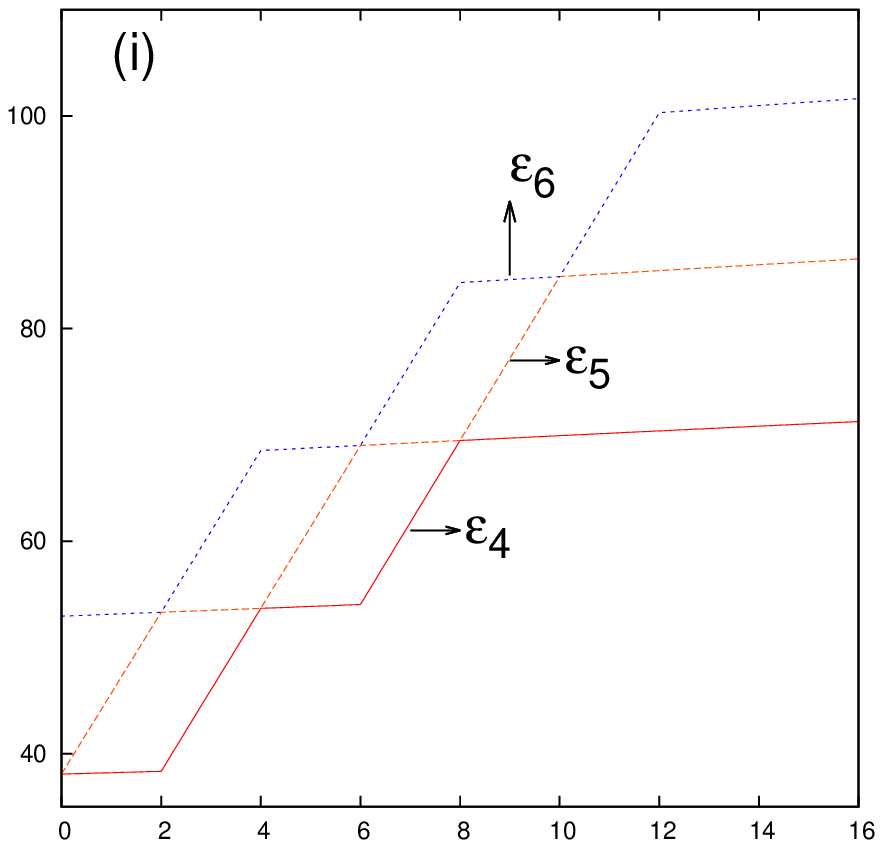}
\end{minipage}
\\[5pt]
\begin{minipage}[c]{0.30\textwidth}\centering
\includegraphics[scale=0.48]{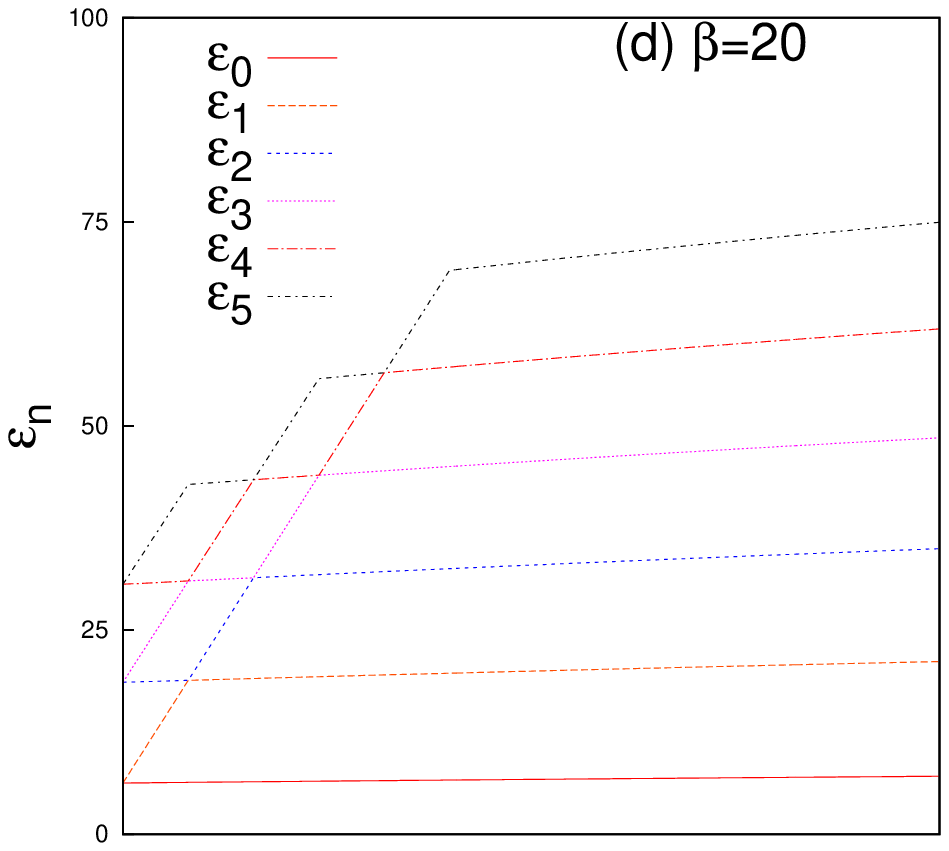}
\end{minipage}\hspace{0.05in}
\begin{minipage}[c]{0.30\textwidth}\centering
\includegraphics[scale=0.48]{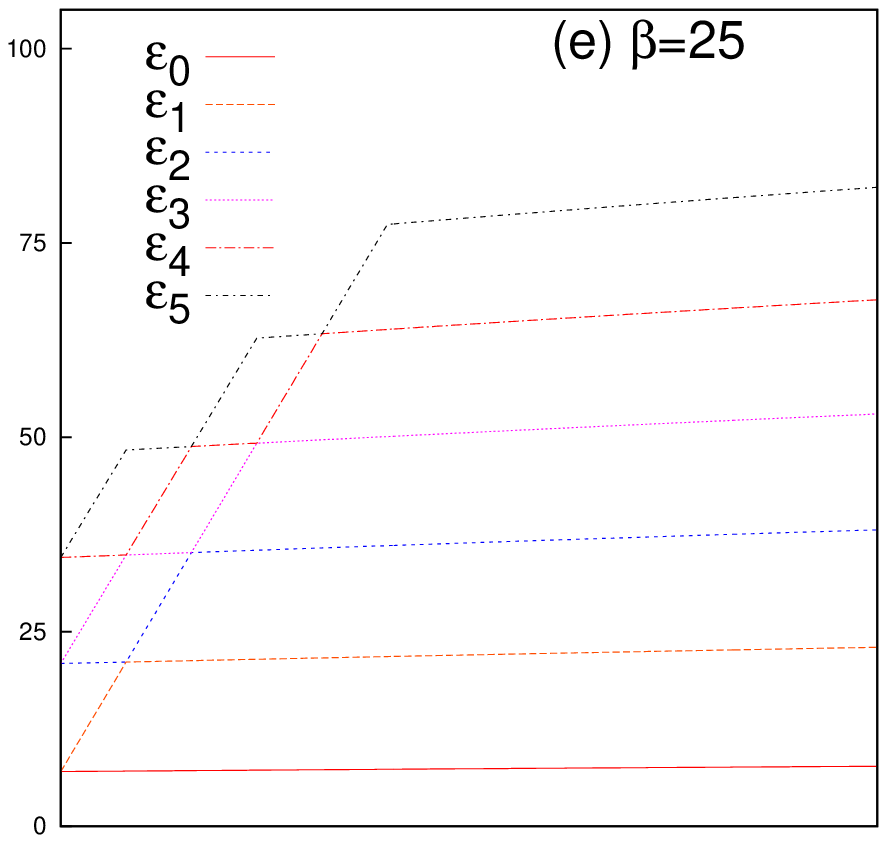}
\end{minipage}\hspace{0.05in}
\begin{minipage}[c]{0.30\textwidth}\centering
\includegraphics[scale=0.48]{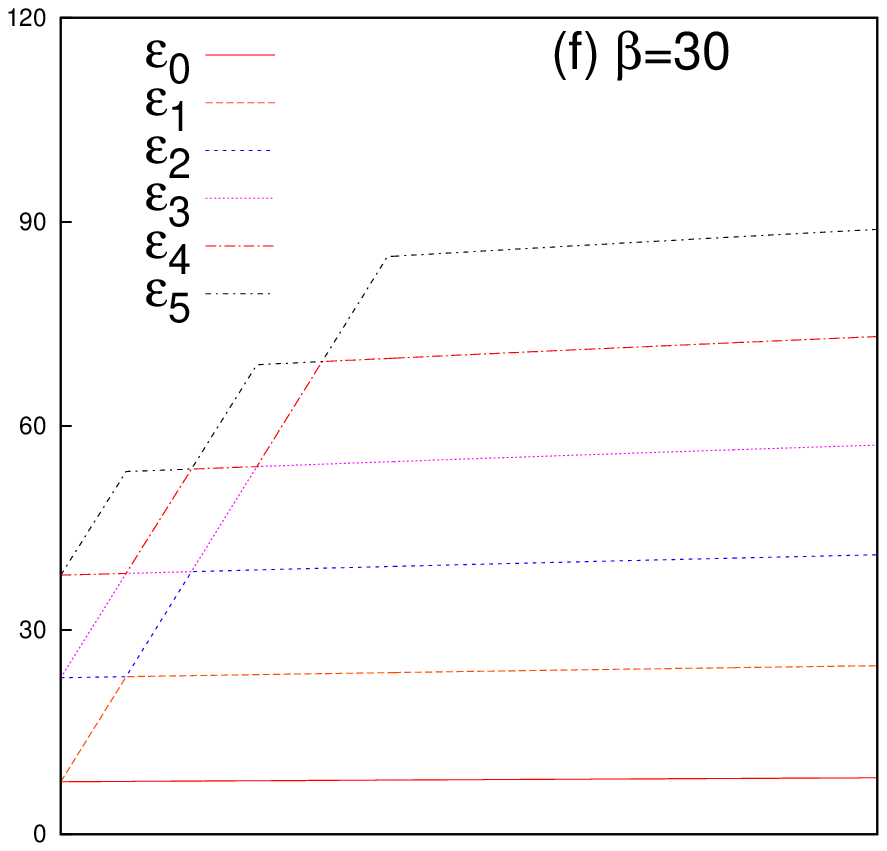}
\end{minipage}
\\[5pt]
\begin{minipage}[c]{0.30\textwidth}\centering
\includegraphics[scale=0.48]{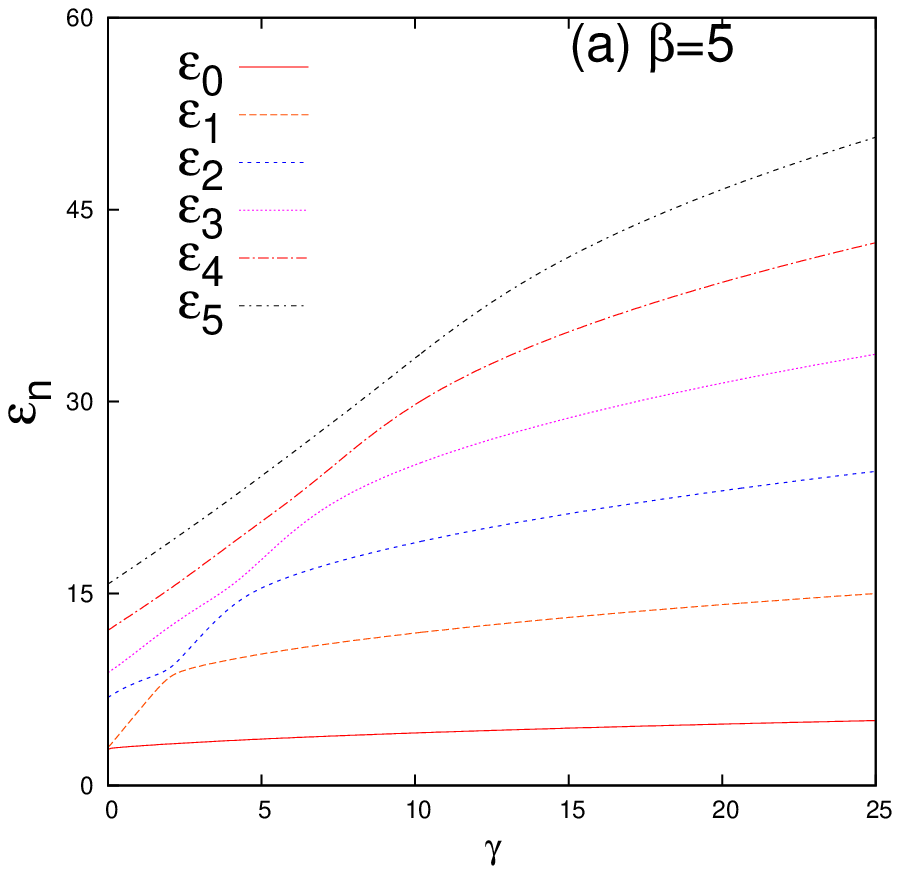}
\end{minipage}\hspace{0.05in}
\begin{minipage}[c]{0.30\textwidth}\centering
\includegraphics[scale=0.48]{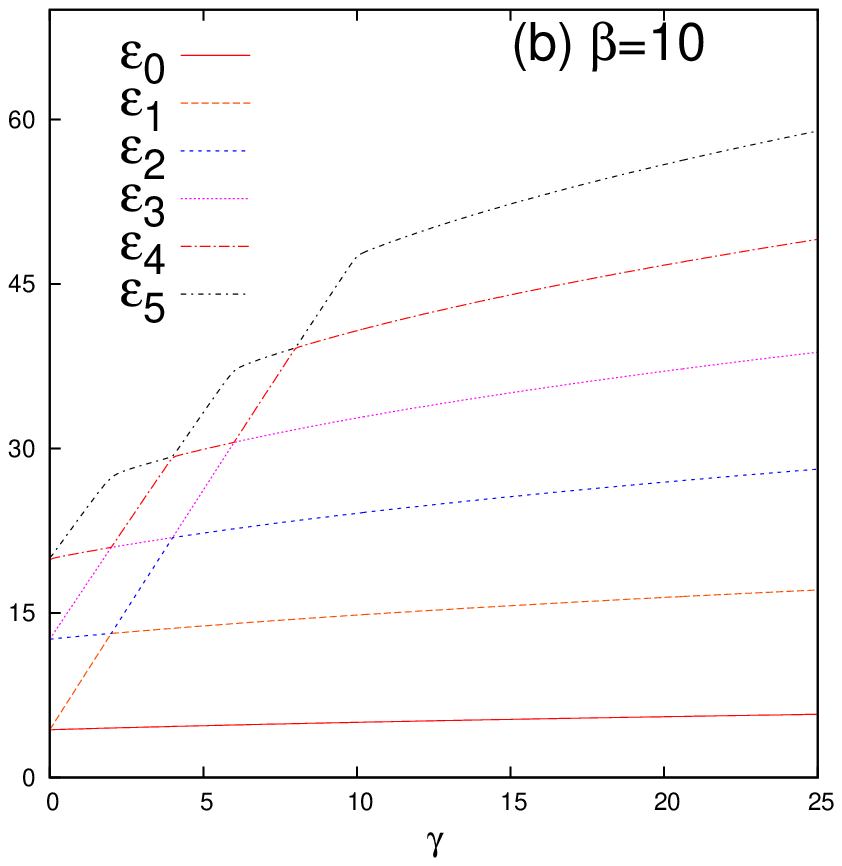}
\end{minipage}\hspace{0.05in}
\begin{minipage}[c]{0.30\textwidth}\centering
\includegraphics[scale=0.48]{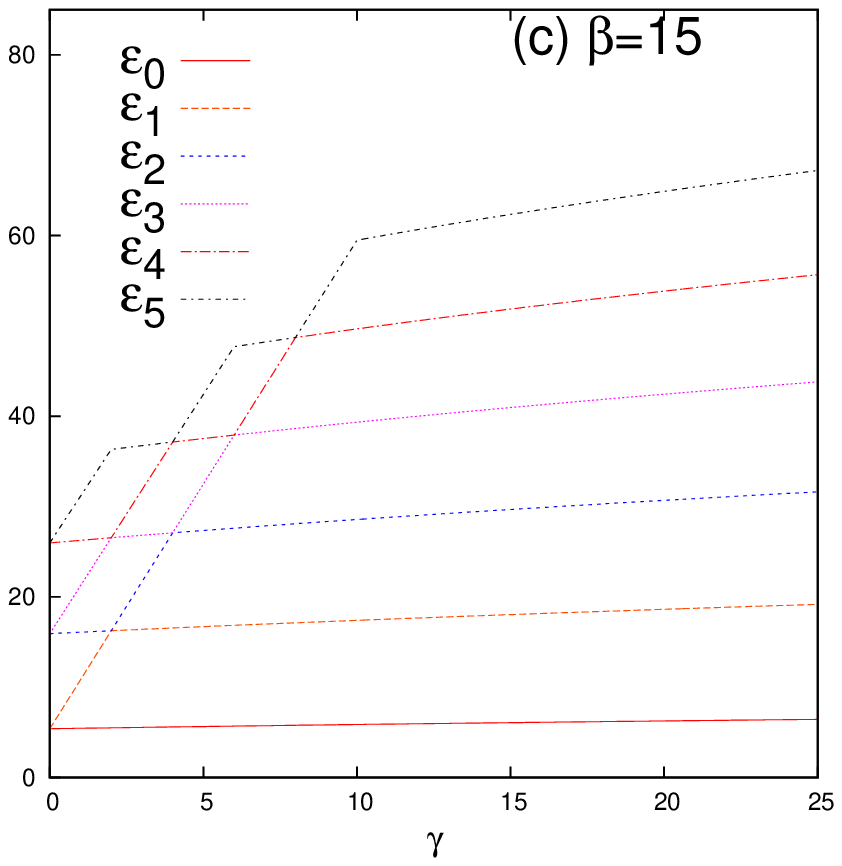}
\end{minipage}
\caption[optional]{Lowest six eigenvalues, plotted against $\gamma$ at six different $\beta$, keeping $\alpha \! = \! 1$, for 
asymmetric DW potential in Eq.~(4). Panels (a)-(f) correspond to $\beta=$5,10,15,20,25,30 respectively. A close-up 
of (f) is reproduced in (g), (h), (i), corresponding to $n \! = \! 0,1,2$; $n \! = \! 2,3,4$ and $n \! = \! 4,5,6$ states.}
\end{figure}

\section{Results and Discussion}
Our primary motivations is to understand effects of variation of $\beta$ and $\gamma$ on behavior of a particle in an 
asymmetric DW potential. As already mentioned, varying $\beta$ causes following things concurrently; it facilitates 
delocalization by increasing spacing between classical turning points, reduces separation between two minima, and promotes 
confinement through increase of barrier height. Whereas the role of $\gamma$ is to bring in asymmetry in to picture.
Therefore, it would be quite fascinating to examine the combined effects of these inter-related and intricate aspects 
together; how these are reflected in various properties, starting from energy, wave function, probability distribution, 
uncertainty, to information-related measures, etc. In the following, these are discussed, one by one. 

Above task is accomplished through two distinct avenues. Initially, we focus on asymmetry effect; for that, co-efficient 
of linear term, $\gamma$ is varied for separate values of parameter $\beta$. Thus, pertinent calculations are done at six 
selected $\beta$, ranging from 5 to 30 at an interval of 5, to cover a wide spectrum. Later, the role of contrasting 
effect of localization-delocalization is probed by altering co-efficient of quadratic term $\beta$ at eight particular
$\gamma$; from 0 to 7, at interval of 1. At first, these effects are monitored through energy. Then, 
distribution of particle within the wells has been investigated quite extensively from a consideration of wave function 
and probability. Finally to explain the extent of particle confinement, recourse is taken to conventional uncertainty
relation as well as a bunch of information measures such as $S,I, E, OS$ (in both position and momentum space). 
For all the calculations that follow, $\alpha$ is kept fixed at 1. Note that, one could think of some variations in 
results with changes in $\alpha$, but that does not alter the qualitative nature of conclusions drawn herein. 

\subsection{Energy}
Energy spectrum of our asymmetric DW potential in Eq.~(4) is analyzed using the procedure outlined in II.A. 
For this, lowest six representative states are displayed, as function of $\gamma$ at six specific $\beta$, in Fig.~(2). 
Note that, $\gamma$ axis remains unaltered in (a)-(f). It is known that, a symmetric DW of the form of Eq.~(2) (with 
$\gamma \! = \! 0$), offers quasi-degeneracy between certain even-odd pairs of states after a threshold $\beta$; furthermore, 
number of such pairs increases with increase in $\beta$ \cite{pedram2010}. Asymmetric term ($\gamma x^r$) promptly lifts 
quasi-degeneracy in energy. However, now a new kind of degeneracy arises at certain characteristic $\gamma$, which will be 
highlighted in our forthcoming discussion. Before we begin, for an easy appreciation of figures, let us draw our attention on 
top three panels (g), (h), (i). They reproduce same energy variation of (f) (all parameters same); three panels referring 
to $n \! = \! 0,1,2$, $n \! = \! 2,3,4$ and $n \! = \! 4,5,6$ states respectively. Clearly a stair-case like structure 
is manifest in all these segments, which is distinctly maintained in remaining panels too. Panel (a) shows that, energy of all 
six states at $\beta \! = \! 5$, are non-degenerate and well-separated. They increase with $\gamma$, extent of which becoming 
larger for higher states. However, from (b)-(f), it is discernible that, after a certain $\beta$ (in this case, within a range 
of 5-10), quasi-degeneracy occurs for some \emph{even} $\gamma$ (in this example, $0,2,4,\cdots$). Moreover, except ground 
state, slope of each individual excited-state energy curve changes after a fixed interval of $\gamma$ (here it is 2) and this 
trend continues up to a certain $\gamma$ (characteristic for a given state). For sake of generality, we denote this interval 
as $\Delta\gamma$, that depends on $\alpha$ only. It is found that, for a given $\alpha$, there always exists a positive real 
number, $k \! = \! \frac{\gamma}{\Delta \gamma}$. For this illustrative example at hand, $\Delta \gamma \! = \! 2$, which also 
happens to be the case for our all future $\beta$ plots, as long as $\alpha$ is fixed at 1. Now, following simple general rules 
can be established. 

\begin{table}      
\caption{Energy differences $\left(\Delta \epsilon_{mn}=|\epsilon_{m}-\epsilon_{n}|\right)$ between several quasi-degenerate 
pairs of asymmetric DW potential, at six selected $\beta$, \emph{viz.}, 5,10,15,20,25,30. See text for more details.}
\centering
\begin{tabular}{l|c|cccccc}
\hline
$\Delta \epsilon_{mn}$ \hspace{0.1in} &  \hspace{0.1in}$\gamma$ \hspace{0.1in} & \hspace{0.1in} $ \beta$=5 
& \hspace{0.2in}$\beta$=10 \hspace{0.1in} & \hspace{0.2in} $\beta$=15 \hspace{0.1in} 
& \hspace{0.2in}$\beta$=20 \hspace{0.1in} & \hspace{0.2in} $\beta$=25 \hspace{0.1in} & \hspace{0.2in}$\beta$=30\\
\hline
$\Delta \epsilon_{12}$ & 2 & \hspace{0.05in} 0.728 & \hspace{0.1in} 0.00035 & \hspace{0.1in} $ 3.3 \times 10^{-9}$ & \hspace{0.1in} $ 5.0 \times 10^{-14}$ & \hspace{0.1in} $ 8.2 \times 10^{-16}$ & \hspace{0.1in} $ 6.3 \times 10^{-18}$  \\
$\Delta \epsilon_{23}$ & 4 & \hspace{0.05in} 1.670 & \hspace{0.1in} 0.00280 & \hspace{0.1in} $ 3.9 \times 10^{-8}$ & \hspace{0.1in} $ 1.0 \times 10^{-13}$ & \hspace{0.1in} $ 7.1 \times 10^{-15}$ & \hspace{0.1in} $ 1.1 \times 10^{-16}$  \\
$\Delta \epsilon_{34}$ & 6 & \hspace{0.05in} 2.616 & \hspace{0.1in} 0.01687 & \hspace{0.1in} $ 3.7 \times 10^{-7}$ & \hspace{0.1in} $ 4.0 \times 10^{-13}$ & \hspace{0.1in} $ 1.4 \times 10^{-14}$ & \hspace{0.1in} $ 3.5 \times 10^{-14}$  \\
$\Delta \epsilon_{45}$ & 8 & \hspace{0.05in} 3.355 & \hspace{0.1in} 0.07994 & \hspace{0.1in} $ 2.9 \times 10^{-6}$ & \hspace{0.1in} $ 4.5 \times 10^{-12}$ & \hspace{0.1in} $ 3.6 \times 10^{-14}$ & \hspace{0.1in} $ 5.7 \times 10^{-14}$  \\
\hline
\end{tabular}
\end{table}

\begin{enumerate}[(i)]      
\item 
When $(n \! \geq \! k)$: three possible outcomes can be envisaged in this scenario.

\begin{enumerate}[(a)]
\item
\emph{$k$ is odd positive integer}: an \emph{odd}-$n$ state will be quasi-degenerate with its immediate higher state.
\item
\emph{$k$ is even positive integer}: an \emph{even}-$n$ state will be quasi-degenerate with its adjacent higher state.
\item
\emph{$k$ is a fraction}: no quasi-degeneracy is possible.                   
\end{enumerate}
\item
 When $k \! > \! n$: $n$-th state \emph{can not} be quasi-degenerate. However, other higher ($\! > \! n$) states may 
show degeneracy depending on $k$, as delineated above in (i.a), (i.b). 
\end{enumerate}

\begin{figure}             
\centering
\begin{minipage}[c]{0.20\textwidth}\centering
\includegraphics[scale=0.38]{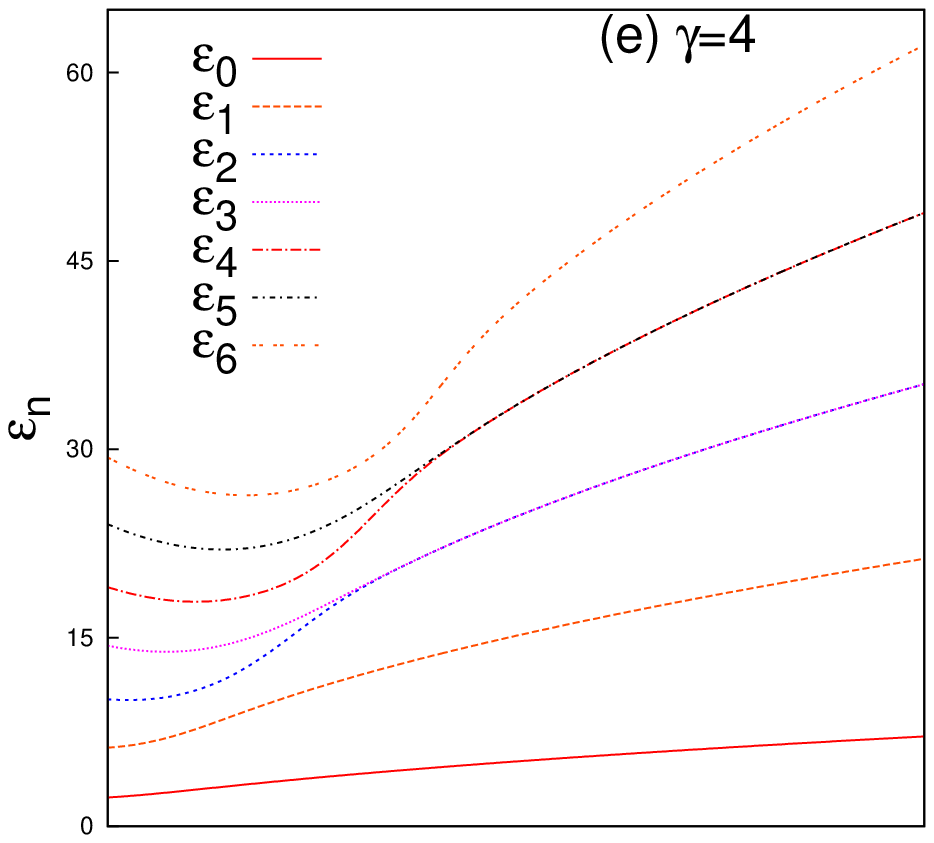}
\end{minipage}\hspace{0.1in}
\begin{minipage}[c]{0.20\textwidth}\centering
\includegraphics[scale=0.38]{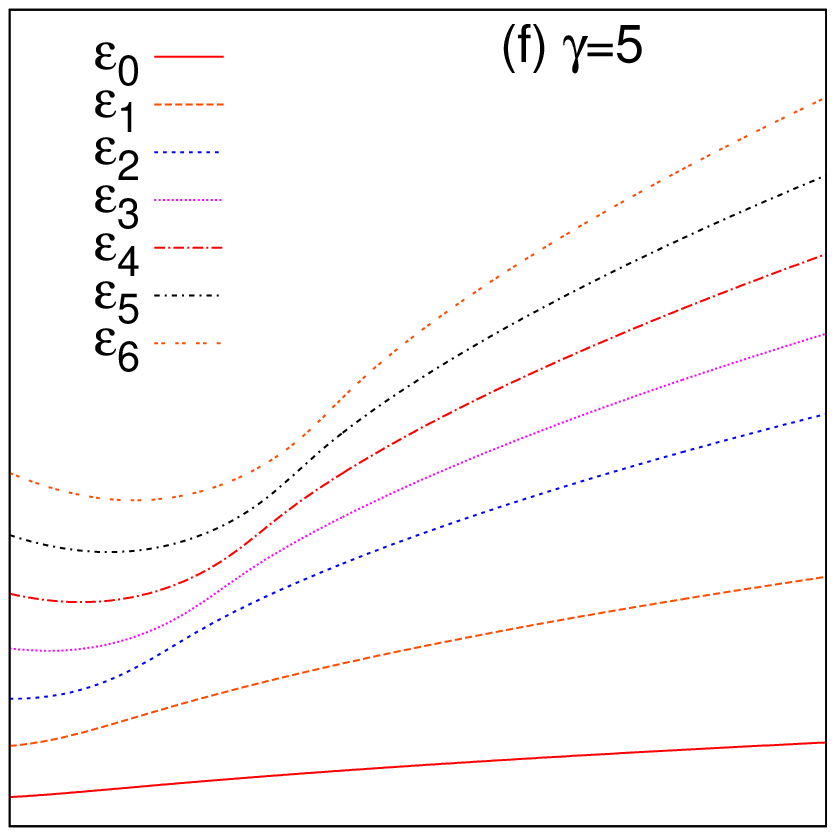}
\end{minipage}\hspace{0.1in}
\begin{minipage}[c]{0.20\textwidth}\centering
\includegraphics[scale=0.38]{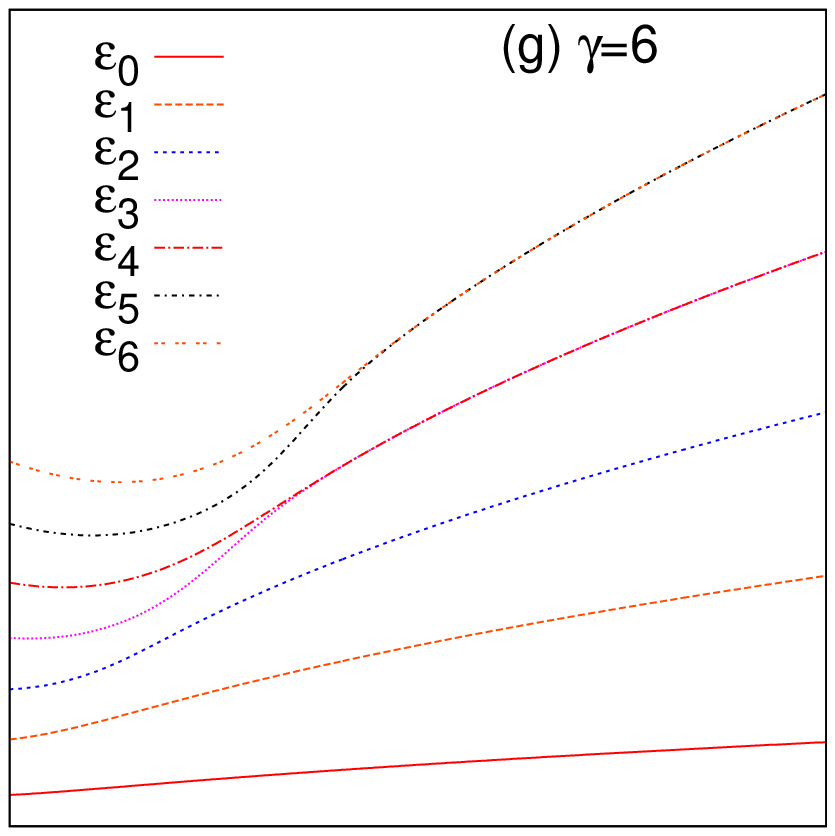}
\end{minipage}\hspace{0.1in}
\begin{minipage}[c]{0.20\textwidth}\centering
\includegraphics[scale=0.38]{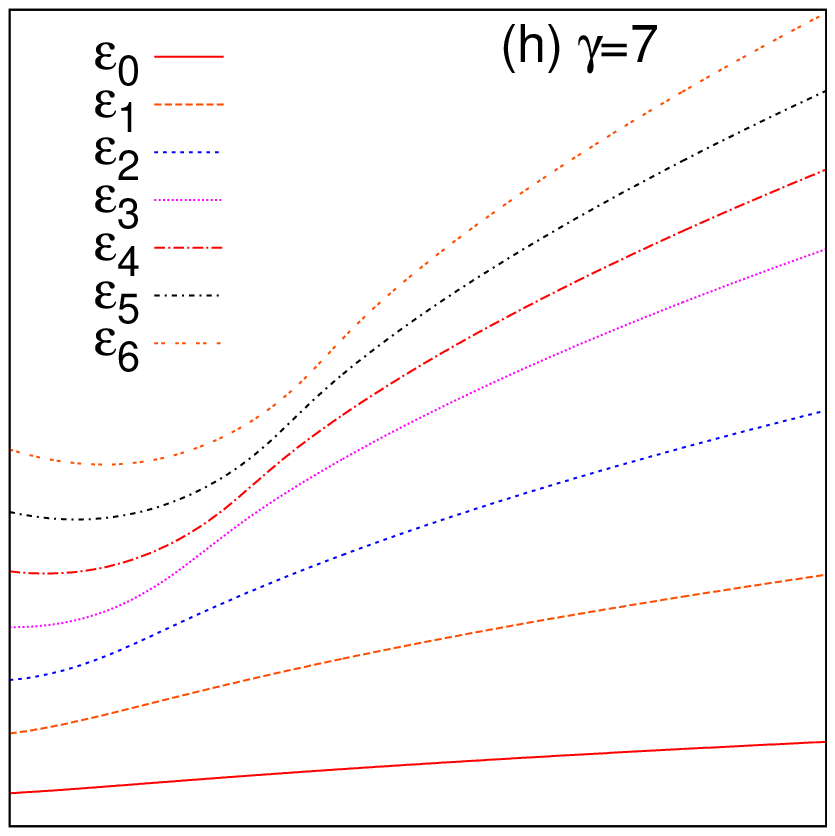}
\end{minipage}
\\[5pt]
\begin{minipage}[c]{0.20\textwidth}\centering
\includegraphics[scale=0.38]{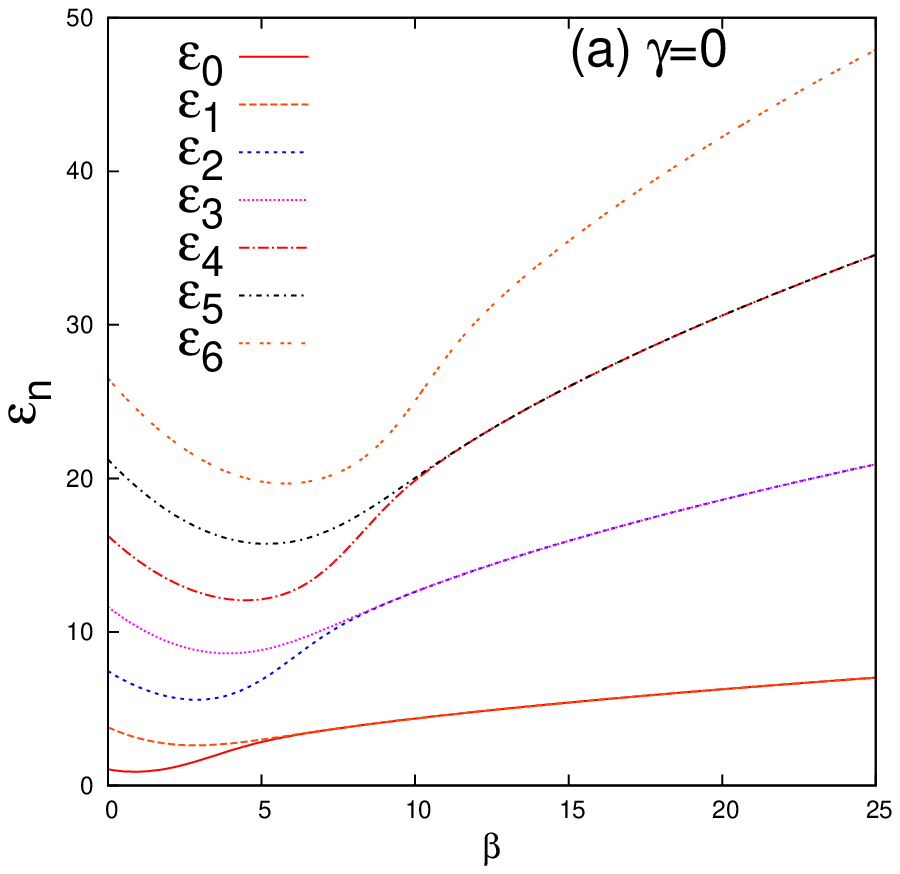}
\end{minipage}\hspace{0.1in}
\begin{minipage}[c]{0.20\textwidth}\centering
\includegraphics[scale=0.38]{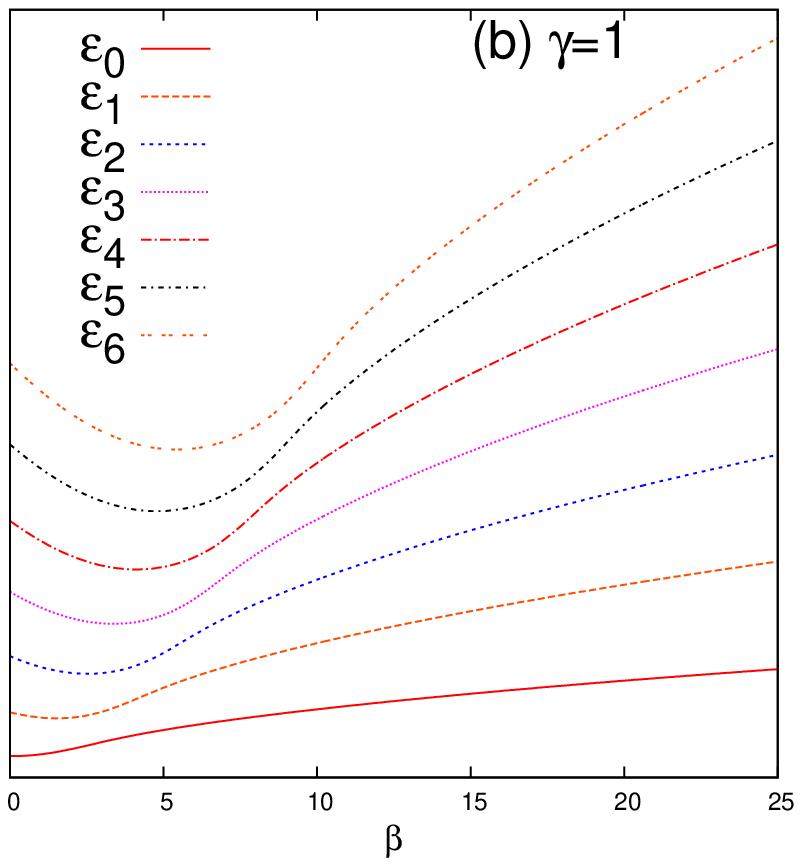}
\end{minipage}\hspace{0.1in}
\begin{minipage}[c]{0.20\textwidth}\centering
\includegraphics[scale=0.38]{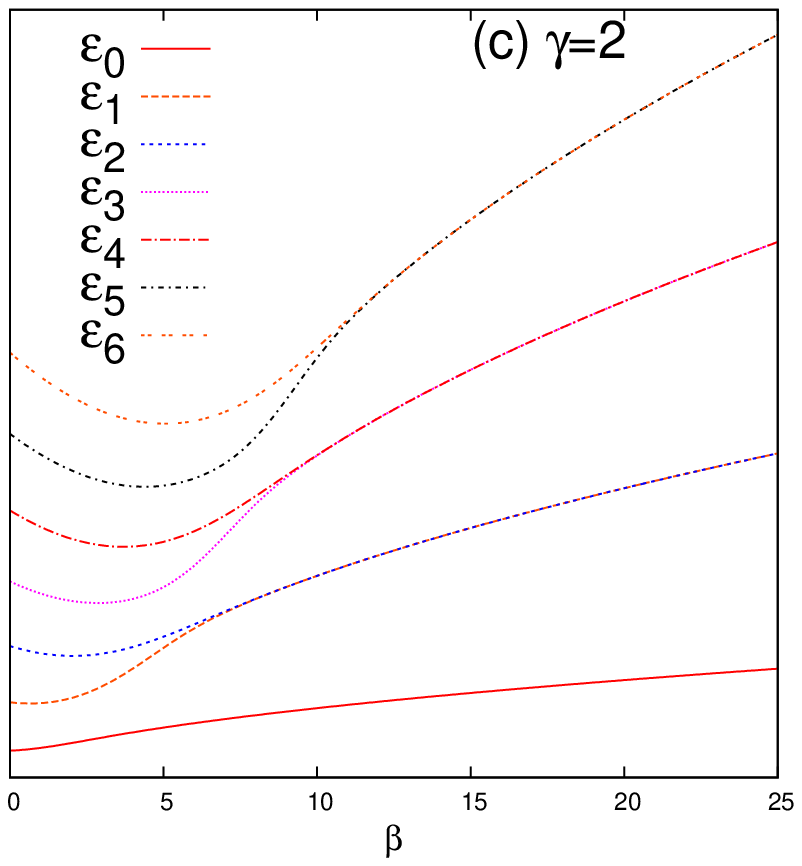}
\end{minipage}\hspace{0.1in}
\begin{minipage}[c]{0.20\textwidth}\centering
\includegraphics[scale=0.38]{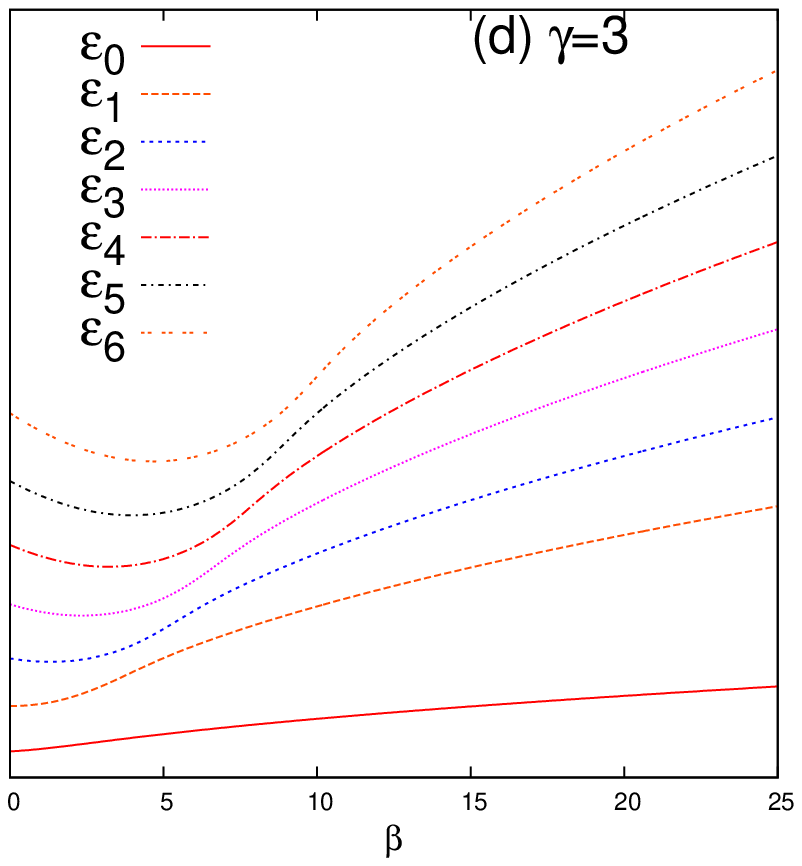}
\end{minipage}
\caption[optional]{Lowest seven eigenvalues, plotted against $\beta$ at eight different $\gamma$, keeping $\alpha \! = \! 1$, 
for asymmetric DW potential in Eq.~(4). Panels (a)-(h) correspond to $\gamma=$0,1,2,3,4,5,6,7 respectively.}
\end{figure}

Therefore, depending on whether $k$ is even or odd, arrangement of quasi-degenerate pair changes. For a fixed $\beta$, 
an increase in integer $k$ reduces number of such quasi-degenerate pairs. As seen from all panels of Fig.~(2) except 
(a), $\epsilon_1, \epsilon_2$ become quasi-degenerate at $\gamma \! = \! 2$; $k \! = \! 1$ and thus (i.a) is valid. 
For any $\gamma$ beyond this, $n \! = \! 1$ always remains non-degenerate, in accordance with (ii). Then at $\gamma \! = \! 4$ 
(correspondingly $k \! = \! 2$), $n \! = \! 2, 3$ become degenerate, consistent with (i.b). Further, from (ii), $n \! = \!2  $ 
will remain non-degenerate for all $\gamma \! > \! 4$. Again at $\gamma \! = \! 6$, $\epsilon_3$ merges with $\epsilon_4$ and 
continues to remain non-degenerate thereafter. Figure (2) displays that at odd $\gamma$ ($k$ is a fraction), there is no 
quasi-degeneracy in any states whatsoever, in agreement with (i.c). 
They all eventually become non-degenerate after a characteristic $\gamma$. Thus, there are $n+1$ such integer $k$ where, 
$n$ and its adjacent state coalesce. Interestingly, at each of those $k$ (for $k \! < \! n$), slope of 
$\frac{d\epsilon_n}{d\gamma}$ curve changes. From above six plots, clearly this phenomenon occurs after a threshold $\beta$ 
(characteristics for a definite state). It is verified that other states also satisfy these rules. 
Table~II now shows effect of $\beta$ on quasi-degeneracy more succinctly.  
It demonstrates that with an increase in $\beta$, separation between respective quasi-degenerate pairs decrease gradually 
and it has no impact on a particular $\Delta \gamma$ so far as $\alpha$ remains constant. At this stage, it is necessary 
to study the role of $\beta$, which is done next. 

\begingroup    
\squeezetable
\begin{table}
\caption{Eigenvalues for first eleven states of potential in Eq.~(4) at five selected $\gamma$, for $\alpha=1, \beta=30$. 
Separate colors indicate sequential quasi-degeneracy.}
\centering
\begin{ruledtabular}
\begin{tabular}{clllll}
$n$  & $\gamma$=0,$k$=0   &  $\gamma$=2,$k$=1   &  $\gamma$=4,$k$=2    &  $\gamma$=6,$k$=3   &   $\gamma$=8,$k$=4 \\
\hline
0   & \color{red} {7.7123035268648}   &  7.7625622475687   &  7.8120692428683  &  7.8608522067983  & 7.9089372344196   \\                       
1   & \color{red} {7.7123035268649}   & \color{red}{23.153053576823}  & 23.304002627698   & 23.452679856354    &  23.599169782772 \\
2   & \color{blue} {22.999742809258} &\color{red}{23.153053576823}  & \color{red}{38.592130067269}   & 38.844503801999   
    & 39.093040002338 \\
3   & \color{blue} {22.999742809258} & \color{blue}{38.335764499909}  & \color{red}{38.592130067269} & \color{red}{54.028636855454}   
    & 54.383230870215 \\
4   & \color{brown} {38.075242534947}  & \color{blue}{38.335764499909}  & \color{blue}{53.668365490156} & \color{red}{54.028636855454}    
    & \color{red}{69.461672176182} \\
5   & \color{brown} {38.075242534947}  & \color{brown}{53.302178296650} & \color{blue}{53.668365490156} & \color{blue}{68.996573867408}  
    & \color{red}{69.461672176182} \\
6   & \color{cyan}{52.929820105577}    & \color{brown}{53.302178296650} & \color{brown}{68.523732643668} & \color{blue}{68.996573867408}
    & \color{blue}{84.319406666242} \\
7   & \color{cyan}{52.929820105577}    & \color{cyan}{68.042807311026}  & \color{brown}{68.523732643668} & \color{brown}{83.738841122037}
    & \color{blue}{84.319406666242}  \\
8   & \color{orange}{67.553431414959}  & \color{cyan}{68.042807311026}  & \color{cyan}{83.148195113819} & \color{brown}{83.738841122037}  
    & \color{brown}{98.946420952478} \\ 
9   & \color{orange}{67.553431414959}  & \color{orange}{82.547000469436} & \color{cyan}{83.148195113819} & \color{cyan}{98.244804553262} 
    & \color{brown}{98.946420952479} \\
10  & 81.934752122370  & \color{orange}{82.547000469436}  & 97.530438011177  & \color{cyan}{98.244804553262}   &  113.33142843137 \\   
\end{tabular}
\end{ruledtabular}
\end{table}
\endgroup 

Figure~(3) exhibits variation of energy of seven lowest states w.r.t. $\beta$, at eight distinct $\gamma$. 
In this instance also, $\Delta \gamma \! = \! 2$ and as usual, one sees that quasi-degeneracy arises for 
both even, odd integer $k$, i.e., $even$ $\gamma$. And there is no trace of it for non-integer $k$, in  (b), (d), 
(f), (h), for the entire range of $\beta$. Reduction in energy separation between quasi-degenerate pairs is again observed (as in 
Table~II) with increase in $\beta$. In case of (a), (e), $k$ is 0, 2; thus (i.b) above is obeyed, as observed from degeneracy of 
($\epsilon_0, \epsilon_1$); $(\epsilon_2, \epsilon_3$); $(\epsilon_4, \epsilon_5$); $\cdots$ pairs in (a) and $(\epsilon_2, \epsilon_3)$;  
($\epsilon_4, \epsilon_5$) pairs in (e). Moreover, in accordance with (ii), there is no degeneracy for $n \! = \! 0,1$ 
states in (e). Panel (c) corresponds to $k \! = \! 1$; thus except for $n \! = \! 0$, where $k \! > \!n$ (hence can not 
contribute to degeneracy, from (ii)), higher states, such as $(\epsilon_1, \epsilon_2)$; $(\epsilon_3, \epsilon_4)$; 
($\epsilon_5, \epsilon_6)$; $\cdots$ are quasi-degenerate, following (i.a). In (g), $k \! = \! 3$; consequently, 
$n \! < 2 \!$ states can not be degenerate, in conformity to (ii); whereas other pairs such as ($\epsilon_3, \epsilon_4$); 
$(\epsilon_5, \epsilon_6$); $\cdots$ will be degenerate with each other, as (i.a) dictates. In (b), (d), (f), (h), 
$k$ is fraction; thus no degeneracy can occur in any state, from (i.c). Now, as a numerical exercise, 
Table~III displays eigenvalues of first 11 states of our asymmetric DW, at five integer $k=0, 1, 2, 3,4,$ keeping 
$\beta, \alpha$ fixed at 30, 1 respectively. This clearly illustrates the satisfaction of such rules for quasi-degenerate levels;
sketched in distinct colors. For example, when $k \! = \! 0$, column 2 shows that, an even-$n$ state is quasi-degenerate
with its immediate higher neighbor, from (i.b). Thus we get $(\epsilon_0, \epsilon_1), \cdots, (\epsilon_8, \epsilon_9)$, 
pairs. Same rule accounts for quasi-degeneracy of $(\epsilon_2, \epsilon_3), \cdots, (\epsilon_8, \epsilon_9)$ pairs in column 4
for $k \! = \! 2$, and of $(\epsilon_4, \epsilon_5), \cdots, (\epsilon_8, \epsilon_9)$ pairs in last column corresponding to 
$k \! = \! 4$, etc. Likewise, when $k$ is odd, an odd-$n$ state becomes degenerate with its adjacent higher neighbor as (i.a) 
applies. Thus in column 3, $(\epsilon_1, \epsilon_2), \cdots, (\epsilon_9, \epsilon_{10})$ pairs and in column 5, 
$(\epsilon_3, \epsilon_4), \cdots, (\epsilon_9, \epsilon_{10})$ pairs show quasi-degeneracy corresponding to $k \! = \! 3$,5
respectively. Moreover, when $k \! = \! 2,3,4$ in columns 4-6, following rule (ii), $n \! = \! 0,1$; $n \! = \! 0,1,2$ and 
$n \! = \! 0,1,2,3$ can not have quasi-degeneracy. In passing, we note that, when $k \! = \! 0$, the potential is actually symmetric
DW, obeying (i.b); treating 0 as \emph{even} integer. From this, one could anticipate that symmetric may be considered as 
a special case of more general asymmetric situation, which is revealed further in results that follow. 

\begin{figure}             
\centering
\begin{minipage}[c]{0.30\textwidth}\centering
\includegraphics[scale=0.48]{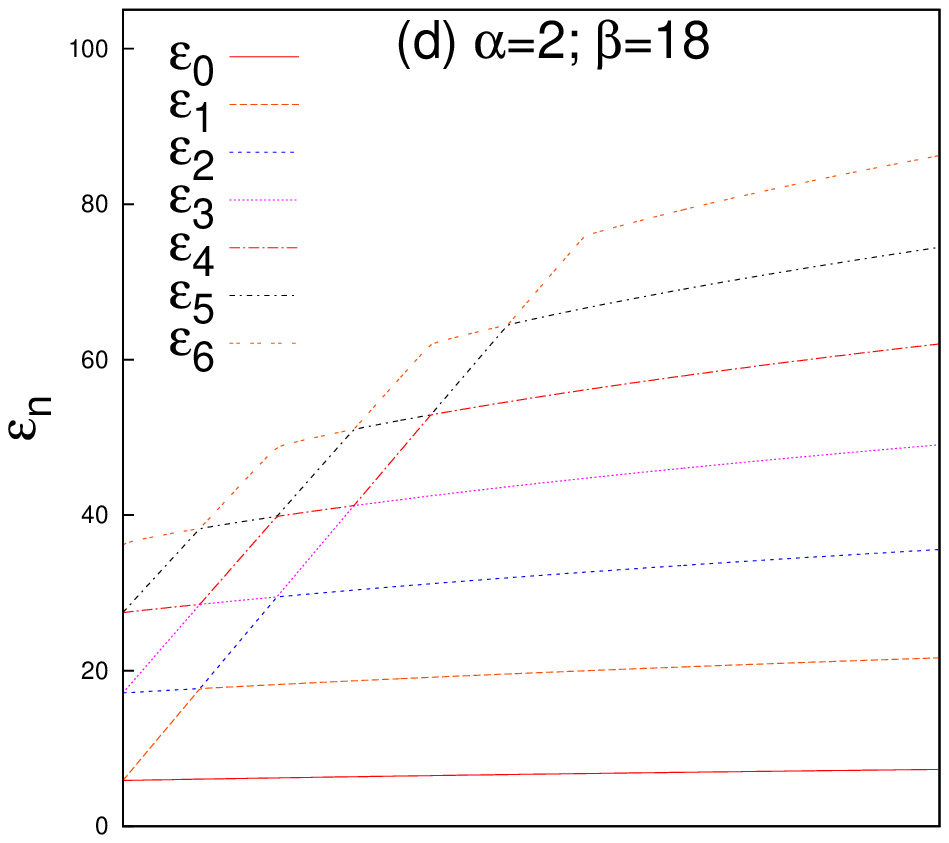}
\end{minipage}\hspace{0.05in}
\begin{minipage}[c]{0.30\textwidth}\centering
\includegraphics[scale=0.48]{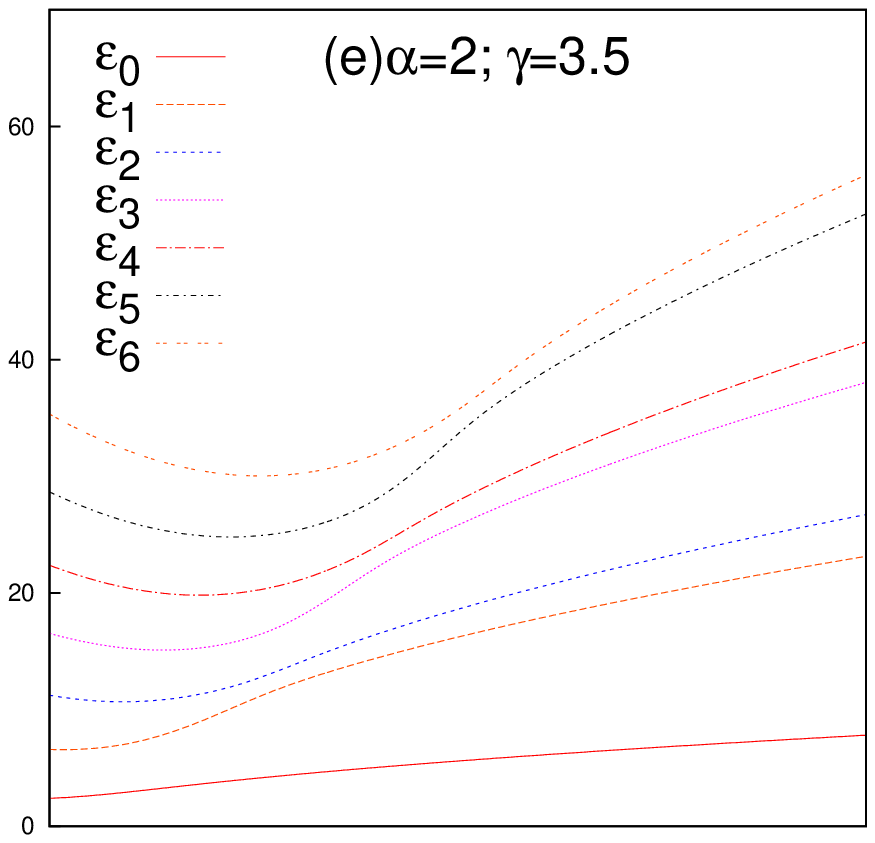}
\end{minipage}\hspace{0.05in}
\begin{minipage}[c]{0.30\textwidth}\centering
\includegraphics[scale=0.48]{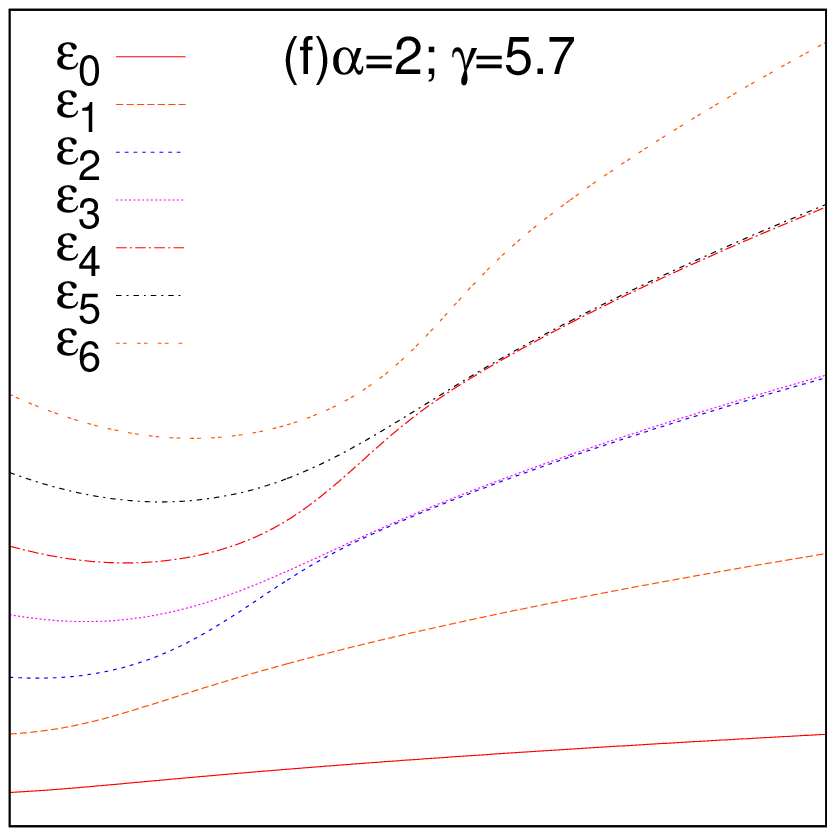}
\end{minipage}
\\[5pt]
\begin{minipage}[c]{0.30\textwidth}\centering
\includegraphics[scale=0.48]{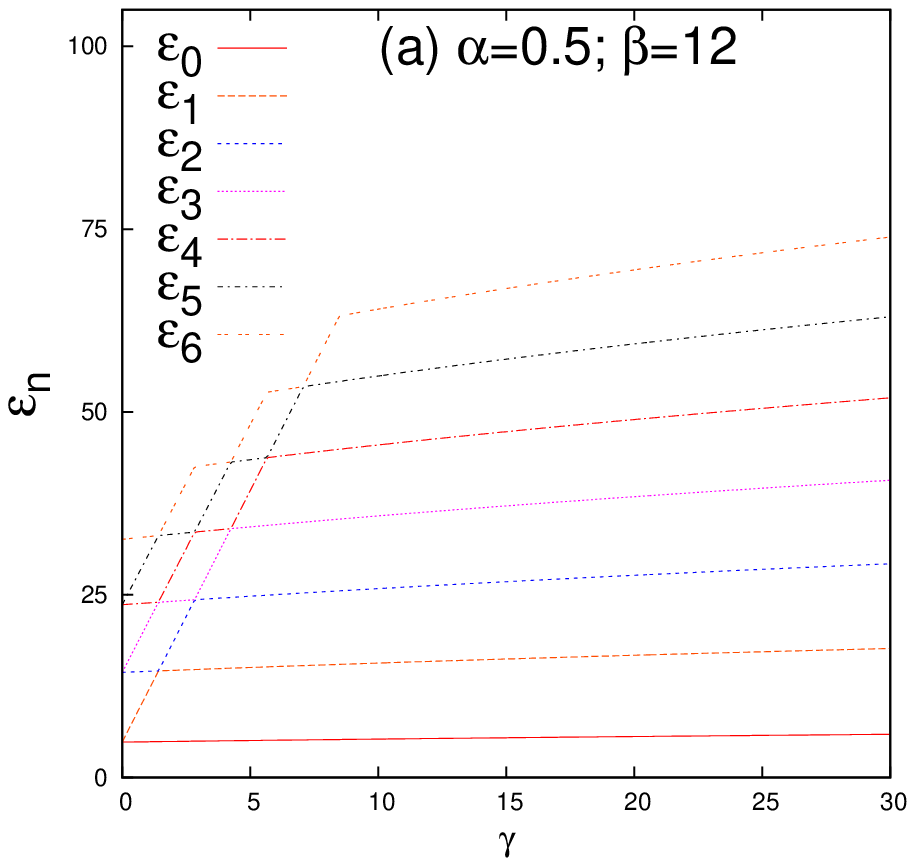}
\end{minipage}\hspace{0.05in}
\begin{minipage}[c]{0.30\textwidth}\centering
\includegraphics[scale=0.48]{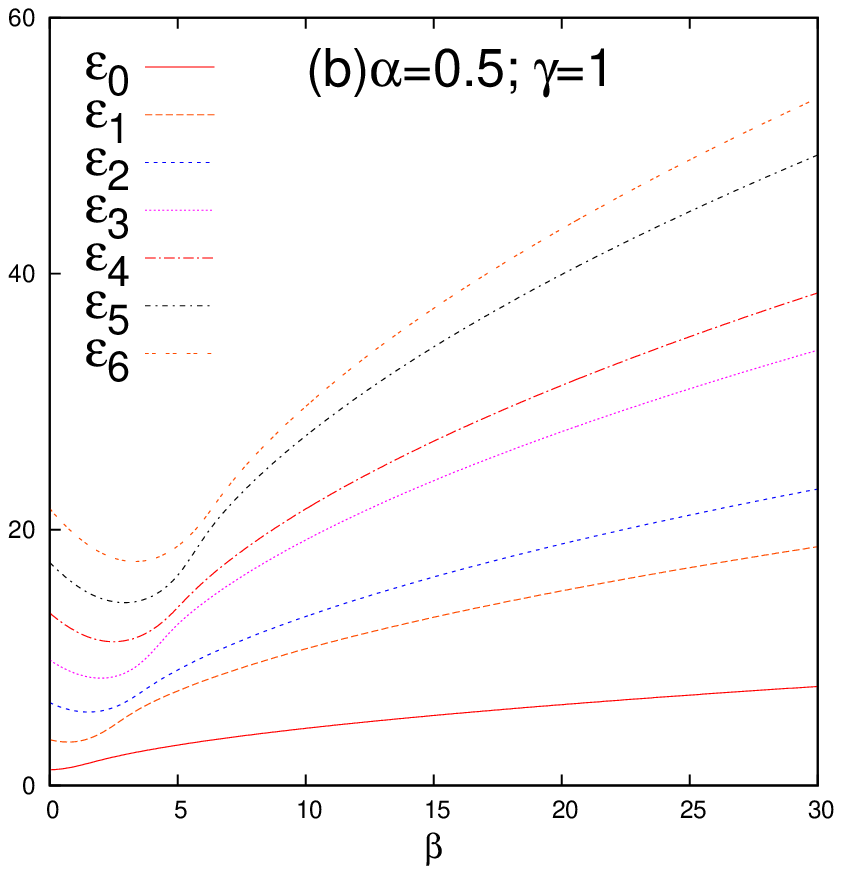}
\end{minipage}\hspace{0.05in}
\begin{minipage}[c]{0.30\textwidth}\centering
\includegraphics[scale=0.48]{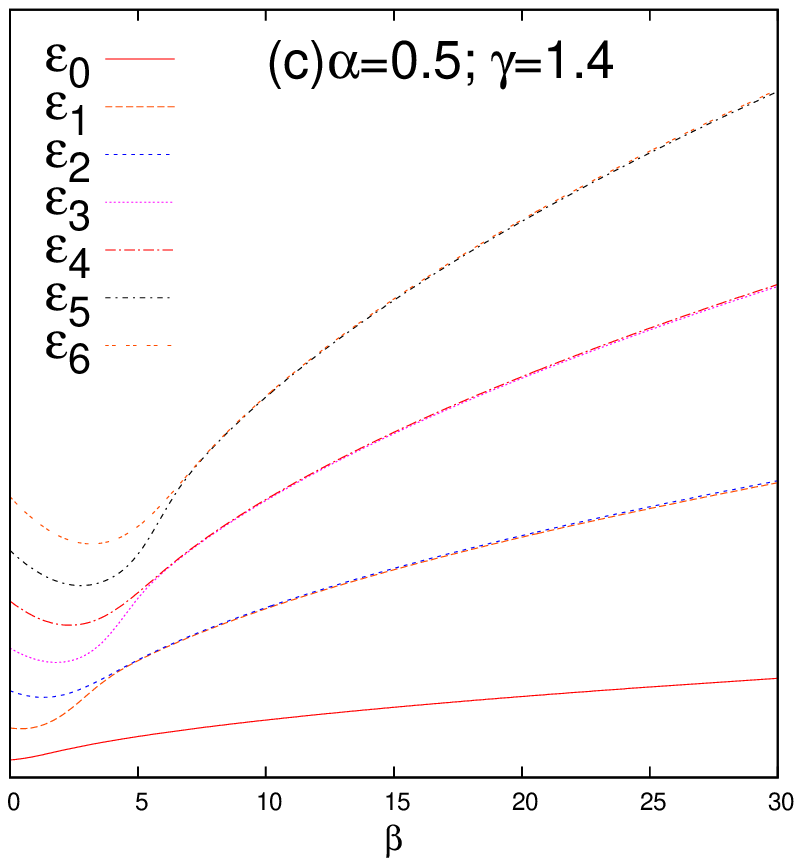}
\end{minipage}
\caption[optional]{First 7 eigenvalues of asymmetric DW, plotted against two $\alpha$ values, \emph{viz.,}
0.5 in (a)-(c) (bottom) and 2 in (d)-(f) (top) panels. Left two panels, (a), (d) show variations as function of 
$\gamma$, while (b), (e) (middle) and (c),(f) (right) give same with $\beta$ for fractional and integer $k$ respectively.}
\end{figure}

Next, Fig.~(4) provides changes in energy as well as $\Delta \gamma$ with $\alpha$. Bottom and top three segments correspond to 
$\Delta \gamma=1.4$ and 2.85 respectively. Two left panels give energy plots for two arbitrary $\alpha$, namely, 0.5 (a) and 2 
(b), with $\gamma$. Two $\beta$ values are selected so as to maintain quasi-degeneracy in spectrum. It is evident that 
qualitative nature of these figures essentially remain rather unaffected for
$\alpha$ changes, which further establishes applicability of those rules for all values of $\alpha$. For $\alpha$=0.5 and 
2, $\Delta \gamma$ corresponds to 1.4 and 2.85 respectively. From (a), (b), one also concludes that $\Delta \gamma$ increases 
as $\alpha$ increases. Manifestly, the degeneracy vanishes in (b),(e) in accordance with (i.c), for both of them have fractional 
$k$. Panels (c),(f) are associated with $k \! = \! 1$, 2 respectively. Consequently, (c) provides degenerate pairs of 
states such as $(\epsilon_1, \epsilon_2)$; ($\epsilon_3, \epsilon_4$); $(\epsilon_5, \epsilon_6)$, $\cdots$ leaving out 
$\epsilon_0$, whereas (f) offers $(\epsilon_2, \epsilon_3)$; $(\epsilon_4, \epsilon_5)$, $\cdots$ pairs barring
$n \! = \! 0$, 1 states; thus markedly follow the rules enumerated above, in both occasions.
To substantiate these rules further and as an illustration, we proceed for generation of arbitrary quasi-degenerate pair of
states from a knowledge of $\gamma$ and $\beta$. Table~IV offers such an example, where energies are reported for five 
selected sets of $\beta$, $\gamma$ values, maintaining $\Delta \gamma \! = \! 2$ and 
$\alpha \! = \! 1$ for all of them. First column demonstrates that only $n \! = \! 1$ and 2 can form quasi-degenerate pair, 
as $k \! = \! 1$. Likewise, it also certifies that only $(\epsilon_3, \epsilon_4)$, $(\epsilon_4, \epsilon_5)$, 
$(\epsilon_5, \epsilon_6)$ and $(\epsilon_6, \epsilon_7)$ pairs can be degenerate in columns 2,3,4,5 respectively; it can not 
occur before these states as well. This follows from our proposed rules, since corresponding $k$ values in these cases are 
3,4,5,6 respectively. For easy appreciation, respective quasi-degenerate pairs are colored. This validates that, it is 
possible to design an arbitrary DW potential with desired number of quasi-degenerate pair by controlling values of $\gamma$
and $\beta$.  

\begingroup                
\squeezetable
\begin{table}
\caption{Eigenvalues of first eleven states of potential, in Eq.~(4) for five sets of $\alpha,~\beta$, $\gamma$.}
\centering
\begin{ruledtabular}
\begin{tabular}{cccccc}
$n$  & $\alpha$=1,$\beta$=11,$\gamma$=2  &  $\alpha$=1,$\beta$=15,$\gamma$=8  &  $\alpha$=1,$\beta$=12,$\gamma$=6   &  
       $\alpha$=1,$\beta$=14,$\gamma$=10 &  $\alpha$=1,$\beta$=20,$\gamma$=12 \\
\hline
0   &  4.7350218201   &  5.7909404284   & 5.1742754901   &  5.7173474702  & 6.1147585445 \\                       
1   & \color{red}{13.823057196}   & 17.135972588  & 15.221243041   & 16.912882601    & 18.136676875 \\
2   & \color{red}{13.823101835}   & 28.101393760  & 24.768065792   & 27.725263393    & 29.830671038    \\
3   & 22.236765069    & 38.645757480  & \color{green}{33.723622648}    & 38.112373583     &     41.169171706 \\
4   & 22.241142559    & \color{blue}{48.713199214}  & \color{green}{33.723974810}     & 48.017352006    &   52.117628809 \\ 
5   & 29.653704124    & \color{blue}{48.713202090}  & 41.921650296  & \color{cyan}{57.357596153}     &      62.631093025 \\ 
6   & 29.822152191    & 58.222739977   & 41.944292391   & \color{cyan}{57.357784321}     &   \color{magenta}{72.647903484} \\
7   & 35.034359639    & 58.223006251   & 48.767803745   & 65.9920127006  &     \color{magenta}{72.647914976} \\
\end{tabular}
\end{ruledtabular}
\end{table}
\endgroup 

\begin{figure}             
\centering
\begin{minipage}[c]{0.15\textwidth}\centering
\includegraphics[scale=0.28]{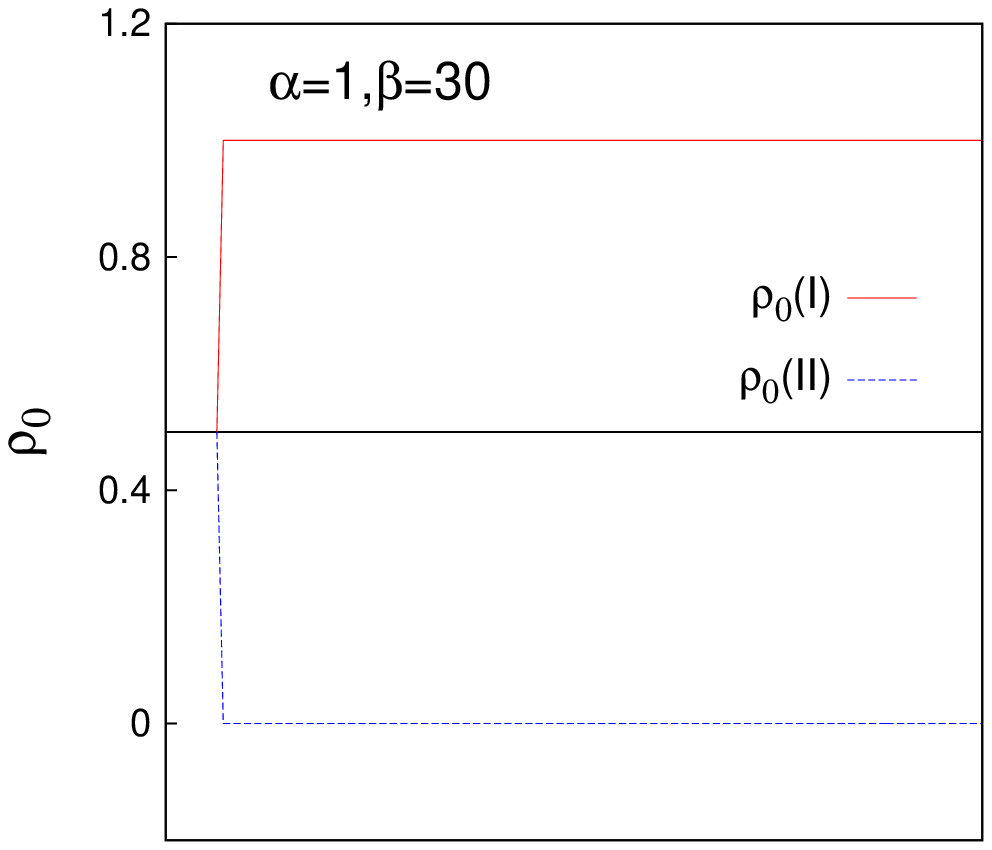}
\end{minipage}\hspace{0.08in}
\begin{minipage}[c]{0.15\textwidth}\centering
\includegraphics[scale=0.28]{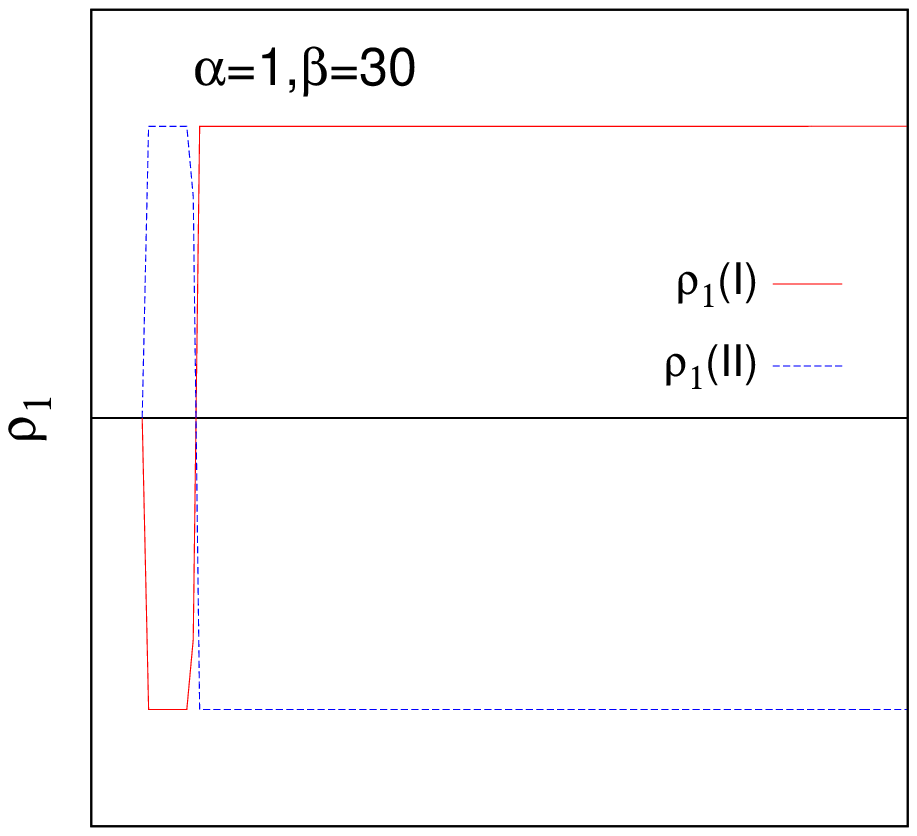}
\end{minipage}\hspace{0.08in}
\begin{minipage}[c]{0.15\textwidth}\centering
\includegraphics[scale=0.28]{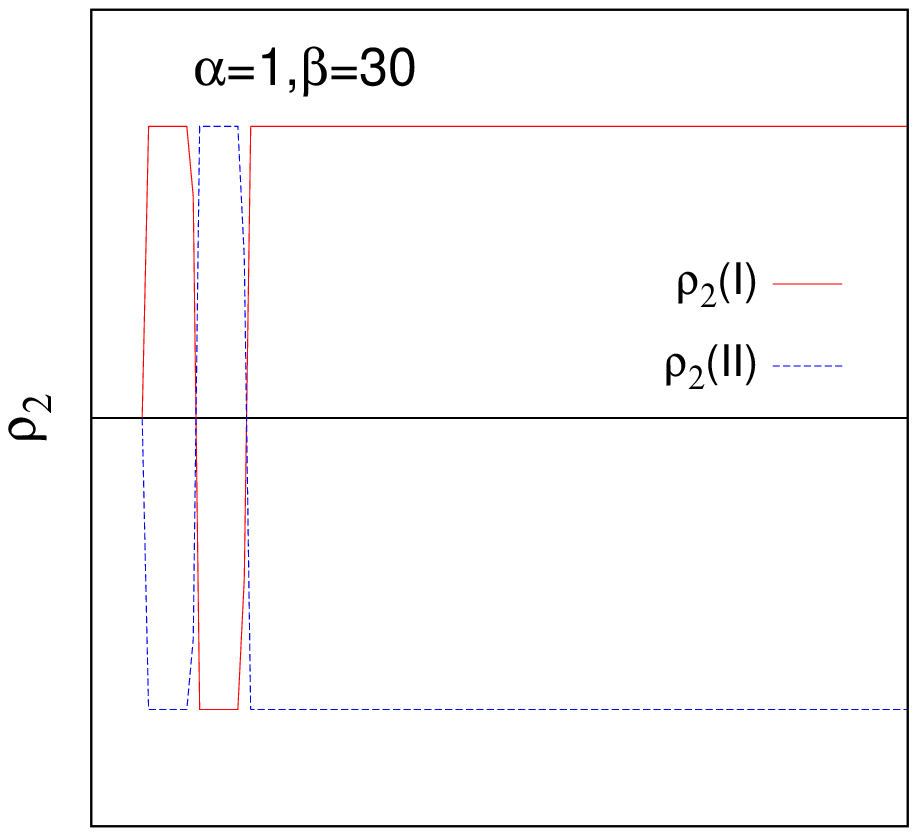}
\end{minipage}\hspace{0.08in}
\begin{minipage}[c]{0.15\textwidth}\centering
\includegraphics[scale=0.28]{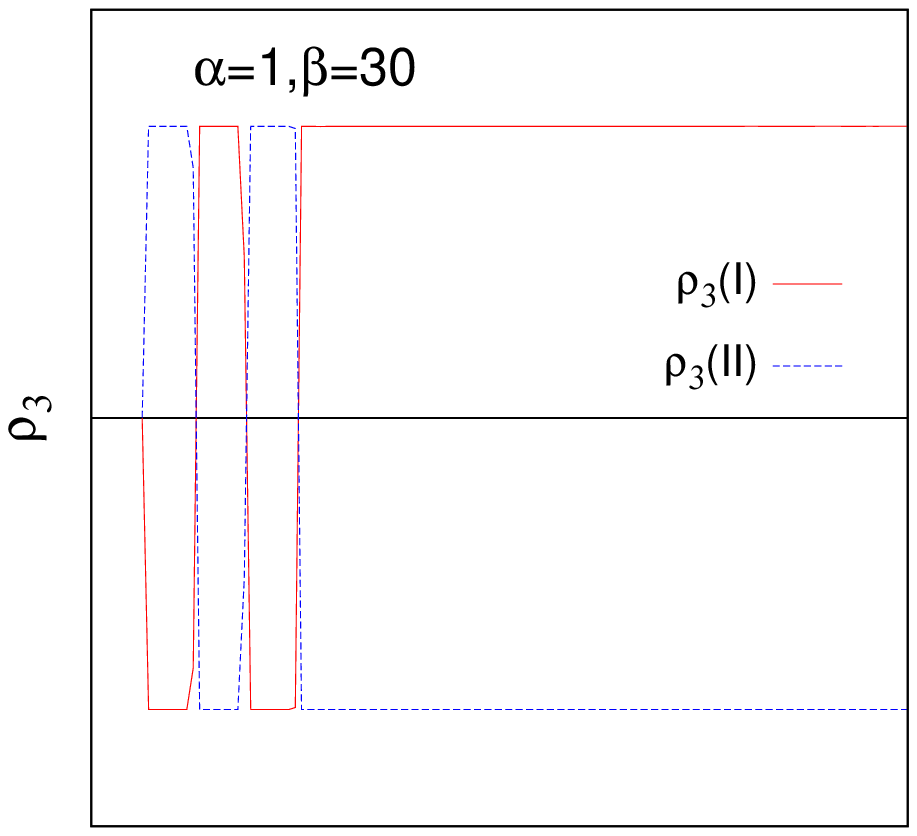}
\end{minipage}\hspace{0.08in}
\begin{minipage}[c]{0.15\textwidth}\centering
\includegraphics[scale=0.28]{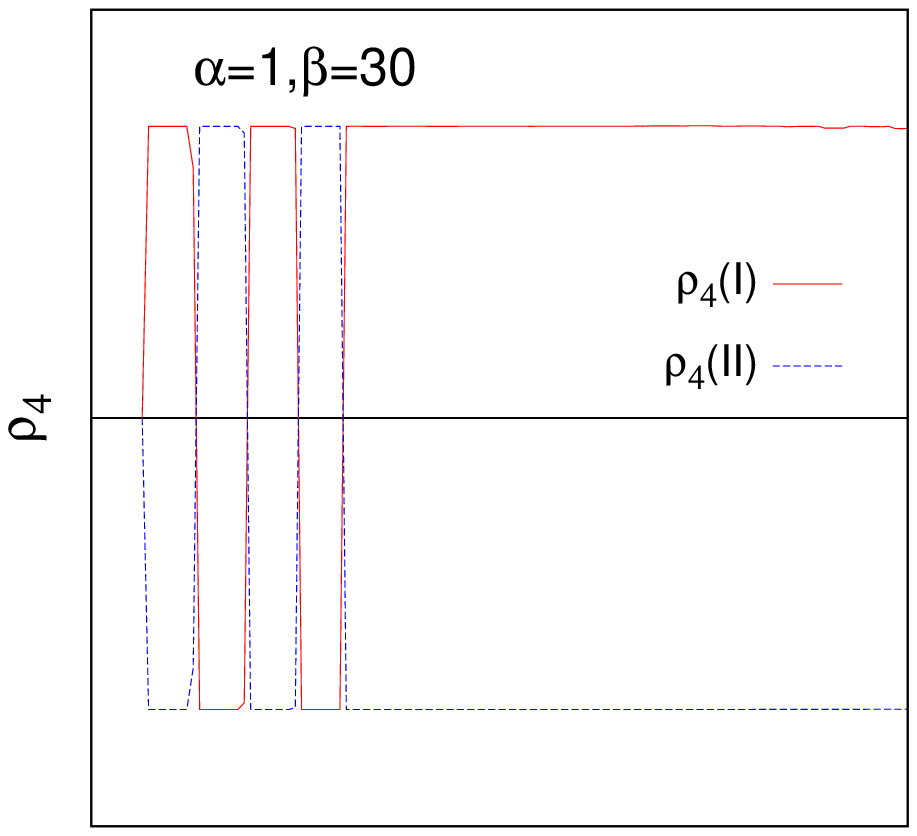}
\end{minipage}\hspace{0.08in}
\begin{minipage}[c]{0.15\textwidth}\centering
\includegraphics[scale=0.28]{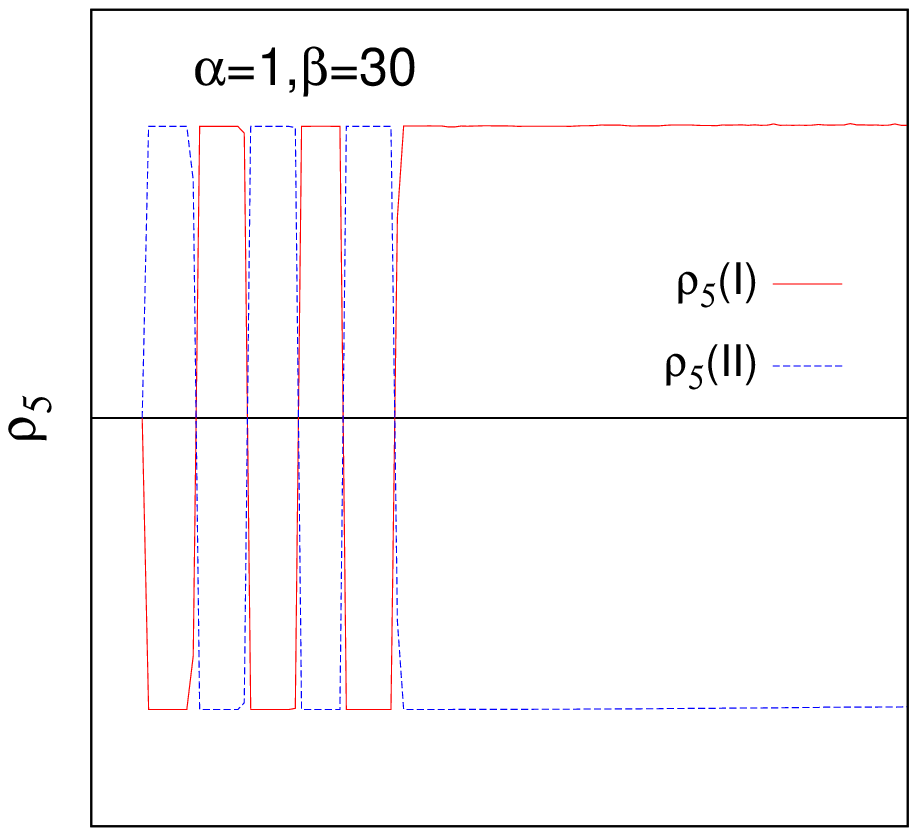}
\end{minipage}
\\[5pt]
\begin{minipage}[c]{0.15\textwidth}\centering
\includegraphics[scale=0.28]{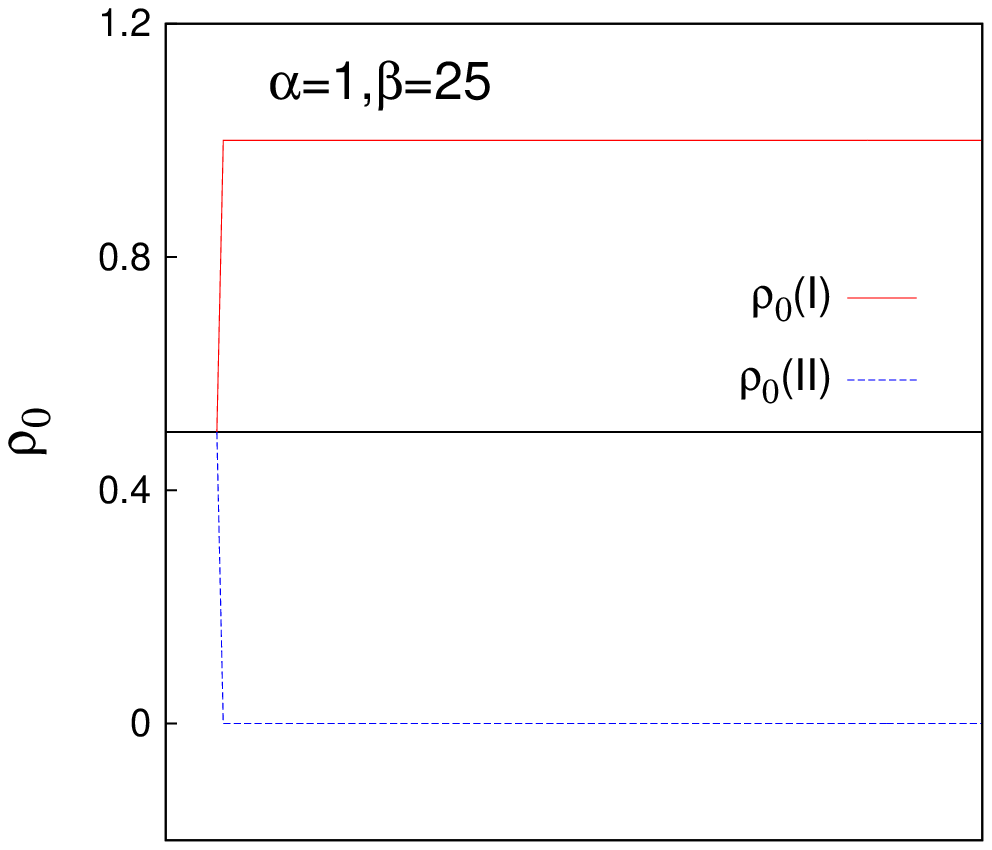}
\end{minipage}\hspace{0.08in}
\begin{minipage}[c]{0.15\textwidth}\centering
\includegraphics[scale=0.28]{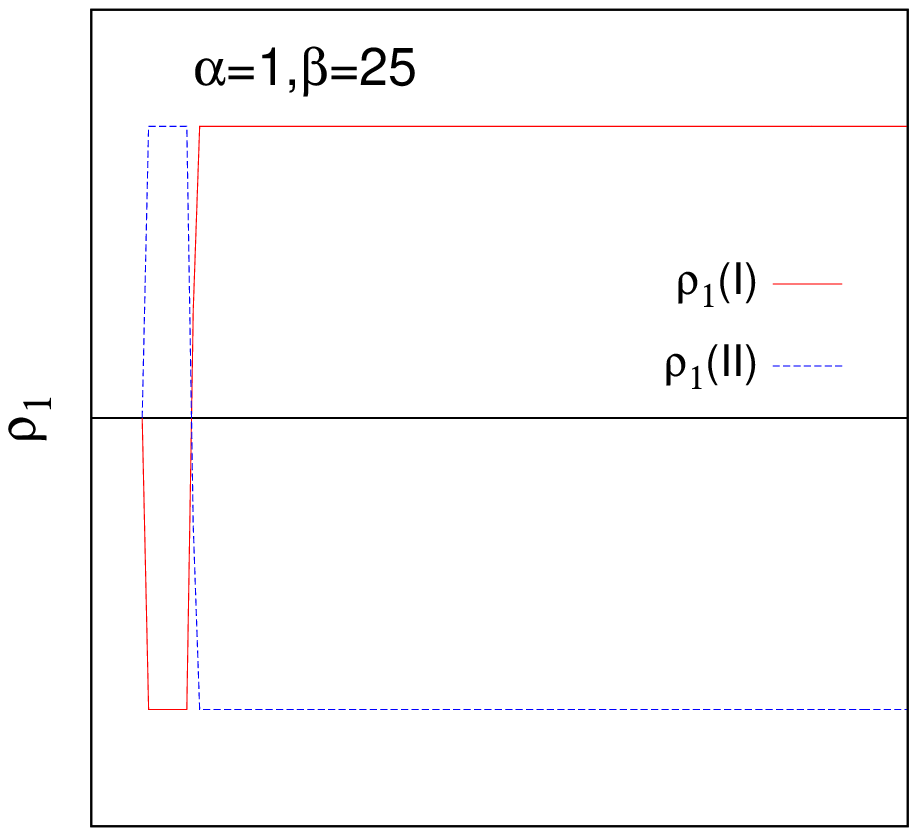}
\end{minipage}\hspace{0.08in}
\begin{minipage}[c]{0.15\textwidth}\centering
\includegraphics[scale=0.28]{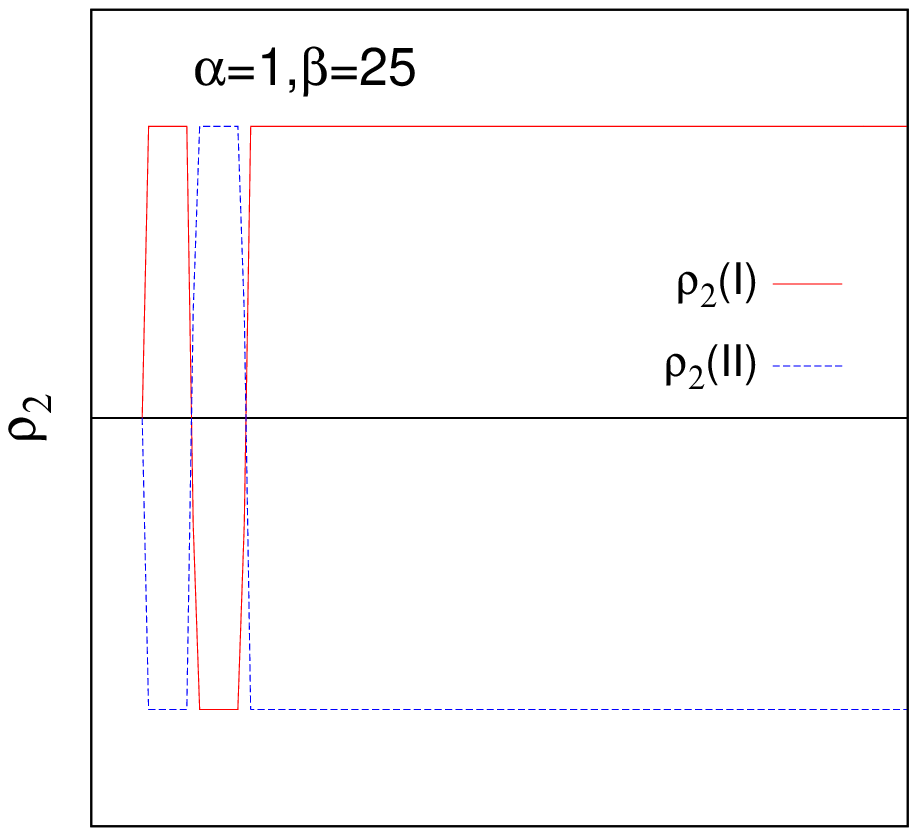}
\end{minipage}\hspace{0.08in}
\begin{minipage}[c]{0.15\textwidth}\centering
\includegraphics[scale=0.28]{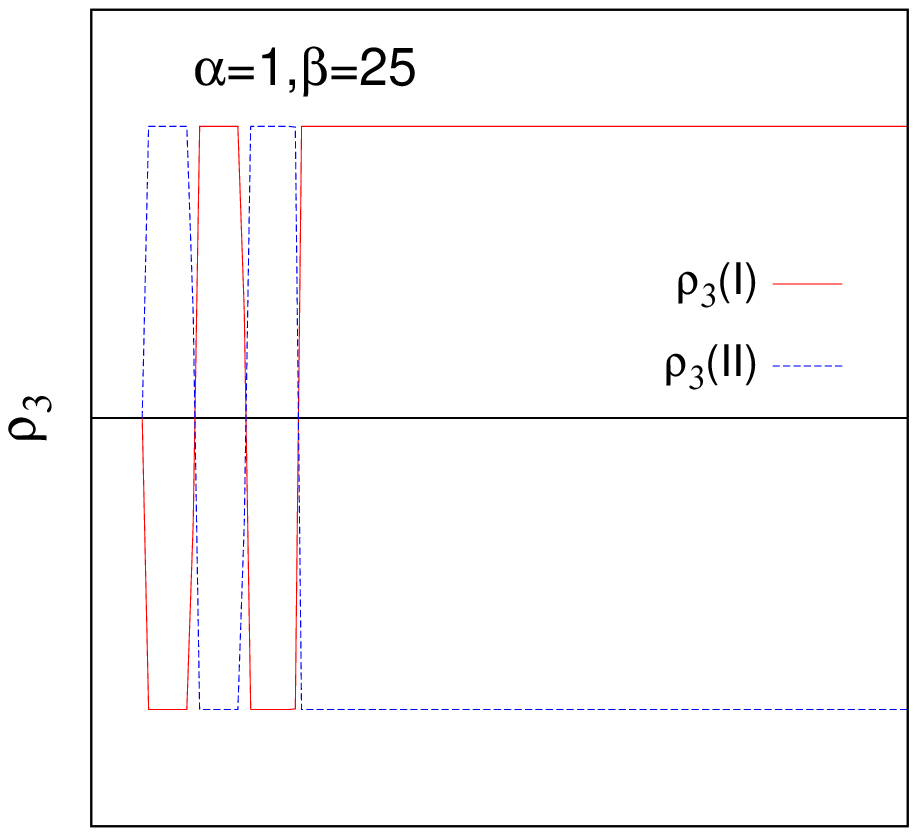}
\end{minipage}\hspace{0.08in}
\begin{minipage}[c]{0.15\textwidth}\centering
\includegraphics[scale=0.28]{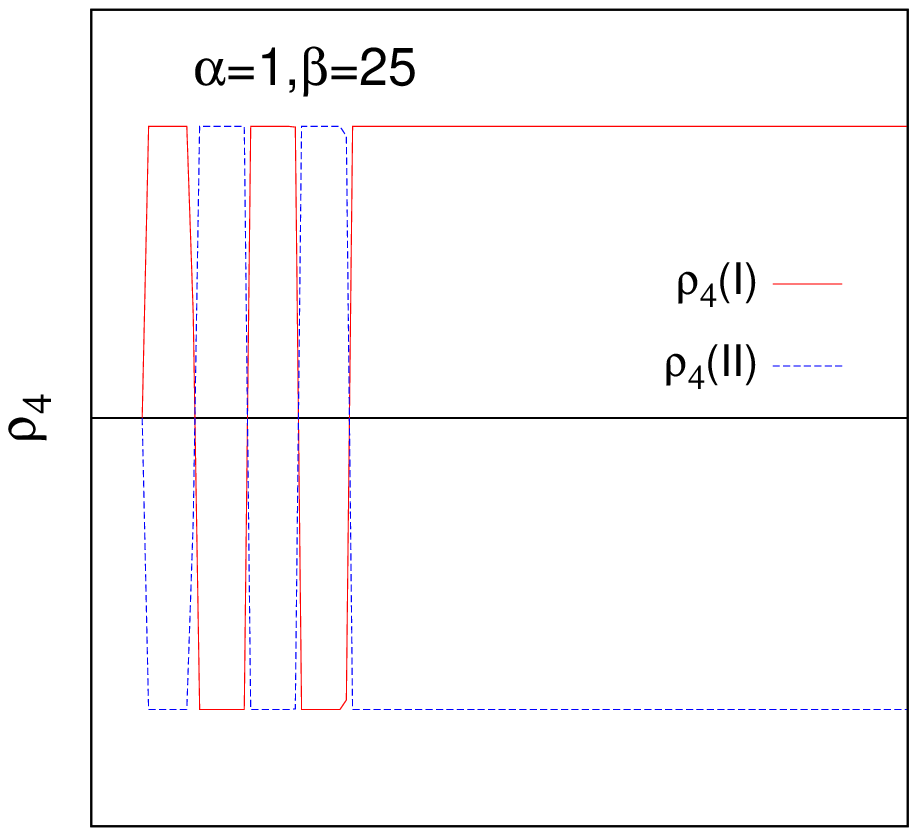}
\end{minipage}\hspace{0.08in}
\begin{minipage}[c]{0.15\textwidth}\centering
\includegraphics[scale=0.28]{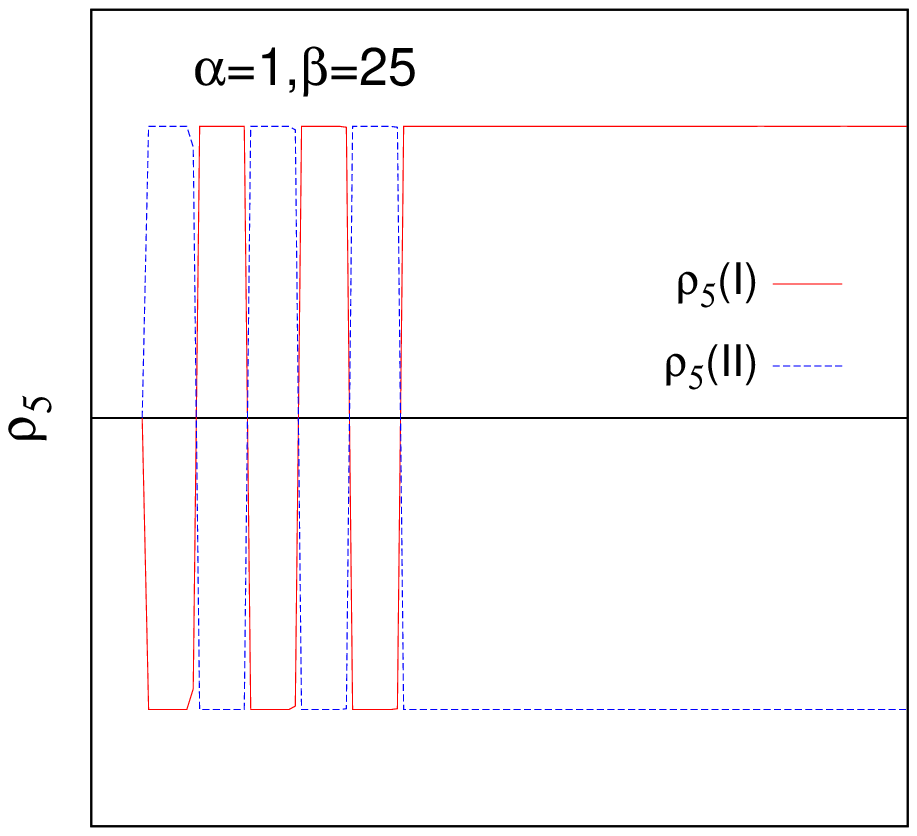}
\end{minipage}
\\[5pt]
\begin{minipage}[c]{0.15\textwidth}\centering
\includegraphics[scale=0.28]{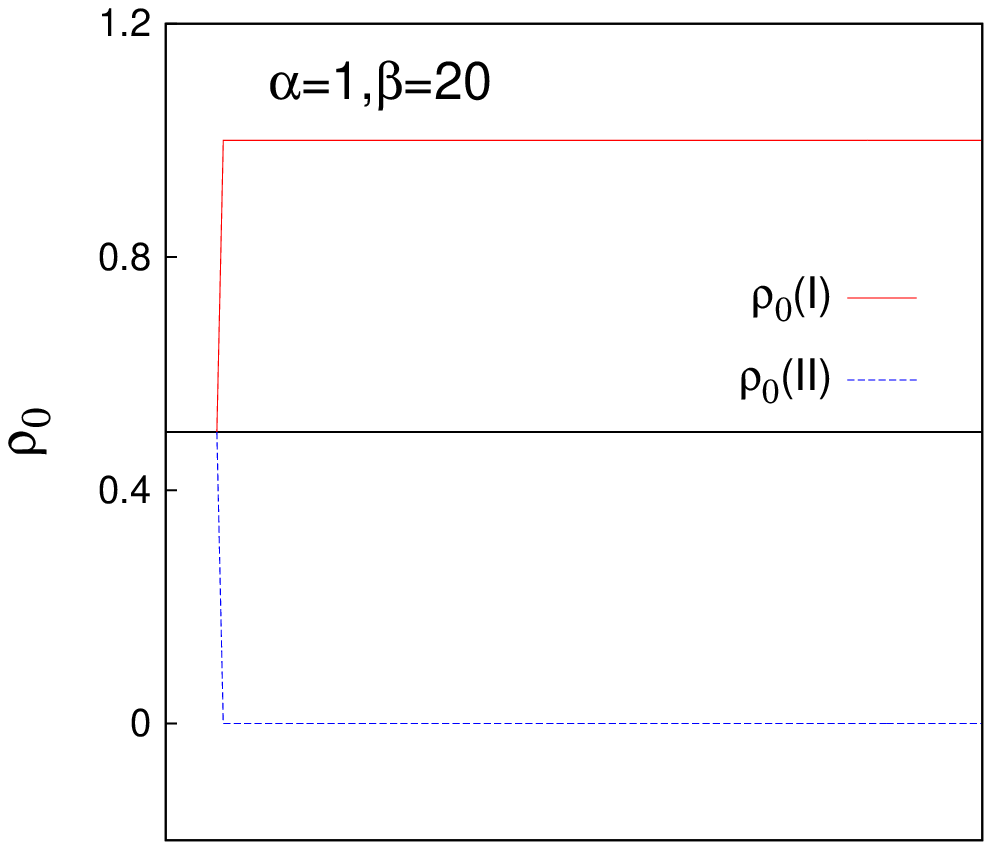}
\end{minipage}\hspace{0.08in}
\begin{minipage}[c]{0.15\textwidth}\centering
\includegraphics[scale=0.28]{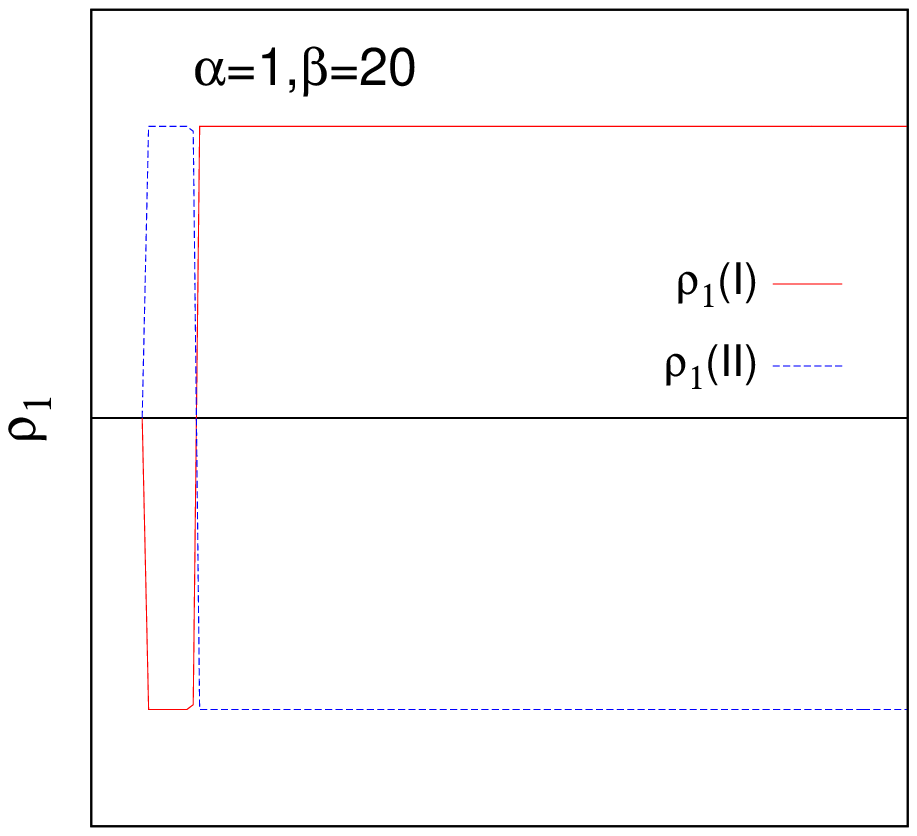}
\end{minipage}\hspace{0.08in}
\begin{minipage}[c]{0.15\textwidth}\centering
\includegraphics[scale=0.28]{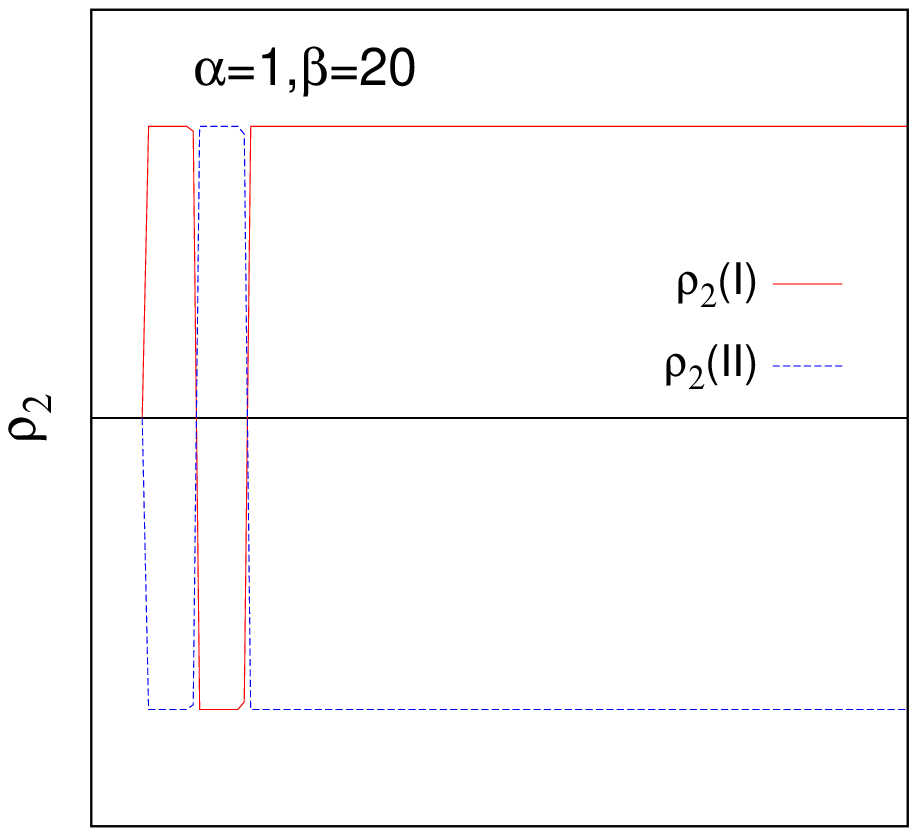}
\end{minipage}\hspace{0.08in}
\begin{minipage}[c]{0.15\textwidth}\centering
\includegraphics[scale=0.28]{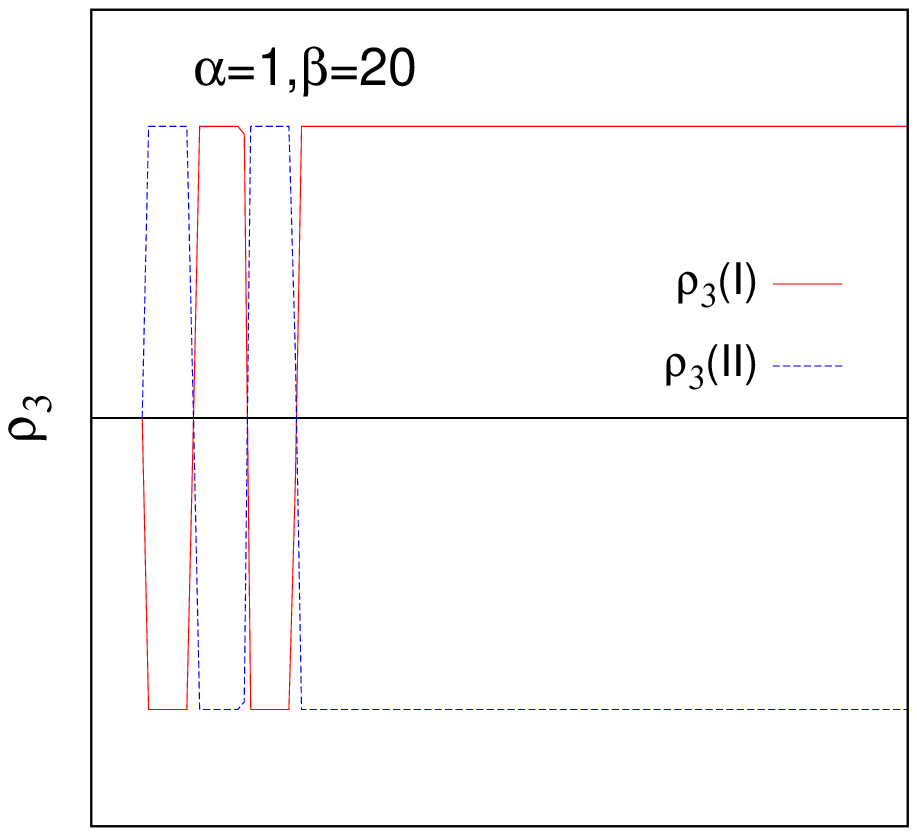}
\end{minipage}\hspace{0.08in}
\begin{minipage}[c]{0.15\textwidth}\centering
\includegraphics[scale=0.28]{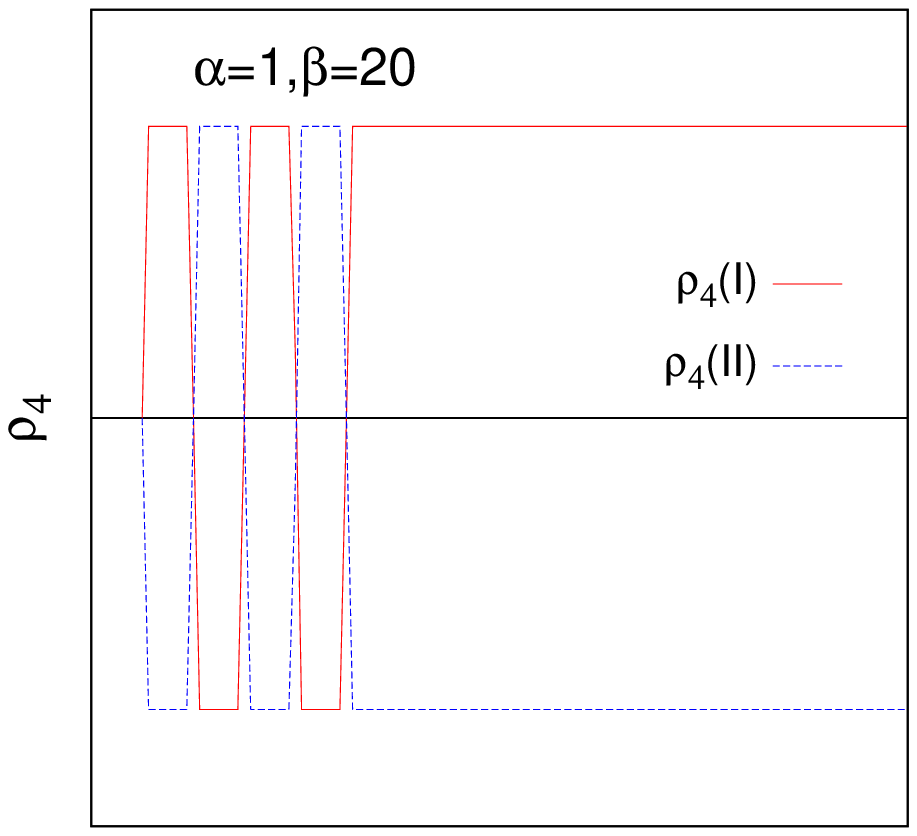}
\end{minipage}\hspace{0.08in}
\begin{minipage}[c]{0.15\textwidth}\centering
\includegraphics[scale=0.28]{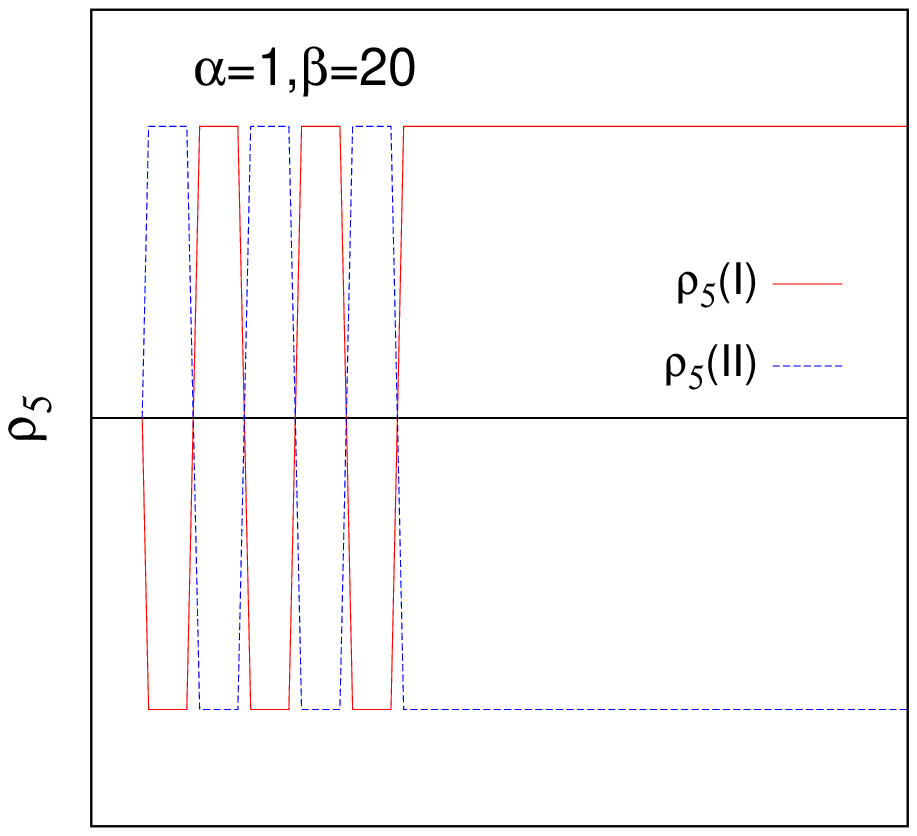}
\end{minipage}
\\[5pt]
\begin{minipage}[c]{0.15\textwidth}\centering
\includegraphics[scale=0.28]{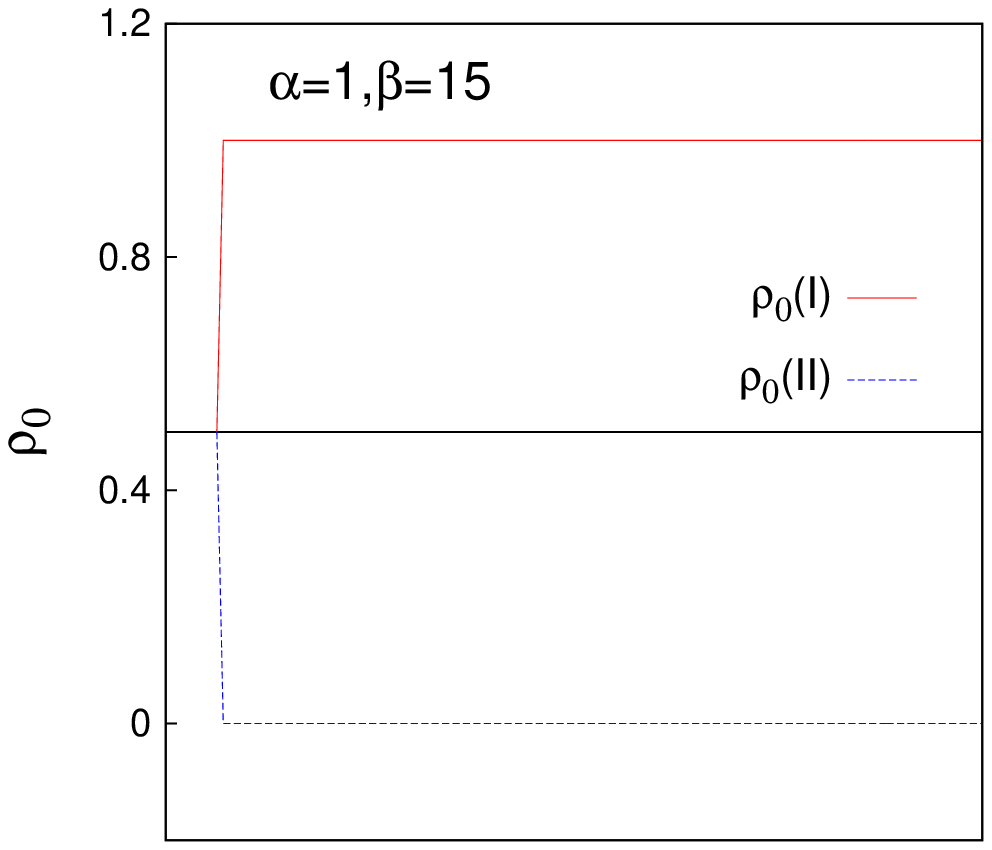}
\end{minipage}\hspace{0.08in}
\begin{minipage}[c]{0.15\textwidth}\centering
\includegraphics[scale=0.28]{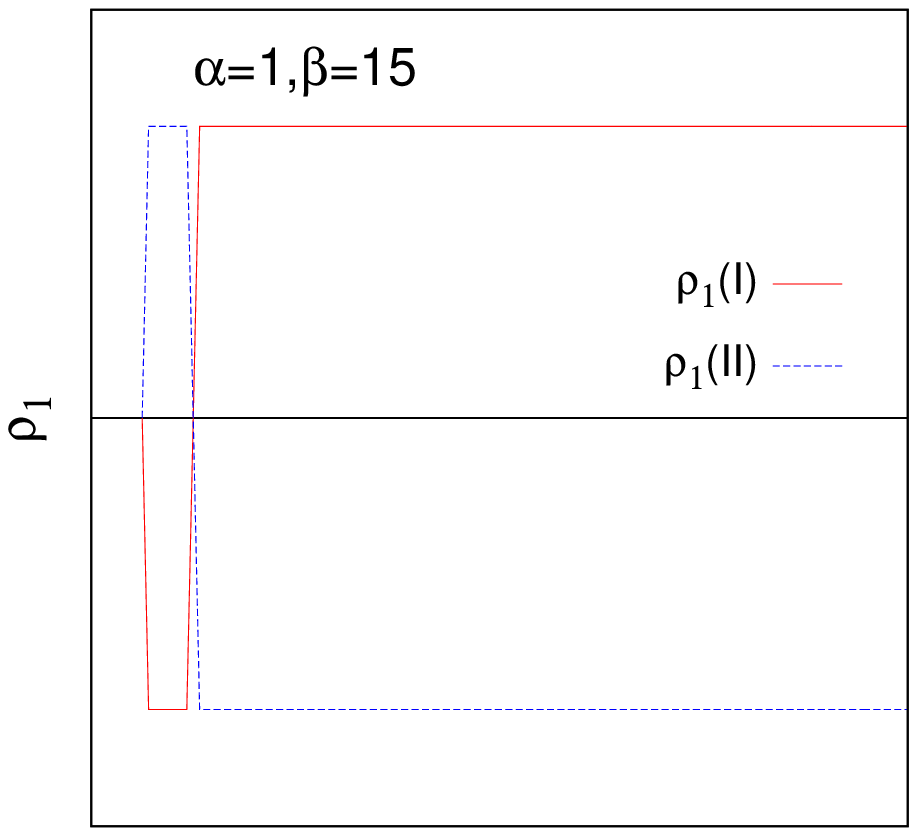}
\end{minipage}\hspace{0.08in}
\begin{minipage}[c]{0.15\textwidth}\centering
\includegraphics[scale=0.28]{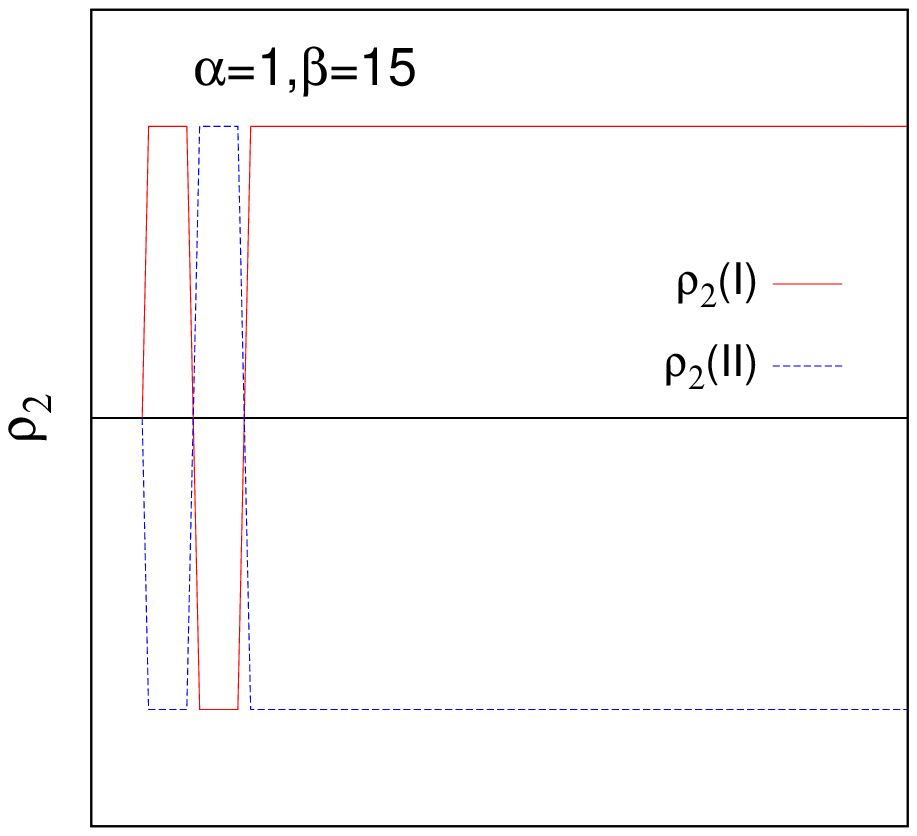}
\end{minipage}\hspace{0.08in}
\begin{minipage}[c]{0.15\textwidth}\centering
\includegraphics[scale=0.28]{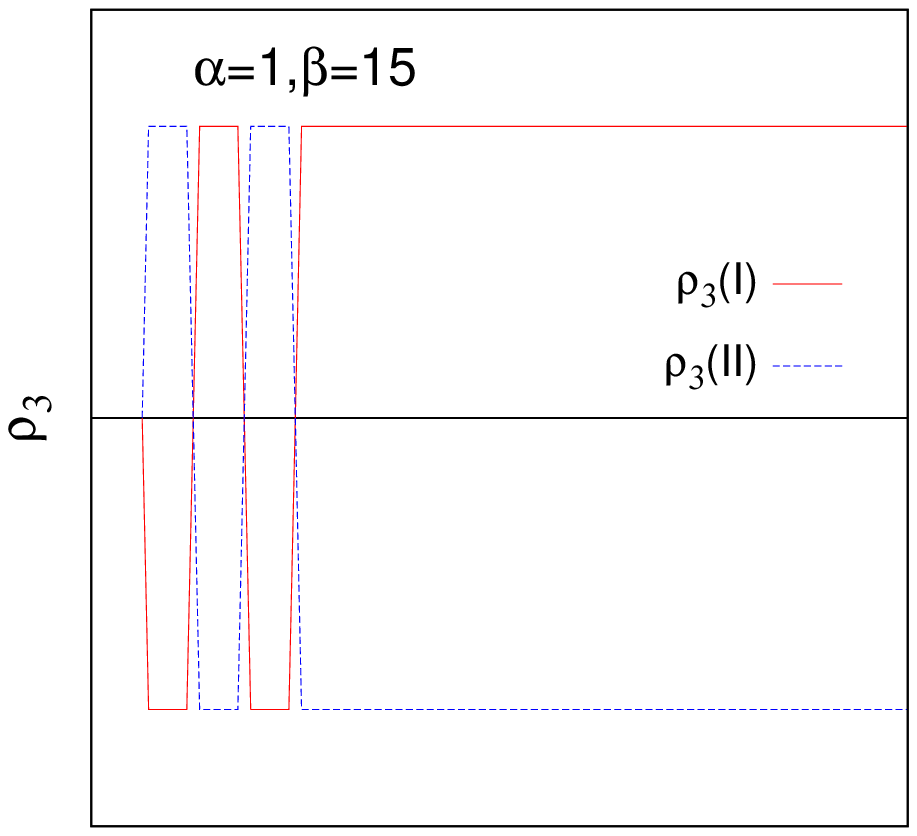}
\end{minipage}\hspace{0.08in}
\begin{minipage}[c]{0.15\textwidth}\centering
\includegraphics[scale=0.28]{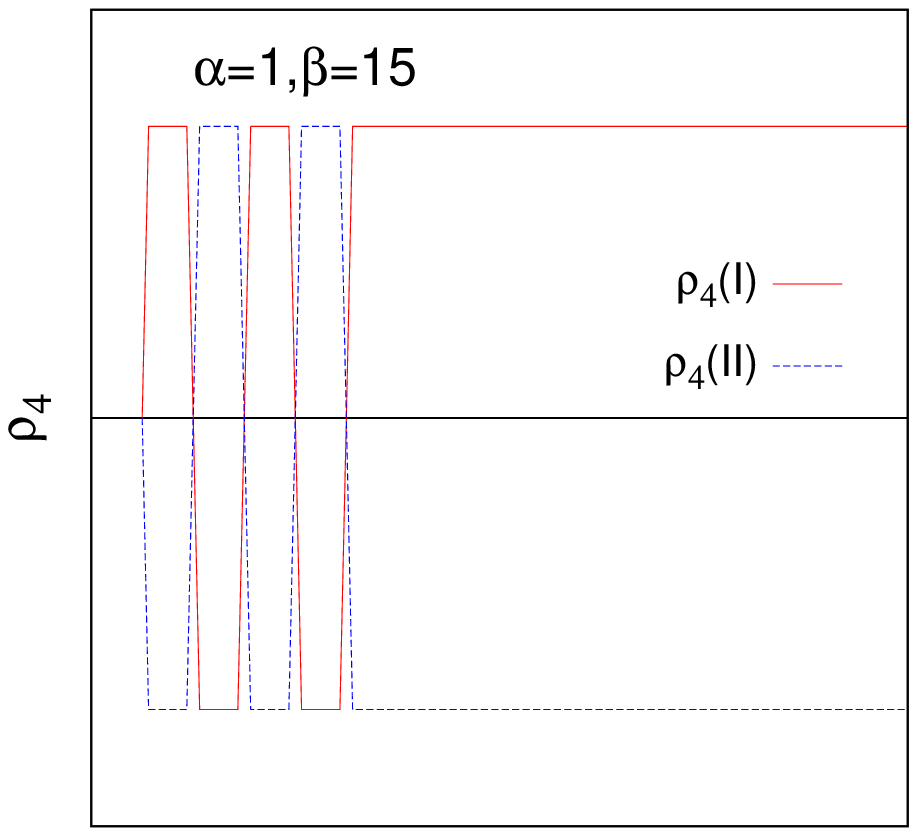}
\end{minipage}\hspace{0.08in}
\begin{minipage}[c]{0.15\textwidth}\centering
\includegraphics[scale=0.28]{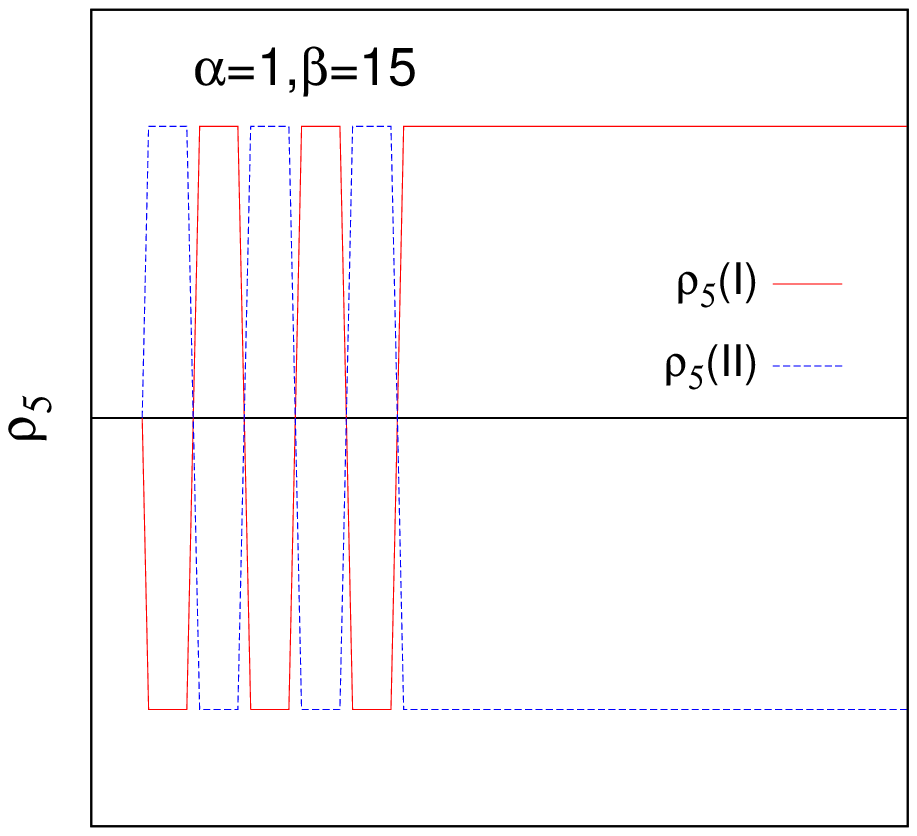}
\end{minipage}
\\[5pt]
\begin{minipage}[c]{0.15\textwidth}\centering
\includegraphics[scale=0.28]{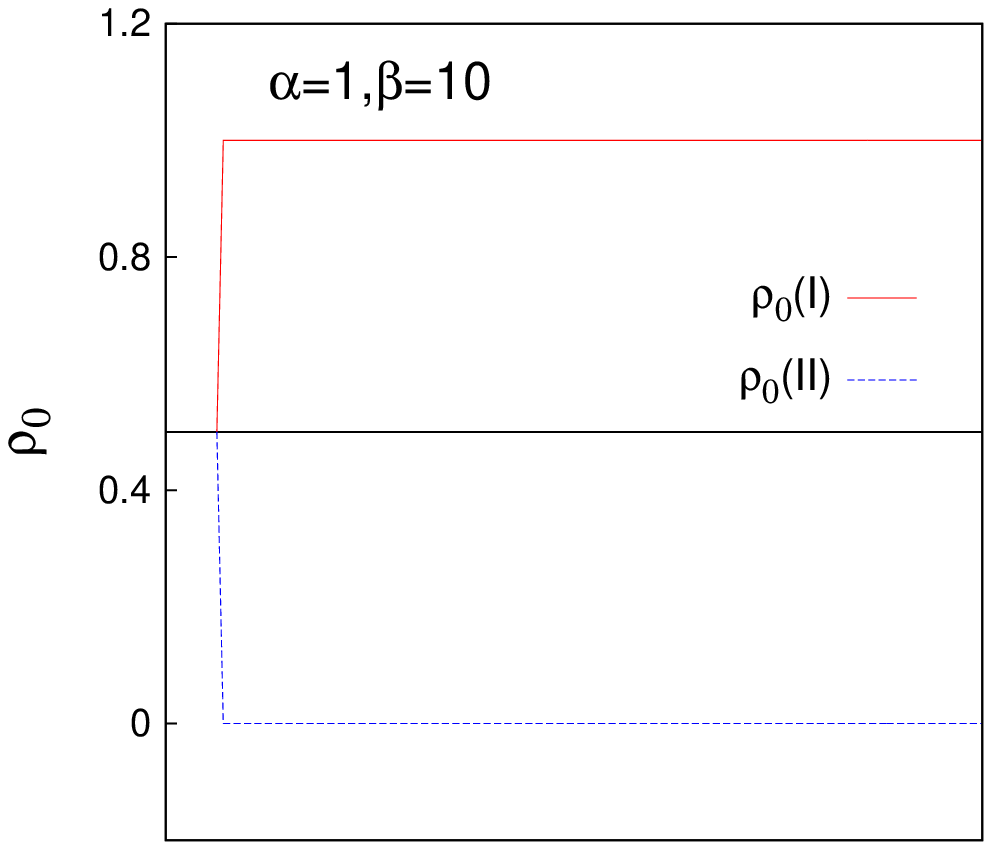}
\end{minipage}\hspace{0.08in}
\begin{minipage}[c]{0.15\textwidth}\centering
\includegraphics[scale=0.28]{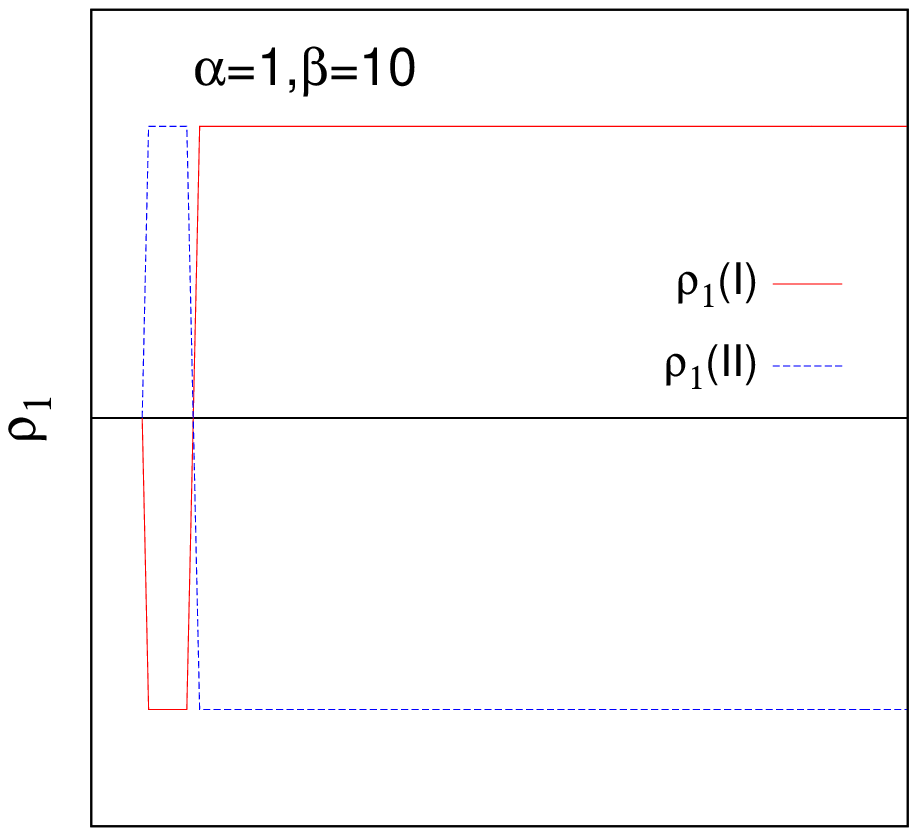}
\end{minipage}\hspace{0.08in}
\begin{minipage}[c]{0.15\textwidth}\centering
\includegraphics[scale=0.28]{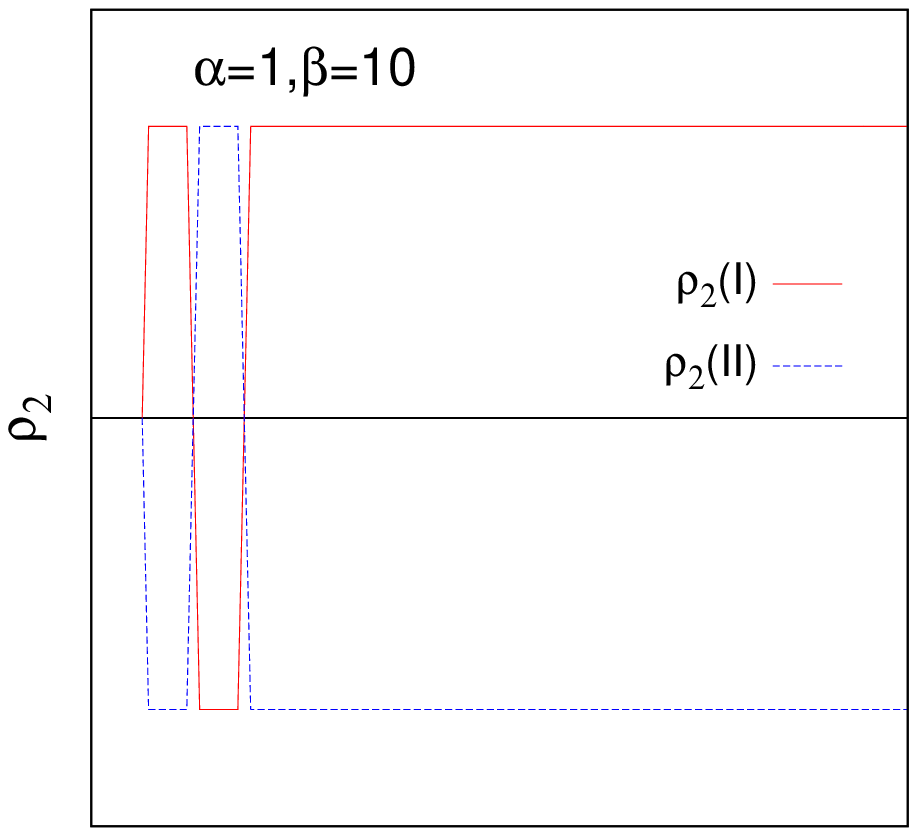}
\end{minipage}\hspace{0.08in}
\begin{minipage}[c]{0.15\textwidth}\centering
\includegraphics[scale=0.28]{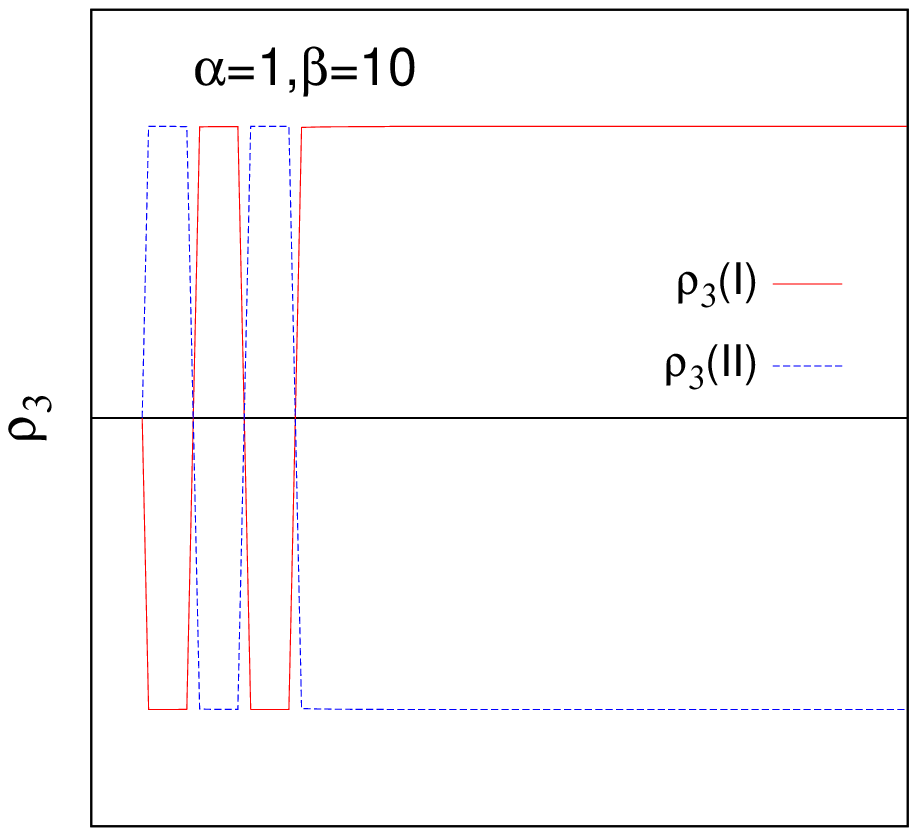}
\end{minipage}\hspace{0.08in}
\begin{minipage}[c]{0.15\textwidth}\centering
\includegraphics[scale=0.28]{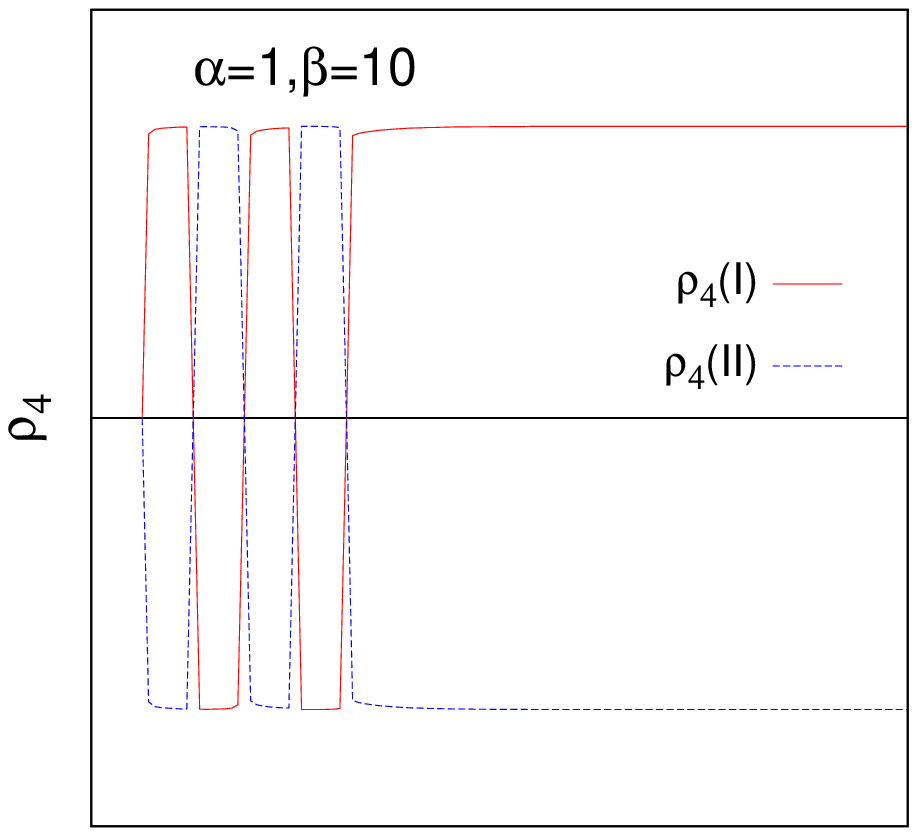}
\end{minipage}\hspace{0.08in}
\begin{minipage}[c]{0.15\textwidth}\centering
\includegraphics[scale=0.28]{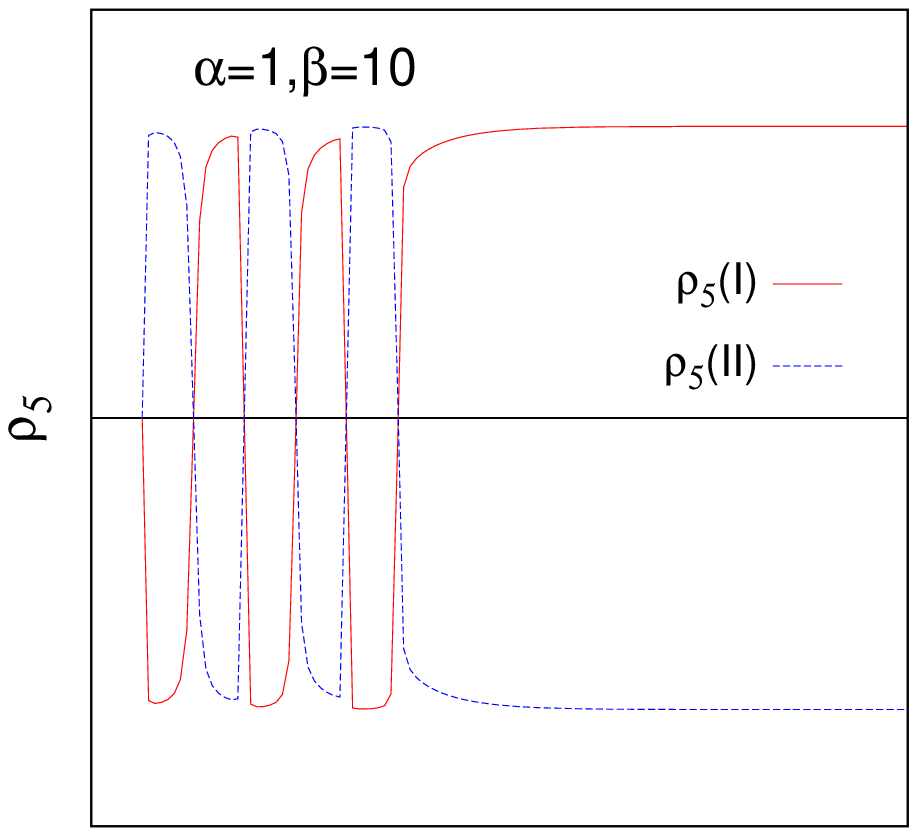}
\end{minipage}
\\[5pt]
\begin{minipage}[c]{0.15\textwidth}\centering
\includegraphics[scale=0.29]{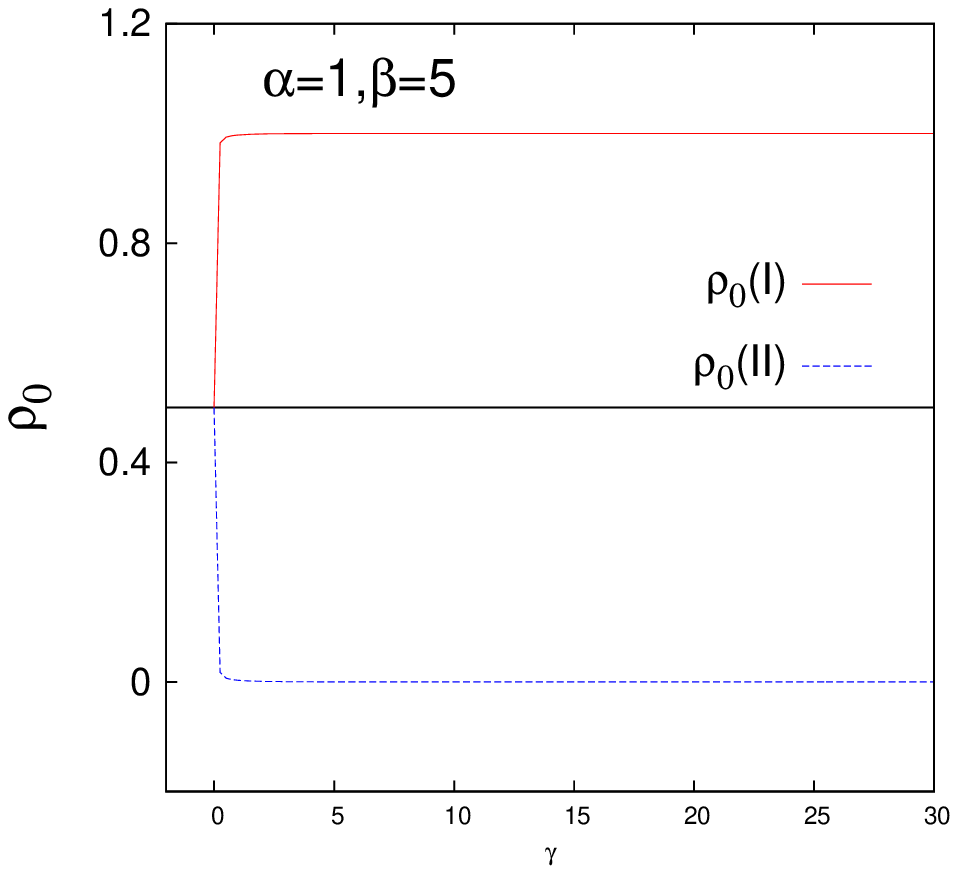}
\end{minipage}\hspace{0.07in}
\begin{minipage}[c]{0.15\textwidth}\centering
\includegraphics[scale=0.29]{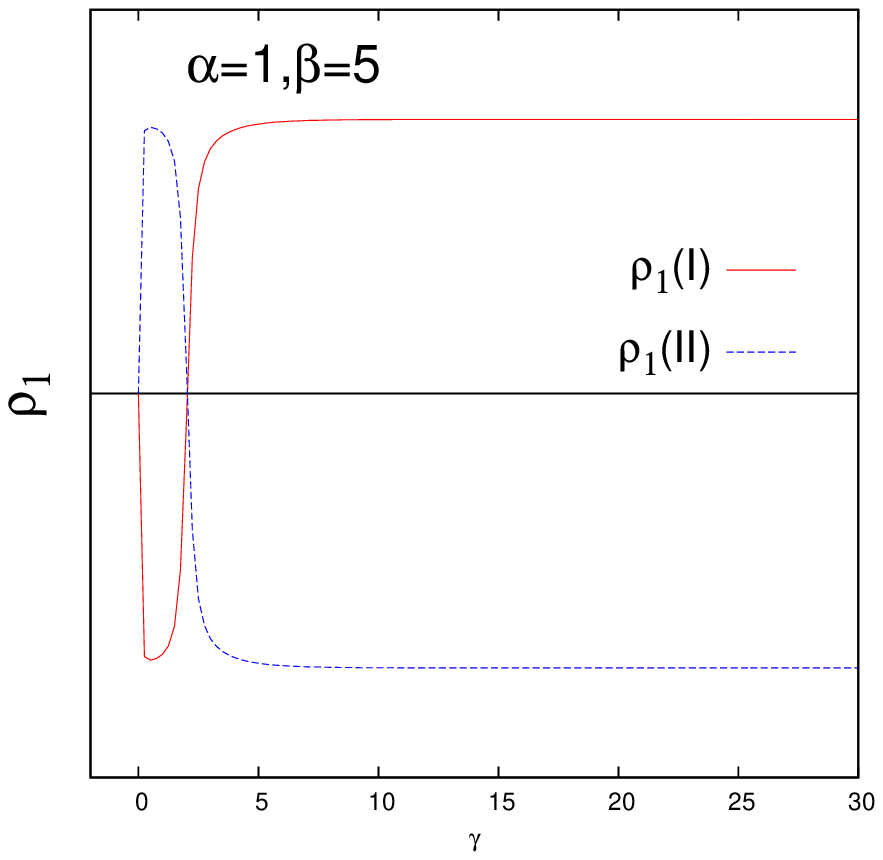}
\end{minipage}\hspace{0.07in}
\begin{minipage}[c]{0.15\textwidth}\centering
\includegraphics[scale=0.29]{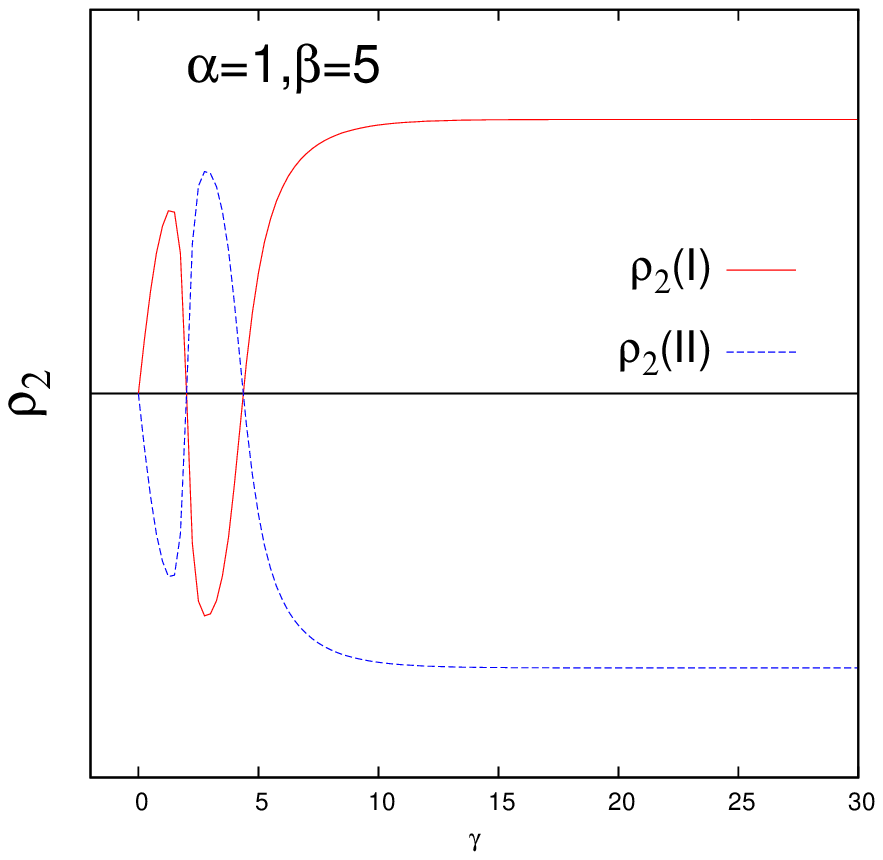}
\end{minipage}\hspace{0.07in}
\begin{minipage}[c]{0.15\textwidth}\centering
\includegraphics[scale=0.29]{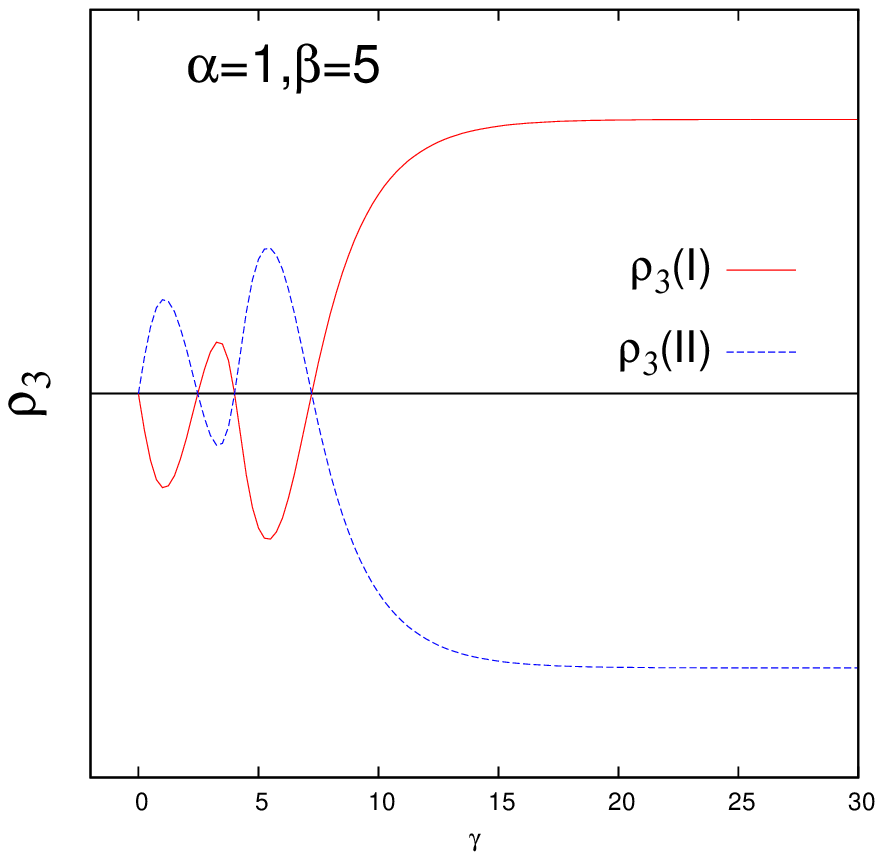}
\end{minipage}\hspace{0.07in}
\begin{minipage}[c]{0.15\textwidth}\centering
\includegraphics[scale=0.29]{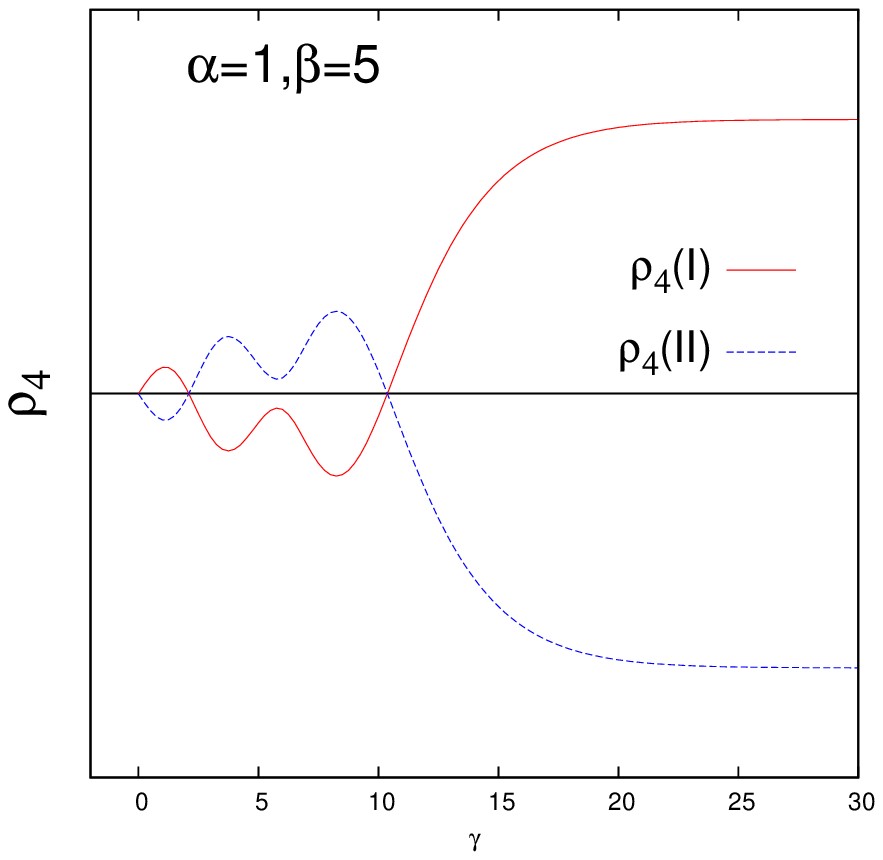}
\end{minipage}\hspace{0.07in}
\begin{minipage}[c]{0.15\textwidth}\centering
\includegraphics[scale=0.29]{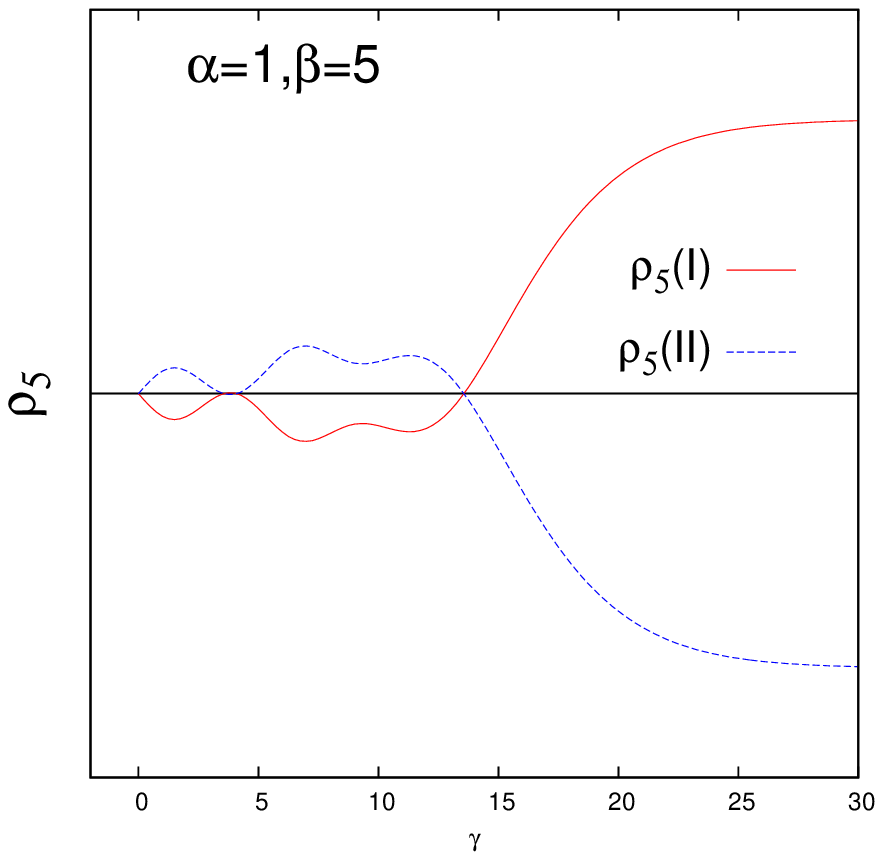}
\end{minipage}
\caption[optional]{Probability of finding the particle inside well~I (red) and II (blue) in an asymmetric DW, as $\gamma$ 
is varied. Six rows from bottom to top represent six $\beta$ (5,10,15,20,25,30); $\alpha$ is fixed at 1 for all. Six columns 
from left to right correspond to six lowest states. For details, see text.}
\end{figure}

\subsection{Wave function and probability distribution}
So far, we have inspected the energy spectrum as function of $\beta$, $\gamma$. Now, it is necessary to explore how these 
parameters impact the distribution of a particle inside the two wells, shown in Fig.~(1). 
Figure~(5) exhibits this for a set of $\gamma$, with $\alpha \! = \! 1$. These are produced in six rows for six $\beta$; 
from bottom to top, they correspond to $\beta \! = \! 5,10,15,20,25,30$, while six columns refer to lowest six states, starting 
from ground state in left side of figure. It leads to the immediate conclusion that, after a certain \emph{characteristic} 
$\beta$ (in this case, somewhere in between 5-10), the particle remains completely localized in either well~I or well~II, 
i.e., respective probabilities are either 0 or 1. Moreover, for a given $\alpha$, $\beta$, there exists some special values 
of $\gamma$, at which there is a finite probability that the particle is distributed in both wells. At these 
\emph{transition points}, particle can travel from one well to the other and these points appear after a fixed interval as 
$\Delta \gamma$ (in this example, it is 2). Finally, beyond a certain $\Delta \gamma$, any particular state eventually 
localizes in the deeper well. After a careful observation, it turns out that it is possible to enumerate a rule (based on 
some modification of energy rule discussed earlier) to explain the distribution pattern inside the wells. Recall that, 
I, II signify deeper and shallow wells respectively. 

Now, in the following, we attempt to quantify these observations in terms of some simple rules to predict 
localization/delocalization of a particle in a given well, in terms of the parameter $k$ and state index $n$. 
For $\epsilon_{n} \! < \!V_0$, following situations could be envisioned. 

\begin{enumerate}[(i)]
\item
$n \! \geq \! k$: two possibilities arise:
\begin{enumerate}[(a)]
\item
$k$ is integer: particle in $n$th state is distributed in both well~I and well~II.
\item
$k$ is fraction: four possibilities need to be considered: 

\begin{enumerate}[(1)]
\item
both $n$ and integer part of $k$ $even$: particle stays in well~I.
\item
both $n$ and integer part of $k$ $odd$: particle stays in well~I.
\item
$n$ $even$ and integer part of $k$ $odd$: particle stays in well~II.
\item
$n$ $odd$ and integer part of $k$ $even$: particle stays in well~II.
\end{enumerate}
\end{enumerate}
\item
$n \! < \! k$: particle always resides in larger (I) well.
\end{enumerate}

Let us now try to explain Fig.~(5) on the basis of above generalizations. At non-zero $\gamma$ ($k \! > \! 0$), particle 
in $n \! = \! 0$ state dwells in deeper well I. This is observed in all sub-figures of column 1; as expected for
$n \! < \! k$, from (ii). Next for $n \! = \! 1$, when $0 \! < \! \gamma \! < \! 2$ ($0 \! < \! k \! < \! 1$), from second 
column, it remains in smaller well~II. This is predicted from (i.b.4) as integer part of $k$ is 0 
($\! < \!n \! = \! 1$). Note that, bottom most panel ($\beta \! = \! 5$) separates out from remaining plots, as complete 
localization is yet to occur (threshold $\beta$ is not reached). At $\gamma \! = \! 2$ ($k \! = \! 1$), (i.a) suggests that 
it resides in both wells; precisely this is observed from figure. For any $\gamma$ higher than 
2 ($ k \! > \! 1$), once again, in agreement with (ii), it stays in well~I. In case of $n \! = \! 2$, third column 
show that, for $0 \! < \! \gamma \! < \! 2$ (or $0 \! < \! k \! < \! 1$), particle inhabits well~I; (i.b.1) applies as 
both $n$ and integer portion of $k$ (0) are even. On the other hand, in $2 \! < \! \gamma \! < \! 4$ region (equivalently 
$1 \! < \! k \! < \! 2$), particle is located in well~II; as integer part of $k$ is odd, (i.b.1) comes in to play. At 
$\gamma \! = \! 2$ or 4 ($k$ is 1 or 2), it stays in both wells, following (i.a). For 
$\gamma \! > \! 4$ ($k \! > \! 2$), it again localizes in well~I. Next in column four, probabilities are given for 
$n \! = \! 3$. When $0 \! < \! \gamma  \! < \! 2$ ($0 \! < \! k < \! 1$) and $4 \! < \! \gamma  \! < \! 6$ 
($2 \! < \! k < \! 3$), it confines in well~II in agreement with (i.b.4), whereas for 
$2 \! < \! \gamma  \! < \! 4$ ($1 \! < \! k < \! 2$), it stays in well~I, as dictated by (i.b.2). At $\gamma \! = \! 2,4,6$
($k \! = \! 1,2,3$), it could be found in both wells following (i). As usual after $\gamma \! > \! 6$ ($k \! > \! 3$), 
it localizes in well~I obeying (ii). Similarly, other higher states in columns five and six also follow these rules
and further establishes their relevance. Thus, one finds that, particle in $n$th excited state localizes to well~I after 
undergoing $n$ number of transitions between the wells. Integer values of $k$ signify transition points. Further, 
it is apparent that for fraction $k$, two wells behave as two separate single-well potentials. Table~V summarizes 
distribution of lowest six states within the wells, for four ranges of $k$, justifying our above discussion. An analysis of
wave function will further authenticate this, which is pursued next. 

\begin{table}     
\caption{Distribution of lowest six states within two wells, as well as number of effective nodes, for asymmetric DW 
potential, in Eq.~(4). Four ranges of $k$ are considered. For details, see text.}
\centering
\begin{tabular}{c|cc|cc|cc|cc}
\hline
\hspace{0.2in} $n$ \hspace{0.2in} &    \multicolumn{2}{c}{\hspace{0.3in}$0 \! < \! k \! < \! 1$ } \hspace{0.3in}  
    &    \multicolumn{2}{c}{\hspace{0.3in}$1 \! < \! k \! < \! 2$} \hspace{0.3in}   
    &    \multicolumn{2}{c}{\hspace{0.3in}$2 \! < \! k \! < \! 3$} \hspace{0.3in} 
    &    \multicolumn{2}{c}{\hspace{0.3in}$3 \! < \! k \! < \! 4$} \hspace{0.3in} \\
\hline
    &    well   &   node   &   well    &   node    &  well    &   node    &   well   &  node           \\
\hline
0   &  I    &  0    &  I    &   0   &    I   &   0    &      I    &     0    \\
1   &  II   &  0    &  I    &   1   &    I   &   1    &      I    &     1    \\
2   &  I    &  1    &  II   &   0   &    I   &   2    &      I    &     2    \\
3   &  II   &  1    &  I    &   2   &    II  &   0    &      I    &     3    \\
4   &  I    &  2    &  II   &   1   &    I   &   3    &      II   &     0    \\
5   &  II   &  2    &  I    &   3   &    II  &   1    &      I    &     4    \\
\hline
\end{tabular}
\end{table}
 
\begin{figure}             
\centering
\begin{minipage}[c]{0.15\textwidth}\centering
\includegraphics[scale=0.28]{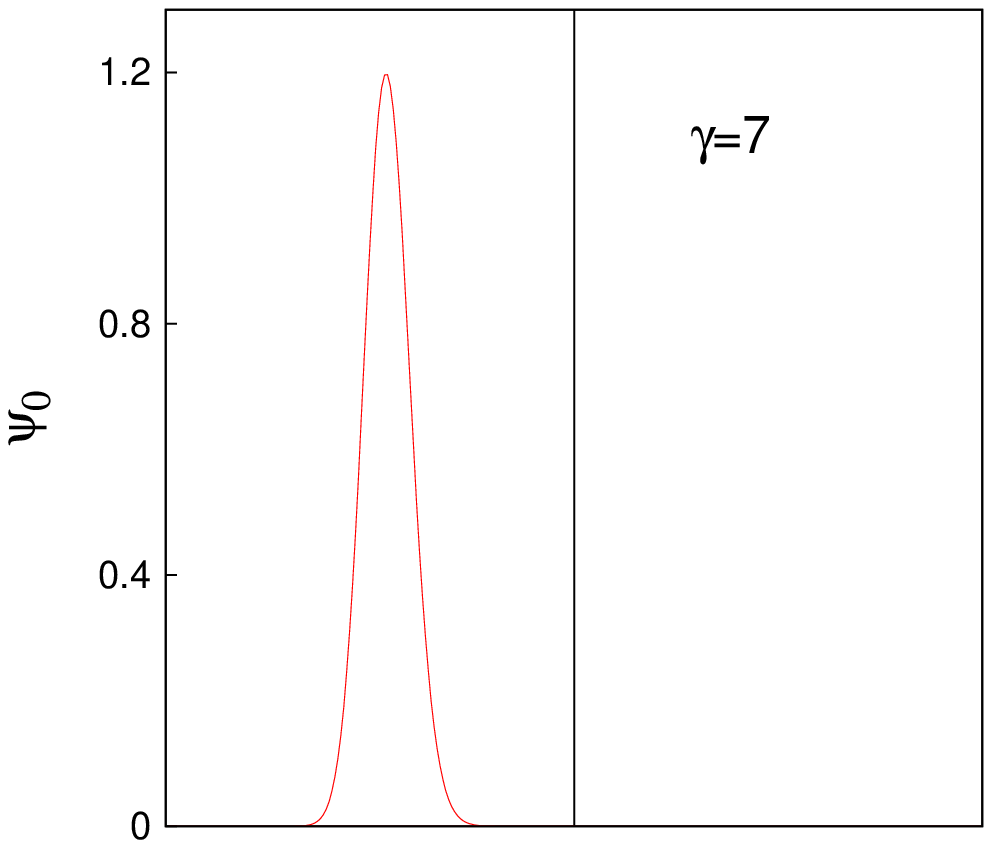}
\end{minipage}\hspace{0.07in}
\begin{minipage}[c]{0.15\textwidth}\centering
\includegraphics[scale=0.28]{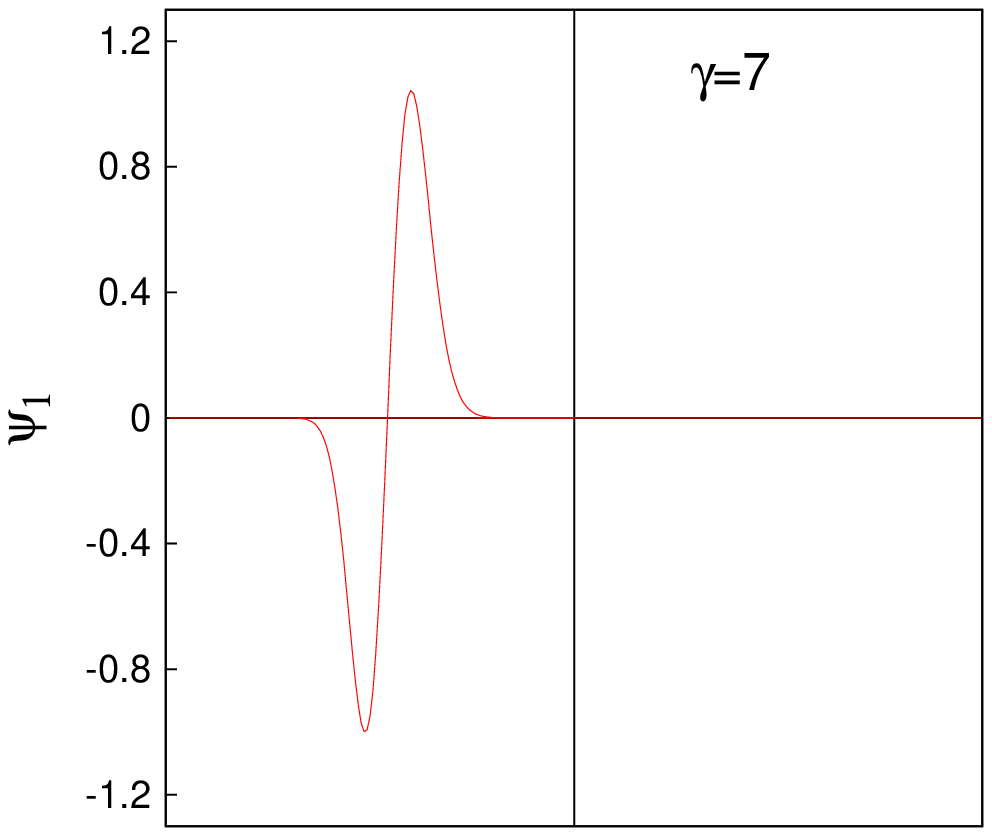}
\end{minipage}\hspace{0.07in}
\begin{minipage}[c]{0.15\textwidth}\centering
\includegraphics[scale=0.28]{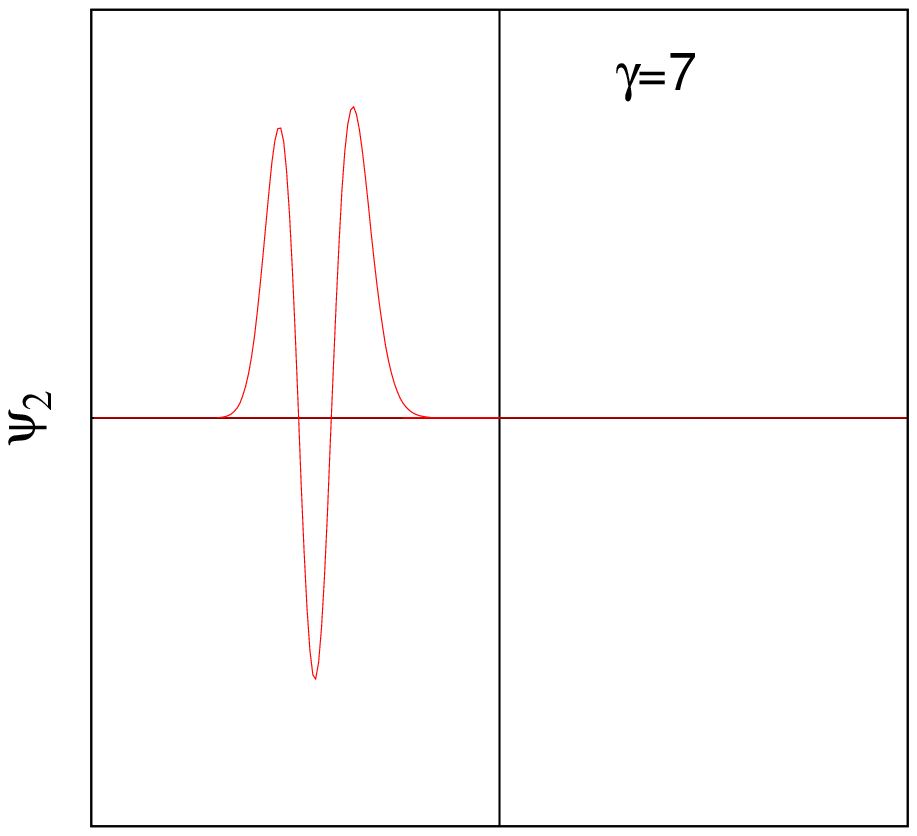}
\end{minipage}\hspace{0.07in}
\begin{minipage}[c]{0.15\textwidth}\centering
\includegraphics[scale=0.28]{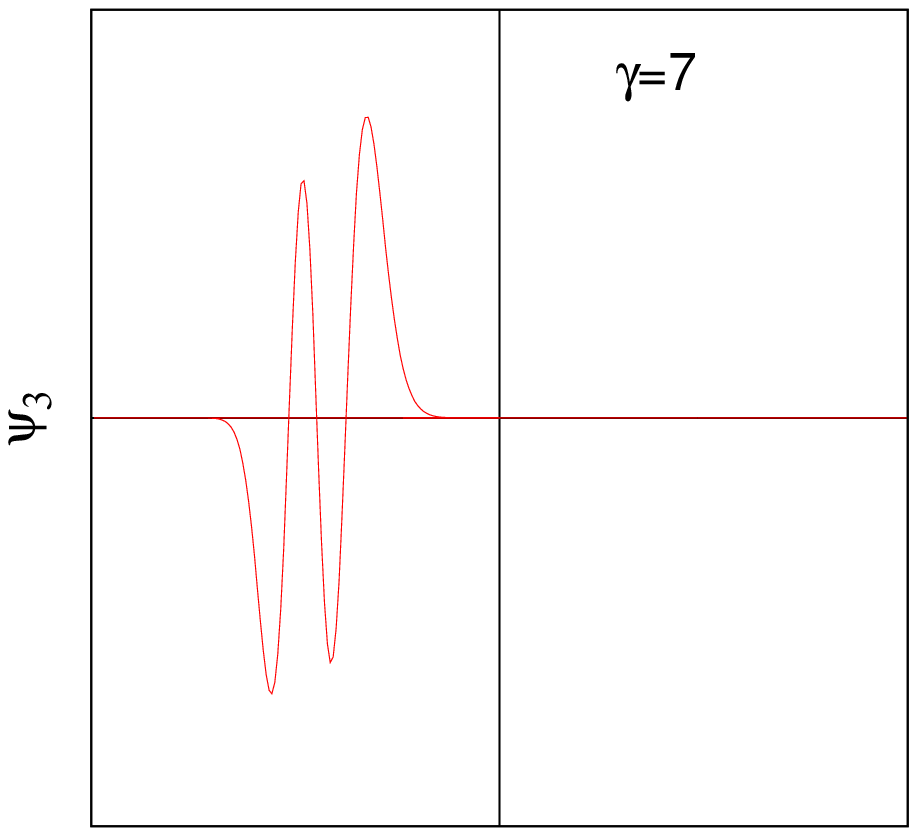}
\end{minipage}\hspace{0.07in}
\begin{minipage}[c]{0.15\textwidth}\centering
\includegraphics[scale=0.28]{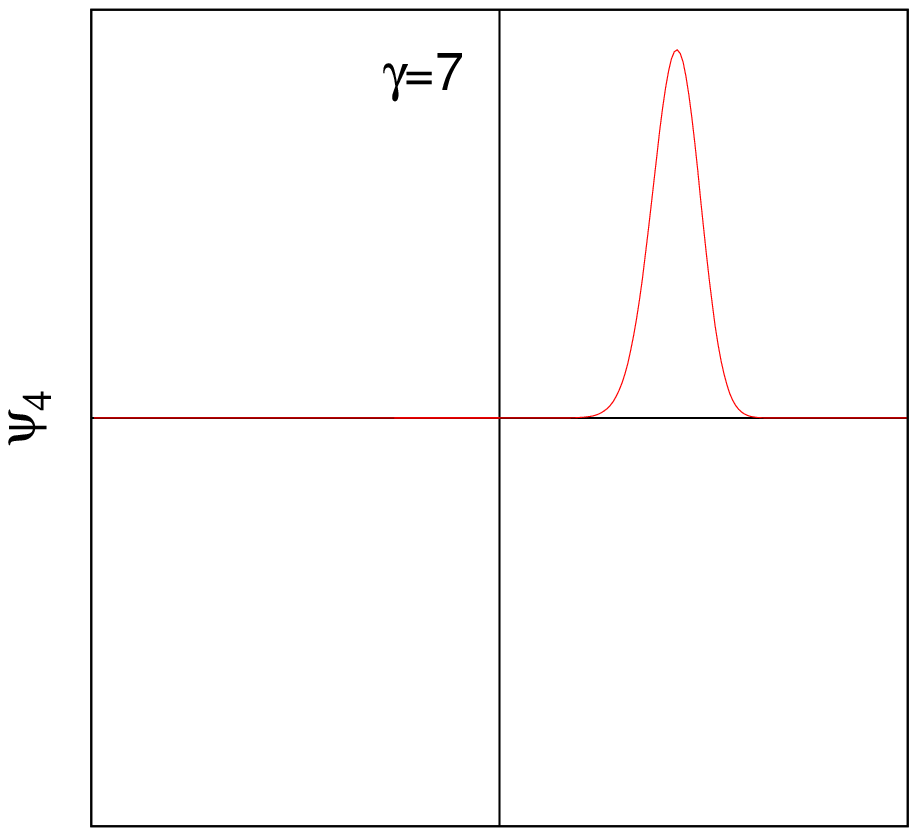}
\end{minipage}\hspace{0.07in}
\begin{minipage}[c]{0.15\textwidth}\centering
\includegraphics[scale=0.28]{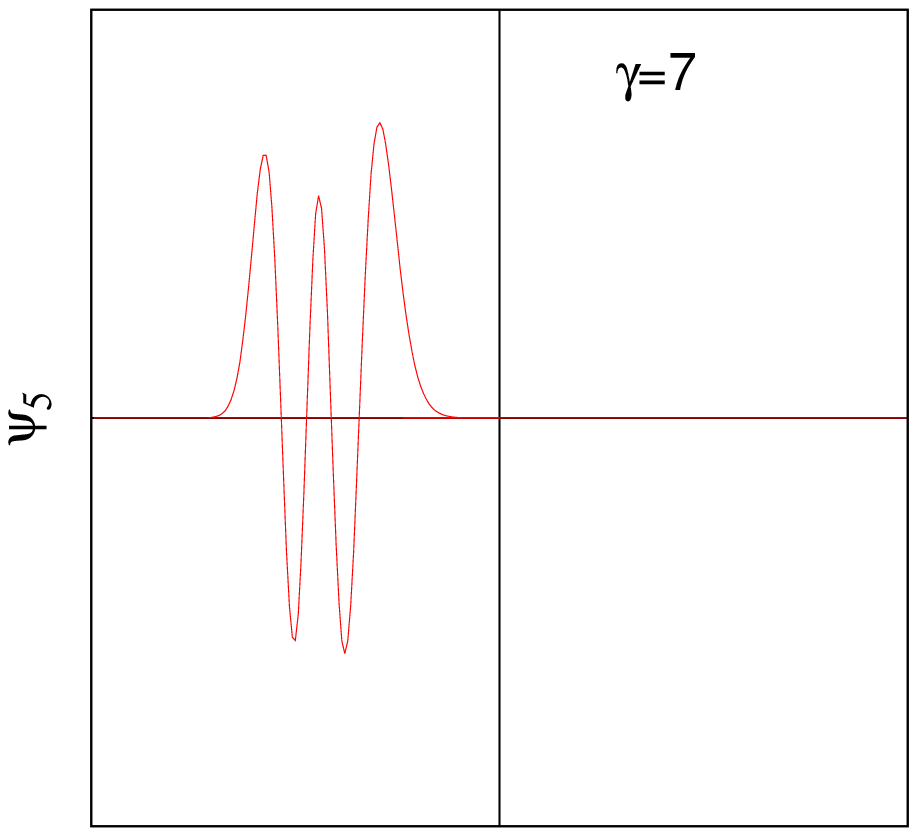}
\end{minipage}
\\[5pt]
\begin{minipage}[c]{0.15\textwidth}\centering
\includegraphics[scale=0.28]{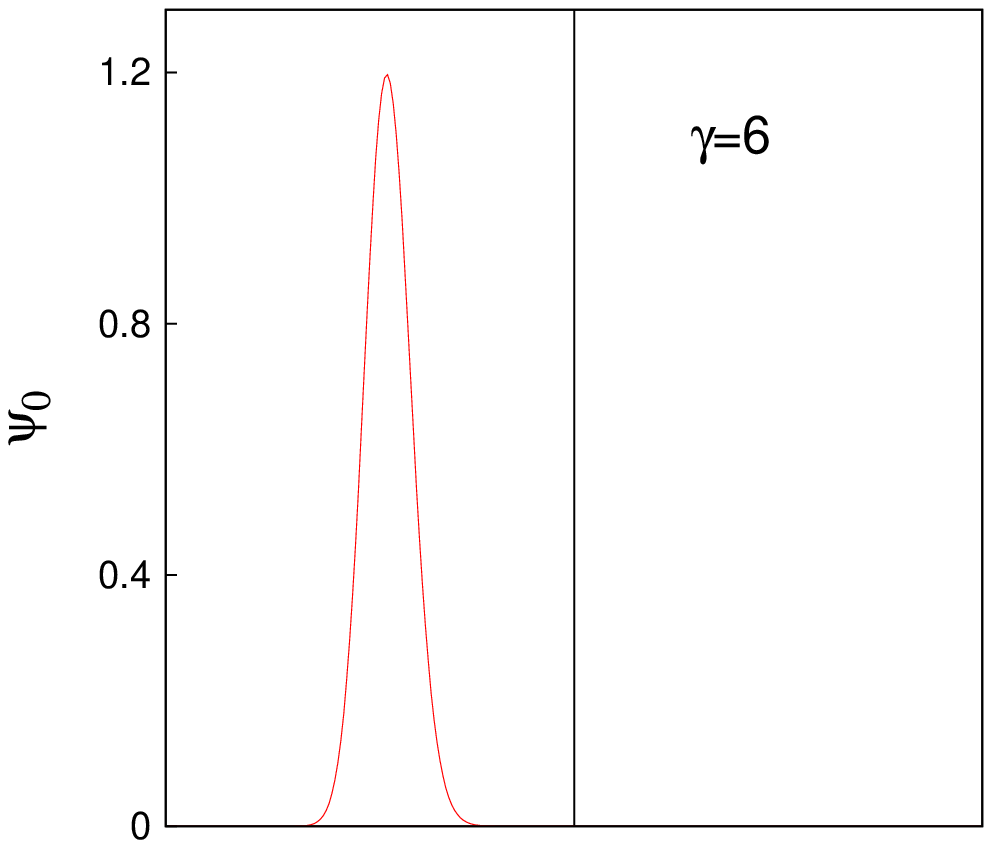}
\end{minipage}\hspace{0.07in}
\begin{minipage}[c]{0.15\textwidth}\centering
\includegraphics[scale=0.28]{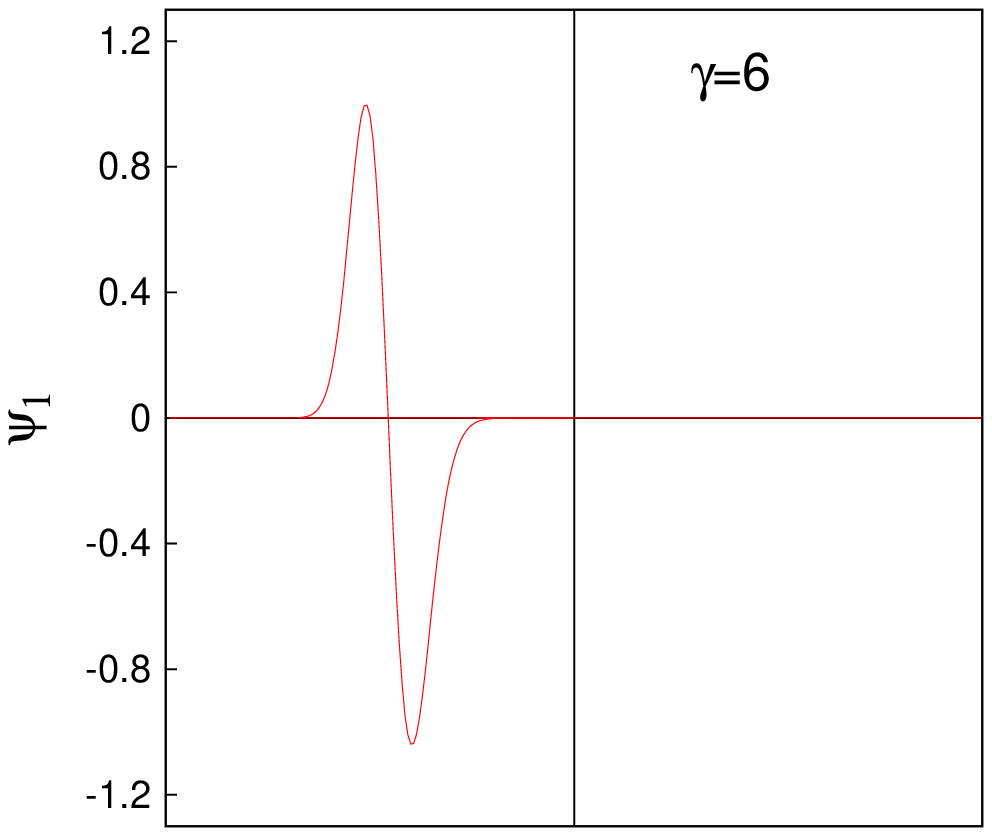}
\end{minipage}\hspace{0.07in}
\begin{minipage}[c]{0.15\textwidth}\centering
\includegraphics[scale=0.28]{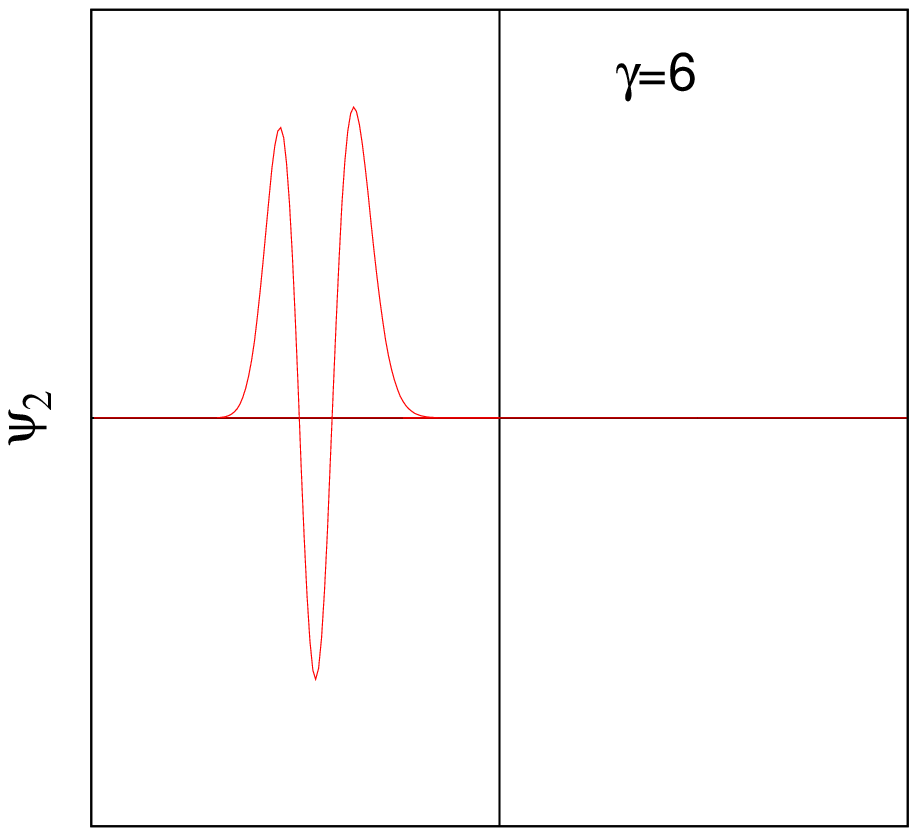}
\end{minipage}\hspace{0.07in}
\begin{minipage}[c]{0.15\textwidth}\centering
\includegraphics[scale=0.28]{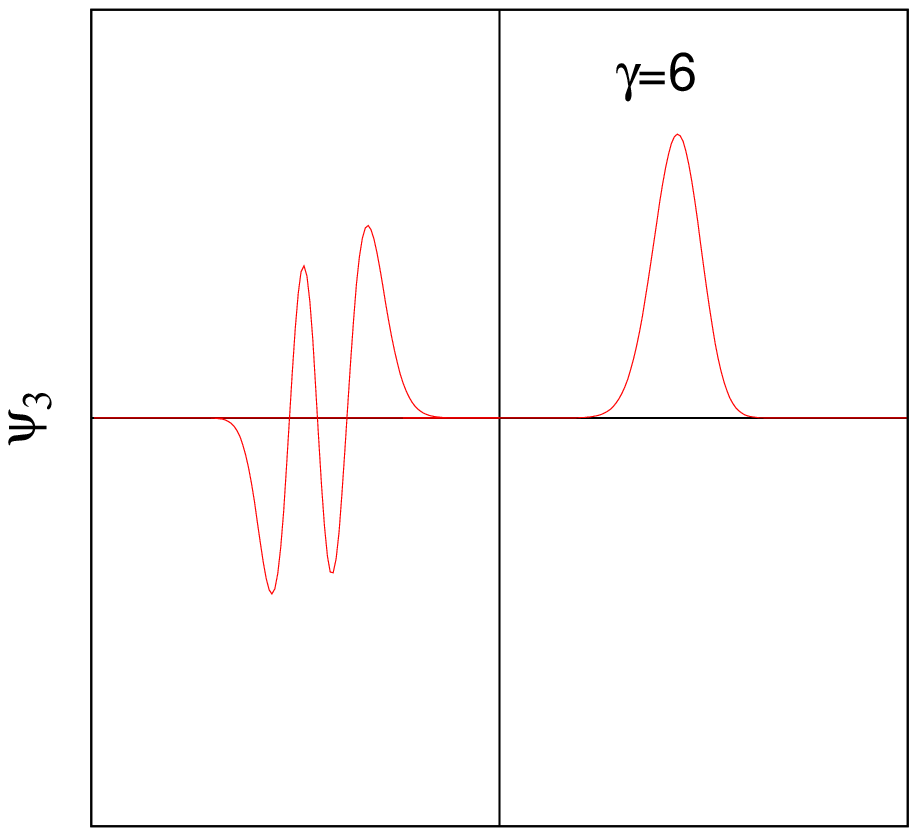}
\end{minipage}\hspace{0.07in}
\begin{minipage}[c]{0.15\textwidth}\centering
\includegraphics[scale=0.28]{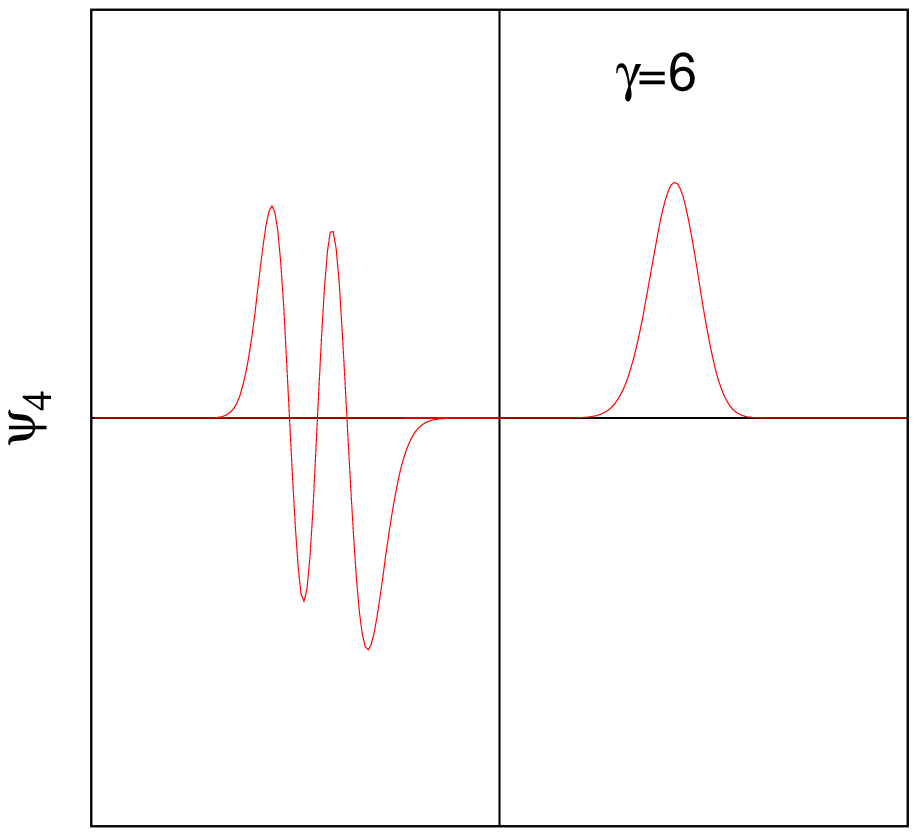}
\end{minipage}\hspace{0.07in}
\begin{minipage}[c]{0.15\textwidth}\centering
\includegraphics[scale=0.28]{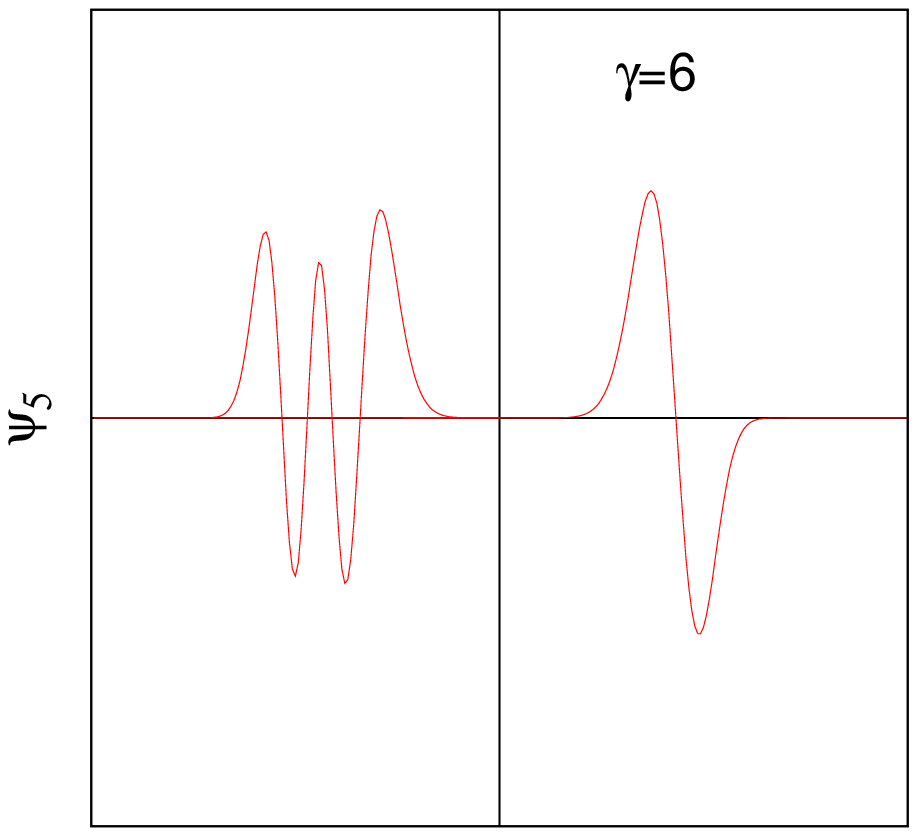}
\end{minipage}
\\[5pt]
\begin{minipage}[c]{0.15\textwidth}\centering
\includegraphics[scale=0.28]{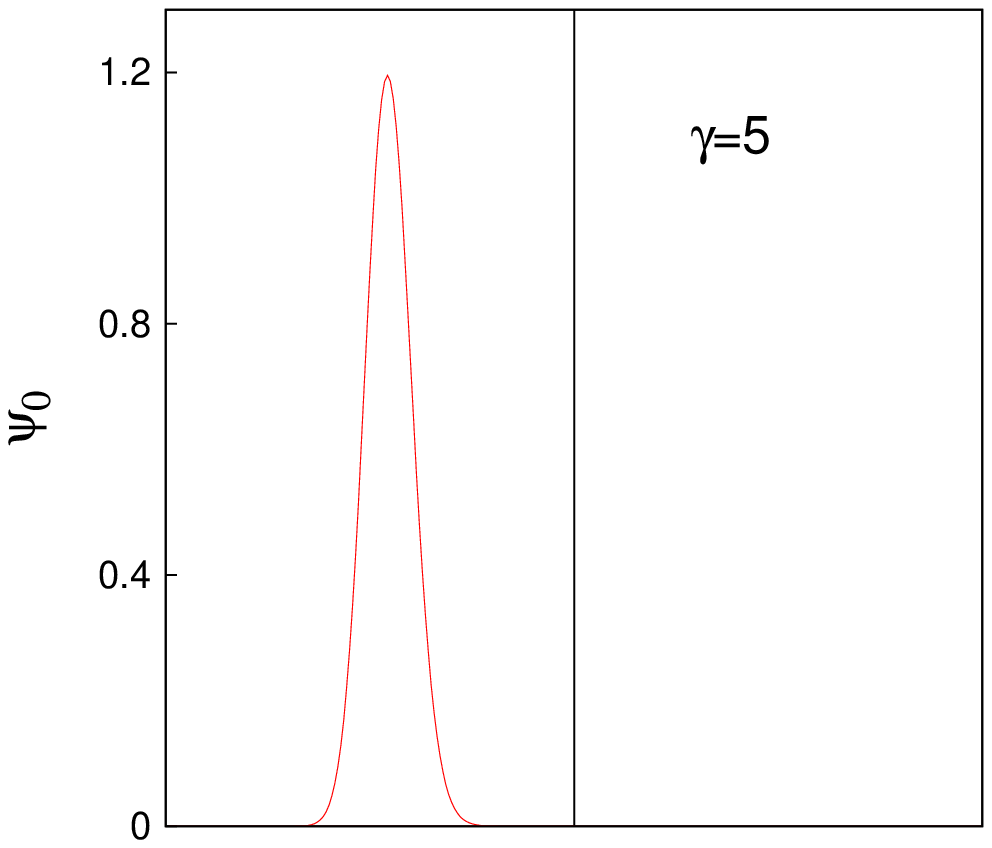}
\end{minipage}\hspace{0.07in}
\begin{minipage}[c]{0.15\textwidth}\centering
\includegraphics[scale=0.28]{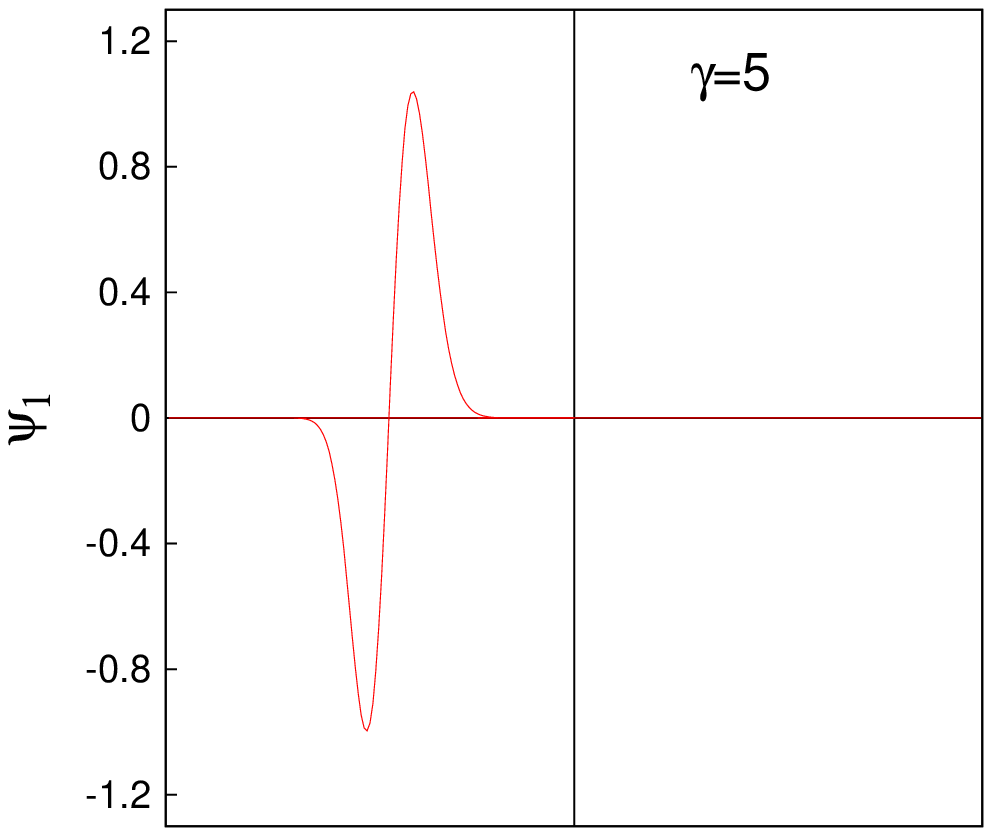}
\end{minipage}\hspace{0.07in}
\begin{minipage}[c]{0.15\textwidth}\centering
\includegraphics[scale=0.28]{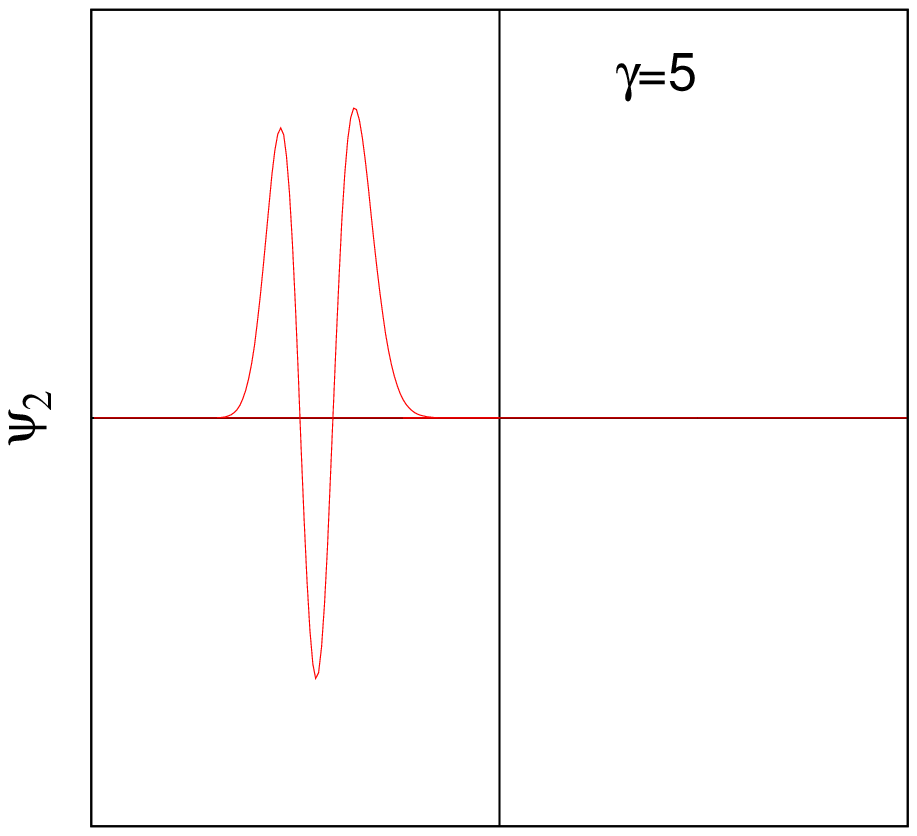}
\end{minipage}\hspace{0.07in}
\begin{minipage}[c]{0.15\textwidth}\centering
\includegraphics[scale=0.28]{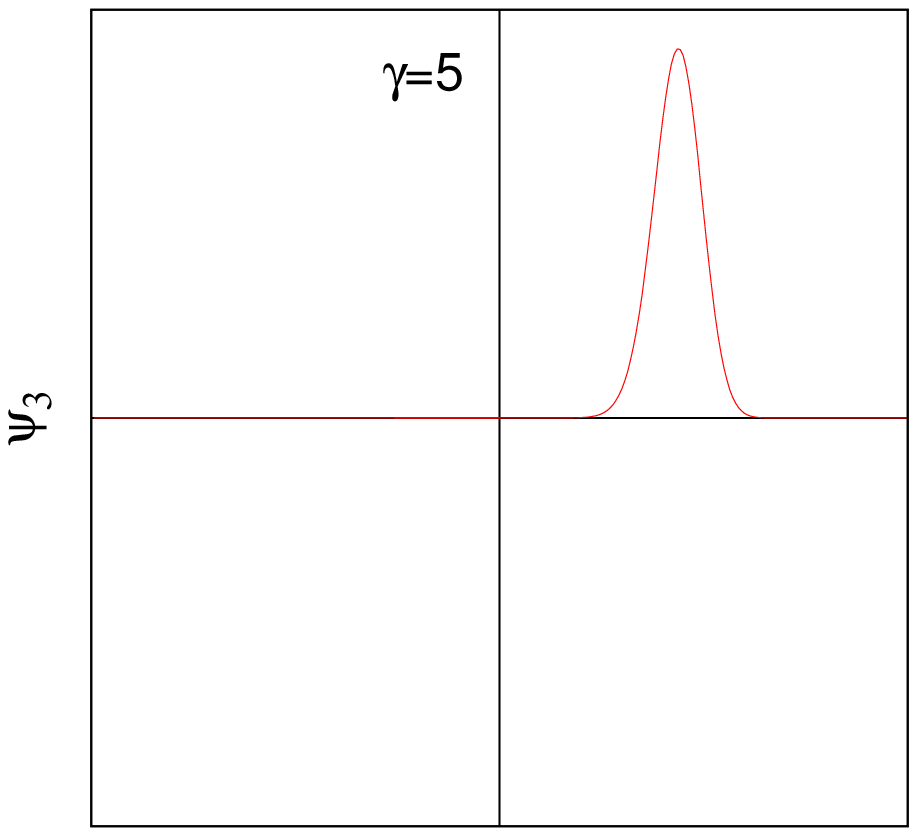}
\end{minipage}\hspace{0.07in}
\begin{minipage}[c]{0.15\textwidth}\centering
\includegraphics[scale=0.28]{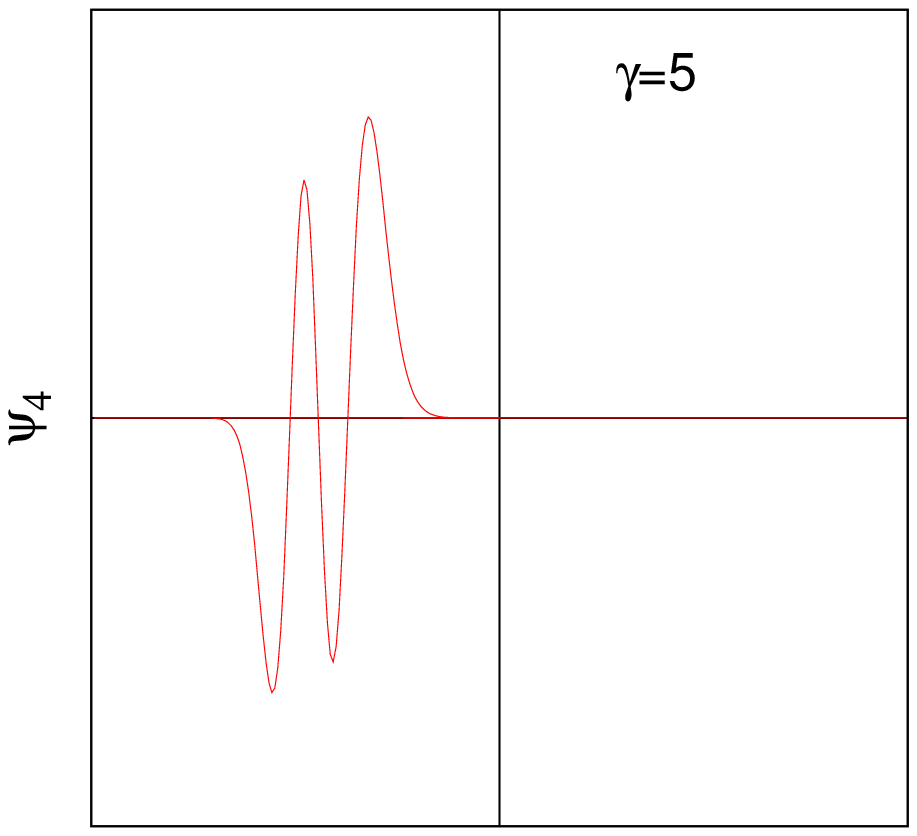}
\end{minipage}\hspace{0.07in}
\begin{minipage}[c]{0.15\textwidth}\centering
\includegraphics[scale=0.28]{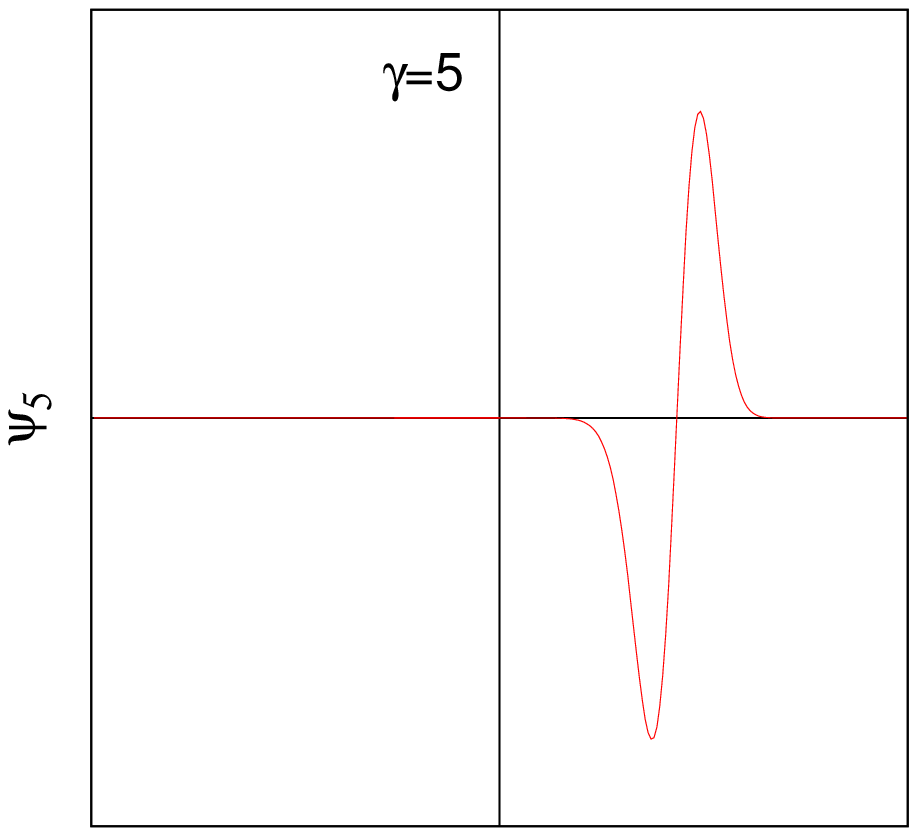}
\end{minipage}
\\[5pt]
\begin{minipage}[c]{0.15\textwidth}\centering
\includegraphics[scale=0.28]{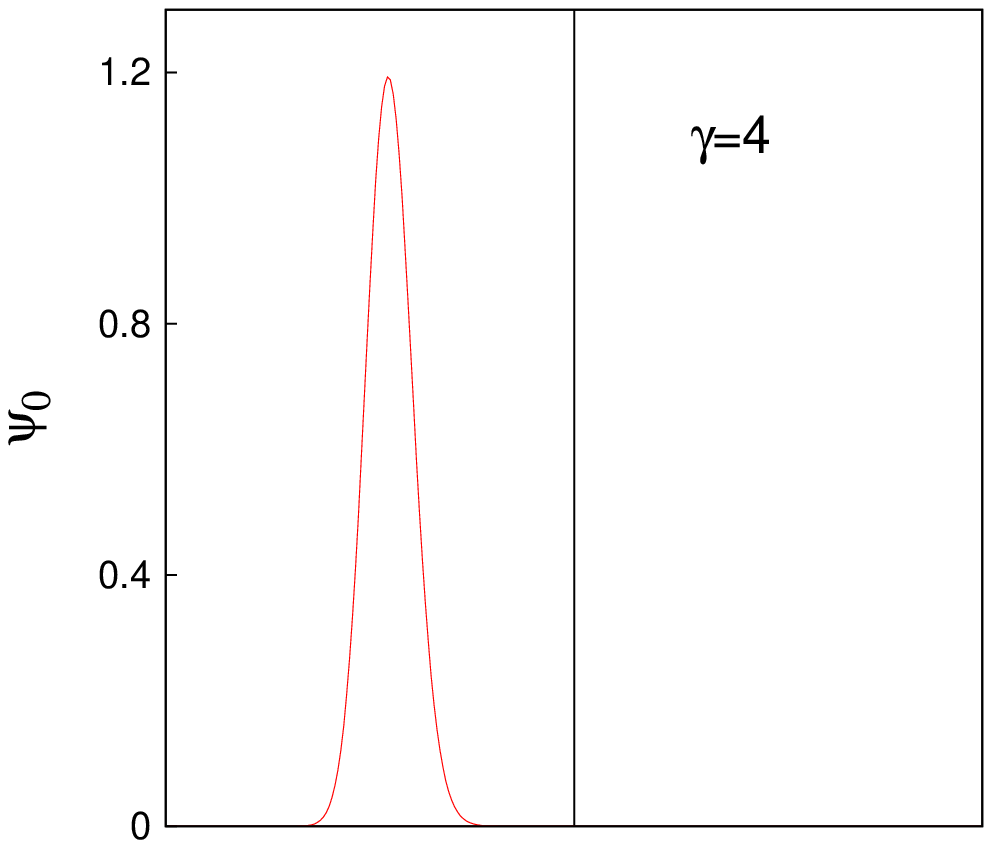}
\end{minipage}\hspace{0.07in}
\begin{minipage}[c]{0.15\textwidth}\centering
\includegraphics[scale=0.28]{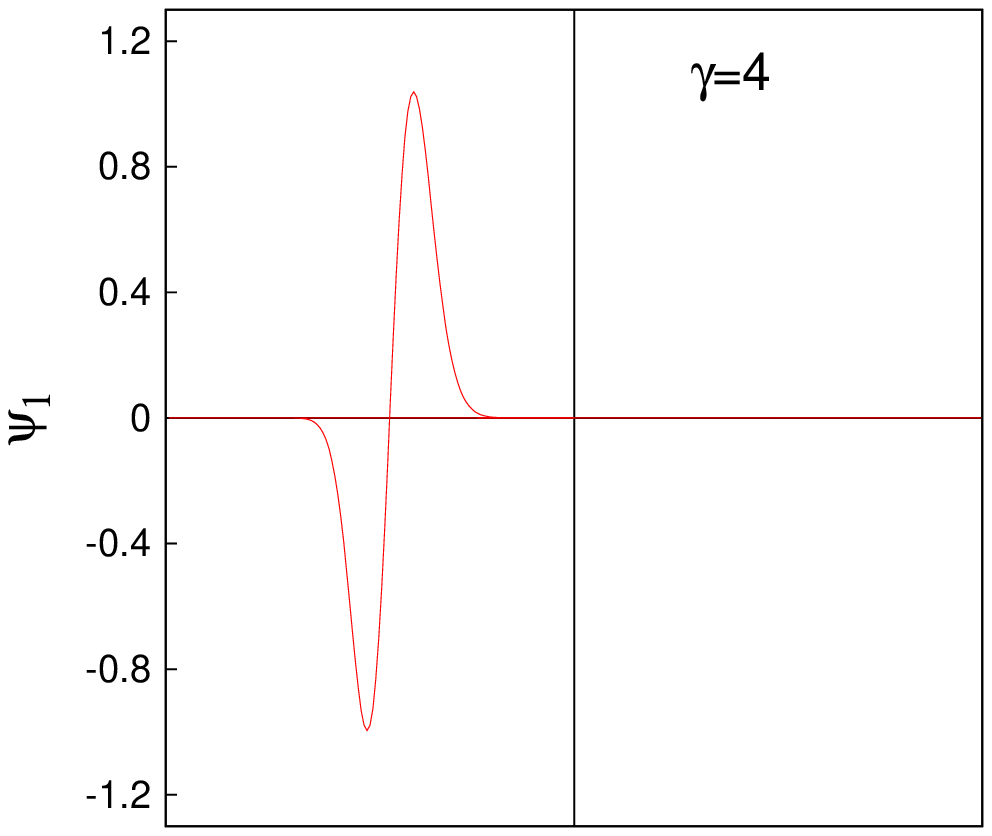}
\end{minipage}\hspace{0.07in}
\begin{minipage}[c]{0.15\textwidth}\centering
\includegraphics[scale=0.28]{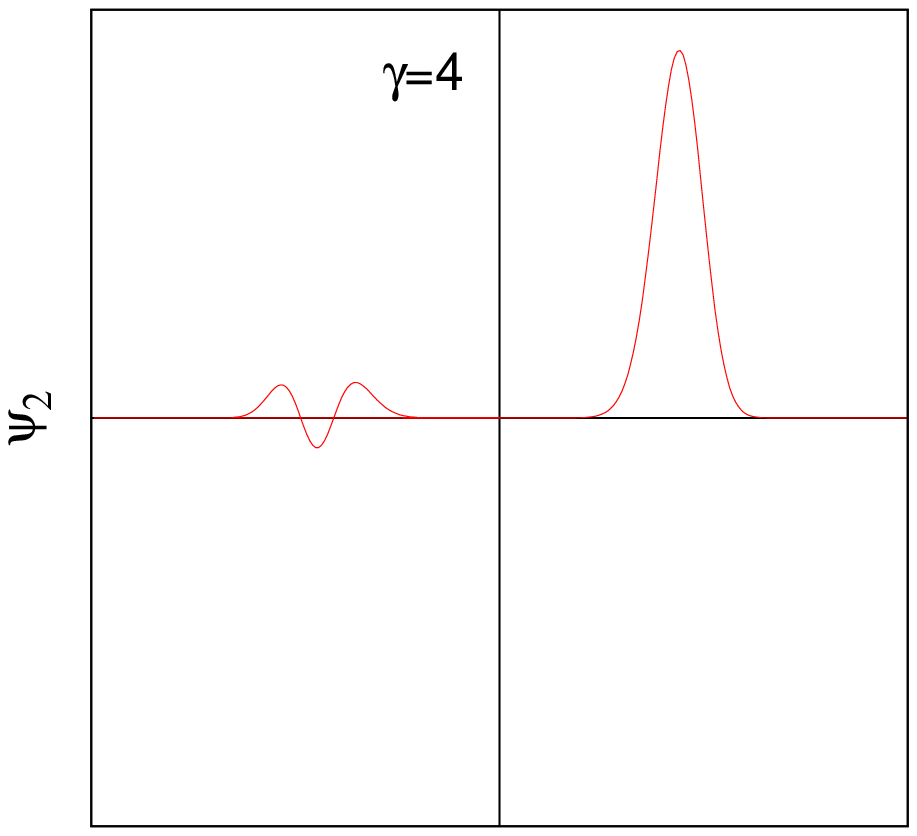}
\end{minipage}\hspace{0.07in}
\begin{minipage}[c]{0.15\textwidth}\centering
\includegraphics[scale=0.28]{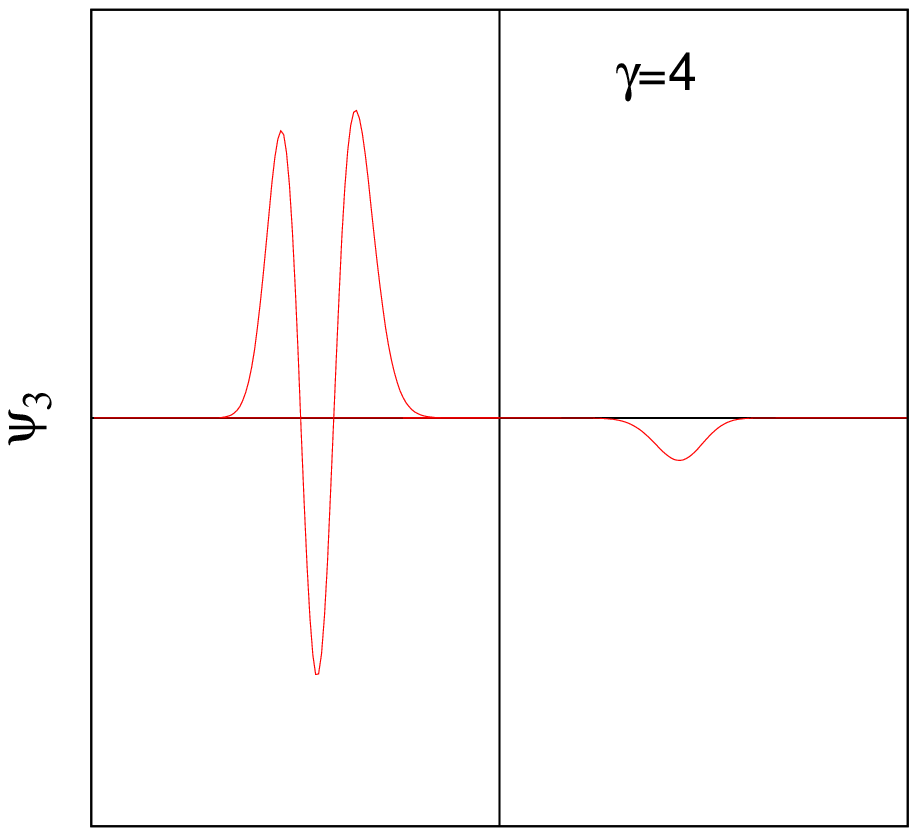}
\end{minipage}\hspace{0.07in}
\begin{minipage}[c]{0.15\textwidth}\centering
\includegraphics[scale=0.28]{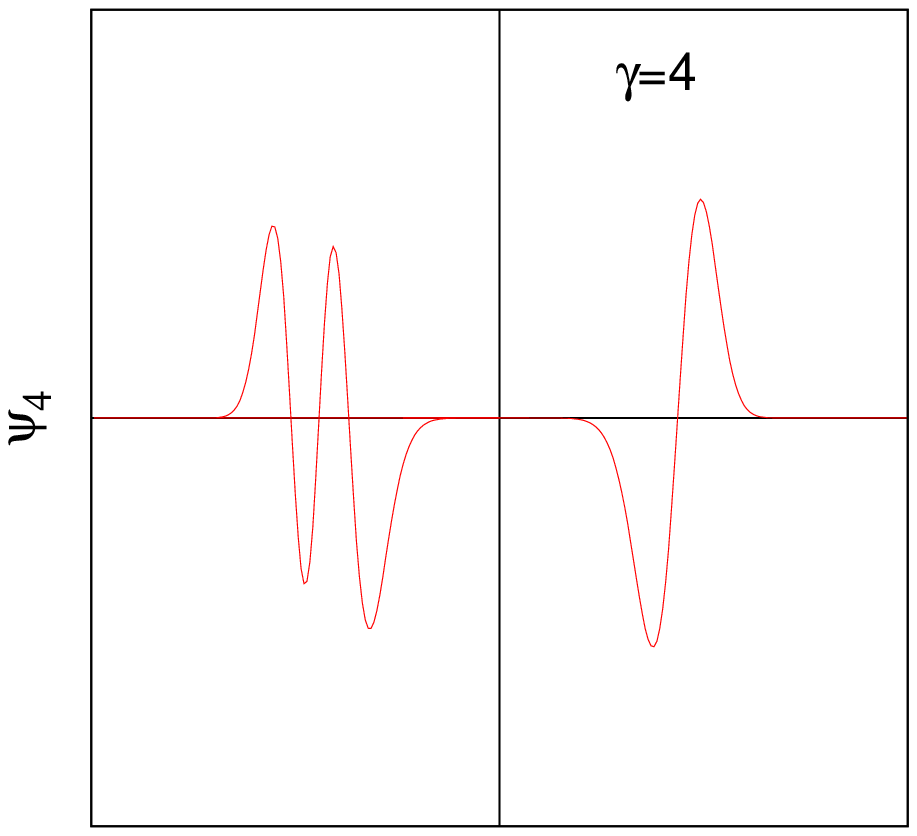}
\end{minipage}\hspace{0.07in}
\begin{minipage}[c]{0.15\textwidth}\centering
\includegraphics[scale=0.28]{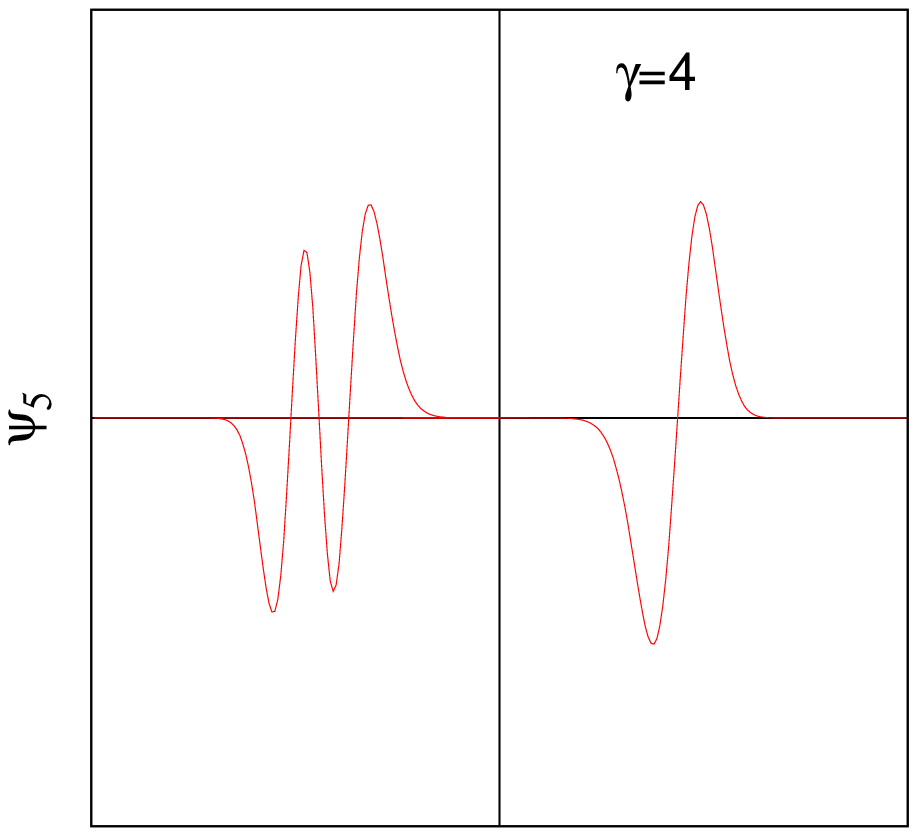}
\end{minipage}
\\[5pt]
\begin{minipage}[c]{0.15\textwidth}\centering
\includegraphics[scale=0.28]{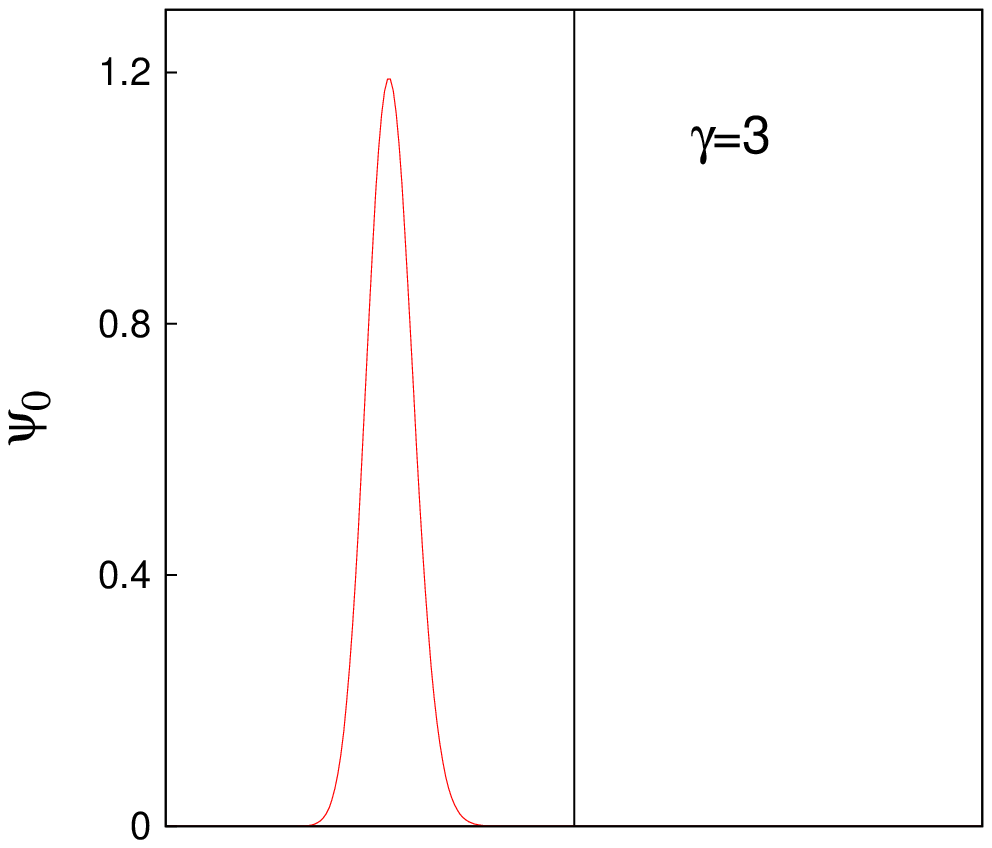}
\end{minipage}\hspace{0.07in}
\begin{minipage}[c]{0.15\textwidth}\centering
\includegraphics[scale=0.28]{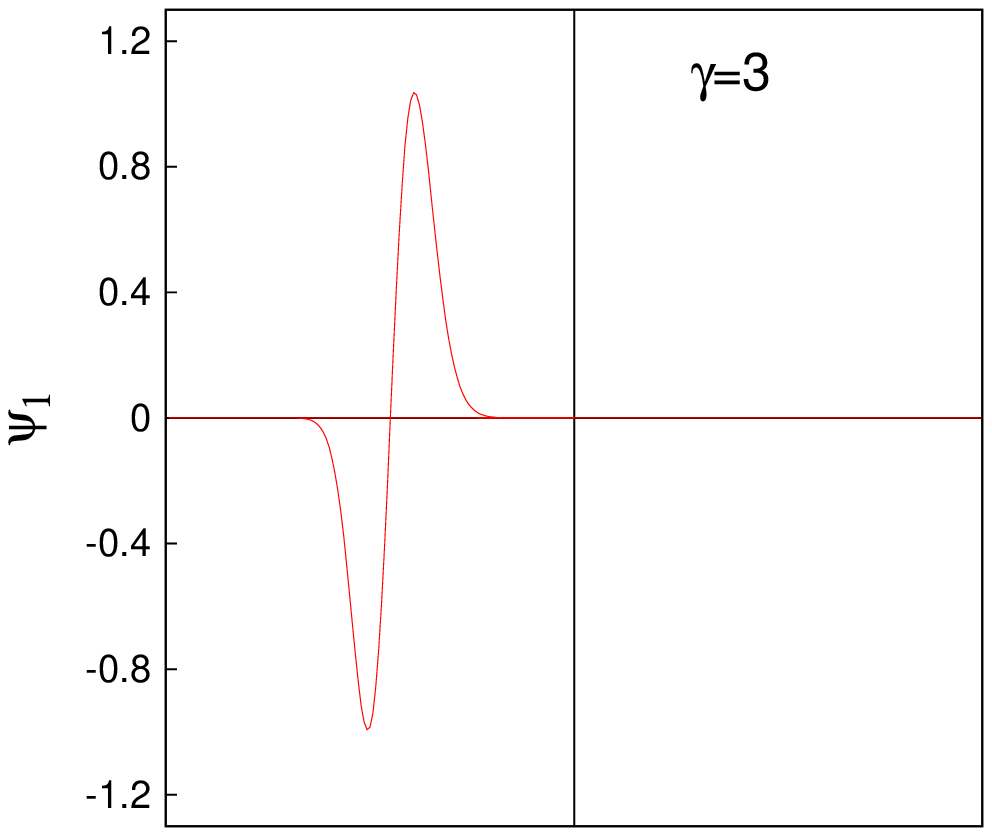}
\end{minipage}\hspace{0.07in}
\begin{minipage}[c]{0.15\textwidth}\centering
\includegraphics[scale=0.28]{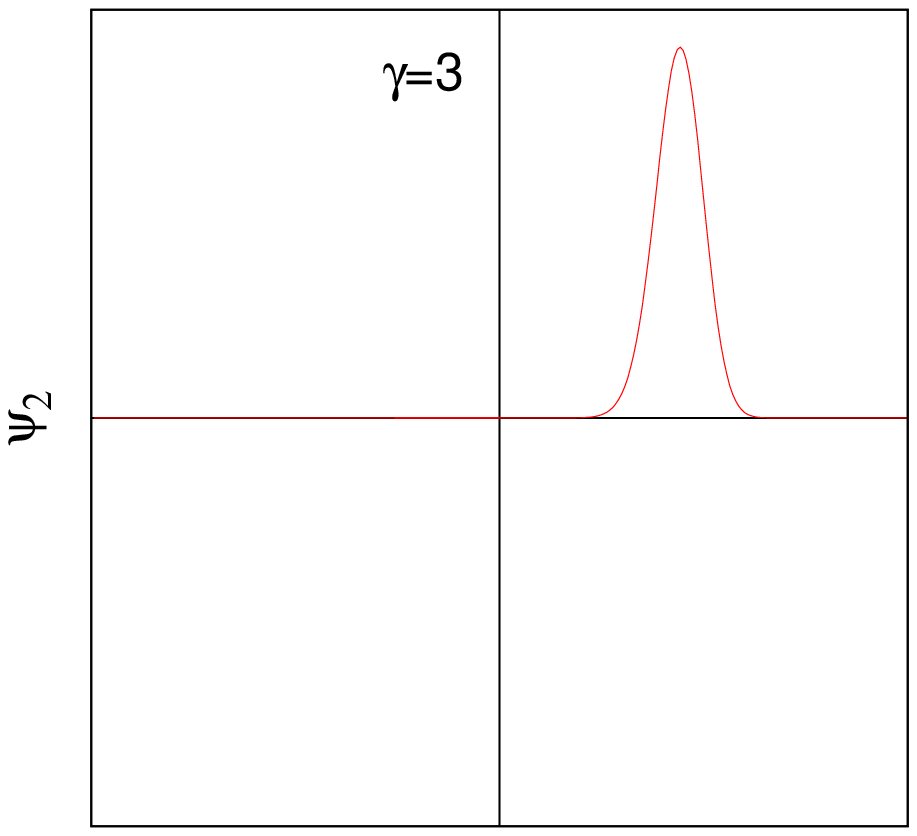}
\end{minipage}\hspace{0.07in}
\begin{minipage}[c]{0.15\textwidth}\centering
\includegraphics[scale=0.28]{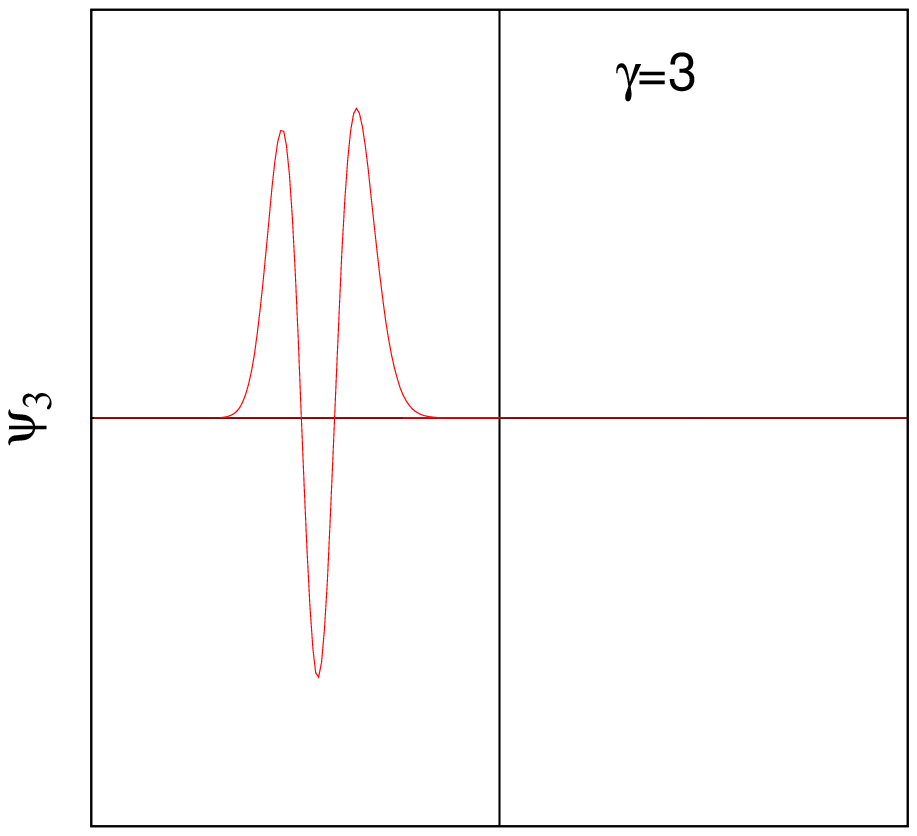}
\end{minipage}\hspace{0.07in}
\begin{minipage}[c]{0.15\textwidth}\centering
\includegraphics[scale=0.28]{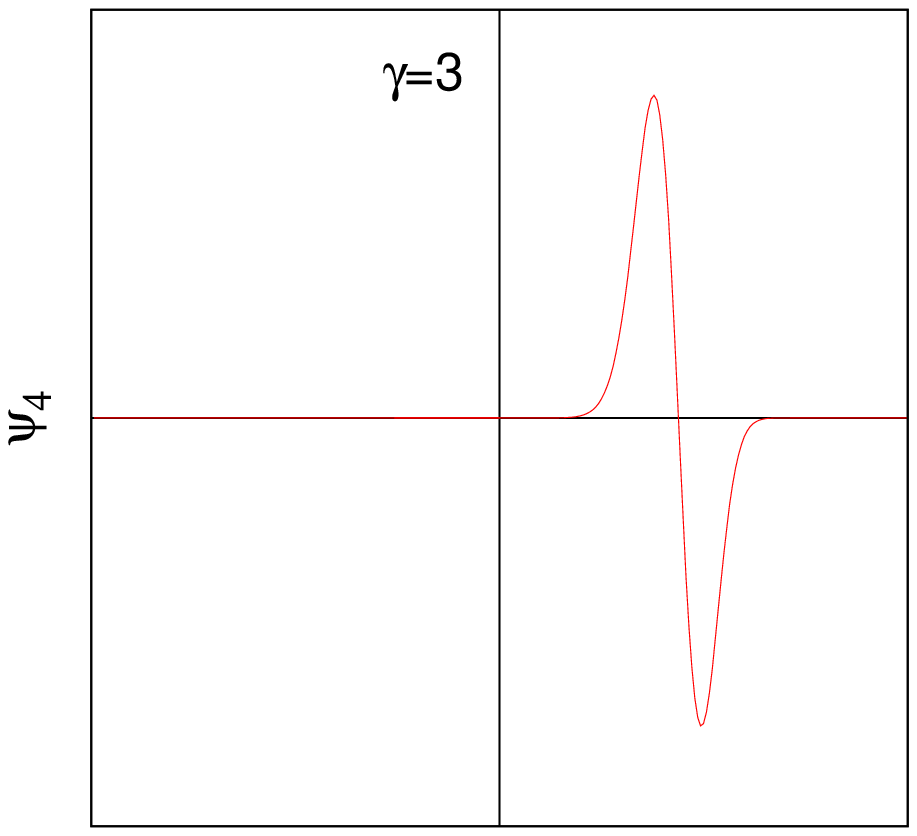}
\end{minipage}\hspace{0.07in}
\begin{minipage}[c]{0.15\textwidth}\centering
\includegraphics[scale=0.28]{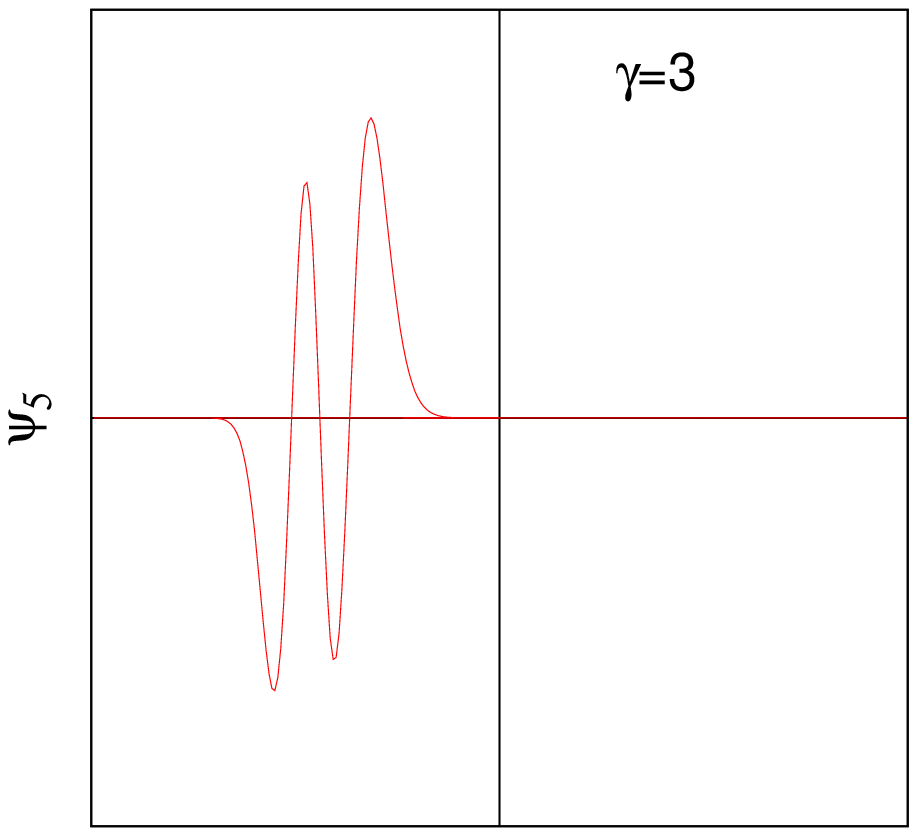}
\end{minipage}
\\[5pt]
\begin{minipage}[c]{0.15\textwidth}\centering
\includegraphics[scale=0.28]{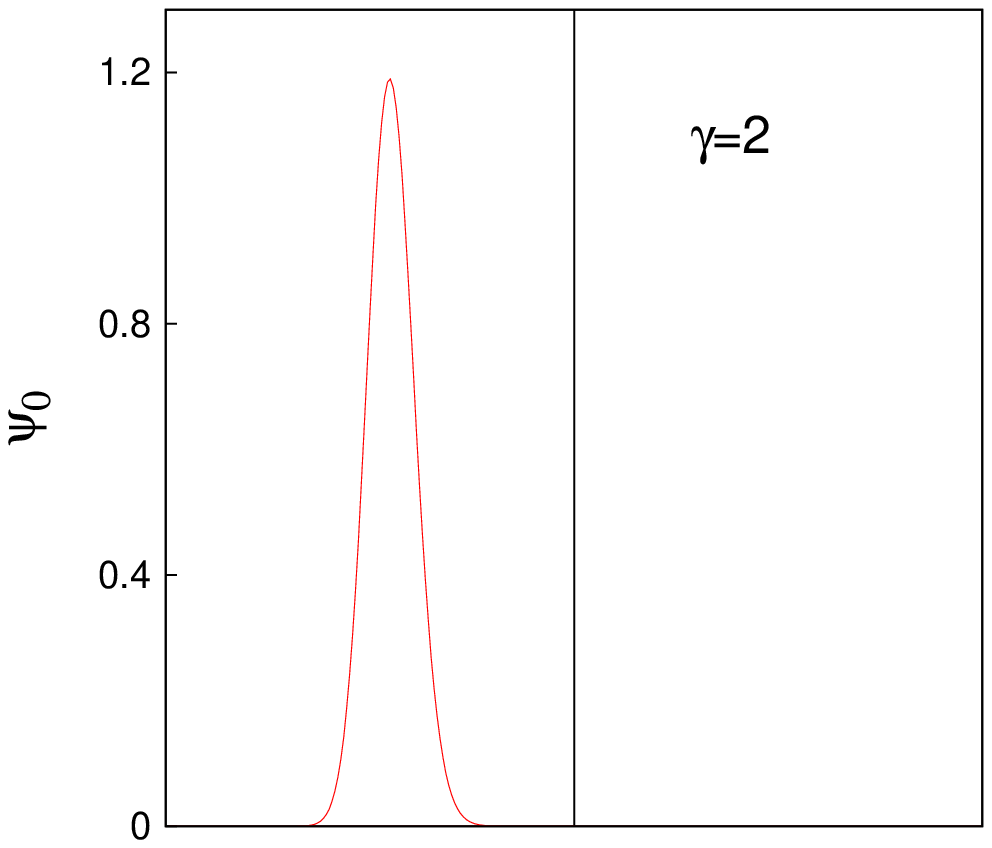}
\end{minipage}\hspace{0.07in}
\begin{minipage}[c]{0.15\textwidth}\centering
\includegraphics[scale=0.28]{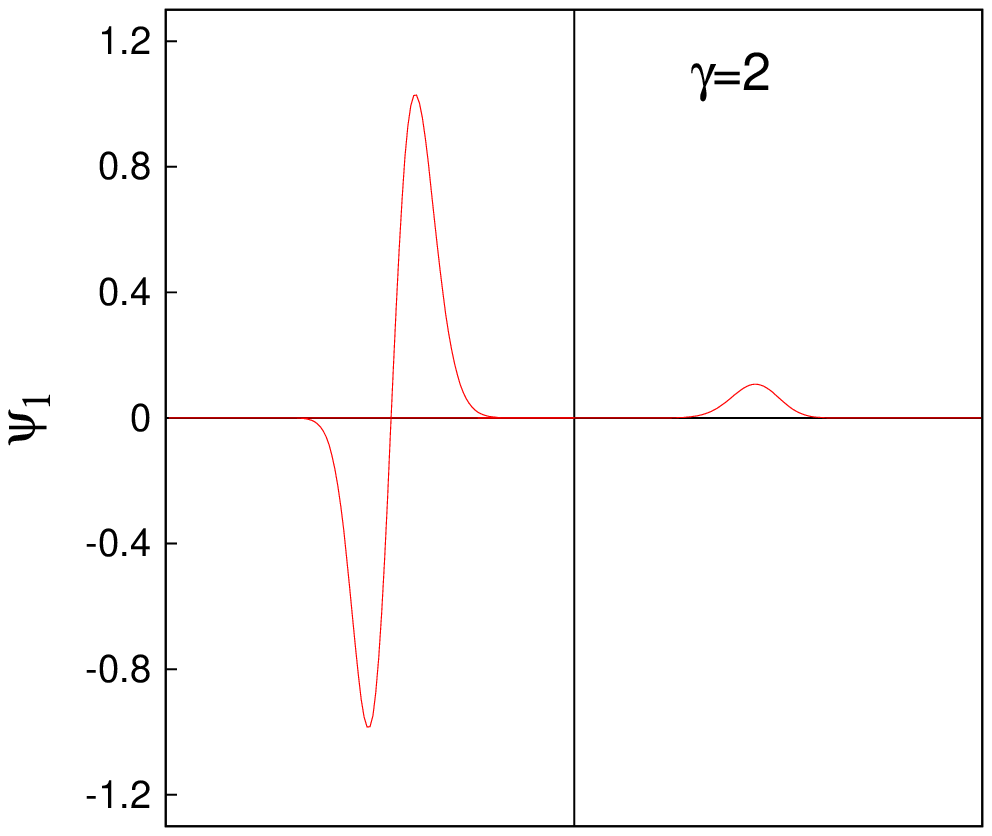}
\end{minipage}\hspace{0.07in}
\begin{minipage}[c]{0.15\textwidth}\centering
\includegraphics[scale=0.28]{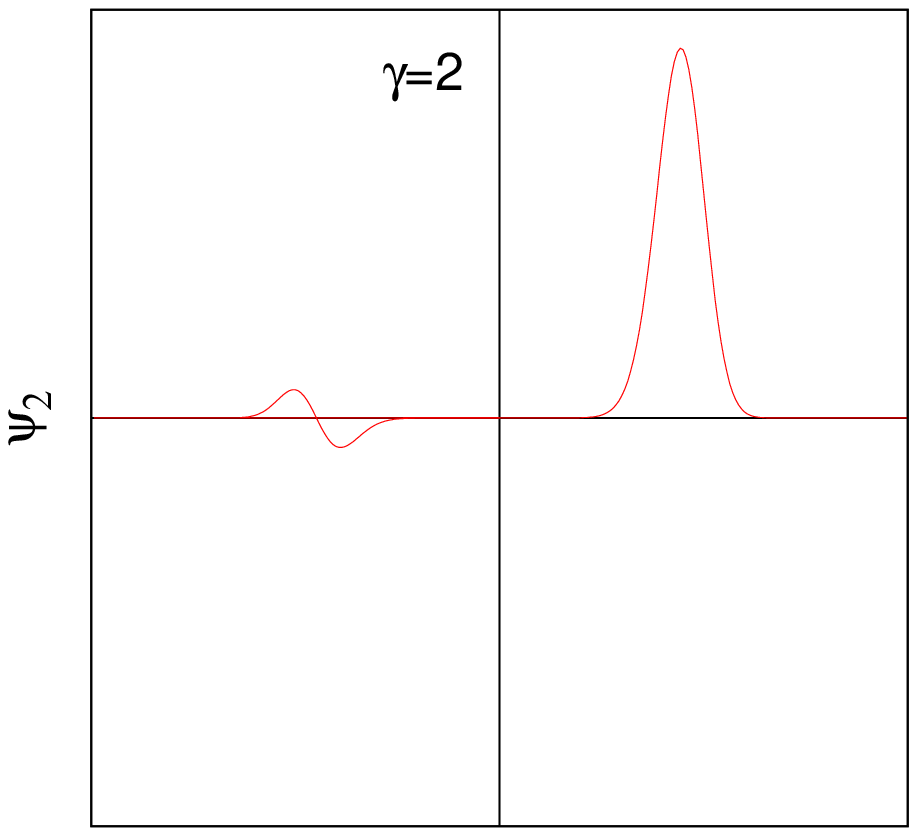}
\end{minipage}\hspace{0.07in}
\begin{minipage}[c]{0.15\textwidth}\centering
\includegraphics[scale=0.28]{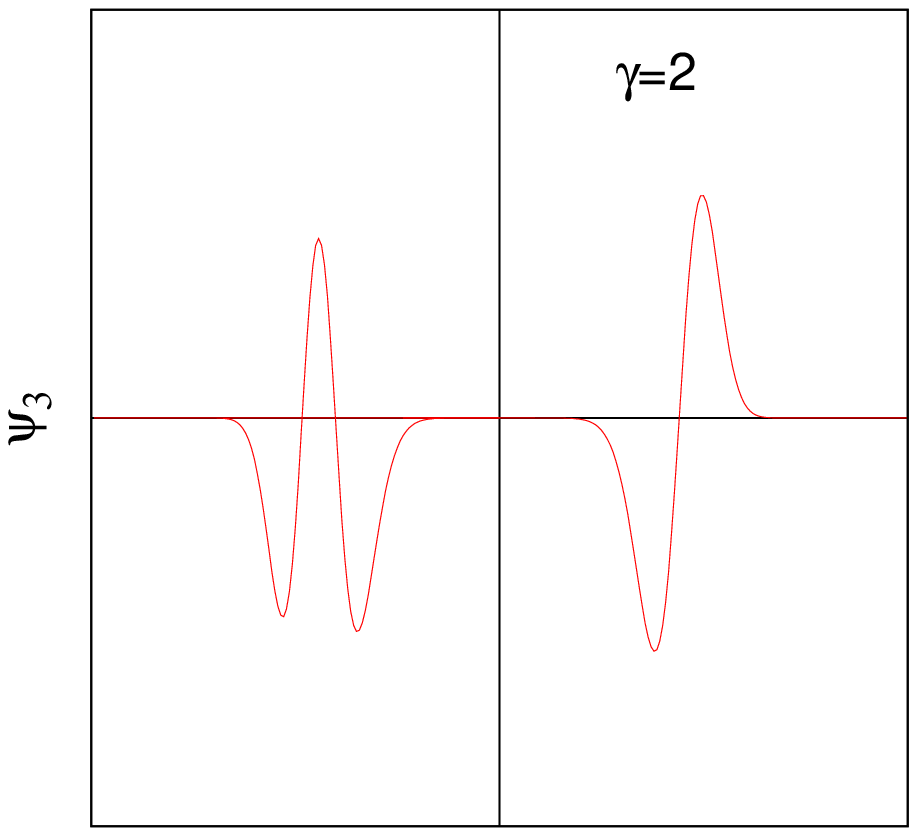}
\end{minipage}\hspace{0.07in}
\begin{minipage}[c]{0.15\textwidth}\centering
\includegraphics[scale=0.28]{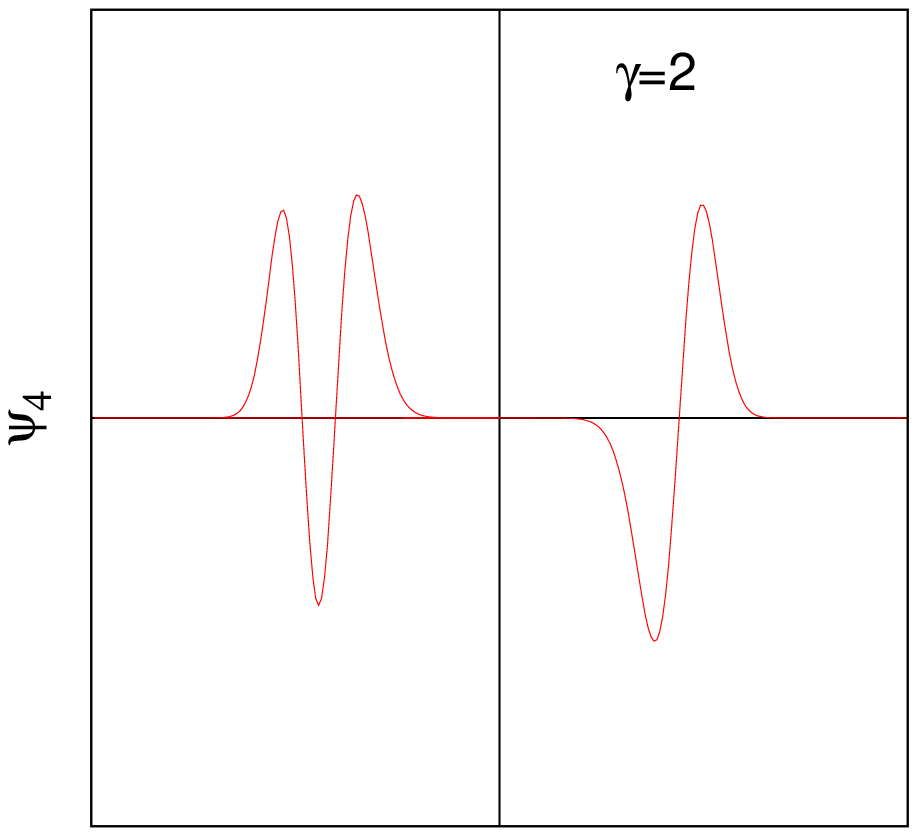}
\end{minipage}\hspace{0.07in}
\begin{minipage}[c]{0.15\textwidth}\centering
\includegraphics[scale=0.28]{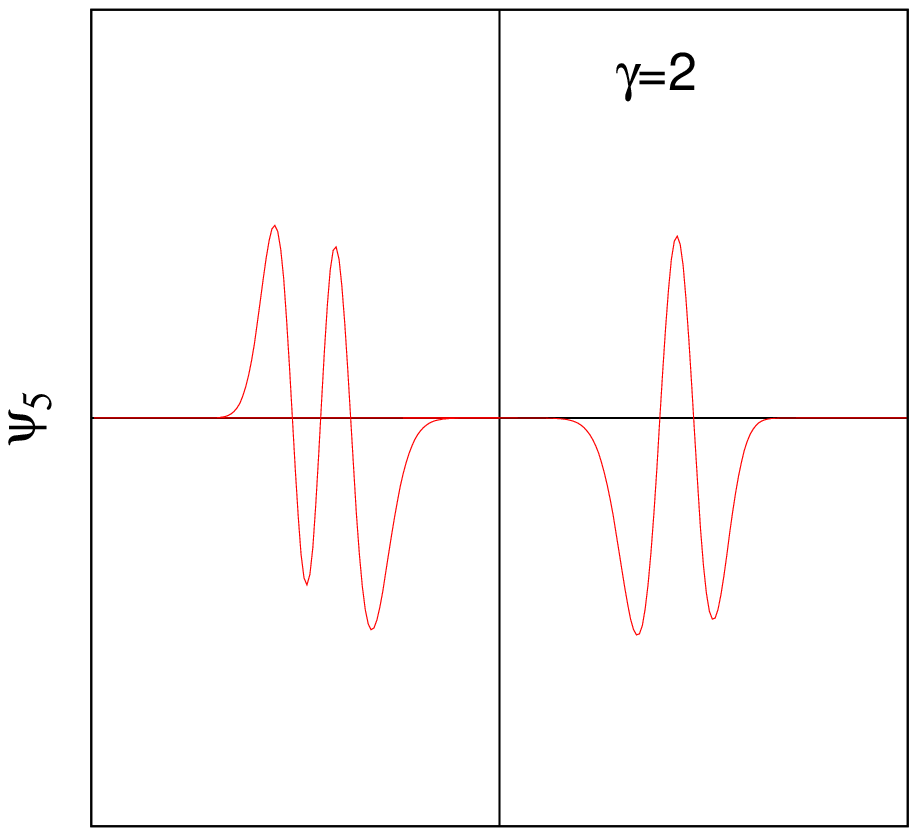}
\end{minipage}
\\[5pt]
\begin{minipage}[c]{0.15\textwidth}\centering
\includegraphics[scale=0.28]{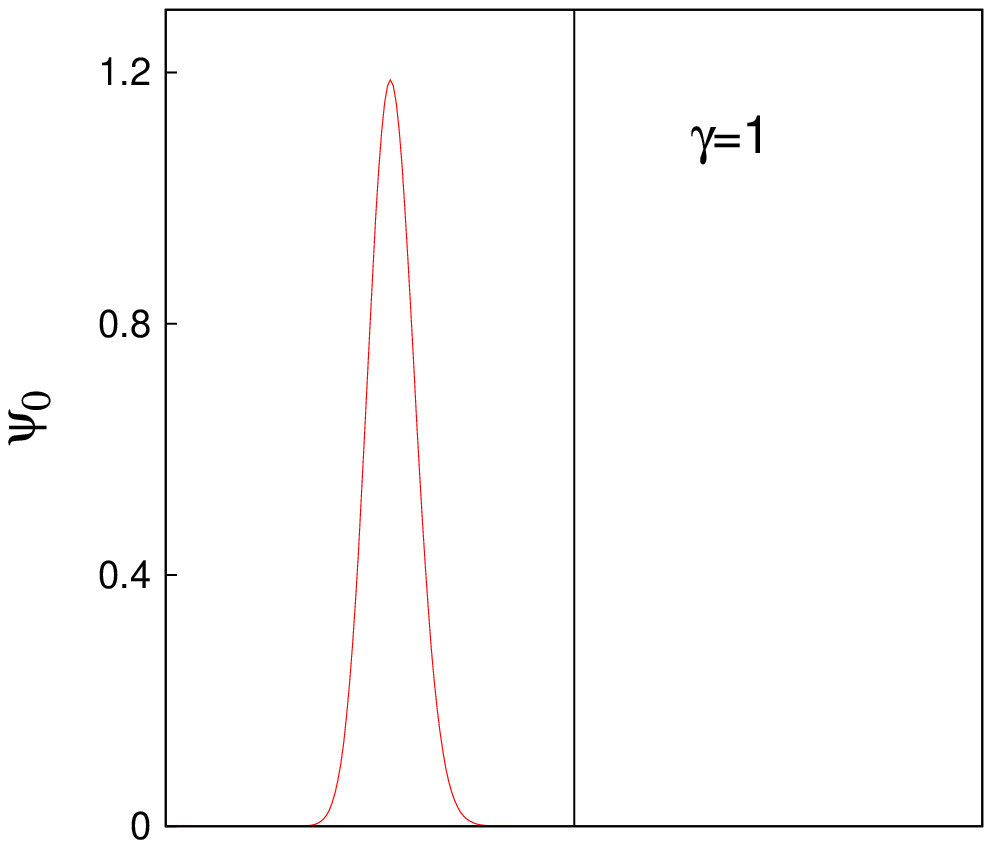}
\end{minipage}\hspace{0.07in}
\begin{minipage}[c]{0.15\textwidth}\centering
\includegraphics[scale=0.28]{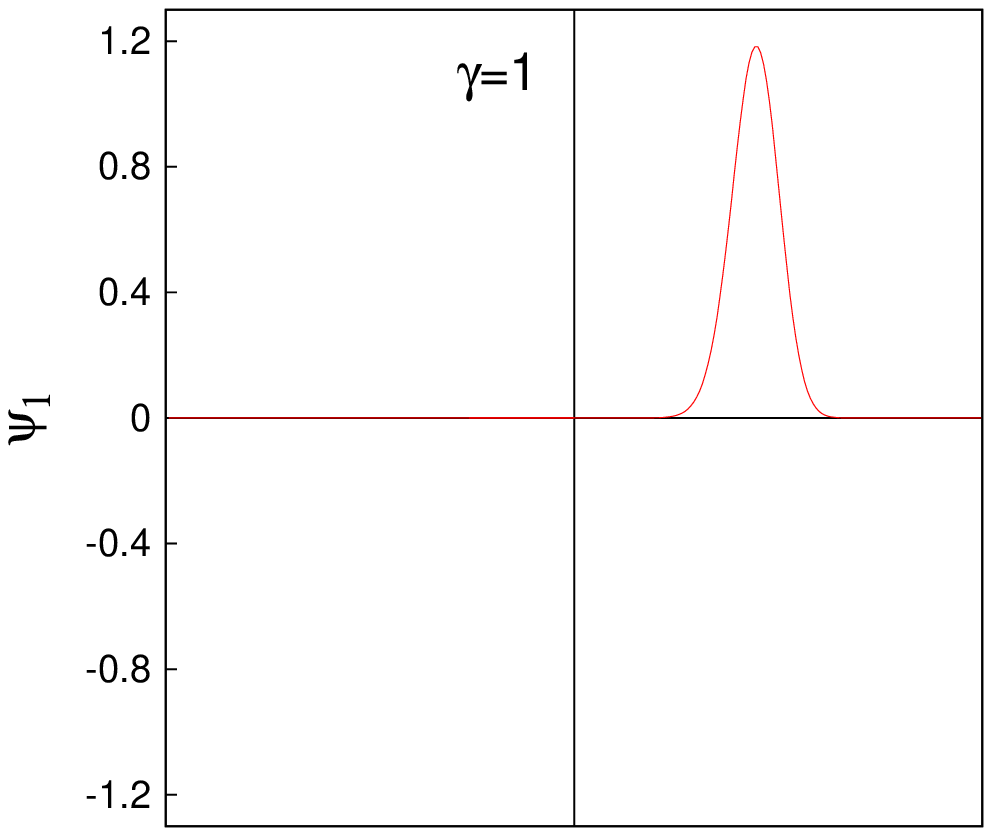}
\end{minipage}\hspace{0.07in}
\begin{minipage}[c]{0.15\textwidth}\centering
\includegraphics[scale=0.28]{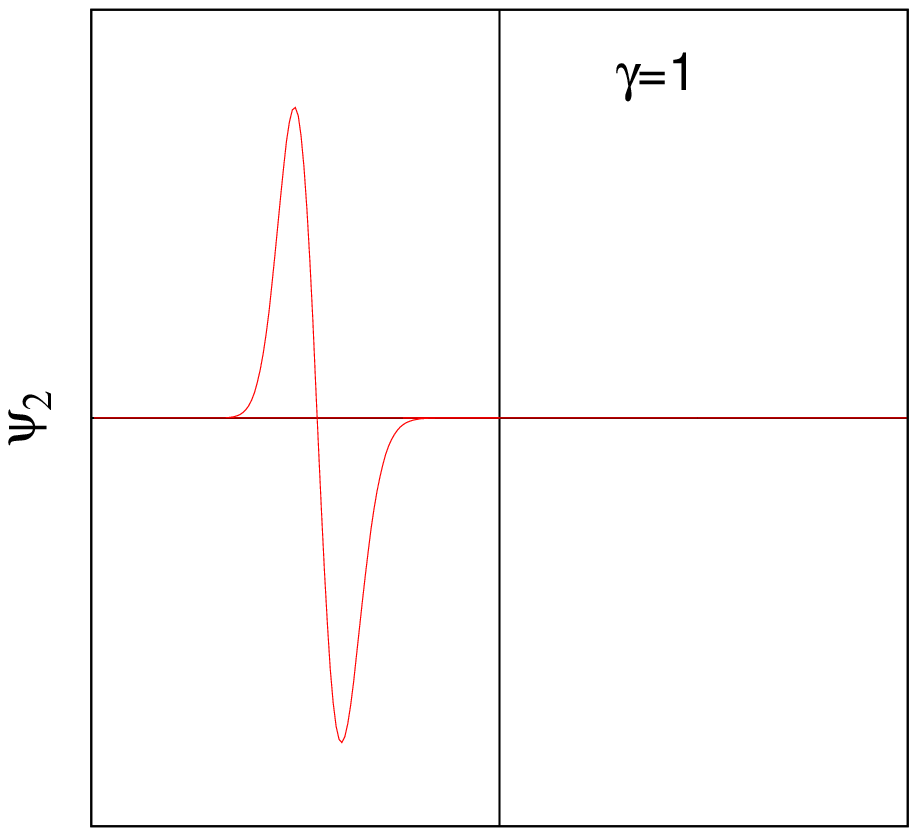}
\end{minipage}\hspace{0.07in}
\begin{minipage}[c]{0.15\textwidth}\centering
\includegraphics[scale=0.28]{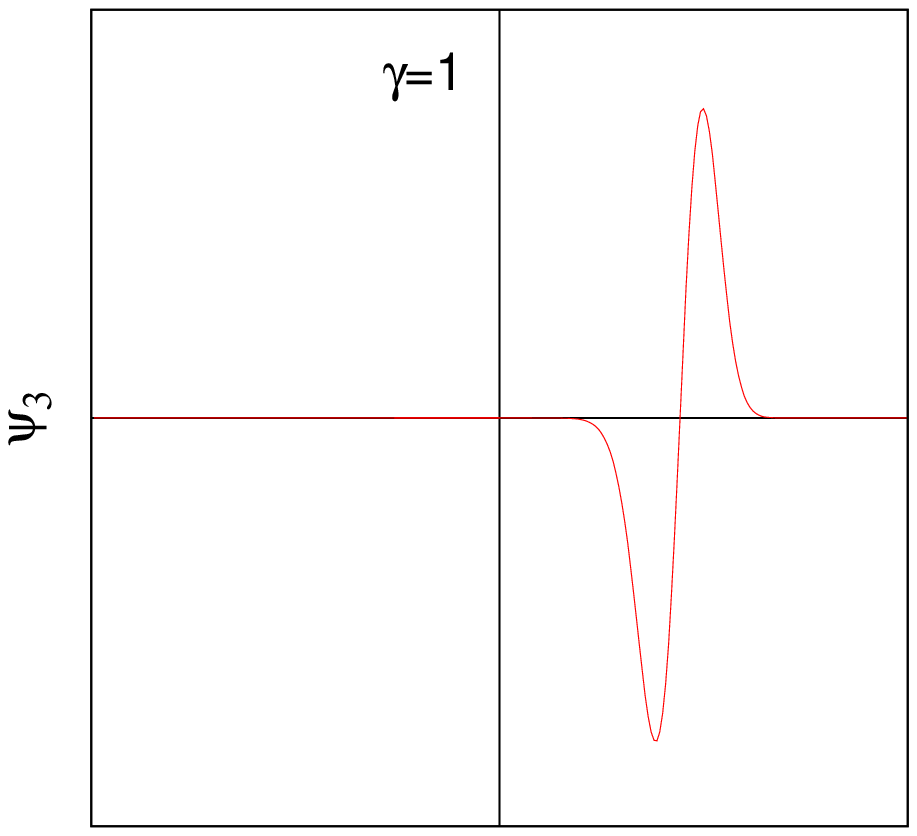}
\end{minipage}\hspace{0.07in}
\begin{minipage}[c]{0.15\textwidth}\centering
\includegraphics[scale=0.28]{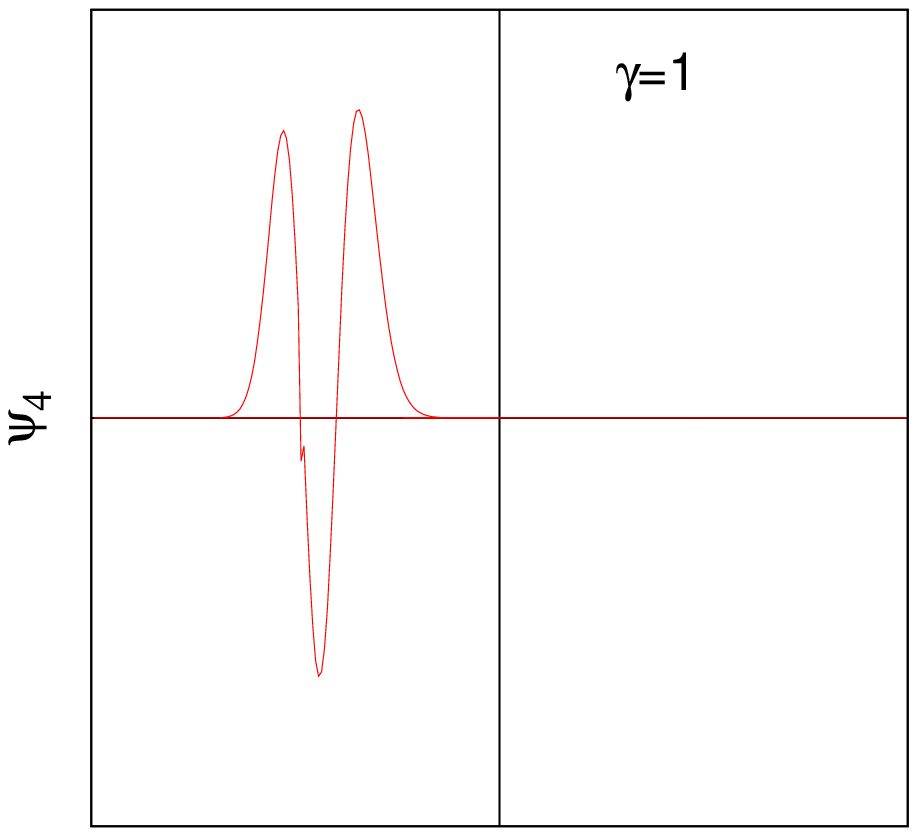}
\end{minipage}\hspace{0.07in}
\begin{minipage}[c]{0.15\textwidth}\centering
\includegraphics[scale=0.28]{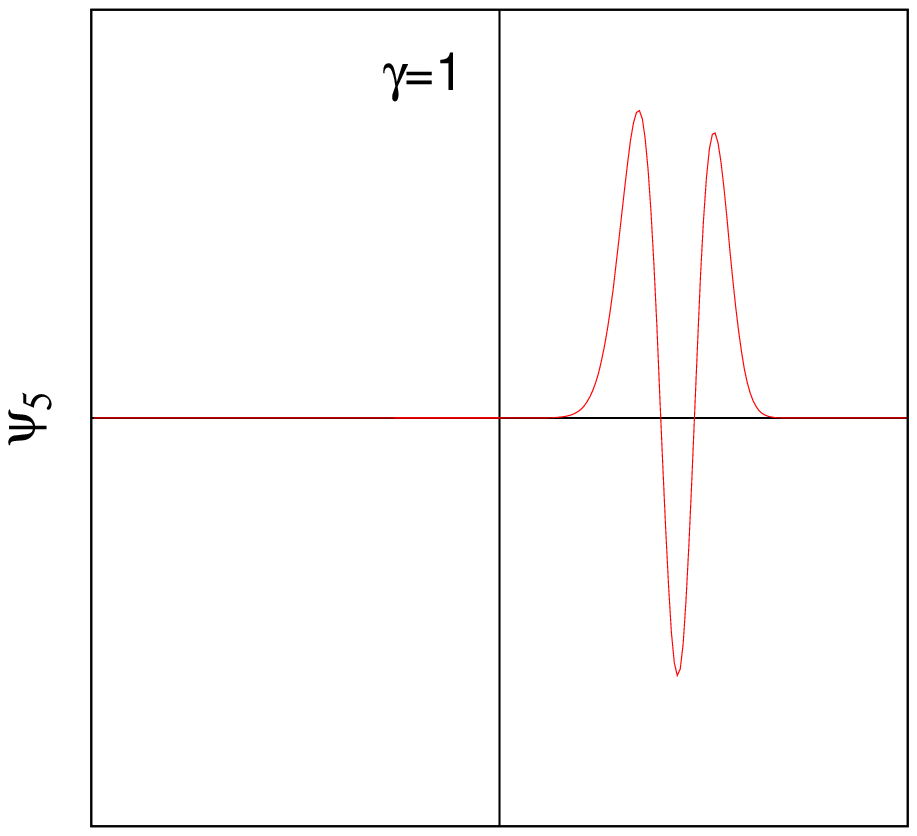}
\end{minipage}
\\[5pt]
\begin{minipage}[c]{0.15\textwidth}\centering
\includegraphics[scale=0.29]{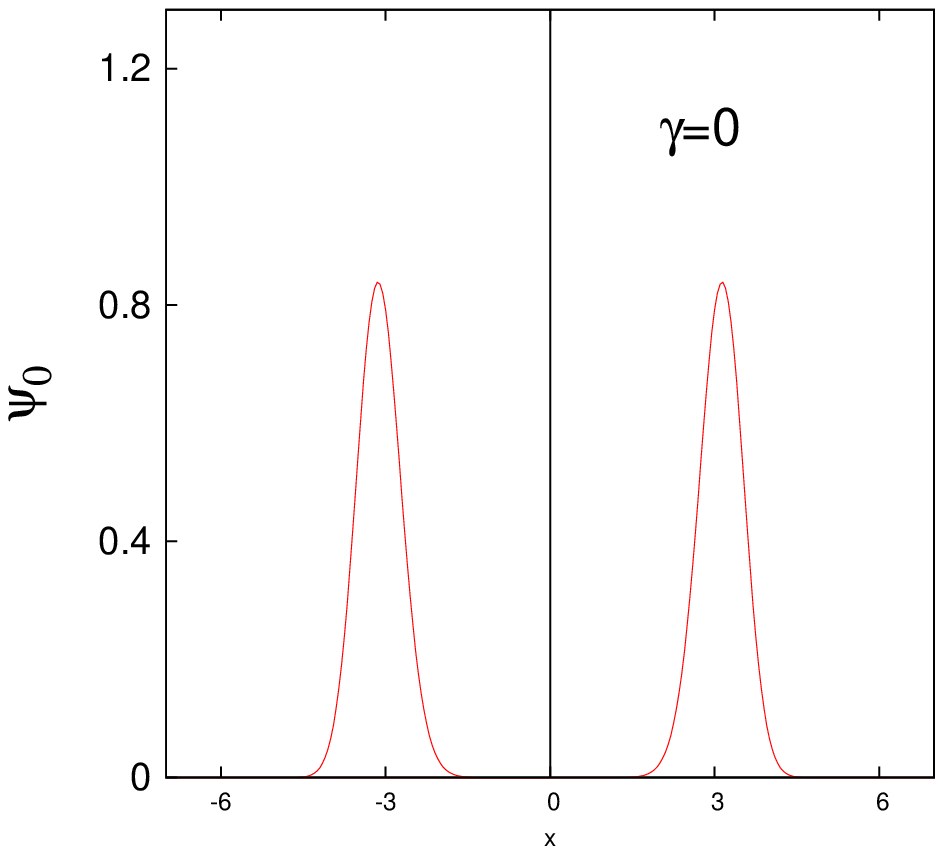}
\end{minipage}\hspace{0.06in}
\begin{minipage}[c]{0.15\textwidth}\centering
\includegraphics[scale=0.29]{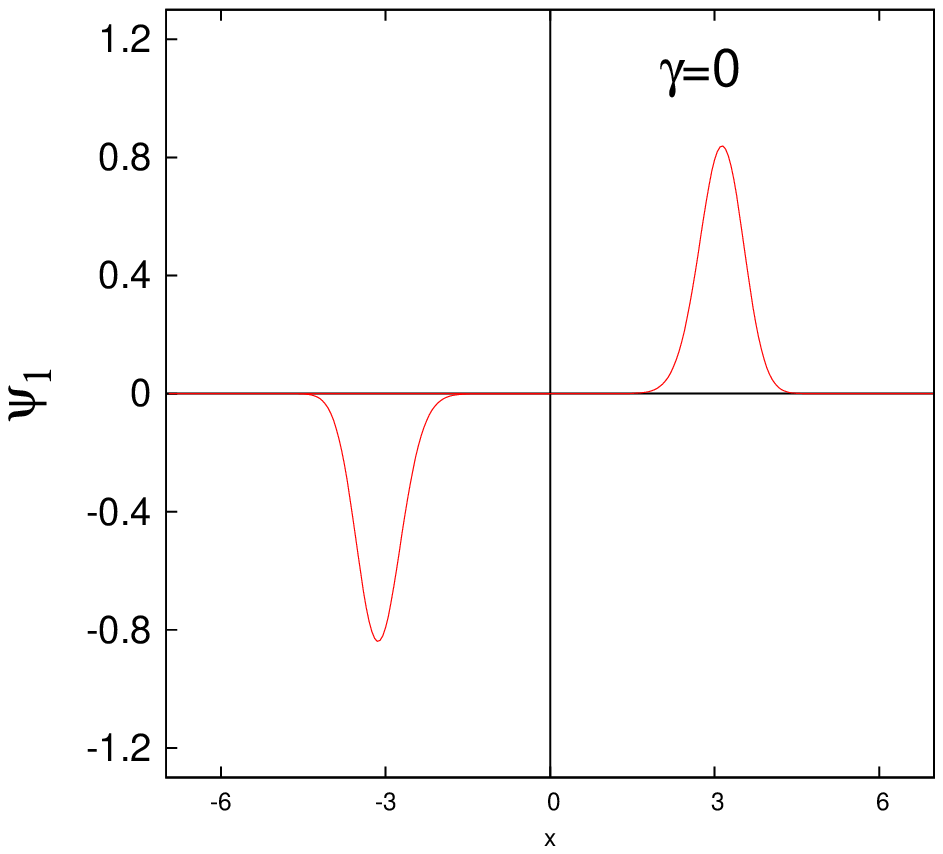}
\end{minipage}\hspace{0.06in}
\begin{minipage}[c]{0.15\textwidth}\centering
\includegraphics[scale=0.29]{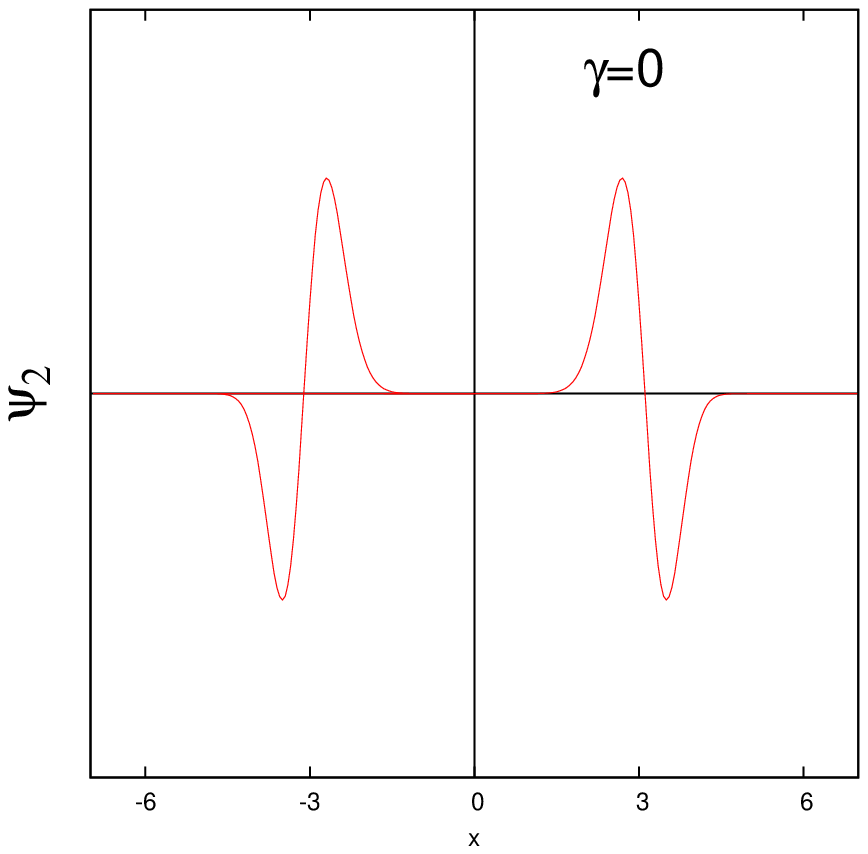}
\end{minipage}\hspace{0.06in}
\begin{minipage}[c]{0.15\textwidth}\centering
\includegraphics[scale=0.29]{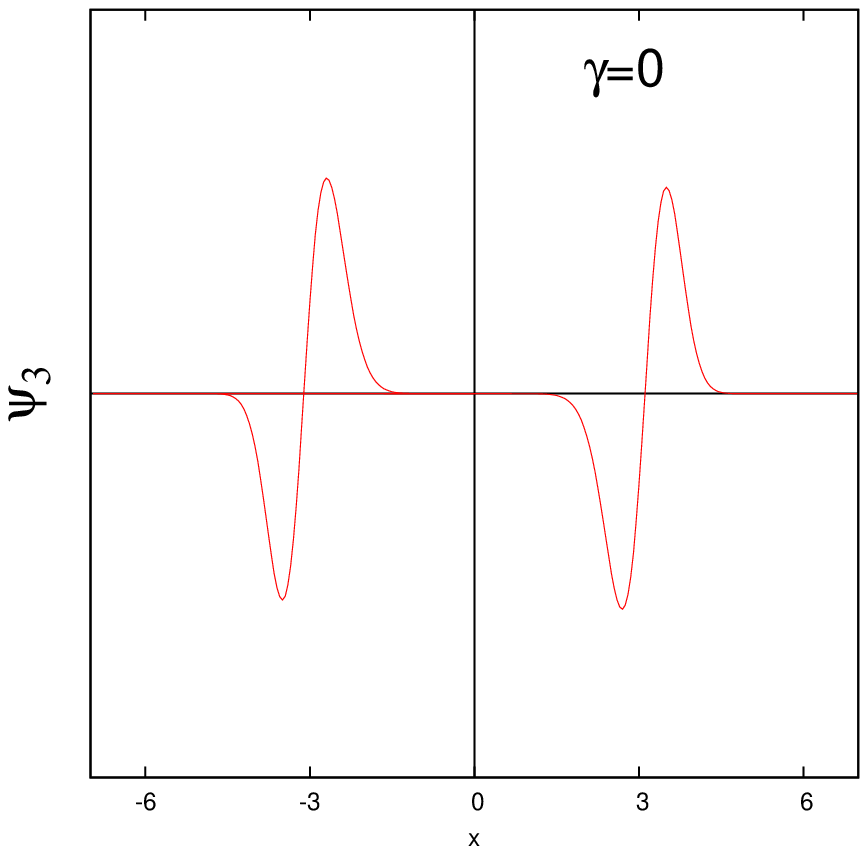}
\end{minipage}\hspace{0.06in}
\begin{minipage}[c]{0.15\textwidth}\centering
\includegraphics[scale=0.29]{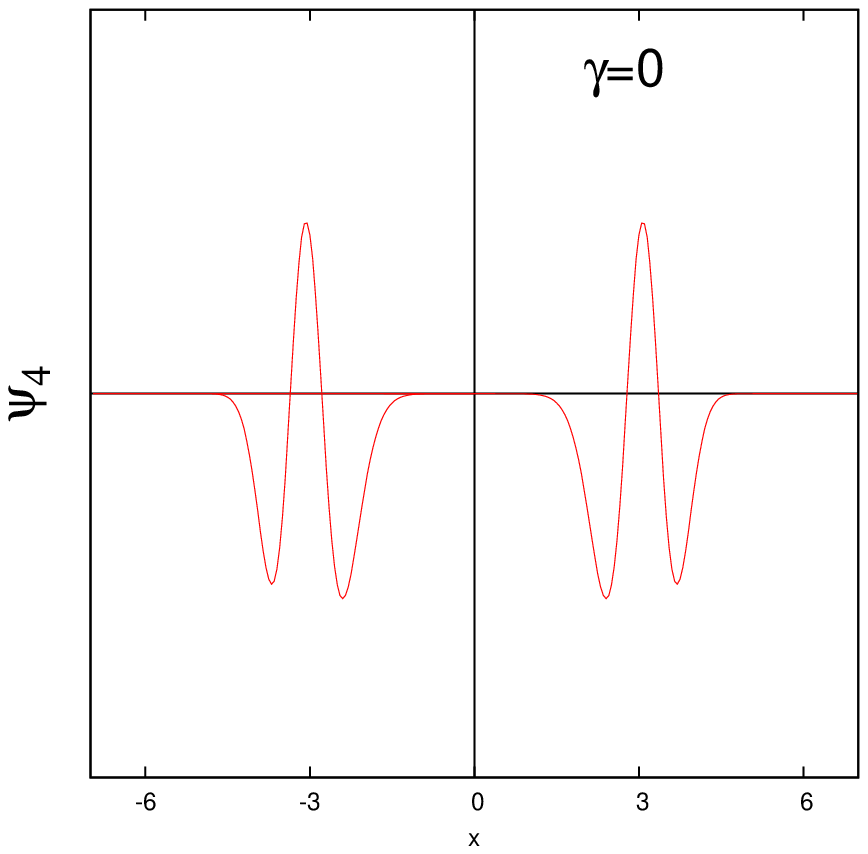}
\end{minipage}\hspace{0.06in}
\begin{minipage}[c]{0.15\textwidth}\centering
\includegraphics[scale=0.29]{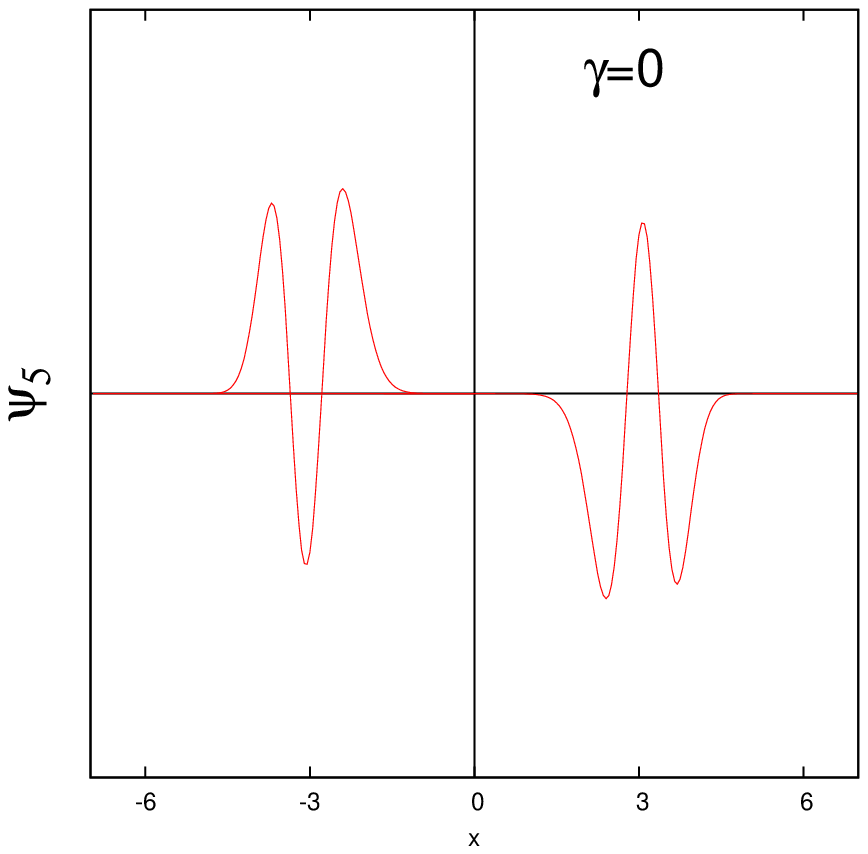}
\end{minipage}
\caption[optional]{Wave function for asymmetric DW potential, given in Eq.~(4), for $\beta \! = \! 20$, $\alpha \! = \! 1$. 
Eight $\gamma$ (0,1,2,3,4,5,6,7) are considered in rows, starting $\gamma \! = \! 0$ at bottom. Six columns present lowest 
six states, beginning with $\psi_0$ in the left. For details, refer to text.}
\end{figure}

Figure~(6) portrays our calculated optimised wave functions for all states below $n \! \leq \! 6$, at eight selected $\gamma$, 
for a constant $\beta \! = \! 20$. Six columns display these for lowest six states; from left to right they correspond 
to $n \! = \! 0$,1,2,3,4,5 respectively. Likewise, eight rows refer to eight chosen $\gamma$ values (0-7, at interval 
of 1); from bottom to top. From first column, we discern that, when $\gamma \! > \! 0$ ($k \! > \! 0$), particle 
in lowest state always dwells in well~I; $k \! > \! n$ and (ii) is satisfied. In second column, at 
first for $\gamma \! = \! 1$ particle in $n \! = \! 1$ prefers to stay in well~II, as suggested from (i.b.4), for integer part 
of $k$ is even (0). At $\gamma \! = 2 \!$, it remains in both wells, in consistence with (i.a), as $k$ is 
integer (1) and $n \! > \! k$. For any $\gamma \! > \! 2$ ($k \! > \! 1$), $k \! > \! n$, and according to 
(ii), it eventually settles in well~I thereafter for all $\gamma$. At this point, it may be prudent to define this 
$\gamma$ (2 in this case) as a \emph{critical transition 
point}, after which it permanently settles in well~I, for all $\gamma$. Consistent with (i.b.1), in $n \! = \! 2$, 
it stays in well~(I) at $\gamma \! = \! 1$, as $n$ and integer part of $k$ are both even. At $\gamma \! = \! 2$, 
$k \! = \! 1$; thus (i.a) says that in $n \! = \! 2$ it may reside in both wells. At $\gamma \! = \! 3$, 
it reverses its position in opposite well from $\gamma \! = \! 1$; in this case since integer part of $k$ is odd, 
(i.b.2) advocates well~I to be the preferential place of stay. At $\gamma \! = \! 4$, it can stay in both wells in 
fulfillment with (i.a); moreover this also indicates a critical transition point, as after that for all $\gamma$, it does 
not move from well~I. For $n \! = \! 3$, it remains in wells~II at $\gamma \! =\! 1$ and 5 in accordance 
with (i.b.4), as integer part of $k$ is even (0, 2 respectively). For $\gamma \! = \! 3$, on the other hand, following 
(i.b.2), it moves to well~I, since integer factor of $k$ is odd. At $\gamma \! = \! 2$, 4, it can occupy both wells, as 
integer $k$ calls for (i.a), whereas $\gamma \! = \! 6$ defines a critical transition point. Thereafter it always remains in 
well~I only. Same procedure could be extended for $n \! = \! 4$, 5 in columns five, six respectively. A quick glance reveals 
that they also follow the same distribution pattern as proposed in the rules. Thus, in essence, even-numbered rows (fractional 
$k$) favor localization in either well~I or II, whereas odd-numbered (integer $k$) ones indicate same in both wells. 

At this stage, it is interesting to note that, in second row from bottom ($k \! = \! 0.5$), $n \! = \! 0,2,4$ states of an 
asymmetric DW well behave as ground, first, second excited states of a single-well potential in well~I, whereas 
$n \! = \! 1,3,5$ act as lowest three states of a potential in well~II. Likewise, fourth row from bottom ($k \! = \! 1.5$) 
suggests that, $n \! = \! 0,1,3,5$ of DW behave as lowest four states of well~I; $n \! = \! 2,4$, on the other hand, appear 
as ground,  first excited state of well~II. Now, in sixth row from bottom ($k \! = \! 2.5$), we notice that, 
$n \! = \! 0,1,2,4$ act as lowest four levels of well~I, while $n \! = \! 3,5$ may be treated as ground, first excited 
states of well~II. Finally, from top-most row ($k \! = \! 3.5$), it is observed that, $n \! = \! 0,1,2,3,5$ of DW represent 
effective lowest five states of well~I, whereas $n \! = \! 4$ may be identified as effective lowest state of well~II. 
Similar trend is found to be valid for other $\gamma$ as well, which further demonstrates the applicability of rules. 
Another interesting feature of these calculations is provided in bottom-most row ($\gamma \! = \! 0, k \! = \! 0$), which 
represents a \emph{symmetric} DW. As expected (from both classical, quantum mechanics), in all states, particle remains 
equally distributed in two wells. This is in accordance with (i.a); this confirms the obvious fact that a \emph{symmetric} 
DW may be treated as a particular case of a more general \emph{asymmetric} DW. 

\begin{figure}             
\centering
\begin{minipage}[c]{0.15\textwidth}\centering
\includegraphics[scale=0.28]{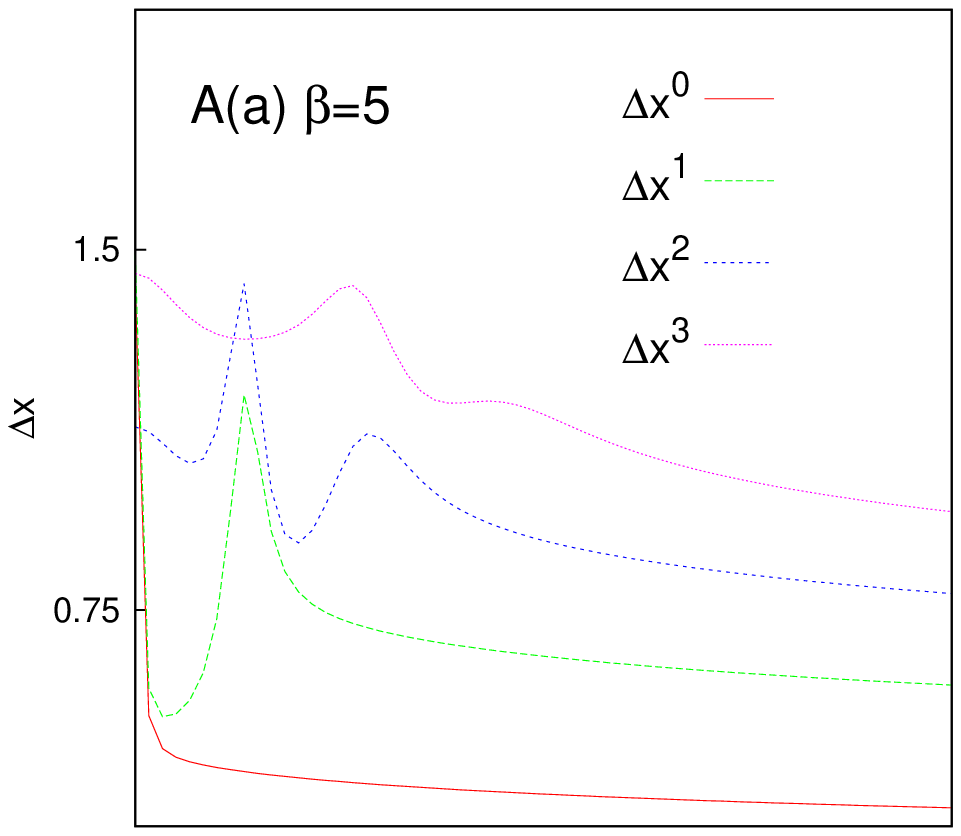}
\end{minipage}\hspace{0.08in}
\begin{minipage}[c]{0.15\textwidth}\centering
\includegraphics[scale=0.28]{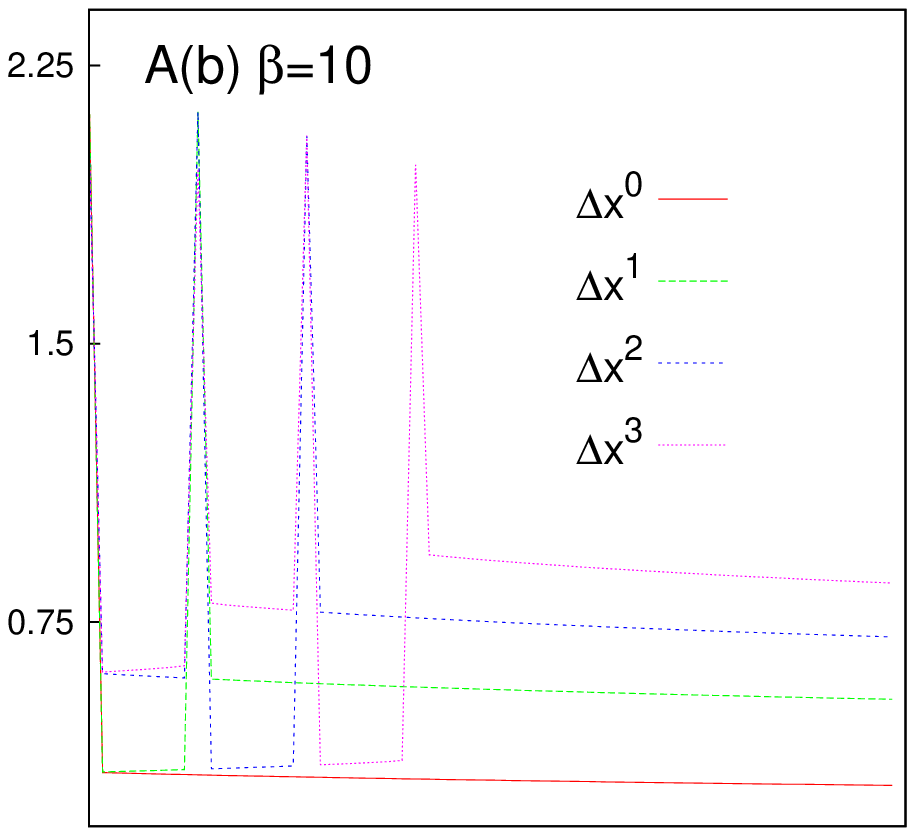}
\end{minipage}\hspace{0.08in}
\begin{minipage}[c]{0.15\textwidth}\centering
\includegraphics[scale=0.28]{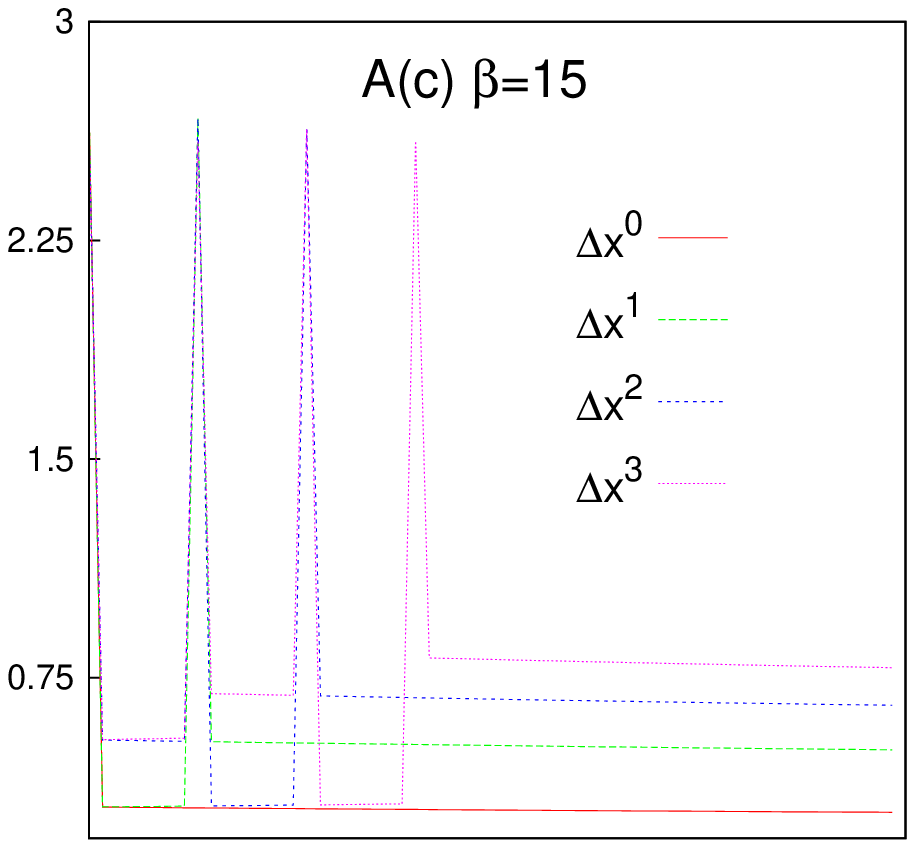}
\end{minipage}\hspace{0.08in}
\begin{minipage}[c]{0.15\textwidth}\centering
\includegraphics[scale=0.28]{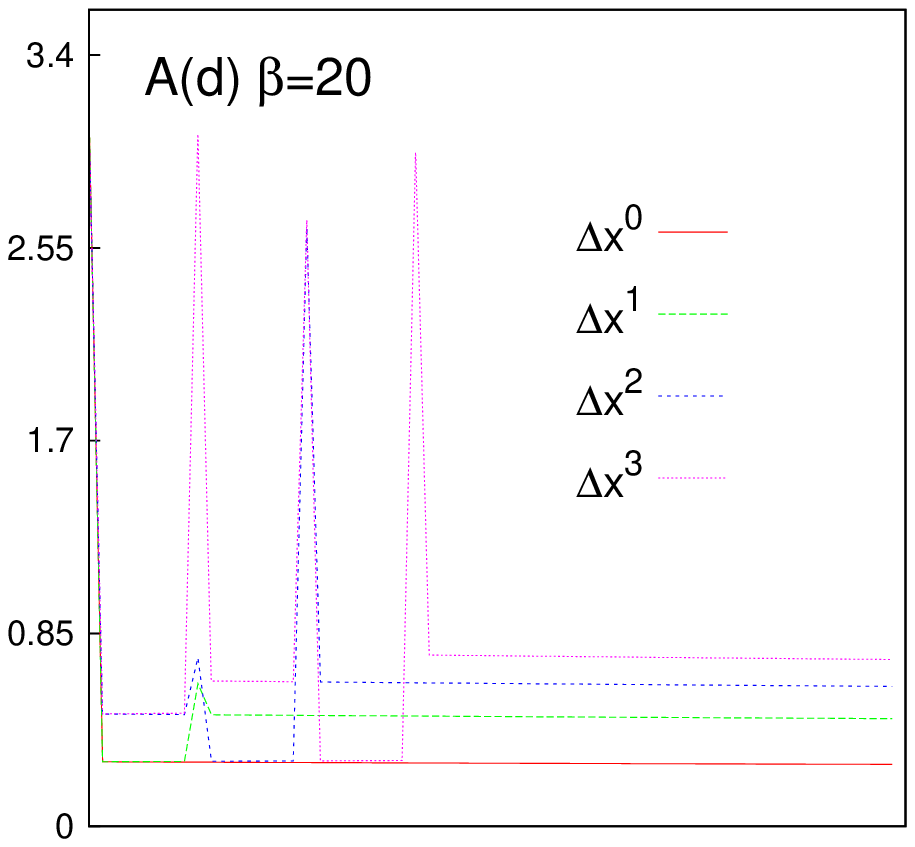}
\end{minipage}\hspace{0.08in}
\begin{minipage}[c]{0.15\textwidth}\centering
\includegraphics[scale=0.28]{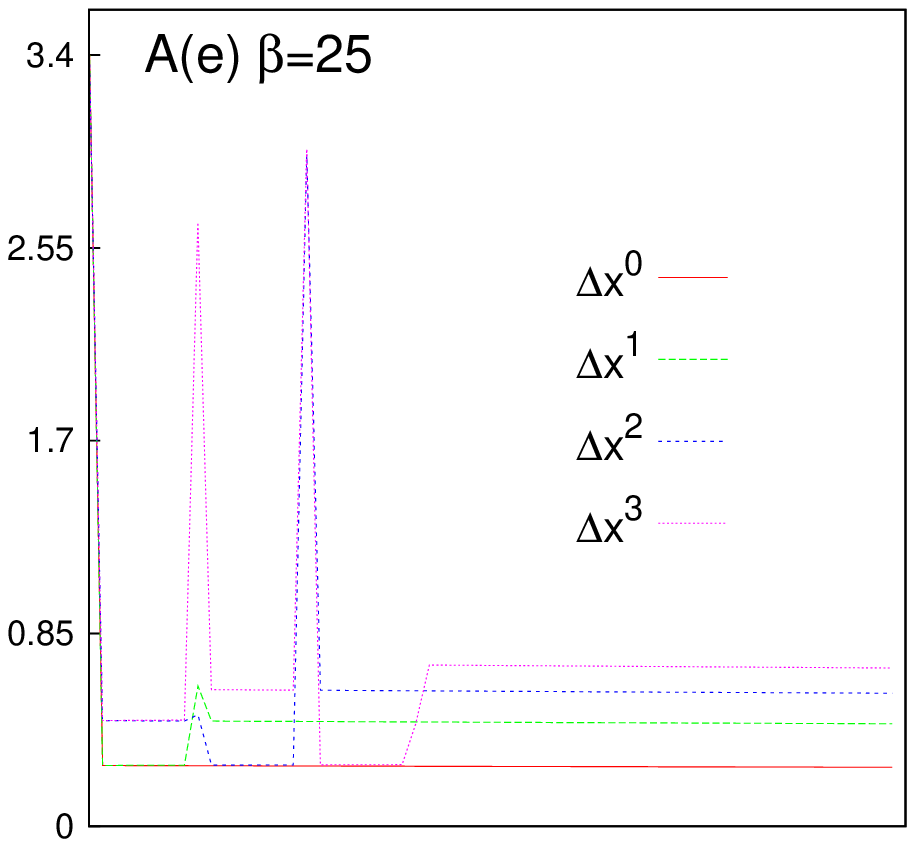}
\end{minipage}\hspace{0.08in}
\begin{minipage}[c]{0.15\textwidth}\centering
\includegraphics[scale=0.28]{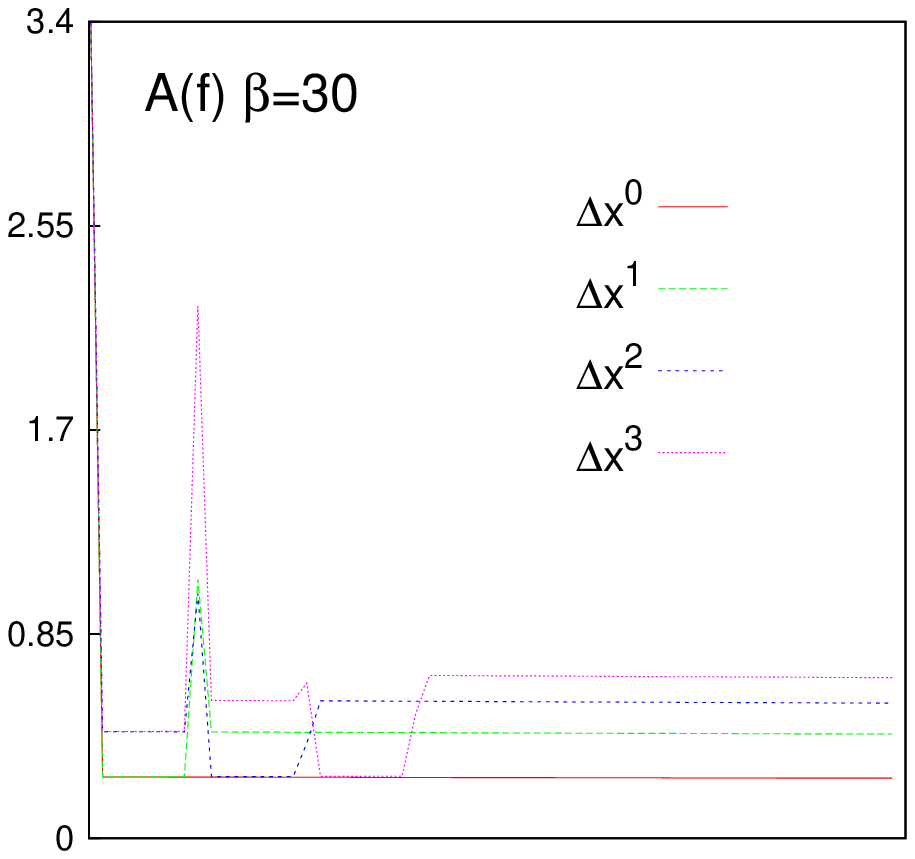}
\end{minipage}
\\[5pt]
\begin{minipage}[c]{0.15\textwidth}\centering
\includegraphics[scale=0.28]{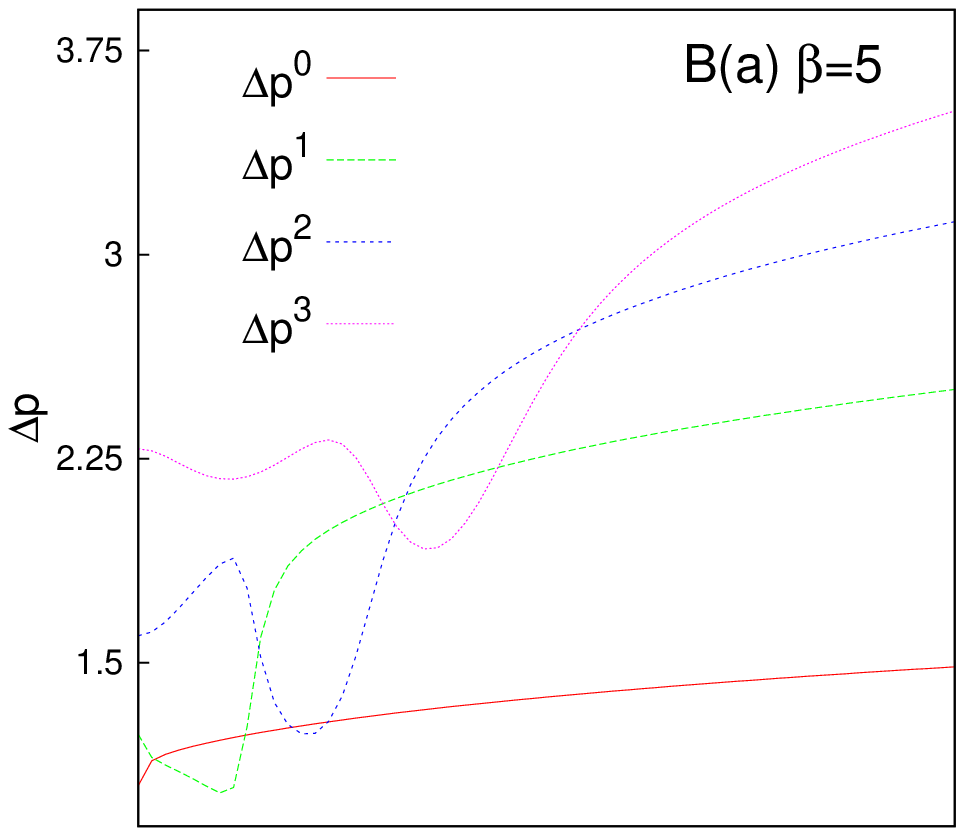}
\end{minipage}\hspace{0.08in}
\begin{minipage}[c]{0.15\textwidth}\centering
\includegraphics[scale=0.28]{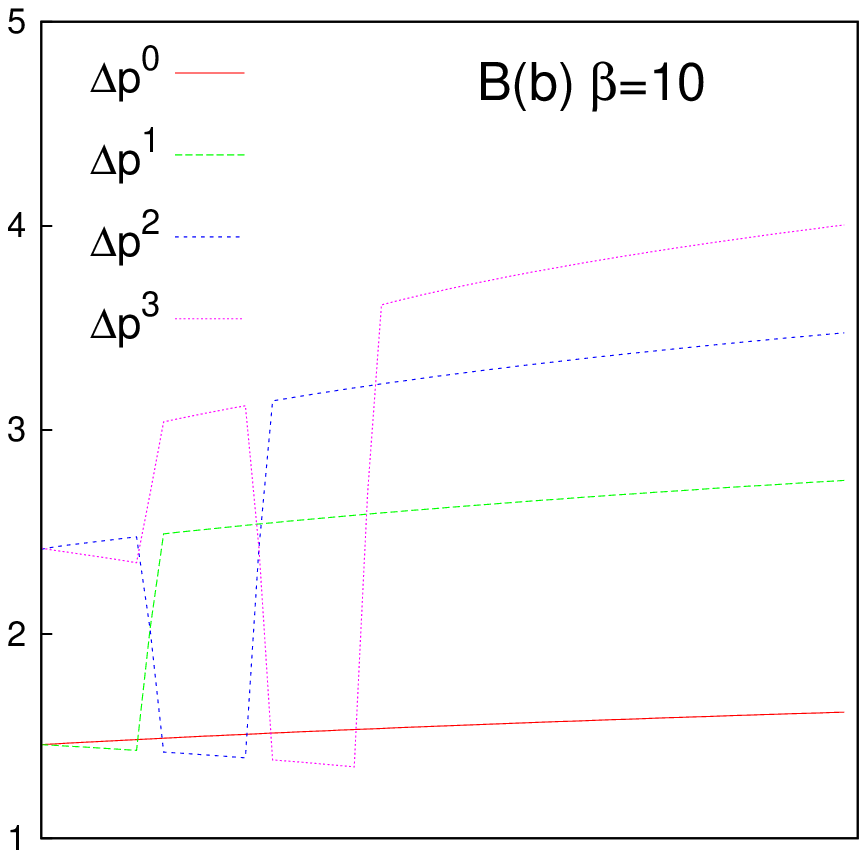}
\end{minipage}\hspace{0.08in}
\begin{minipage}[c]{0.15\textwidth}\centering
\includegraphics[scale=0.28]{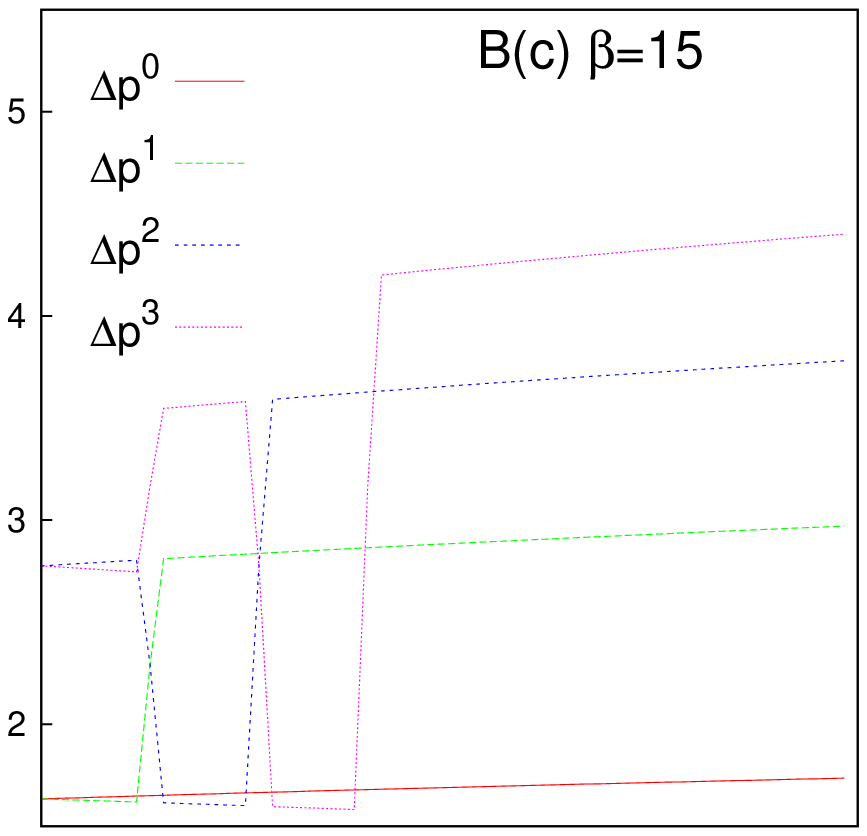}
\end{minipage}\hspace{0.08in}
\begin{minipage}[c]{0.15\textwidth}\centering
\includegraphics[scale=0.28]{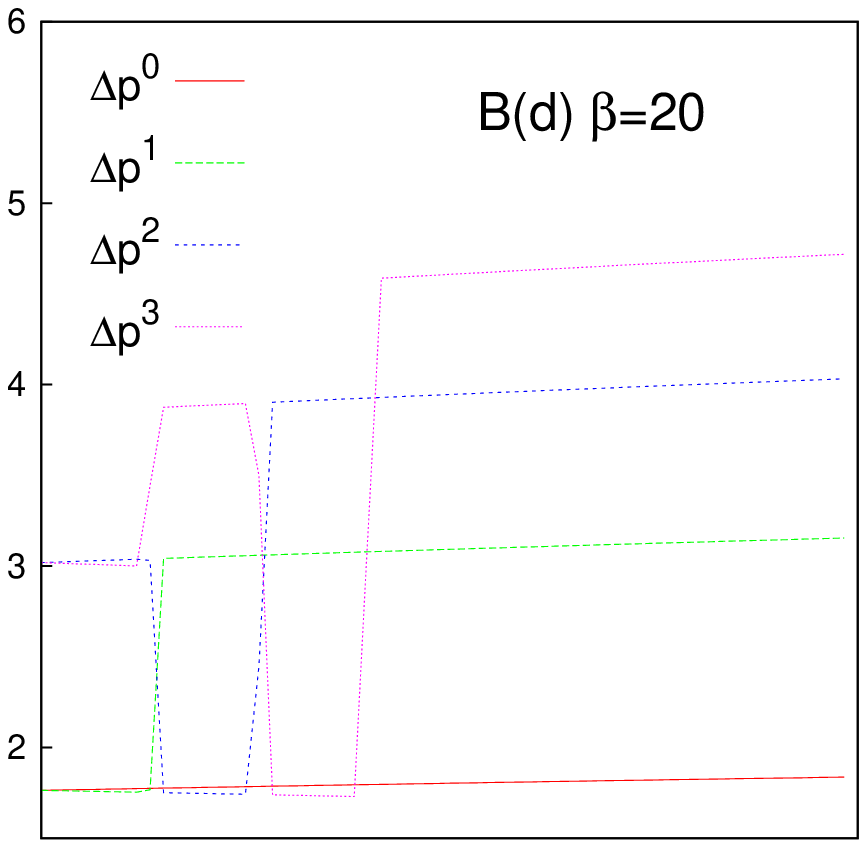}
\end{minipage}\hspace{0.08in}
\begin{minipage}[c]{0.15\textwidth}\centering
\includegraphics[scale=0.28]{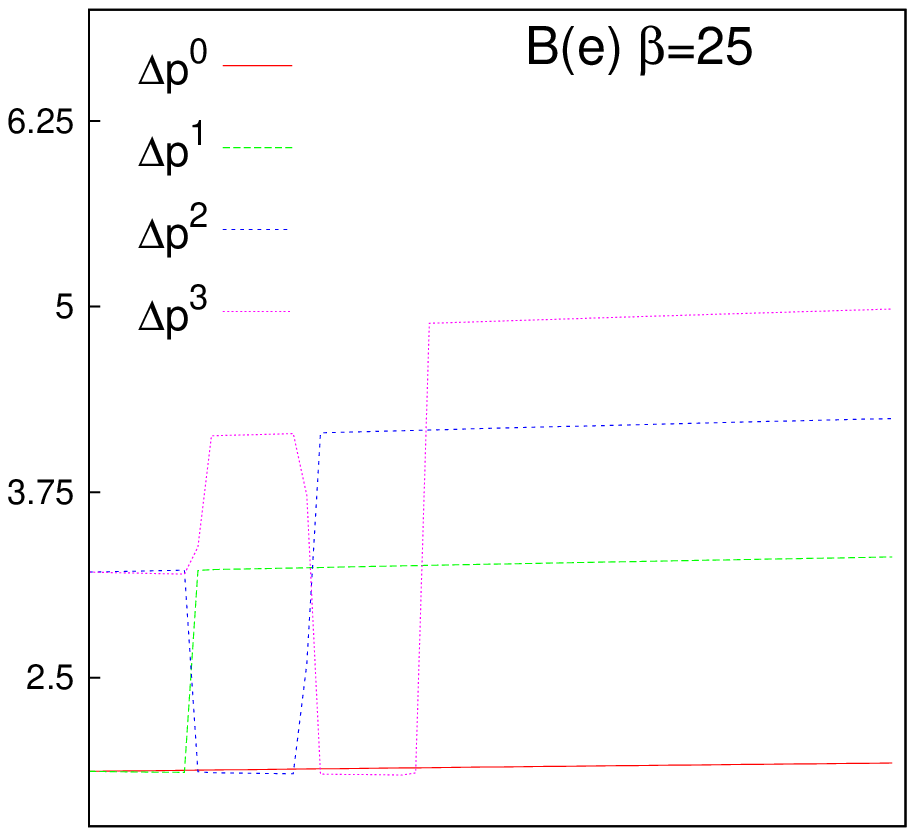}
\end{minipage}\hspace{0.08in}
\begin{minipage}[c]{0.15\textwidth}\centering
\includegraphics[scale=0.28]{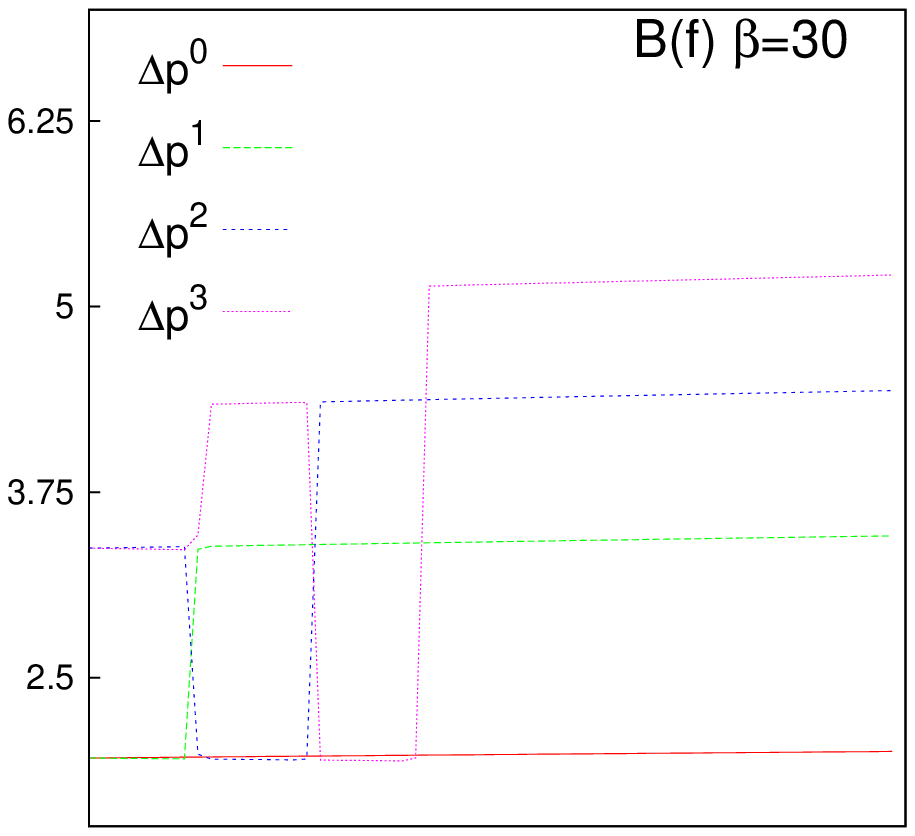}
\end{minipage}
\\[5pt]
\begin{minipage}[c]{0.15\textwidth}\centering
\includegraphics[scale=0.3]{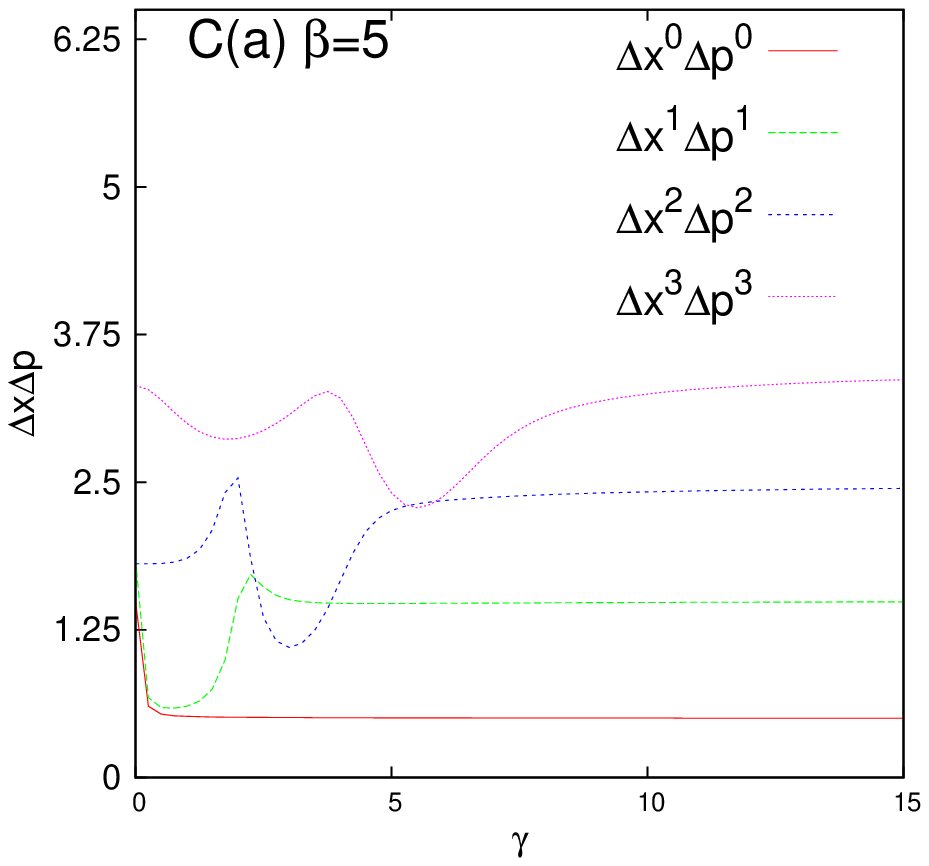}
\end{minipage}\hspace{0.08in}
\begin{minipage}[c]{0.15\textwidth}\centering
\includegraphics[scale=0.3]{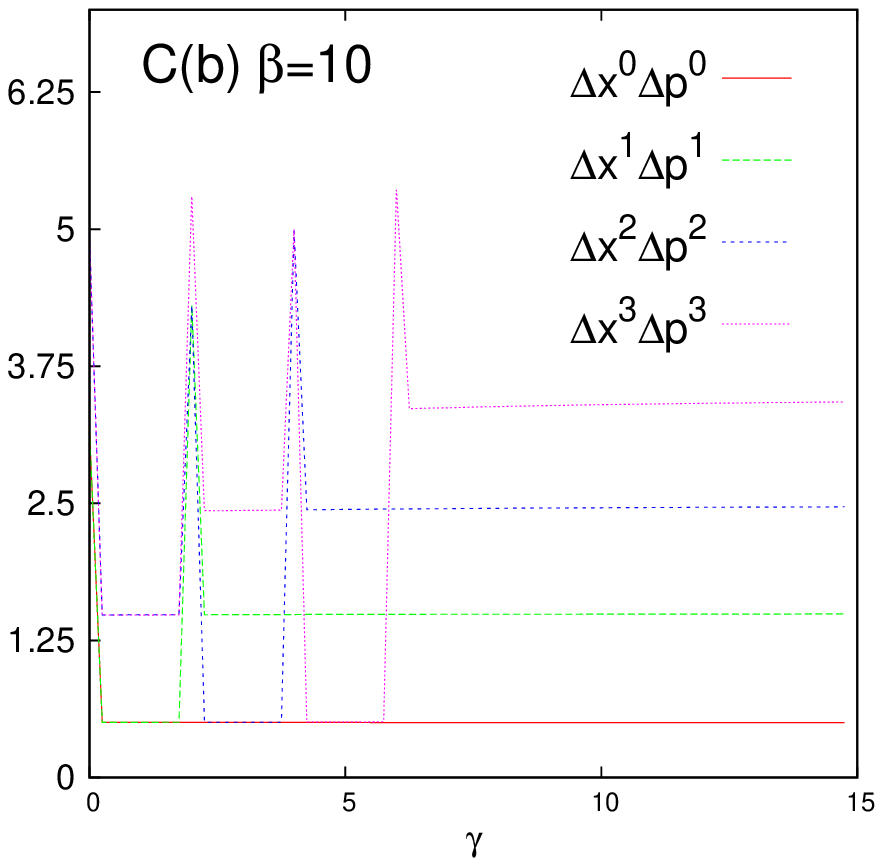}
\end{minipage}\hspace{0.08in}
\begin{minipage}[c]{0.15\textwidth}\centering
\includegraphics[scale=0.3]{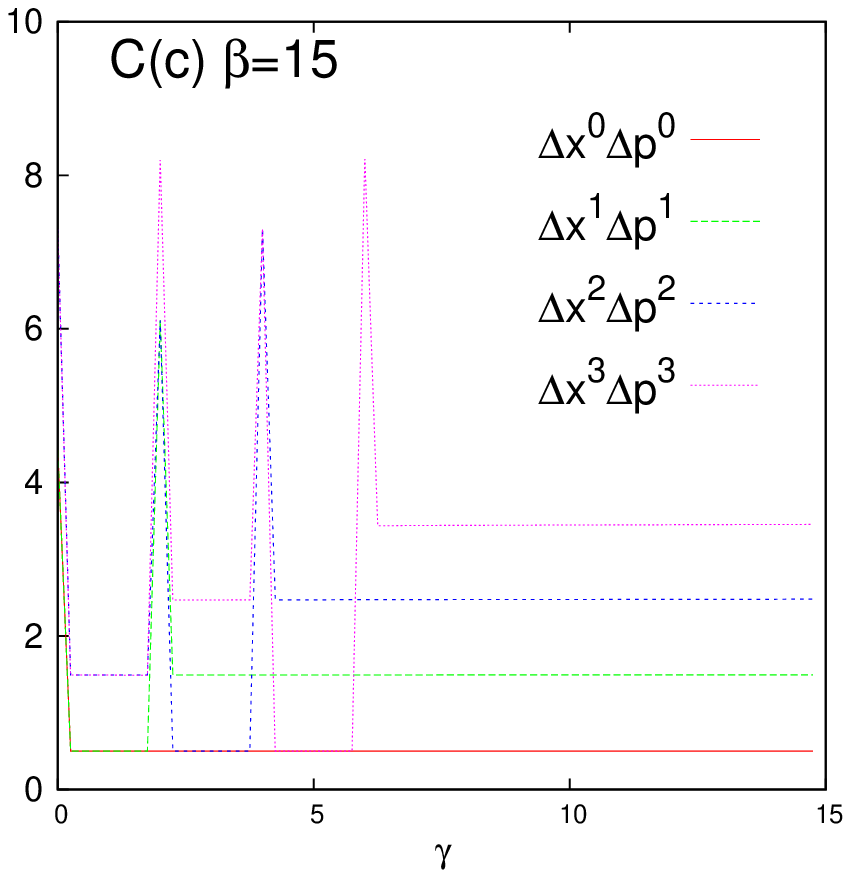}
\end{minipage}\hspace{0.08in}
\begin{minipage}[c]{0.15\textwidth}\centering
\includegraphics[scale=0.3]{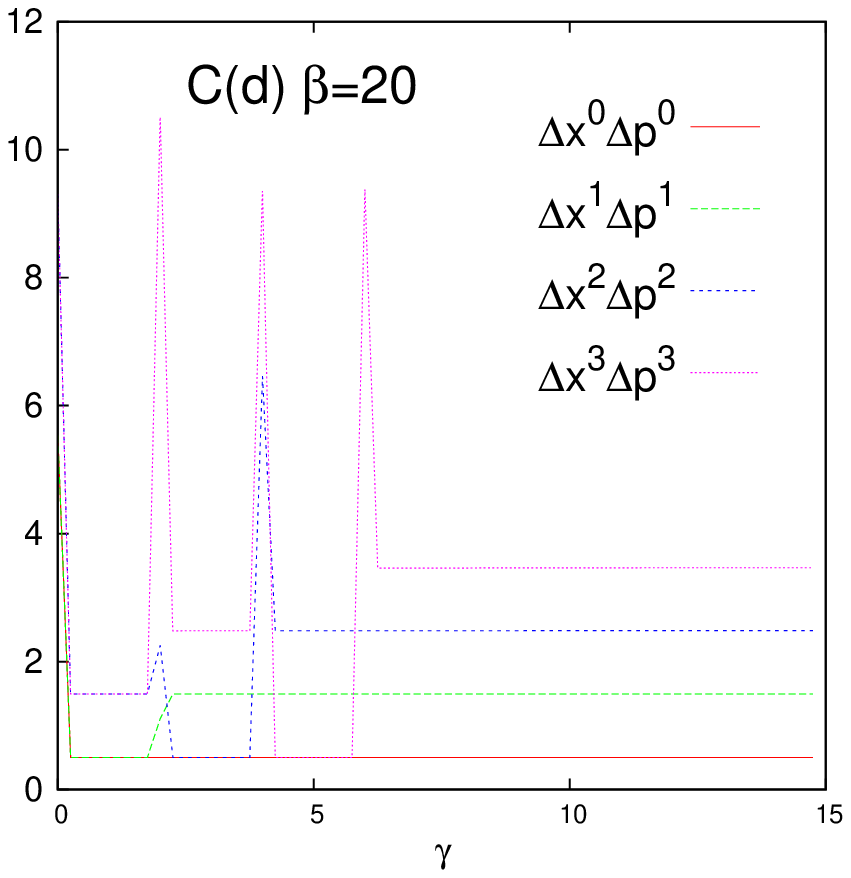}
\end{minipage}\hspace{0.08in}
\begin{minipage}[c]{0.15\textwidth}\centering
\includegraphics[scale=0.3]{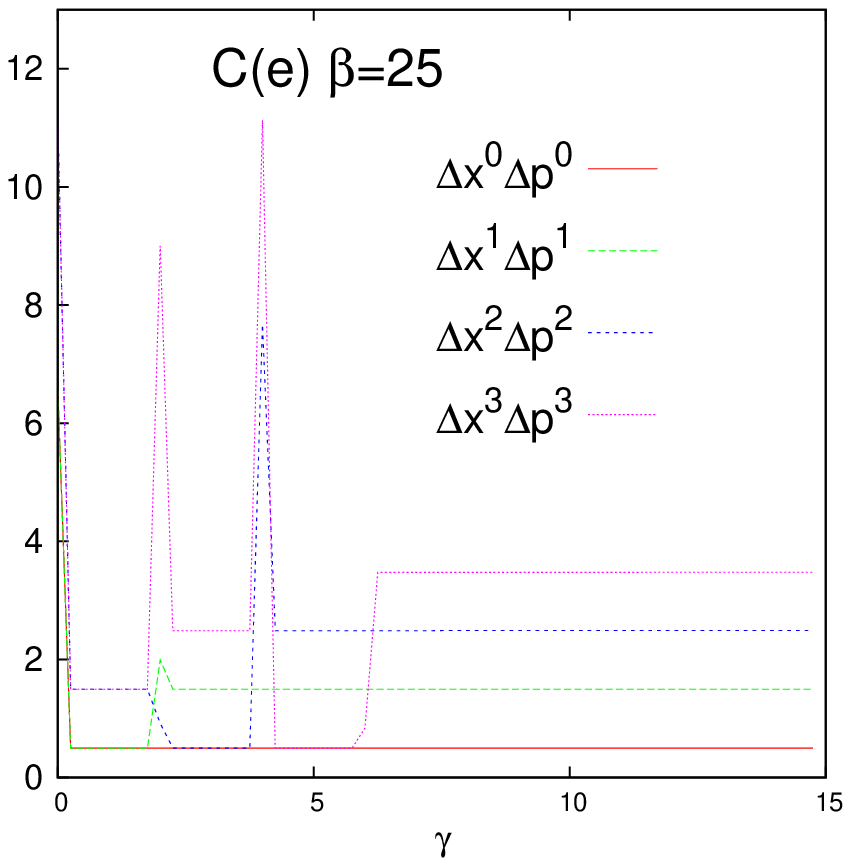}
\end{minipage}\hspace{0.08in}
\begin{minipage}[c]{0.15\textwidth}\centering
\includegraphics[scale=0.3]{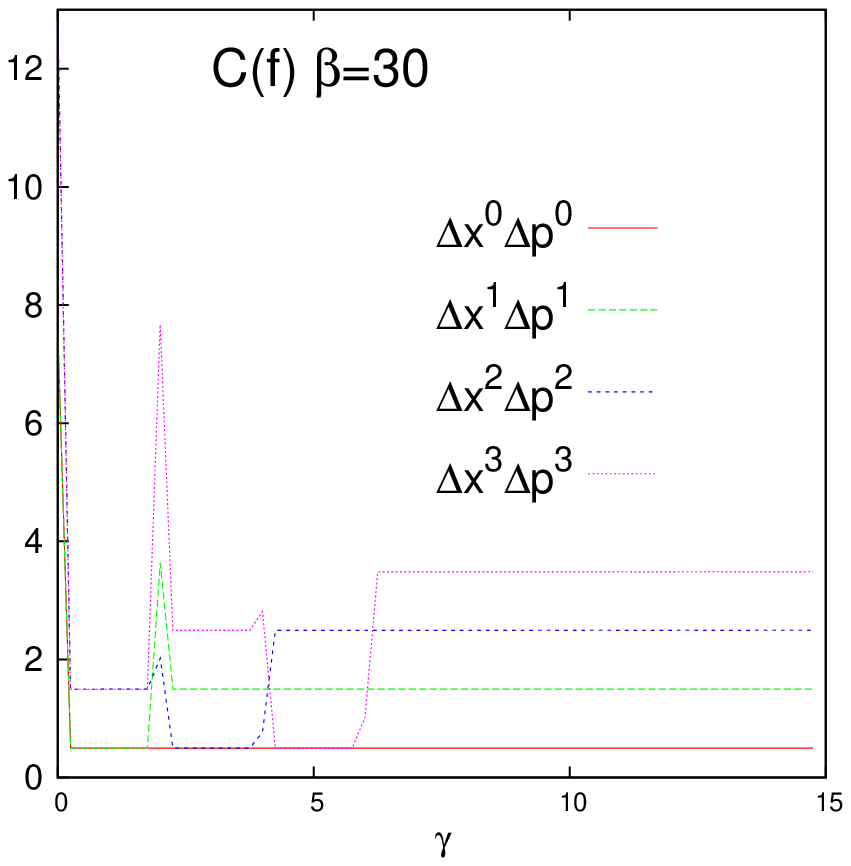}
\end{minipage}
\caption[optional]{$\Delta x$ (A), $\Delta p$ (B) and $\Delta x \Delta p$ (C) for first four states, in top, middle and 
bottom rows, plotted against $\gamma$, for asymmetric DW potential, in Eq~(4) keeping $\alpha$ fixed at 1. Six panels 
(a)-(f), in each row refer to six $\beta$, namely, 5,10,15,20,25,30 respectively. For more details, see text.}
\end{figure}

Now some remarks may be made about nodes in wave function plots. Table~V offers number of effective nodes present in a 
particular state of asymmetric DW, at same selected range of $k$. Actually, this increases with state index ($n$th state
has $n$ nodes lying within classical turning points). However, due to occurrence of two separate wells as discussed 
above, certain nodes become insignificant. In range $0 \! < k \! < 1$, for example, $n \! = \! 0$,1 are effectively nodeless
as they behave as ground state of wells~I, II respectively. Similarly, $n \! = \! 2$,3 possess single node, for they represent
first excited state of wells~I, II; whereas $n \! = \! 4$,5 contain two nodes corresponding to second excited states 
of wells~I, II. These numbers are provided in third column of this table. Next, fifth column suggests that, within 
$1 \! < \! k \! < 2$,
$n \! = \! 0$,2 have zero node; thus they appear as lowest two states of wells~I, II. Also, $n \! = \! 1$,4 have single node, as 
they act as first excited states of wells~I, II. And $n \! = \! 3$,5 possess 2,3 nodes respectively, for they relate to second 
and third excited states of well~I. In a similar fashion, in range $2 \! < k \! < \! 3$, $n \! = \! 0$,1,2,4 of DW appear as 
lowest four states of well~I, whereas $n \! = \! 3$,5 act as lowest two states of well~II. These are evident from column seven. 
Finally for range $ 3 \! < \! k \! < \! 4$, last column confirms that $n \! = \! 0$,1,2,3,5 of our asymmetric DW serve as lowest
five states of well~I, giving rise to 0,1,2,3,4 nodes respectively, whereas $n \! = \! 4$ remains effectively nodeless, as 
it is the ground state of well~II. Thus, one concludes that at fractional $k$, wells~I and II 
virtually function as two different potentials (after a certain threshold $\beta$). 

\begin{figure}              
\centering
\begin{minipage}[c]{0.18\textwidth}\centering
\includegraphics[scale=0.30]{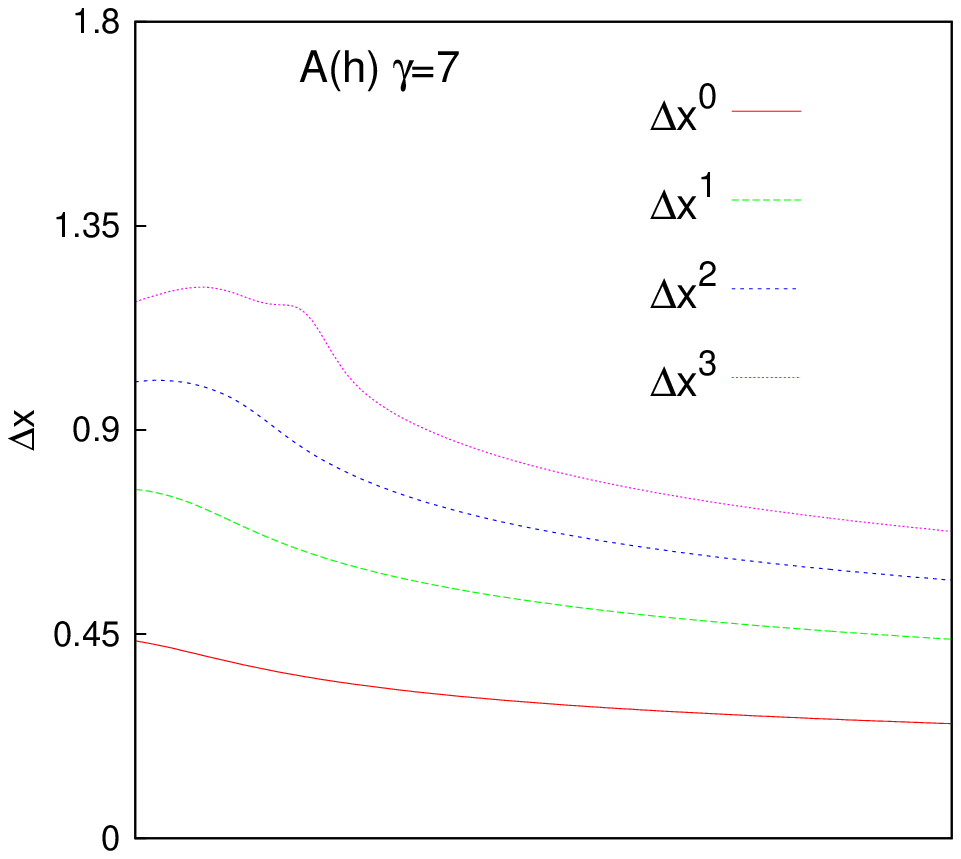}
\end{minipage}\hspace{0.10in}
\begin{minipage}[c]{0.18\textwidth}\centering
\includegraphics[scale=0.30]{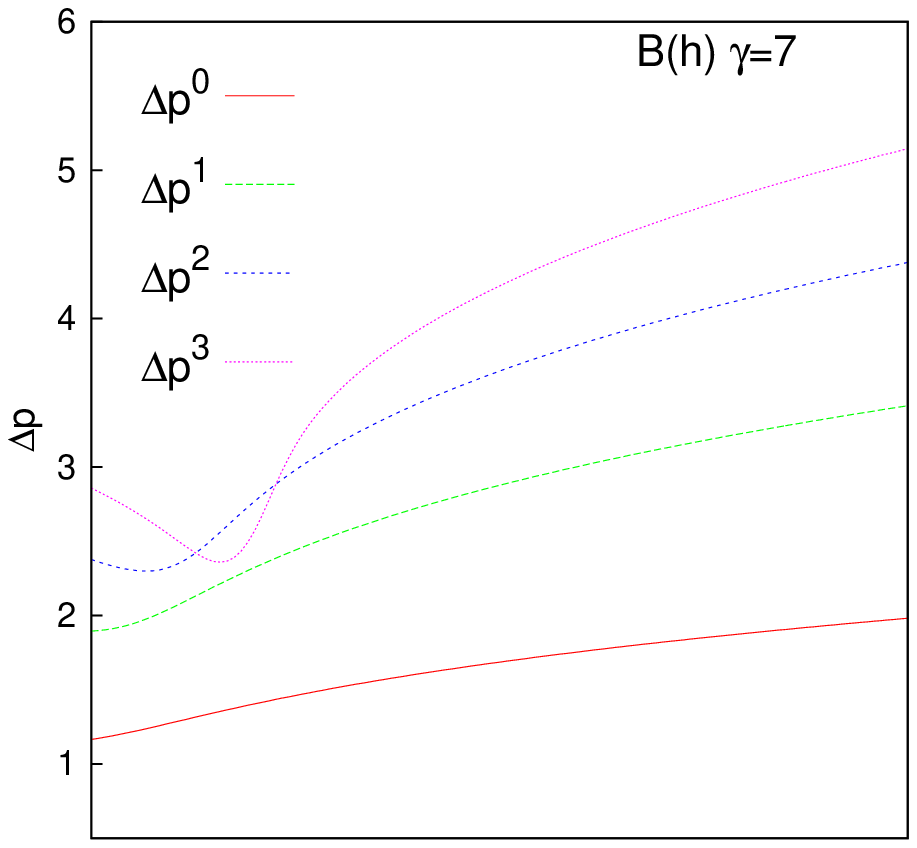}
\end{minipage}\hspace{0.10in}
\begin{minipage}[c]{0.18\textwidth}\centering
\includegraphics[scale=0.30]{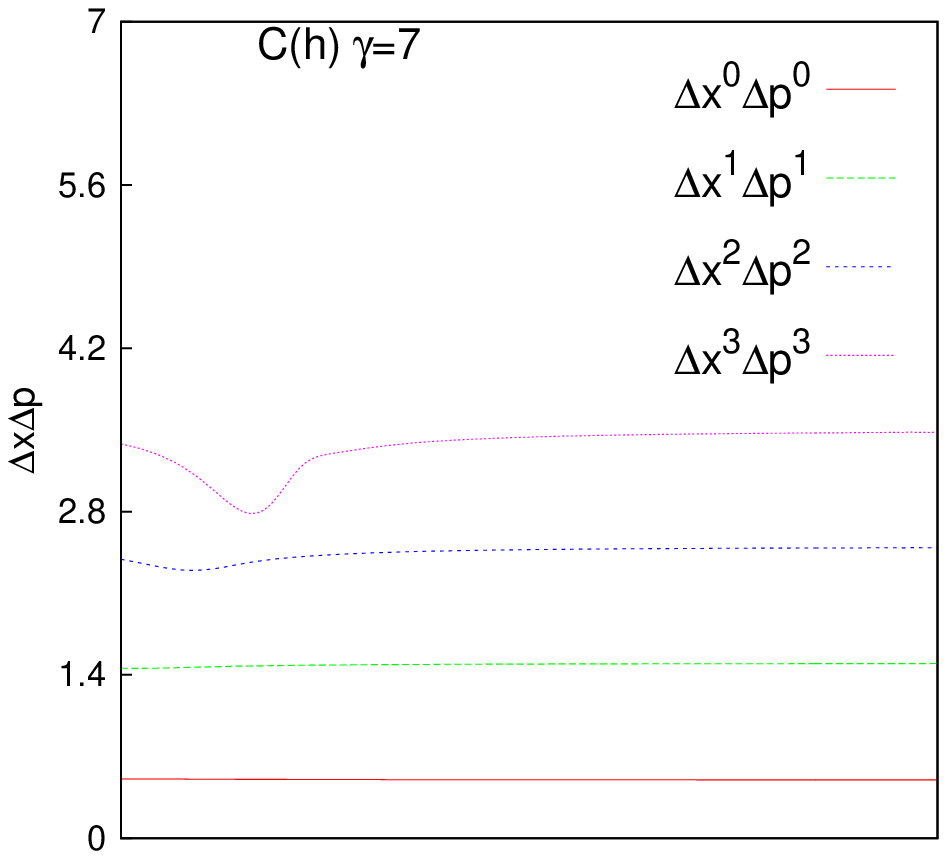}
\end{minipage}\hspace{0.10in}
\\[1pt]
\begin{minipage}[c]{0.18\textwidth}\centering
\includegraphics[scale=0.30]{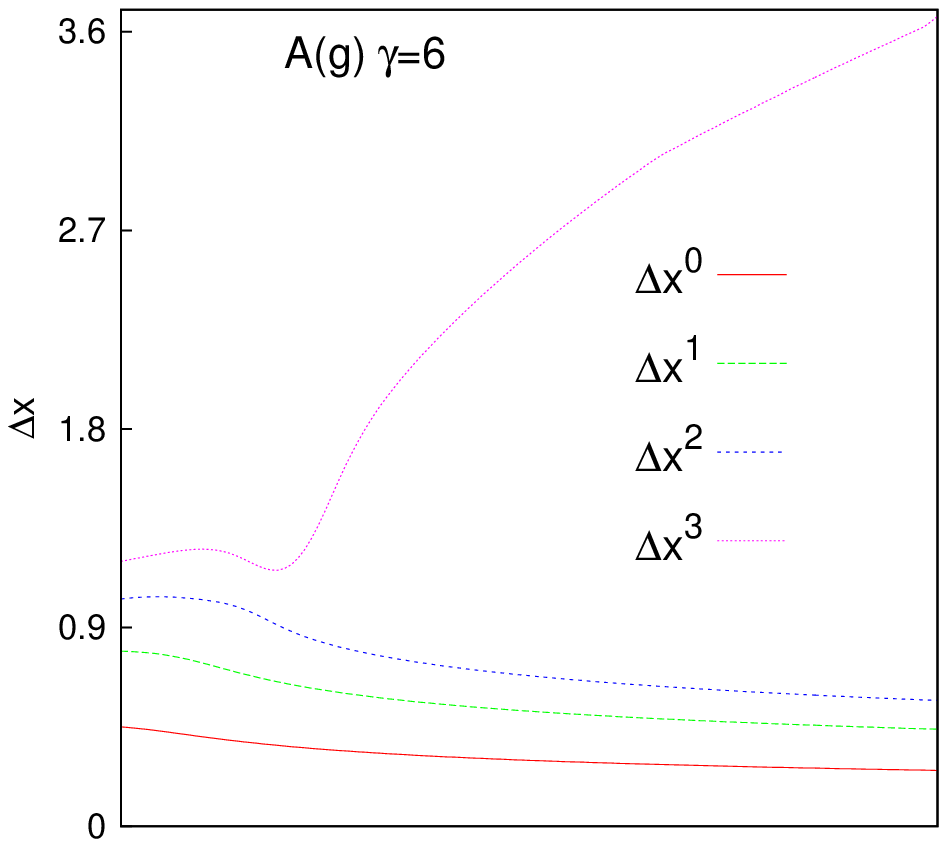}
\end{minipage}\hspace{0.10in}
\begin{minipage}[c]{0.18\textwidth}\centering
\includegraphics[scale=0.30]{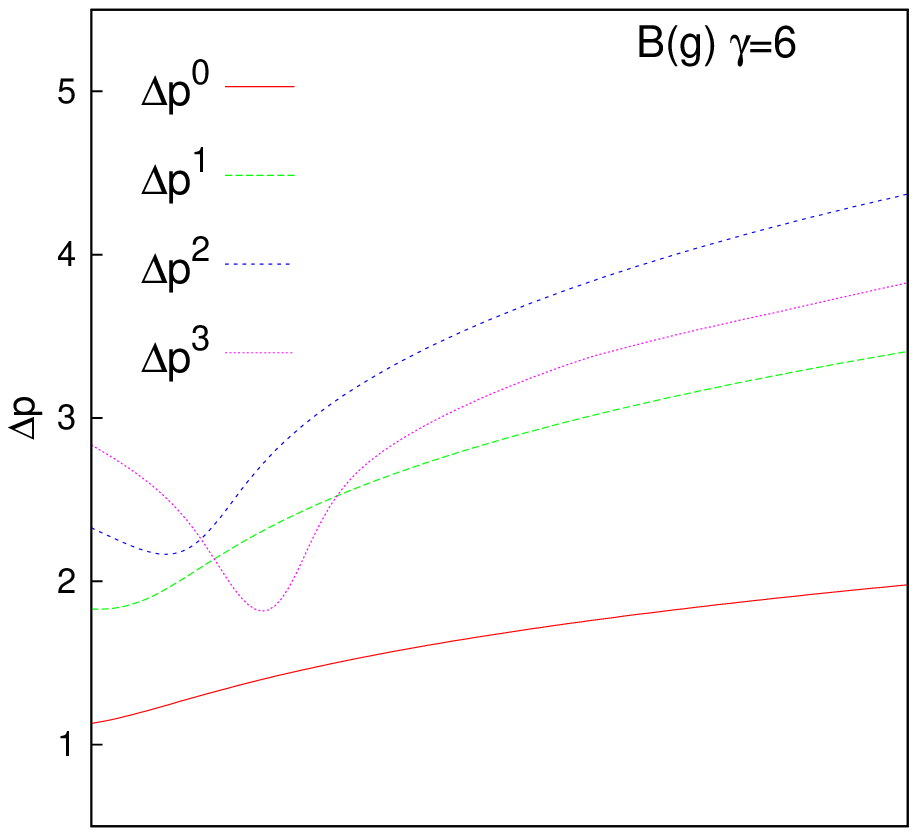}
\end{minipage}\hspace{0.10in}
\begin{minipage}[c]{0.18\textwidth}\centering
\includegraphics[scale=0.30]{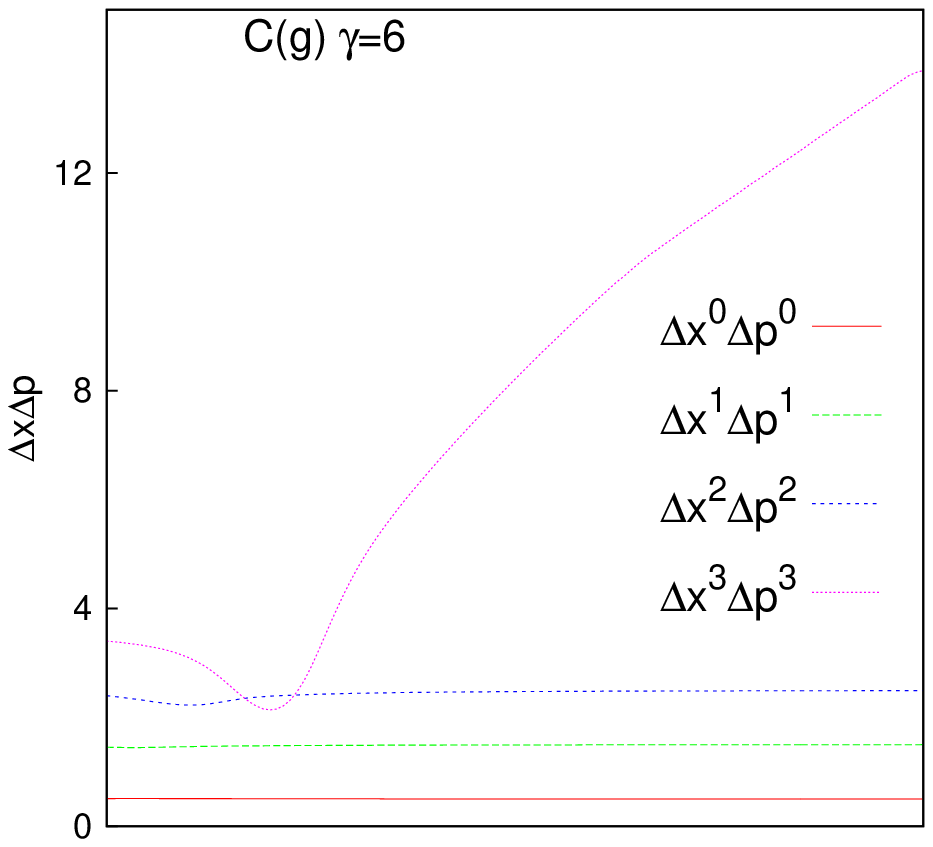}
\end{minipage}\hspace{0.10in}
\\[1pt]
\begin{minipage}[c]{0.18\textwidth}\centering
\includegraphics[scale=0.30]{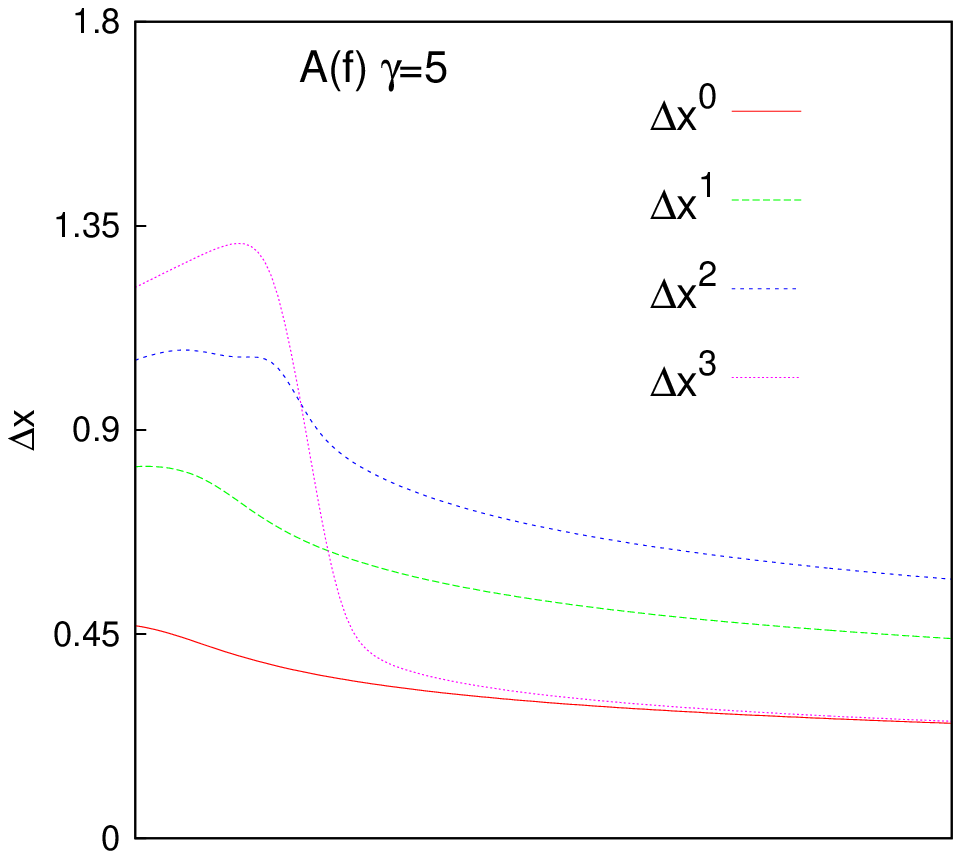}
\end{minipage}\hspace{0.10in}
\begin{minipage}[c]{0.18\textwidth}\centering
\includegraphics[scale=0.30]{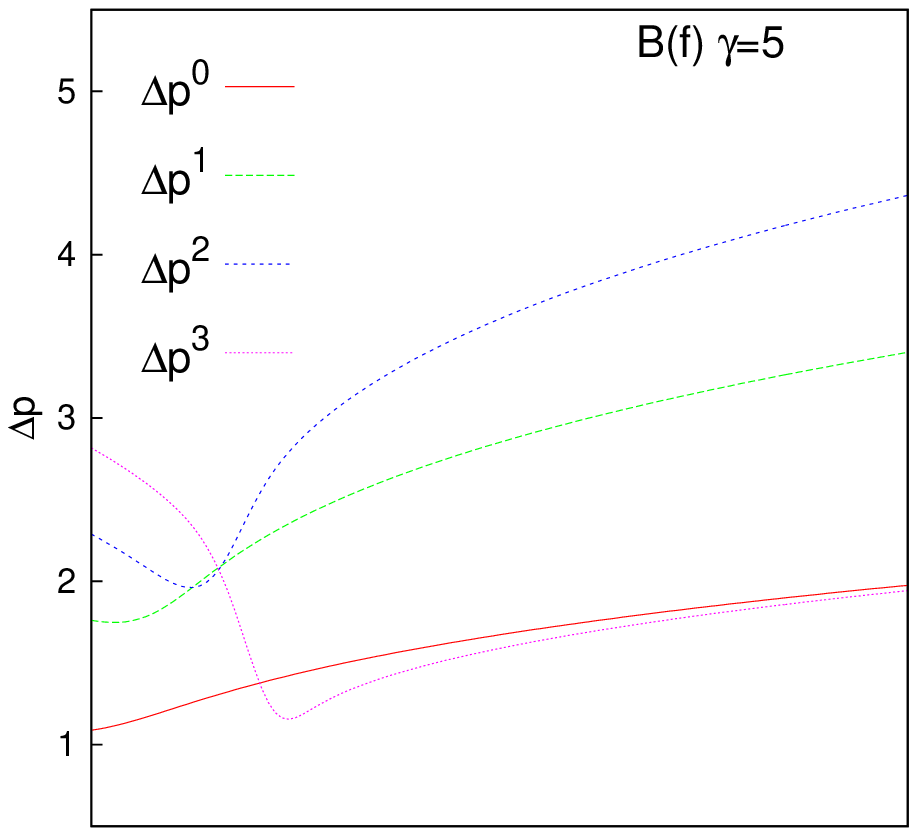}
\end{minipage}\hspace{0.10in}
\begin{minipage}[c]{0.18\textwidth}\centering
\includegraphics[scale=0.30]{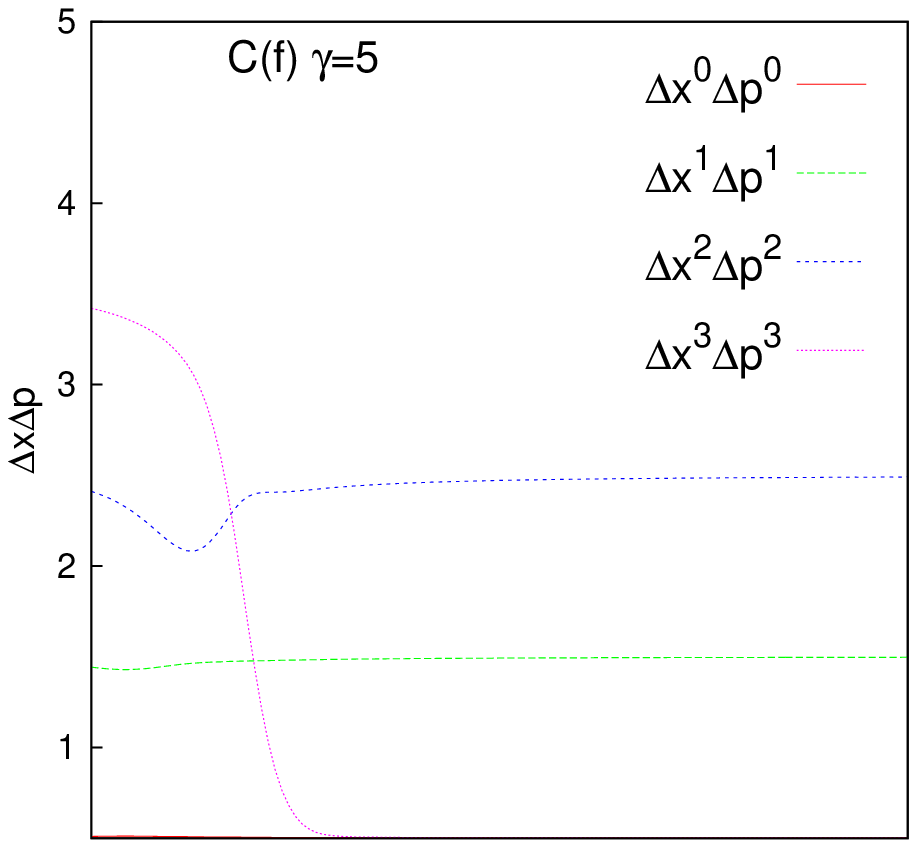}
\end{minipage}\hspace{0.10in}
\\[1pt]
\begin{minipage}[c]{0.18\textwidth}\centering
\includegraphics[scale=0.30]{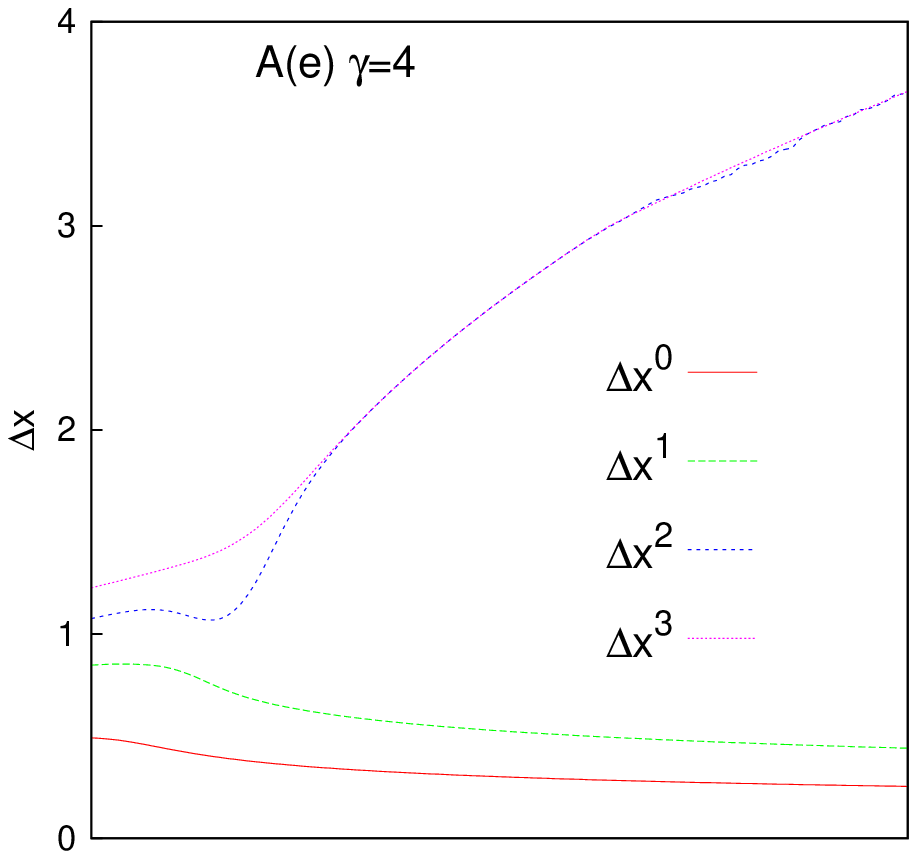}
\end{minipage}\hspace{0.10in}
\begin{minipage}[c]{0.18\textwidth}\centering
\includegraphics[scale=0.30]{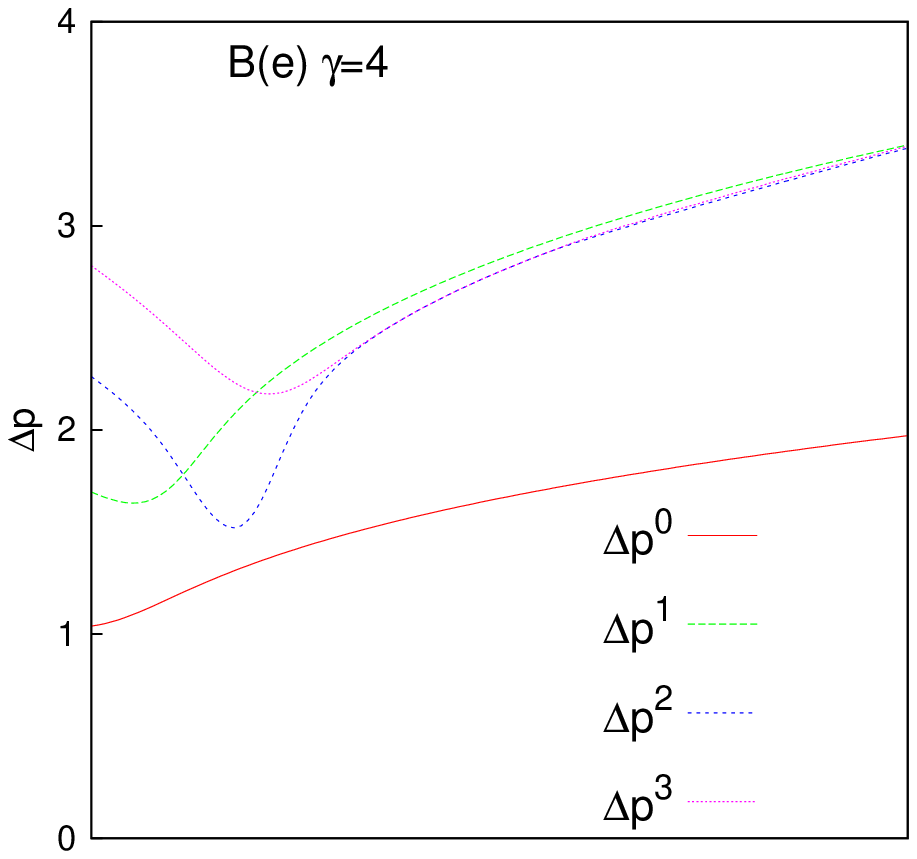}
\end{minipage}\hspace{0.10in}
\begin{minipage}[c]{0.18\textwidth}\centering
\includegraphics[scale=0.30]{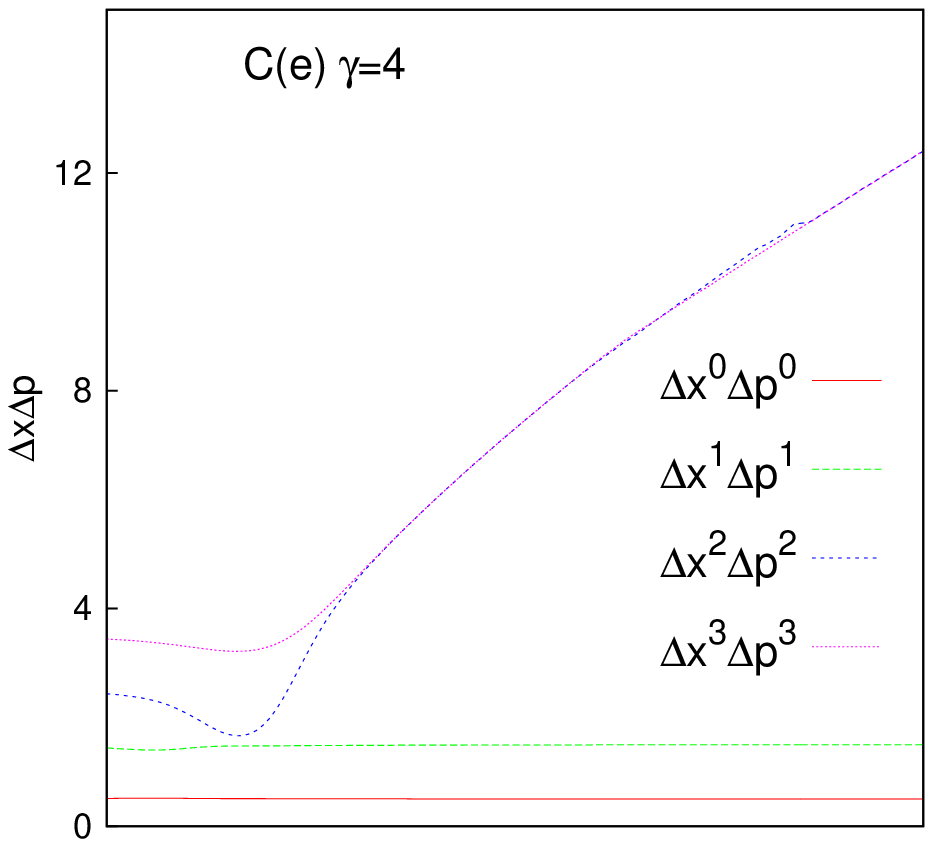}
\end{minipage}\hspace{0.10in}
\\[1pt]
\begin{minipage}[c]{0.18\textwidth}\centering
\includegraphics[scale=0.30]{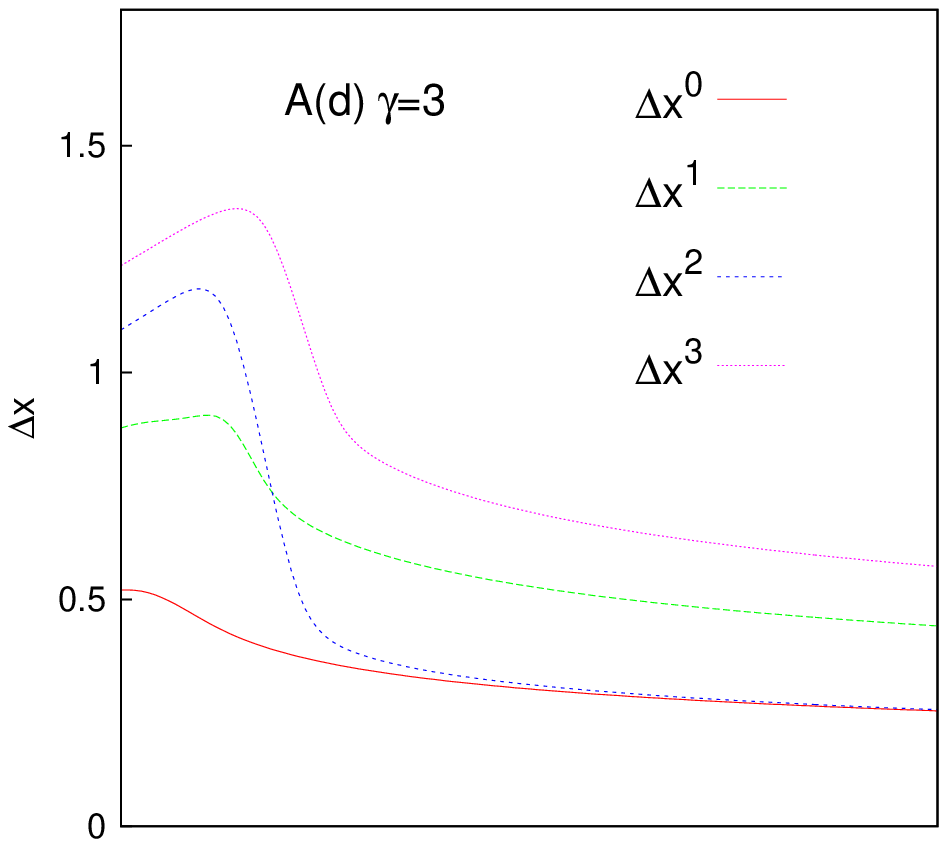}
\end{minipage}\hspace{0.10in}
\begin{minipage}[c]{0.18\textwidth}\centering
\includegraphics[scale=0.30]{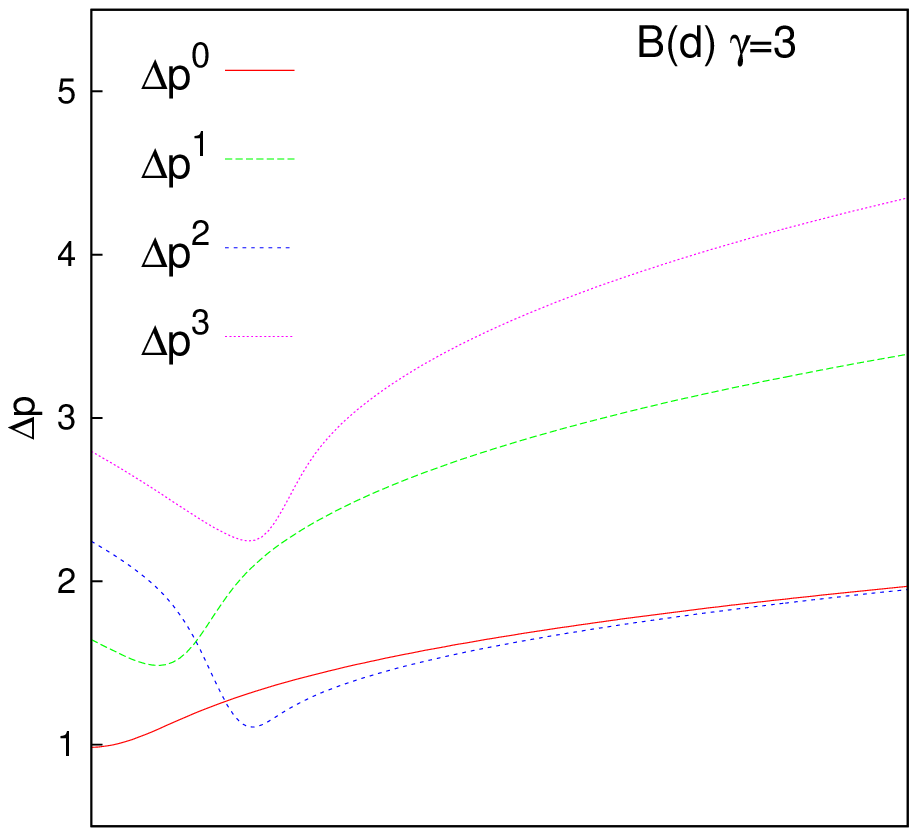}
\end{minipage}\hspace{0.10in}
\begin{minipage}[c]{0.18\textwidth}\centering
\includegraphics[scale=0.30]{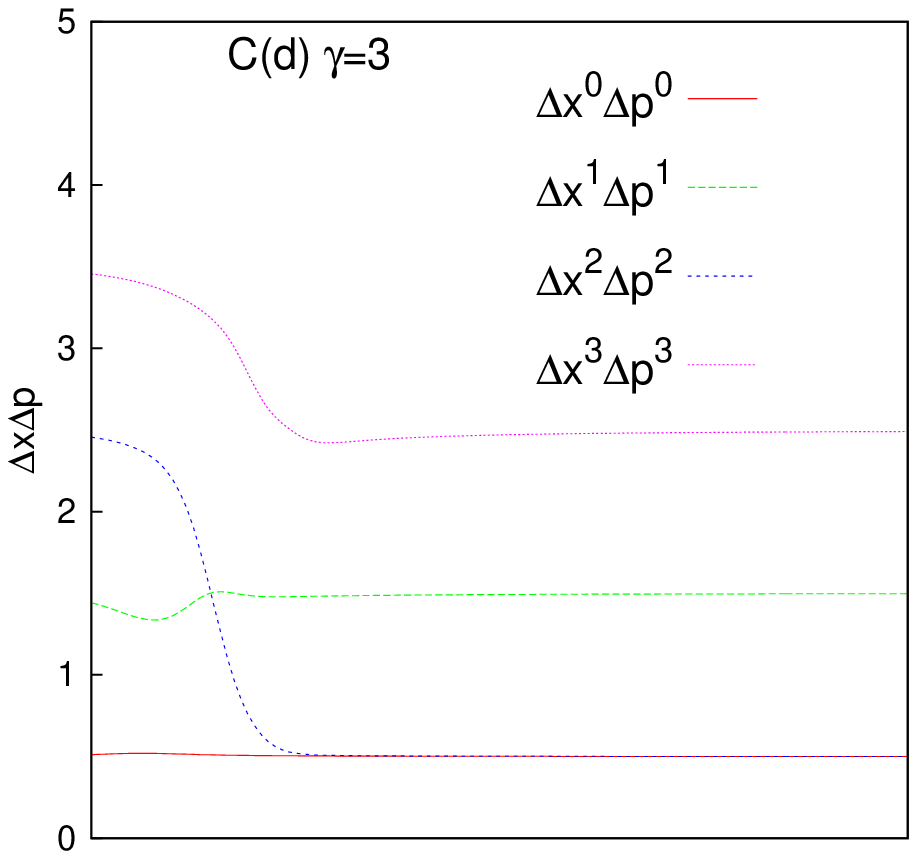}
\end{minipage}\hspace{0.10in}
\\[1pt]
\begin{minipage}[c]{0.18\textwidth}\centering
\includegraphics[scale=0.30]{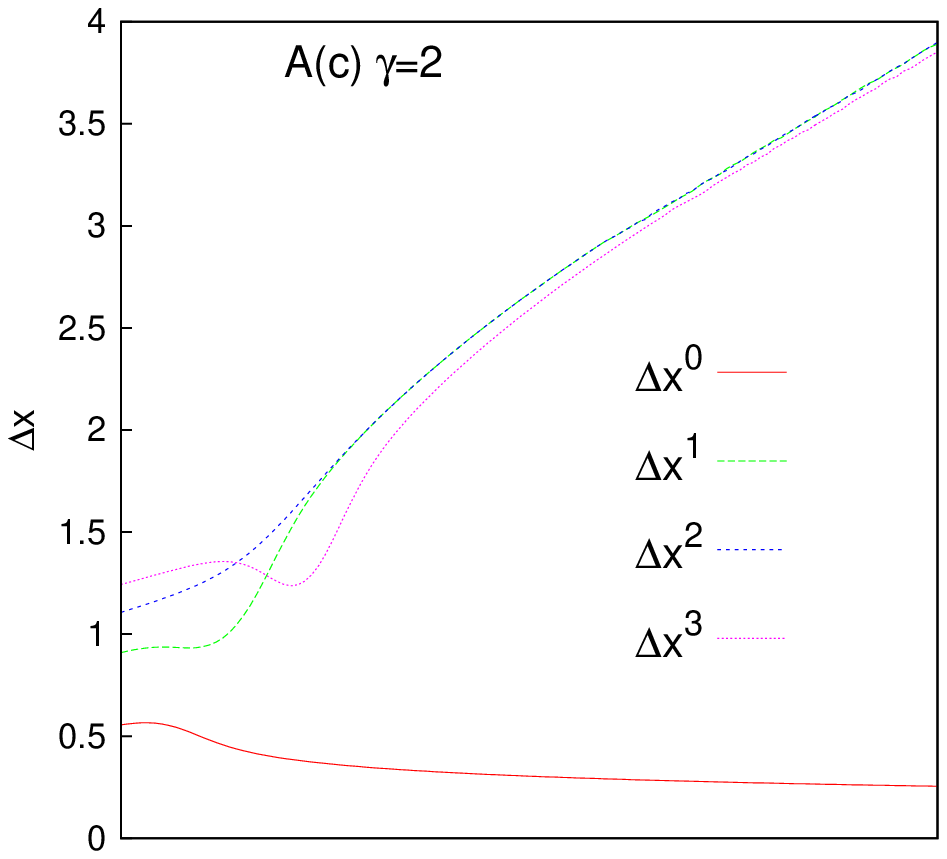}
\end{minipage}\hspace{0.1in}
\begin{minipage}[c]{0.18\textwidth}\centering
\includegraphics[scale=0.30]{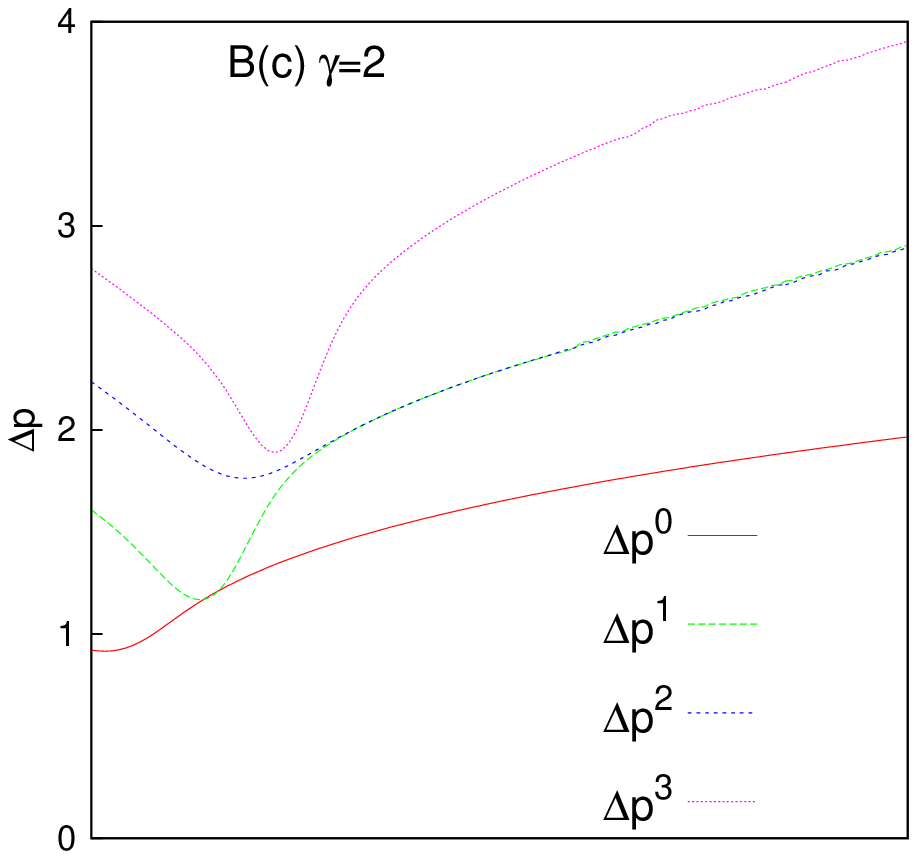}
\end{minipage}\hspace{0.1in}
\begin{minipage}[c]{0.18\textwidth}\centering
\includegraphics[scale=0.30]{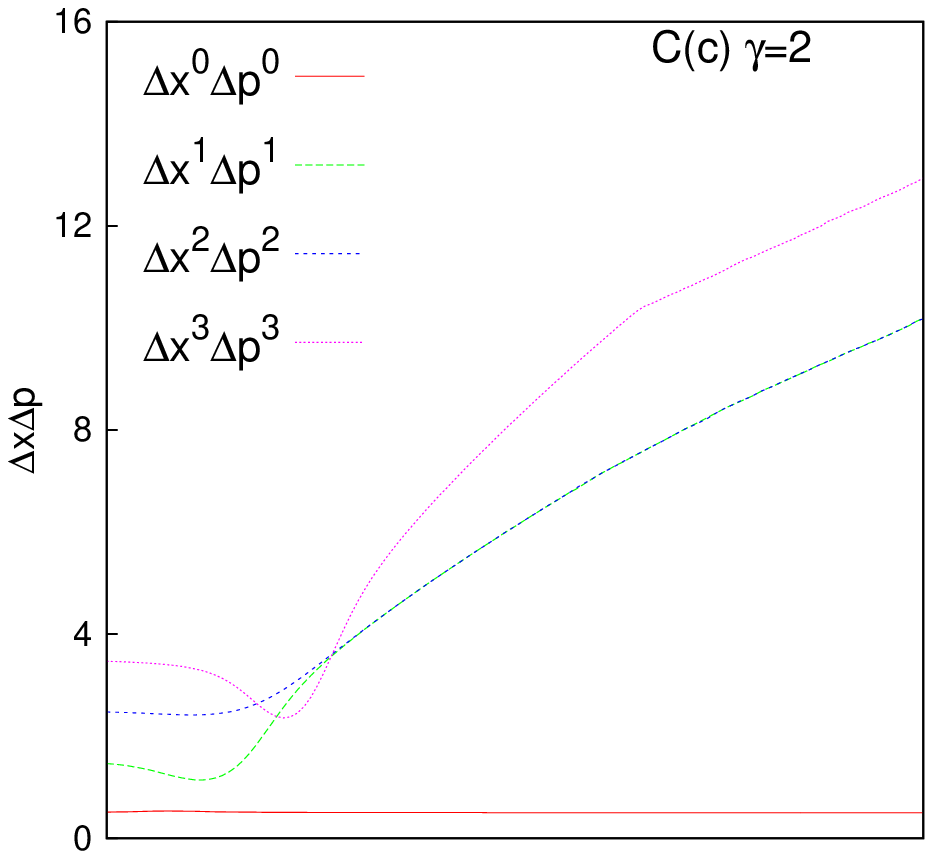}
\end{minipage}\hspace{0.1in}
\\[1pt]
\begin{minipage}[c]{0.18\textwidth}\centering
\includegraphics[scale=0.30]{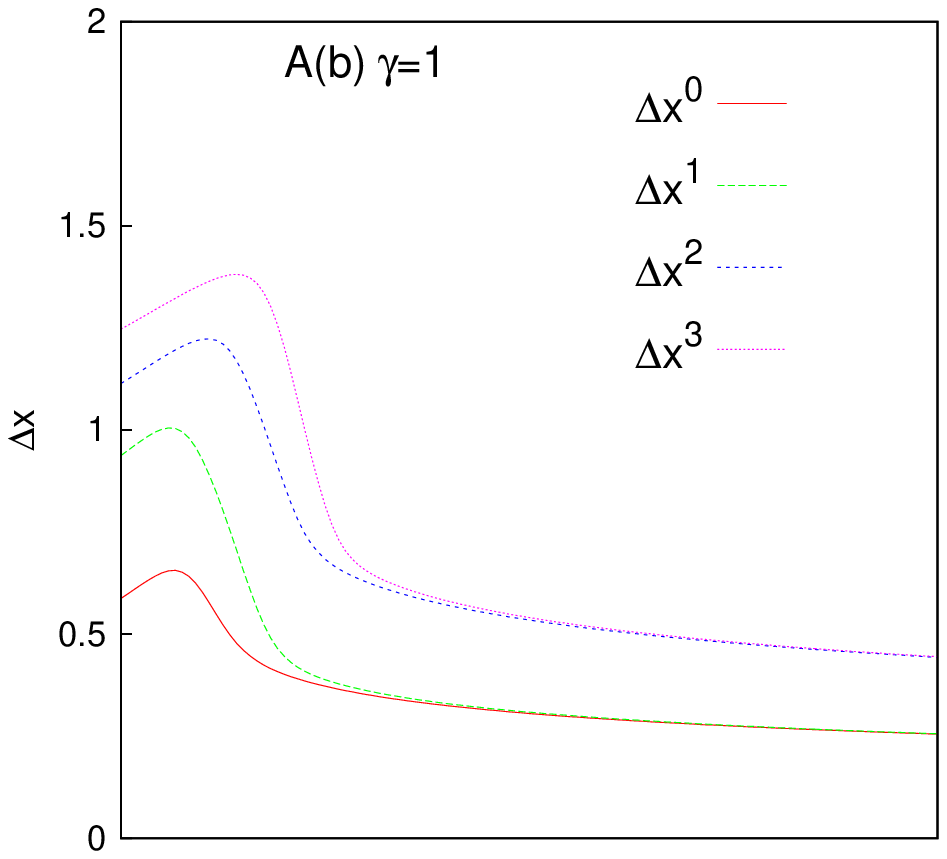}
\end{minipage}\hspace{0.1in}
\begin{minipage}[c]{0.18\textwidth}\centering
\includegraphics[scale=0.30]{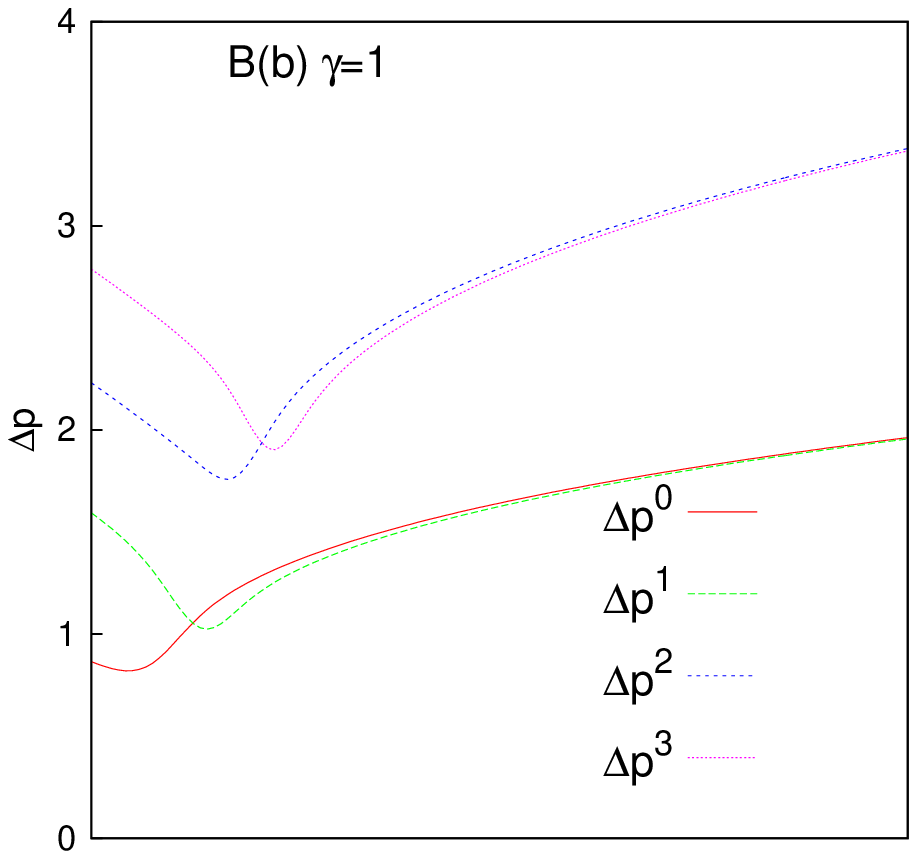}
\end{minipage}\hspace{0.1in}
\begin{minipage}[c]{0.18\textwidth}\centering
\includegraphics[scale=0.30]{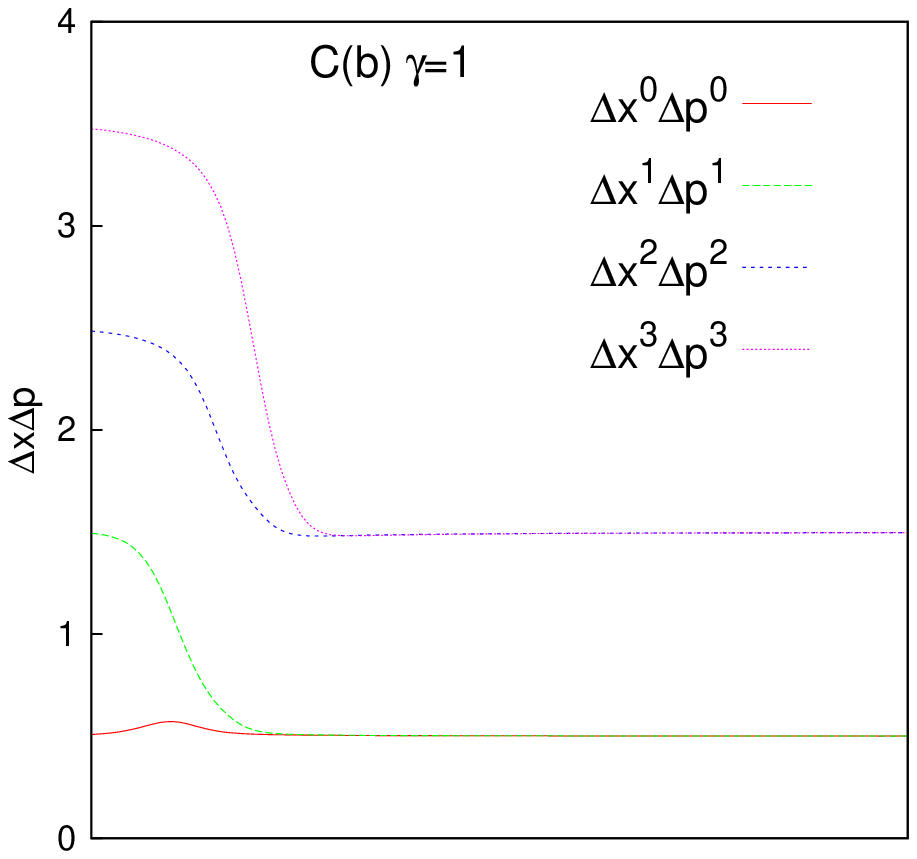}
\end{minipage}\hspace{0.1in}
\\[1pt]
\begin{minipage}[c]{0.18\textwidth}\centering
\includegraphics[scale=0.30]{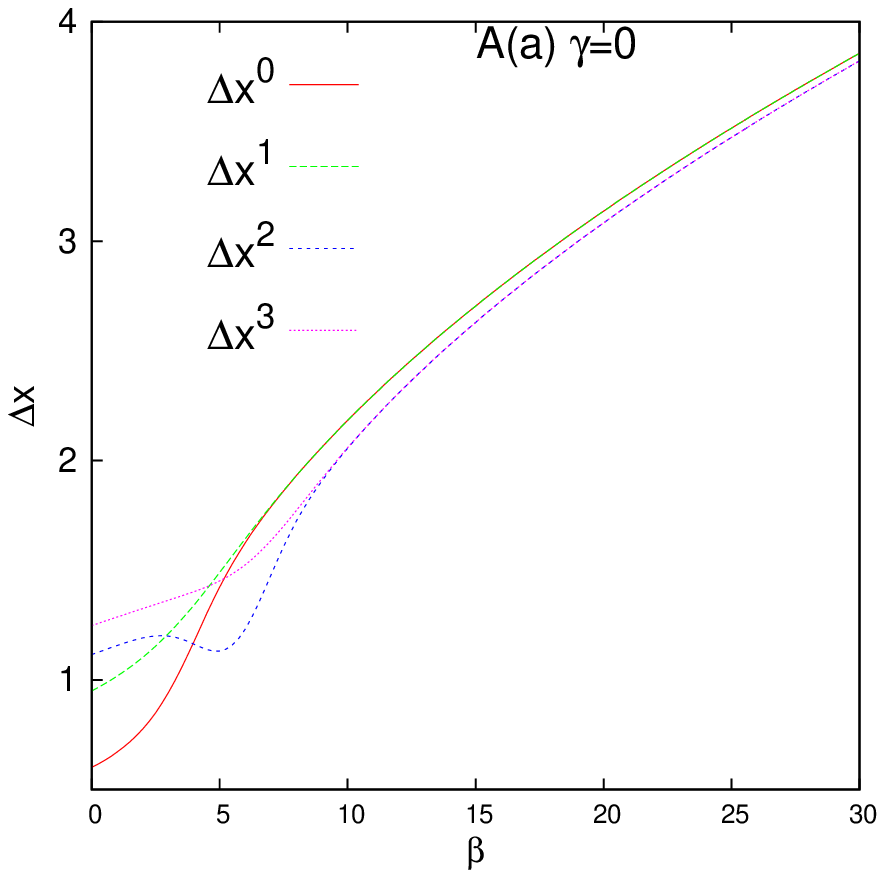}
\end{minipage}\hspace{0.1in}
\begin{minipage}[c]{0.18\textwidth}\centering
\includegraphics[scale=0.30]{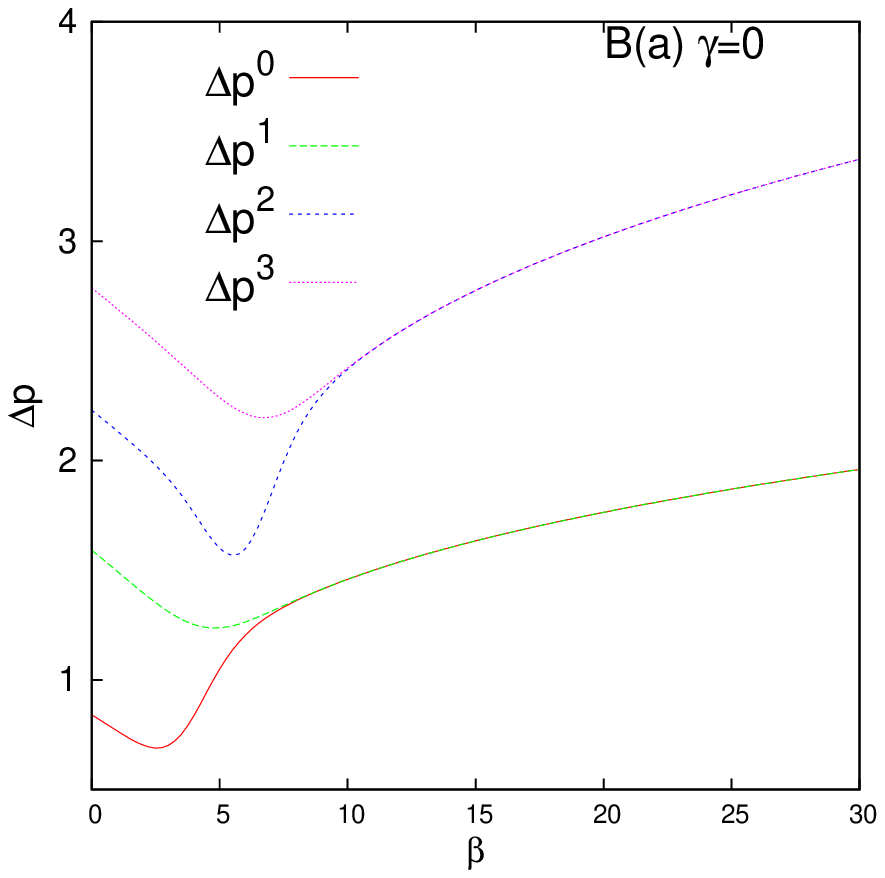}
\end{minipage}\hspace{0.1in}
\begin{minipage}[c]{0.18\textwidth}\centering
\includegraphics[scale=0.30]{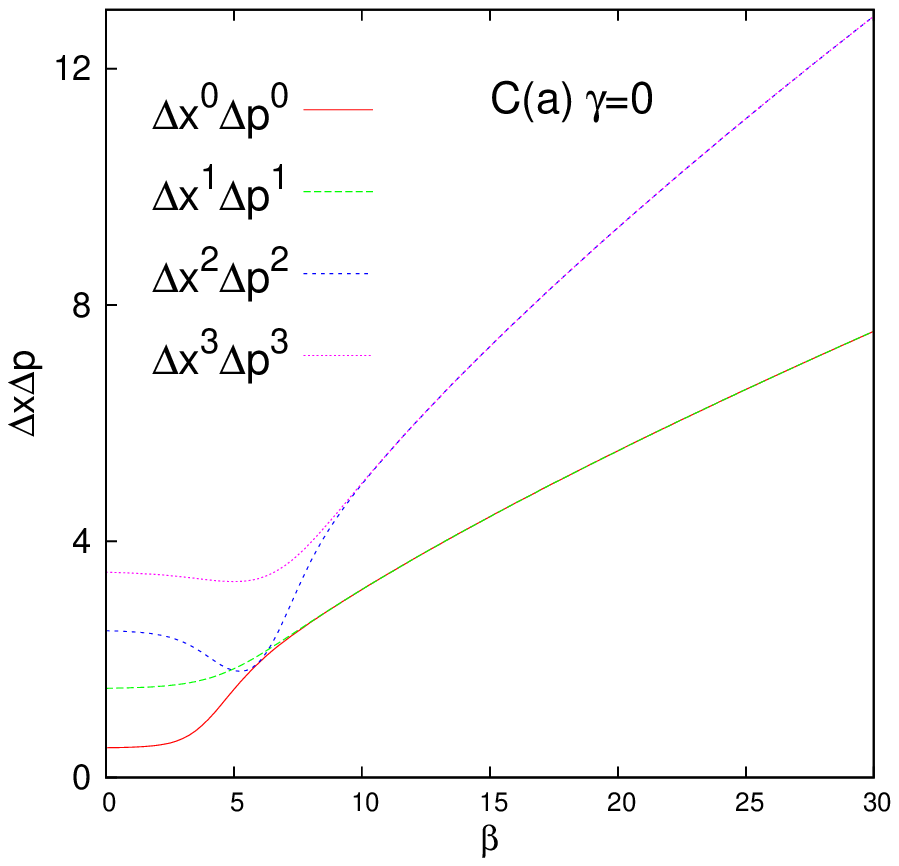}
\end{minipage}\hspace{0.1in}
\caption[optional]{$\Delta x$ (A), $\Delta p$ (B), $\Delta x \Delta p$ (C) for first four states, in left, middle, right
columns, plotted against $\beta$, for asymmetric DW potential, in Eq~(4) keeping $\alpha$ fixed at 1. Eight panels 
(a)-(h), in each column refer to eight $\gamma$, namely, 0,1,2,3,4,5,6,7 respectively. For more details, see text.}
\end{figure}

\subsection{Conventional uncertainty product}
Now we turn to an analysis of $\Delta x$ (A), $\Delta p$ (B) and $ \Delta x \Delta p$ (C) in top, middle and bottom  
rows in Fig.~(7). Four lowest states are depicted as functions of $\gamma$ at six $\beta$, 
in panels (a)-(f) for each row. It is easily noticed that in all excited 
states, slope for all these three properties change in an abrupt manner at an interval of $\Delta \gamma$ (integer $k$)
in consonance with the rule of distribution described earlier. After each interval of $\Delta \gamma$ the particle changes
position from one well to another; hence causing sudden variation in these measures. 

Variation of $\Delta x$ with $\gamma$ in first row implies confinement; at fractional $k$, particle is 
localized in a particular well, whereas integer $k$ signals distribution in both wells. Further it also indicates the  
existence of a critical transition point after which it settles in a specific well and never comes out. Extent of 
confinement is not same in all intervals of $\Delta \gamma$. In first interval $0 \! < \! \gamma \! < \! 2$ (or equivalently
at $0 \! < \! k \! < \! 1 $), $\Delta x$ for $n \! = \! 0$,1 is very close. Whereas in second interval 
$ \! 2 \! < \! \gamma \! < \! 4$ (or $ \! 1 \! < \! k \! < \! 2$), this closeness happens for $n \! = \! 0$,2. Likewise, 
in third interval $ 4 \! < \! \gamma \! < \! 6$ (or $ 2 \! < \! k \! < \! 3$), $\Delta x$ for $n \! = \! 0$ and 3 states
approach each other closely. This is due to the change in number of effective nodes in higher states. Reference to 
Table~V shows that, in these pair of states they are same (0,1,2 respectively) in three mentioned intervals (other nodes 
also lie within classical turning points; but they are located on other well where probability of finding the particle 
is insignificant). Every integer $k$ is characterized by a sharp peak in $\Delta x$. In the lowest state, critical
transition point is at $\gamma \! = \! 0$; therefore, 
only one such discontinuity in slope is observed. First and second excited state, on the other hand, show two and three such 
points where slope changes suddenly. This, again confirms the general rule: \emph{n-th excited state has (n+1)
such points where slope changes because particle makes n+1 such jumps before localizing in the larger well}. 

Change in $\Delta p$ (second row) with increase in $\gamma$ is also characterized by jumps at integer $k$, but there is no 
such spike as in $\Delta x$. Clearly, $n$th state has $n$ points where slope of $\Delta p$ vs $\gamma$ curve changes 
suddenly. In each state, however, there is a critical transition point after which $\Delta p$ assumes continuity. For 
all states, when slope of 
$\Delta x$ vs $\gamma$ plot increases, same of $\Delta p$ curve decreases and vice-versa. Thus, when the extent of confinement 
increases, positional uncertainty decreases and kinetic energy increases. 
Next, bottom row shows uncertainty product plots with respect to $\gamma$. Analogous to $\Delta x$ plots, here also,
at integer $k$, slope of the curve changes sharply indicating a transition of particle from one well to another, and 
$n$th state goes through $(n+1)$ points before complete localization in deeper well. Thus, in an asymmetric DW, traditional
uncertainty relations are capable of explaining confinement. 

\begin{figure}             
\centering
\begin{minipage}[c]{0.15\textwidth}\centering
\includegraphics[scale=0.28]{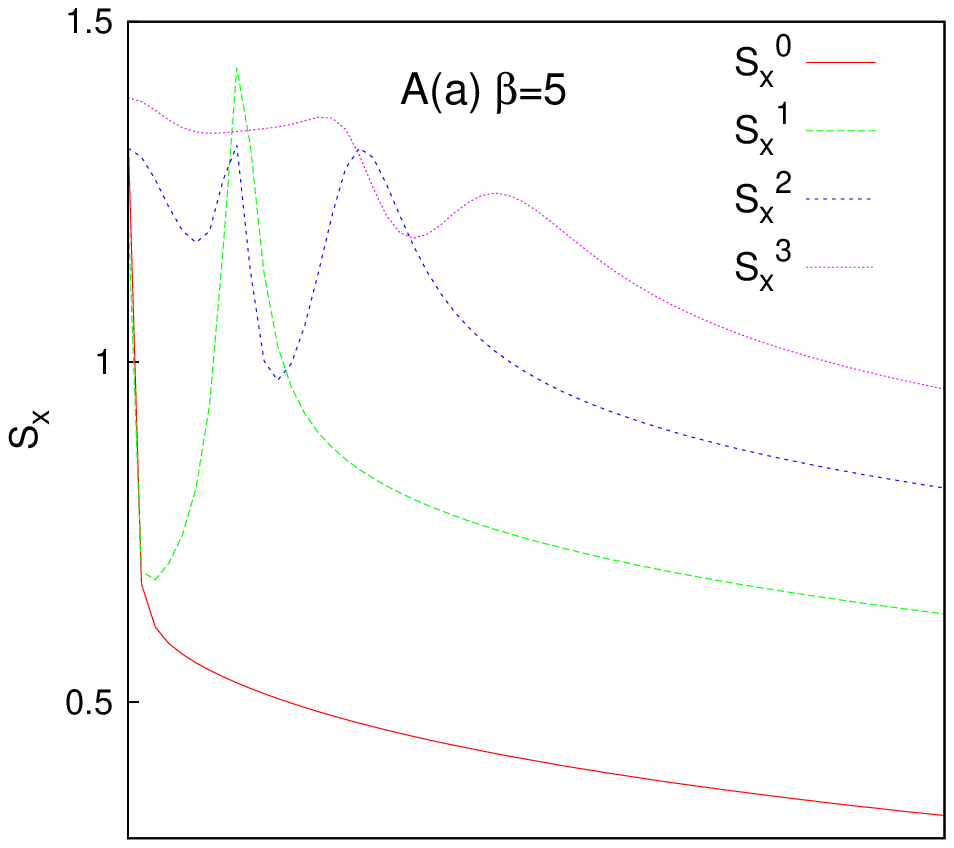}
\end{minipage}\hspace{0.08in}
\begin{minipage}[c]{0.15\textwidth}\centering
\includegraphics[scale=0.28]{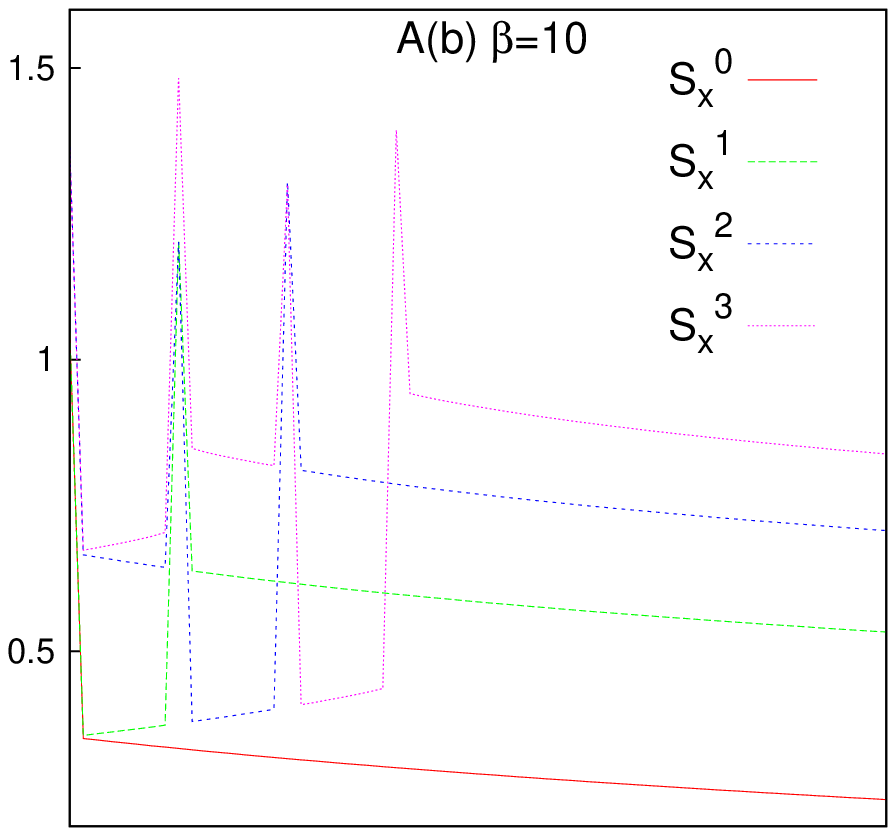}
\end{minipage}\hspace{0.08in}
\begin{minipage}[c]{0.15\textwidth}\centering
\includegraphics[scale=0.28]{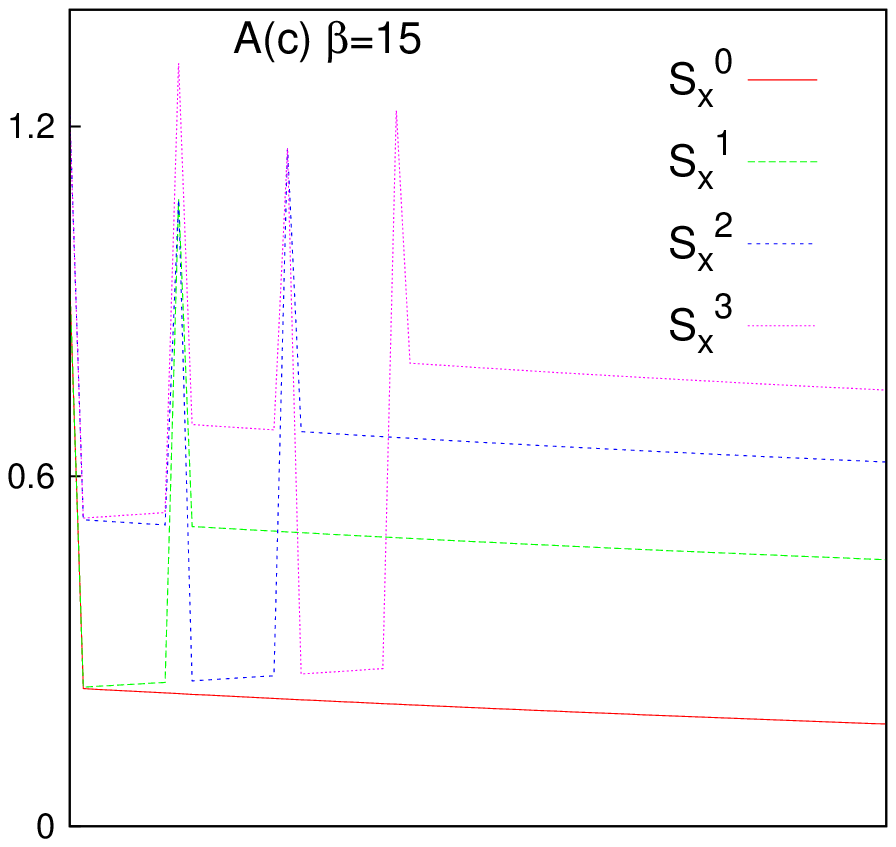}
\end{minipage}\hspace{0.08in}
\begin{minipage}[c]{0.15\textwidth}\centering
\includegraphics[scale=0.28]{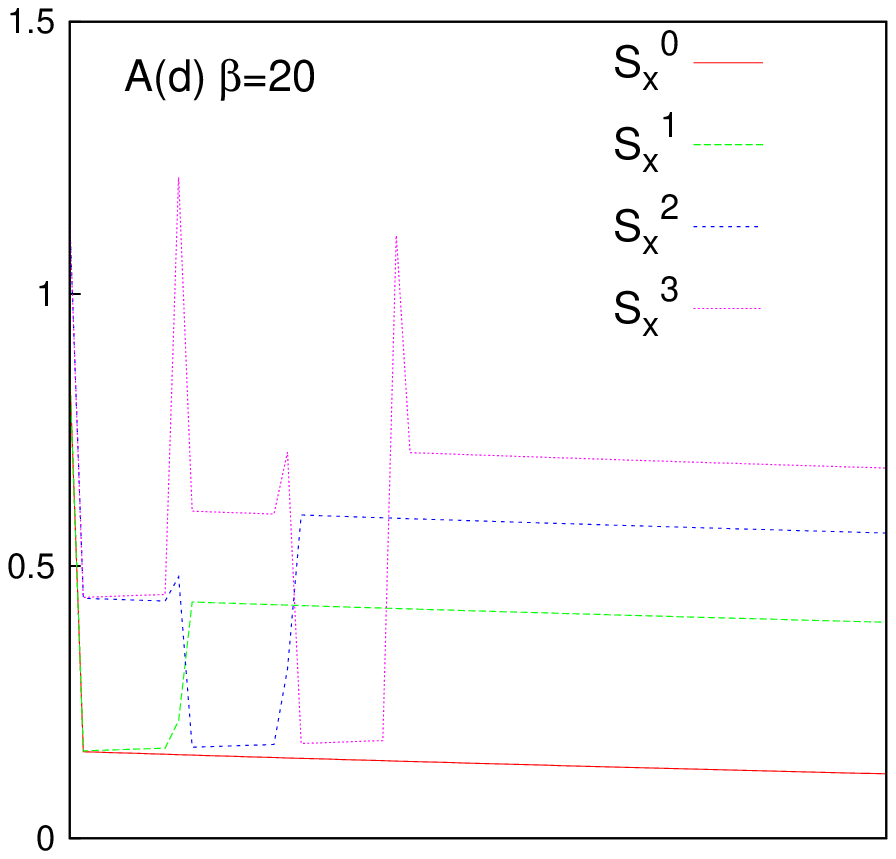}
\end{minipage}\hspace{0.08in}
\begin{minipage}[c]{0.15\textwidth}\centering
\includegraphics[scale=0.28]{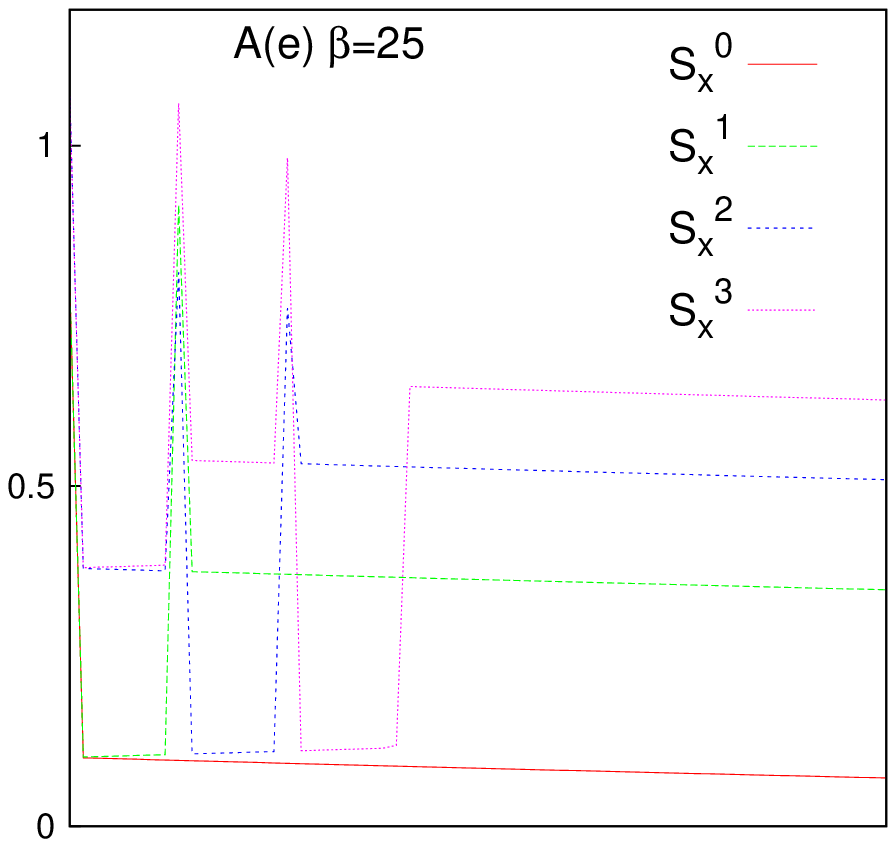}
\end{minipage}\hspace{0.08in}
\begin{minipage}[c]{0.15\textwidth}\centering
\includegraphics[scale=0.28]{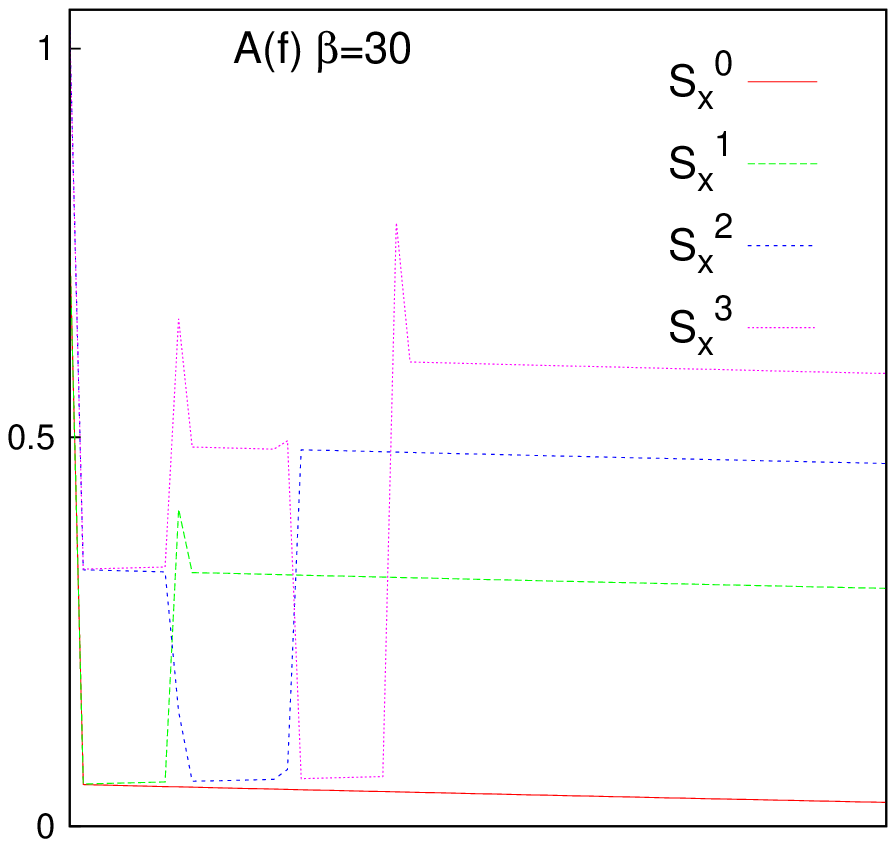}
\end{minipage}
\\[5pt]
\begin{minipage}[c]{0.15\textwidth}\centering
\includegraphics[scale=0.28]{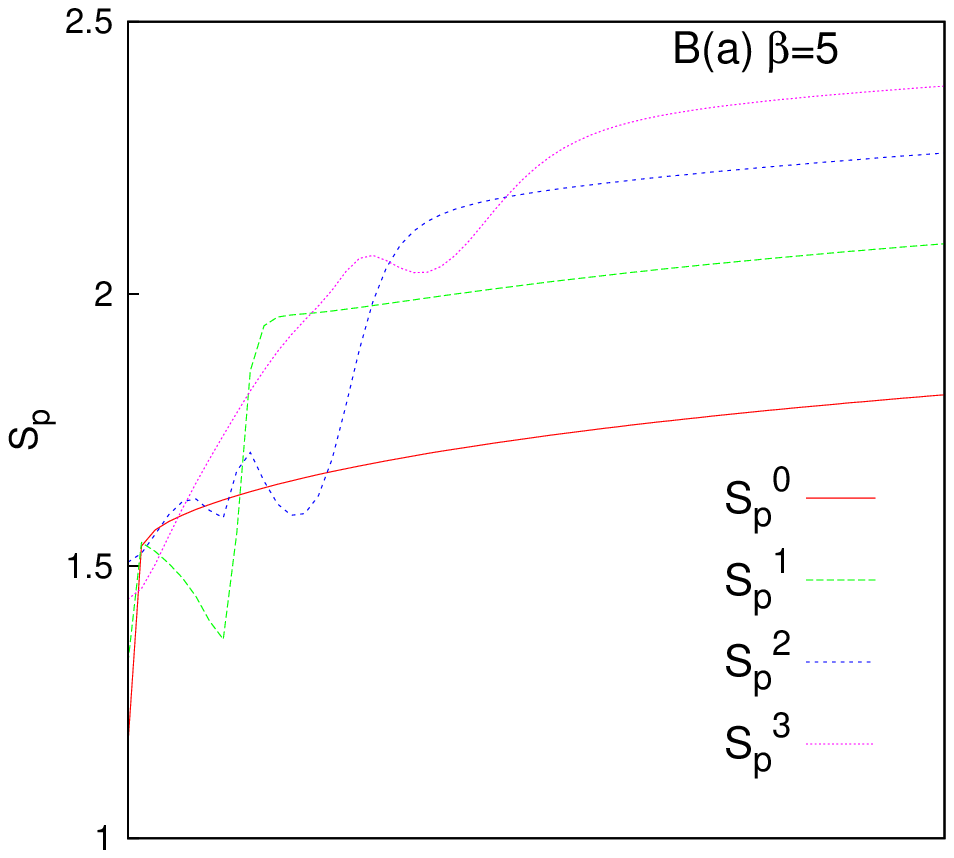}
\end{minipage}\hspace{0.08in}
\begin{minipage}[c]{0.15\textwidth}\centering
\includegraphics[scale=0.28]{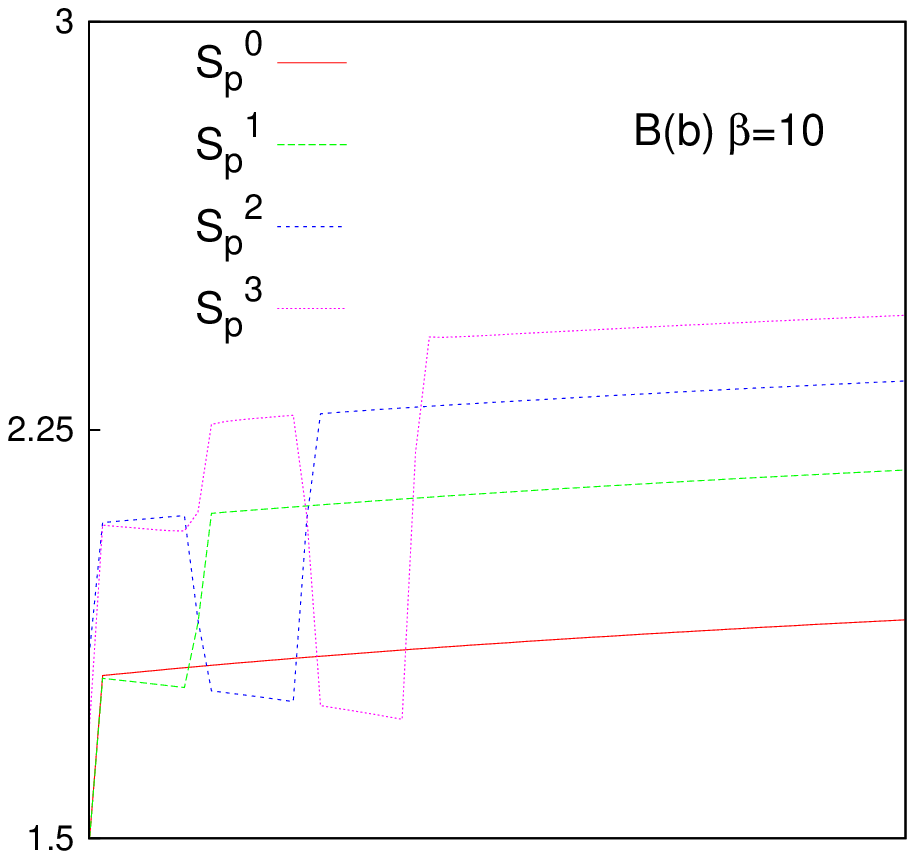}
\end{minipage}\hspace{0.08in}
\begin{minipage}[c]{0.15\textwidth}\centering
\includegraphics[scale=0.28]{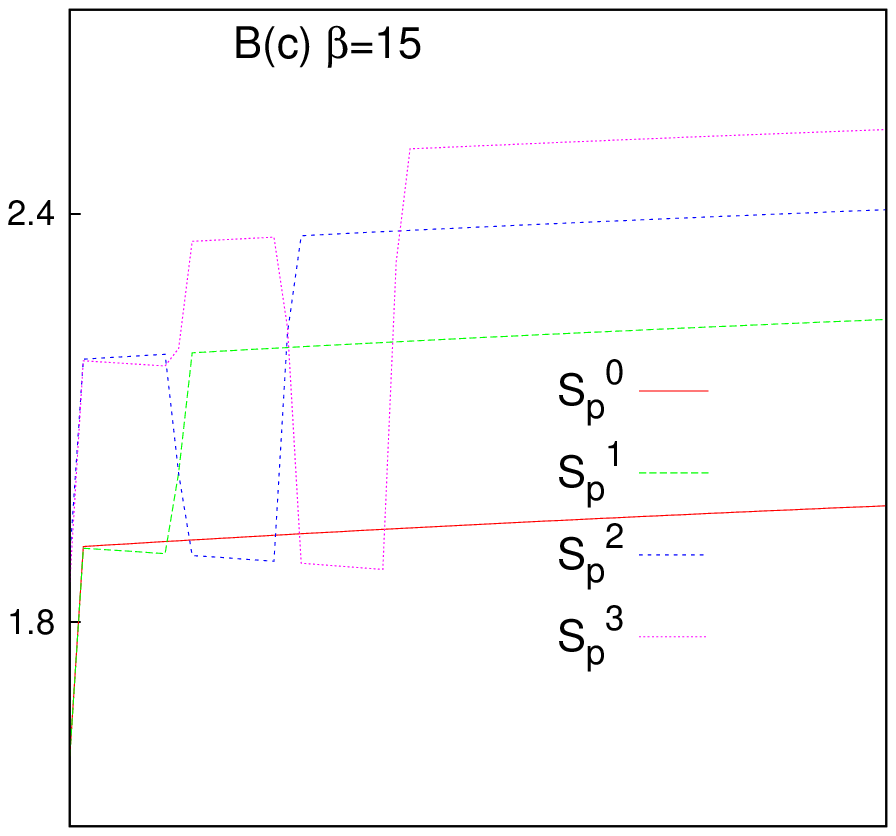}
\end{minipage}\hspace{0.08in}
\begin{minipage}[c]{0.15\textwidth}\centering
\includegraphics[scale=0.28]{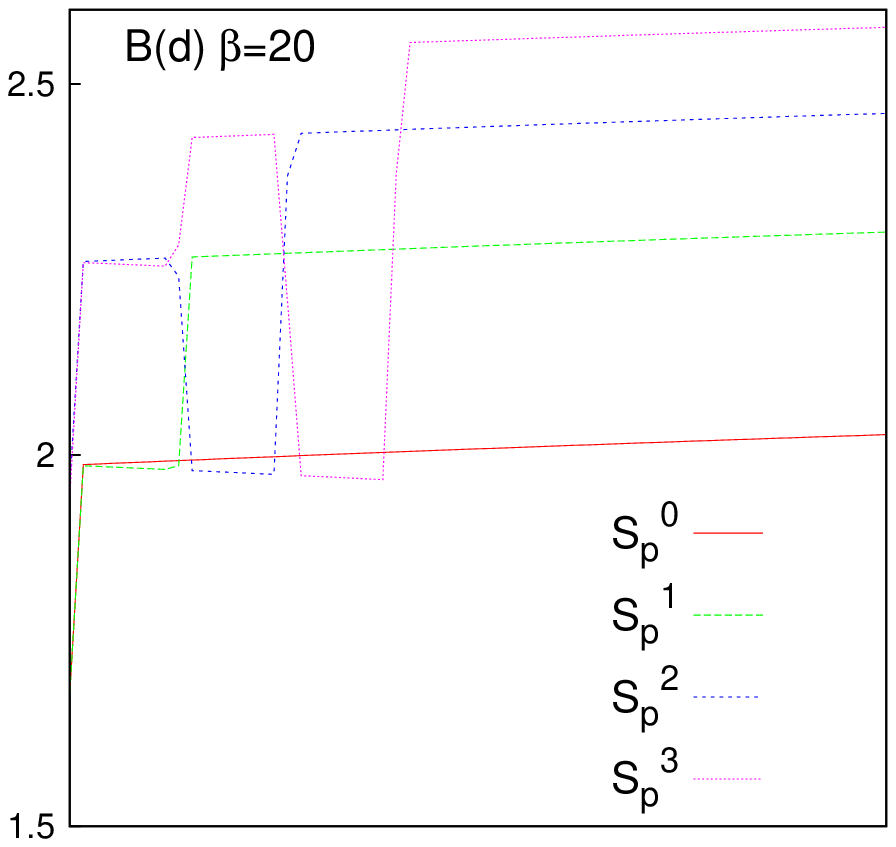}
\end{minipage}\hspace{0.08in}
\begin{minipage}[c]{0.15\textwidth}\centering
\includegraphics[scale=0.28]{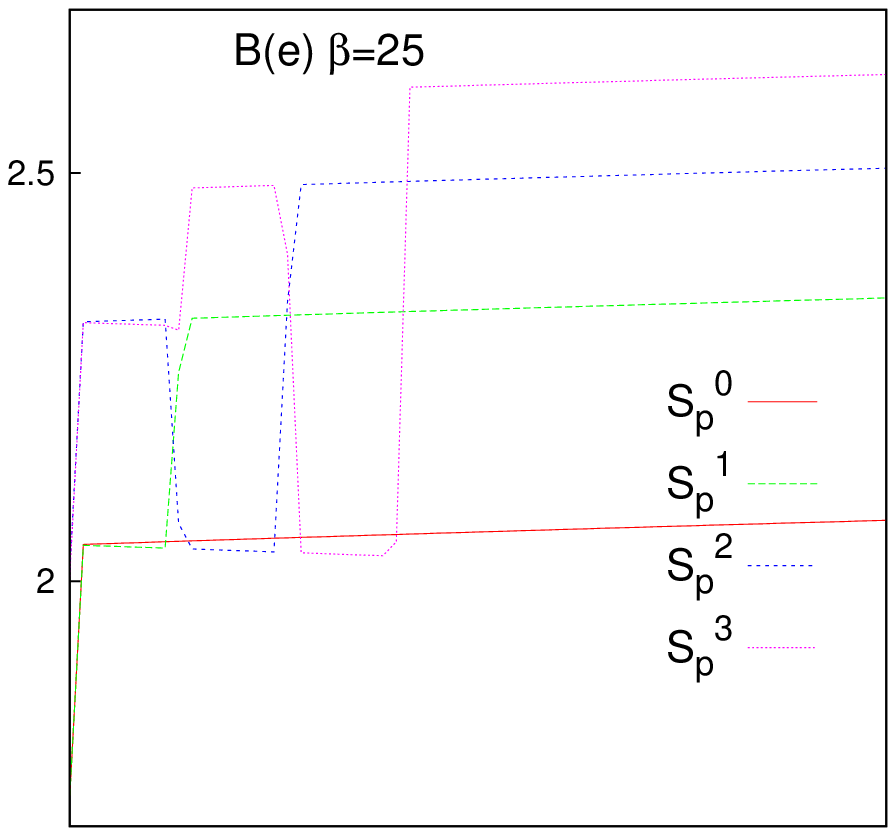}
\end{minipage}\hspace{0.08in}
\begin{minipage}[c]{0.15\textwidth}\centering
\includegraphics[scale=0.28]{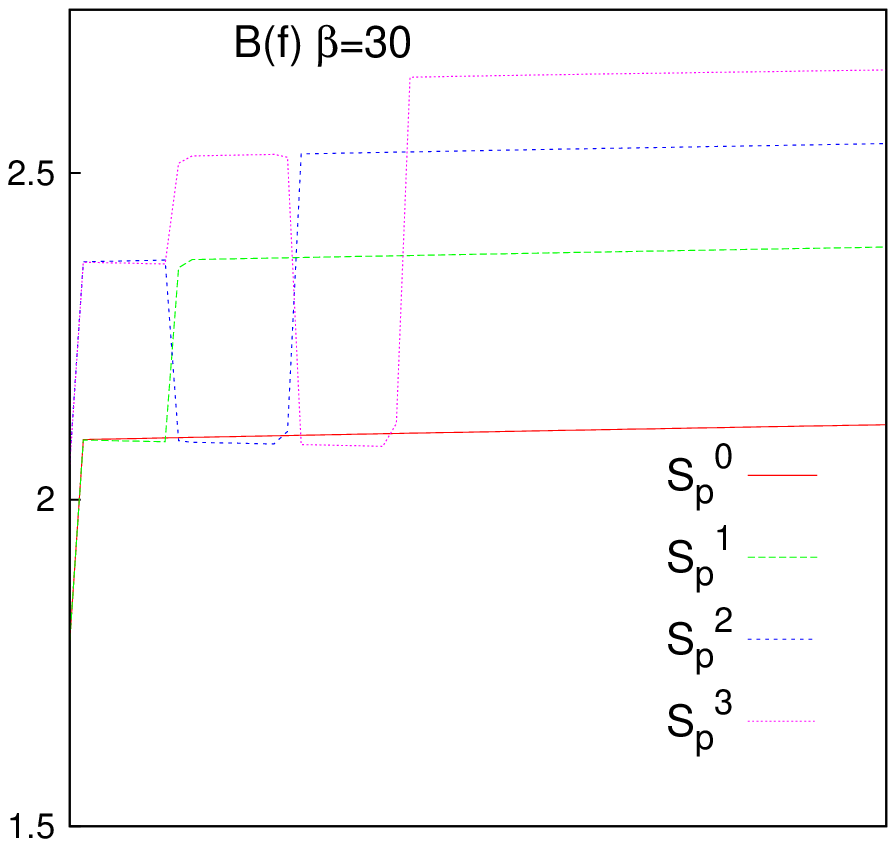}
\end{minipage}
\\[5pt]
\begin{minipage}[c]{0.15\textwidth}\centering
\includegraphics[scale=0.3]{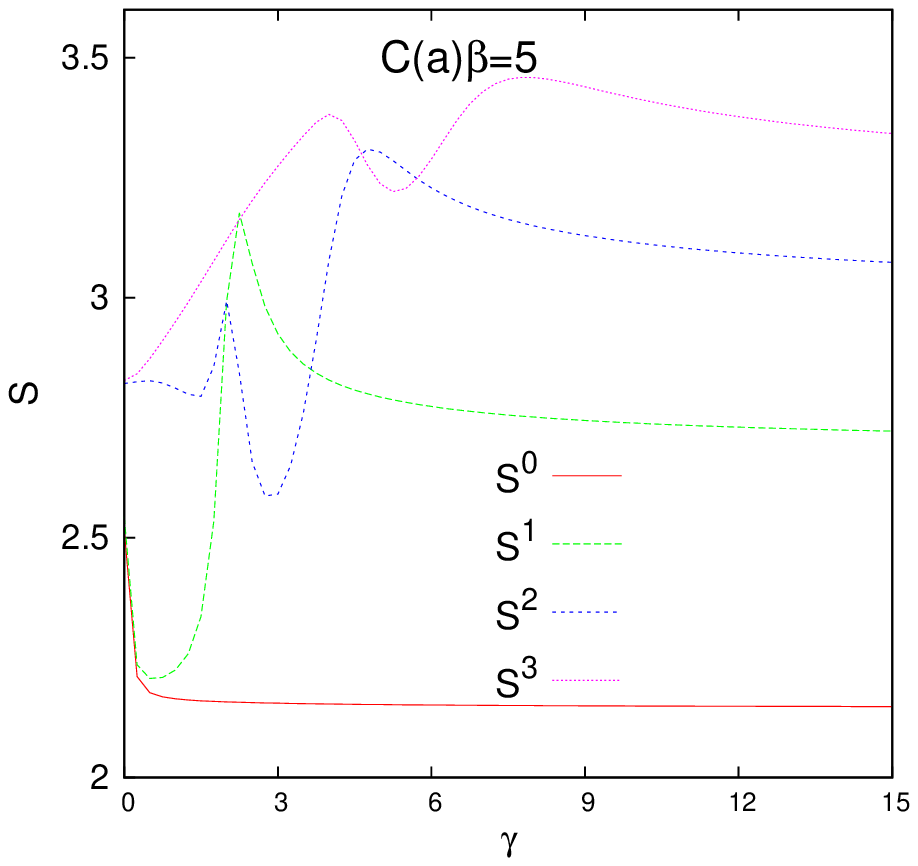}
\end{minipage}\hspace{0.08in}
\begin{minipage}[c]{0.15\textwidth}\centering
\includegraphics[scale=0.3]{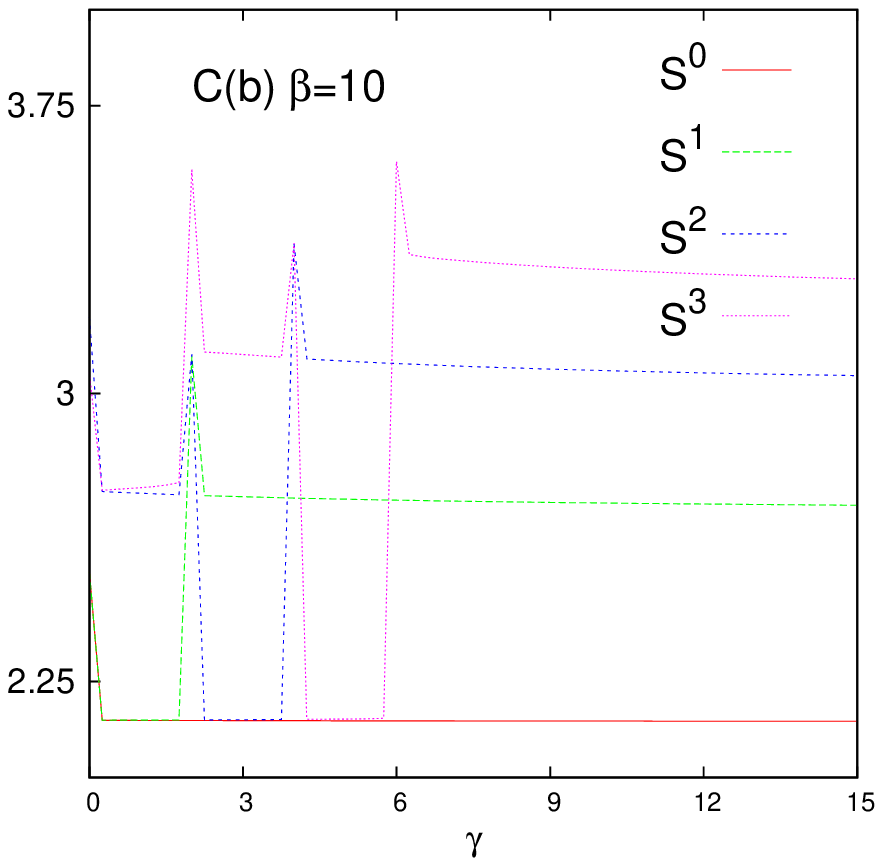}
\end{minipage}\hspace{0.08in}
\begin{minipage}[c]{0.15\textwidth}\centering
\includegraphics[scale=0.3]{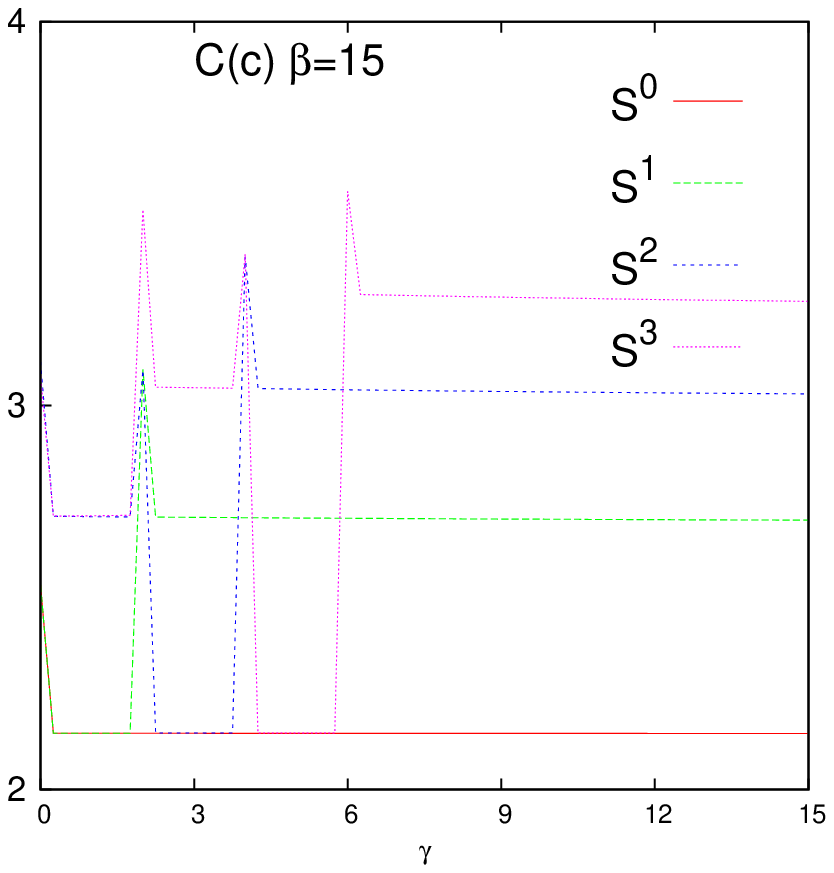}
\end{minipage}\hspace{0.08in}
\begin{minipage}[c]{0.15\textwidth}\centering
\includegraphics[scale=0.3]{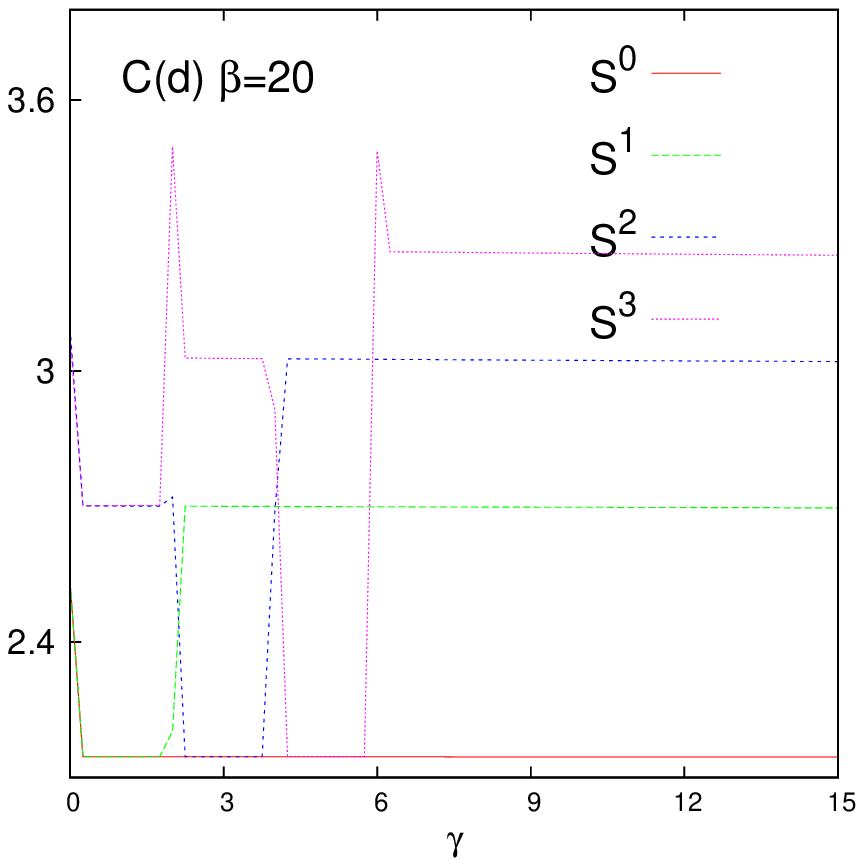}
\end{minipage}\hspace{0.08in}
\begin{minipage}[c]{0.15\textwidth}\centering
\includegraphics[scale=0.3]{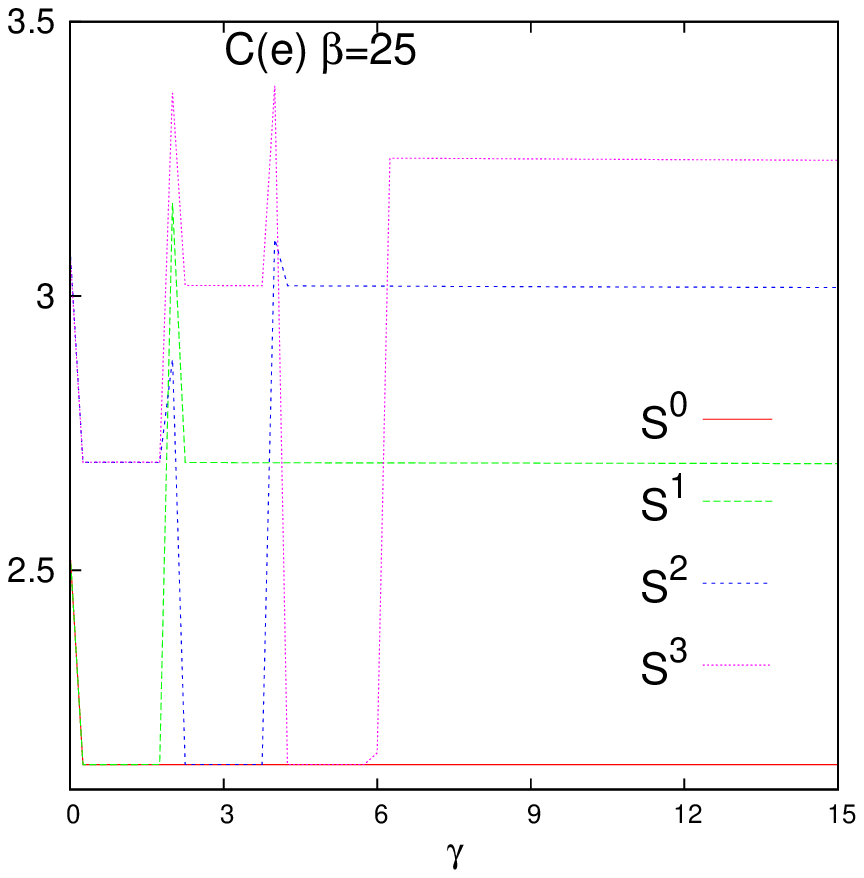}
\end{minipage}\hspace{0.08in}
\begin{minipage}[c]{0.15\textwidth}\centering
\includegraphics[scale=0.3]{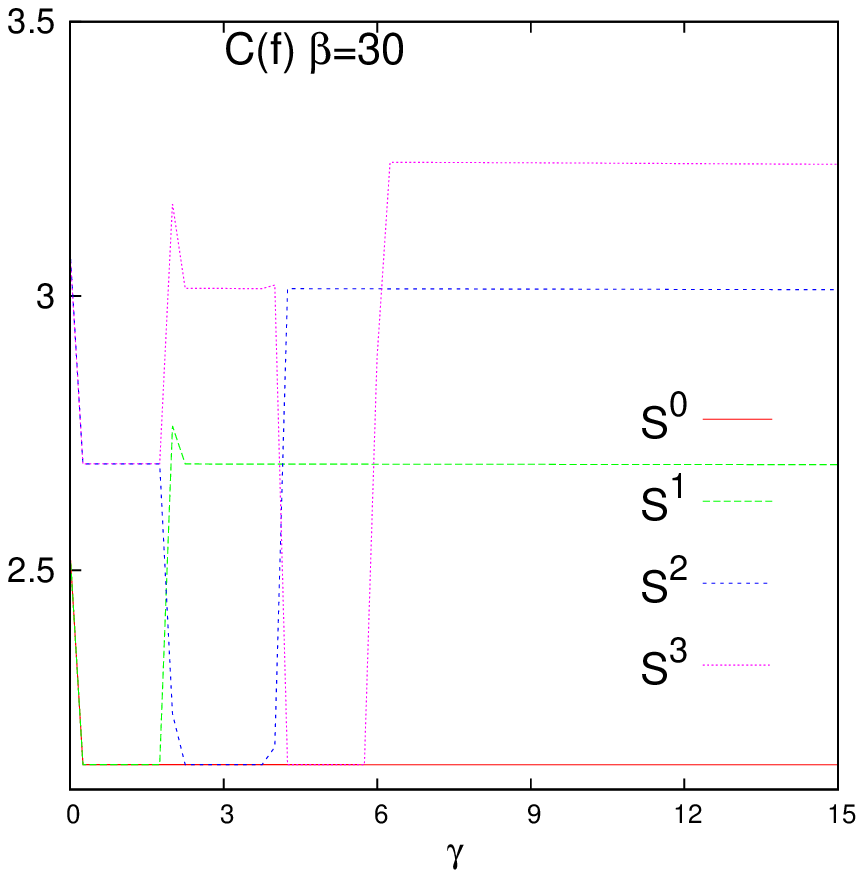}
\end{minipage}
\caption[optional]{$S_x$ (A), $S_p$ (B) and $S_x S_p$ (C) for first four states, in top, middle and 
bottom rows, plotted against $\gamma$, for asymmetric DW potential, in Eq~(4) keeping $\alpha$ fixed at 1. Six panels 
(a)-(f), in each row refer to six $\beta$, namely, 5,10,15,20,25,30 respectively. For more details, see text.}
\end{figure}

We now examine variation of uncertainty measures $\Delta x $ (A), $\Delta p$ (B), $\Delta x \Delta p$ (C) as functions 
of $\beta$ in left, middle, right columns in Fig.~(8). These are considered for eight $\gamma$ in panels (a)-(h). It is seen 
that, at fractional $k$ (odd $\gamma$), i.e., panels A(b), A(d), A(f), A(h), $\Delta x$ increases at first, attains a 
maximum and then decreases. Position of these maxima progressively shifts to right 
as $n$ increases; this also causes peak heights to increase. Moreover, these plots tend to flatten with $\gamma$. 
After initial delocalization, the two opposing effects of $\beta$ balance at the maximum and then localization is favored.   
At $0 \! < \! k \! < \! 1$, i.e., panel A(b), $\Delta x$ merge for $n \! = \! 0$,1 and $n \! = \! 2$,3 after certain 
$\beta$. Here, each well functions as two different potentials. In this interval $n \! = \! 1$ acts as ground state of 
well~II whereas $n \! = \! 2$,3 operate as first excited state of wells~I,II respectively (refer to columns 
2,3 in Table~V).  Similarly, at $1 \! < \! k \! < \! 2$, $\Delta x$ for $n \! = \! $0,2 coalesce in A(d); this again occurs 
for $n \! = \! 0$,3 in range $ 2 \! < \! k \! < \! 3$, in A(e). Because, from Table~V, at these 
respective intervals, $n \! = \! 2$,3 act as lowest states of well~II. Interestingly, however, at integer $k$, 
$\Delta x$ behaves differently. These correspond to transition of a state from one well to 
another causing distribution in two wells. In this occasion, $\Delta x$ of two different states merge due to quasi-degeneracy. 
At $k \! = \! 0$, $\Delta x$ for $n \! = \! 0$,1 and $n \! = \! 2$,3 unite separately in A(a). In an analogous way, 
at $k \! = \! 1$, in A(c), $\Delta x$ for $n \! = \! 1$,2 coalesce, while at $k \! = \! 2$, 
$\Delta x$ for $n \! = \! 2$, 3 join in A(e). Now from middle row, we see that, $\Delta p$ decreases with $\beta$ 
at first, attains a minimum and then increases. For a given states, positions of these minima get left shifted with $\gamma$.   
For a fixed $\gamma$, they shift to right as state index increases. For higher $\gamma$, the minima tends to disappear; this 
$\gamma$ grows larger with $n$. On the contrary, $\Delta x \Delta p$ at integer $k$ ($\gamma \! = \! 0,2,4,6$) in C(a), C(c),
C(e), C(f) initially decrease until reaching a minimum and then continues to grow steadily. However, at fraction $k$ 
($\gamma \! = \! 1,3,5,7$) in C(b), C(d), C(f), C(g) it decreases, finally becoming flat after certain $\beta$. In latter 
case, one notices that $n$th 
state attains minima after ($n$+1)th transition points. For example, $n \! = \! 1$, after one transition point at C(c) for
$\gamma \! = \! 2$ ($k \! = \! 1$), localizes in C(d) for $k \! > \! 1$; $n \! = \! 2$ gives the minimum in C(f) after two 
transition points at $k \! = \! 1,2$; $n \! = \! 3$ shows minimum in C(h) after 3 such points for $k \! = \! 1,2,3$. 

Thus, on the basis of Figs.~(7), (8), it is clear that at fractional $k$, traditional uncertainties can   
explain confinement, where asymmetric effect dominates. But, at integer $k$, situation remains unclear; combined 
response of $\alpha$, $\beta$ overshadows asymmetric effect of $\gamma$ then.

\begin{figure}            
\centering
\begin{minipage}[c]{0.18\textwidth}\centering
\includegraphics[scale=0.30]{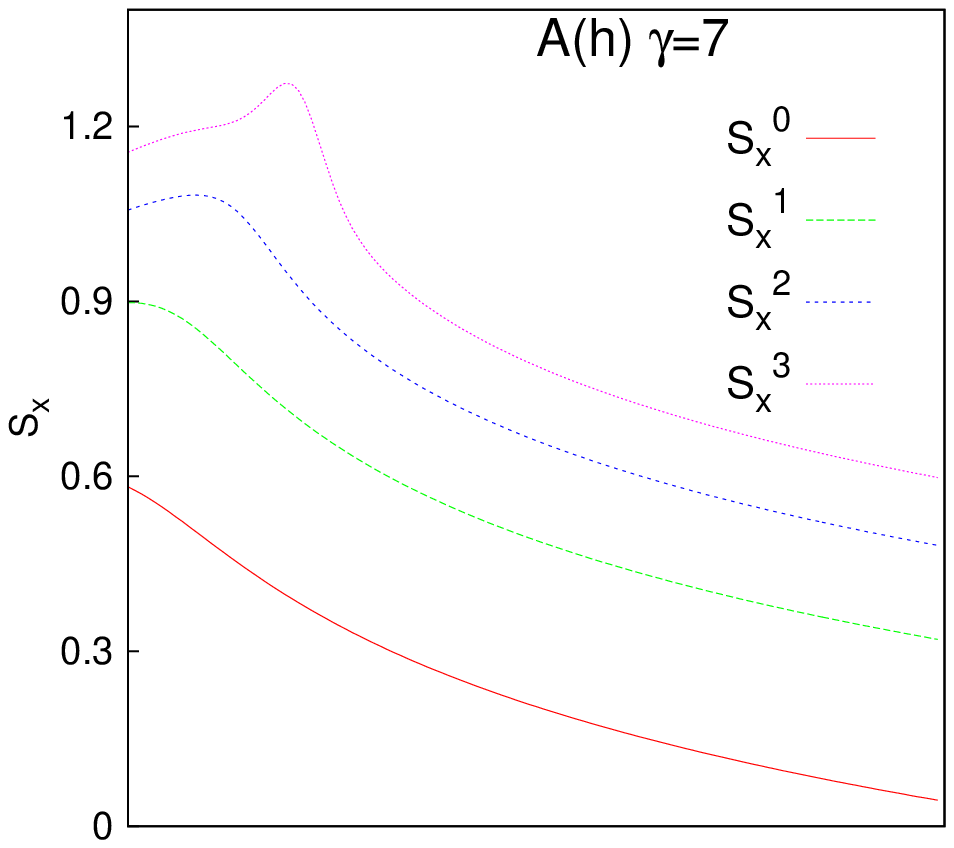}
\end{minipage}\hspace{0.10in}
\begin{minipage}[c]{0.18\textwidth}\centering
\includegraphics[scale=0.30]{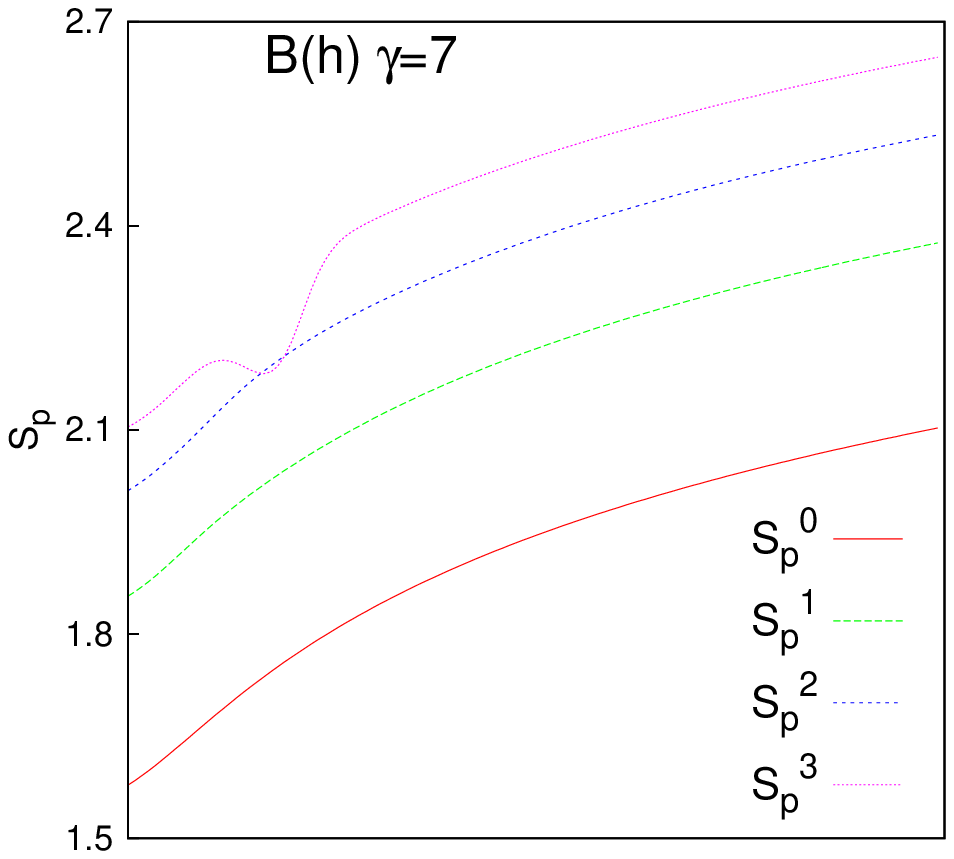}
\end{minipage}\hspace{0.10in}
\begin{minipage}[c]{0.18\textwidth}\centering
\includegraphics[scale=0.30]{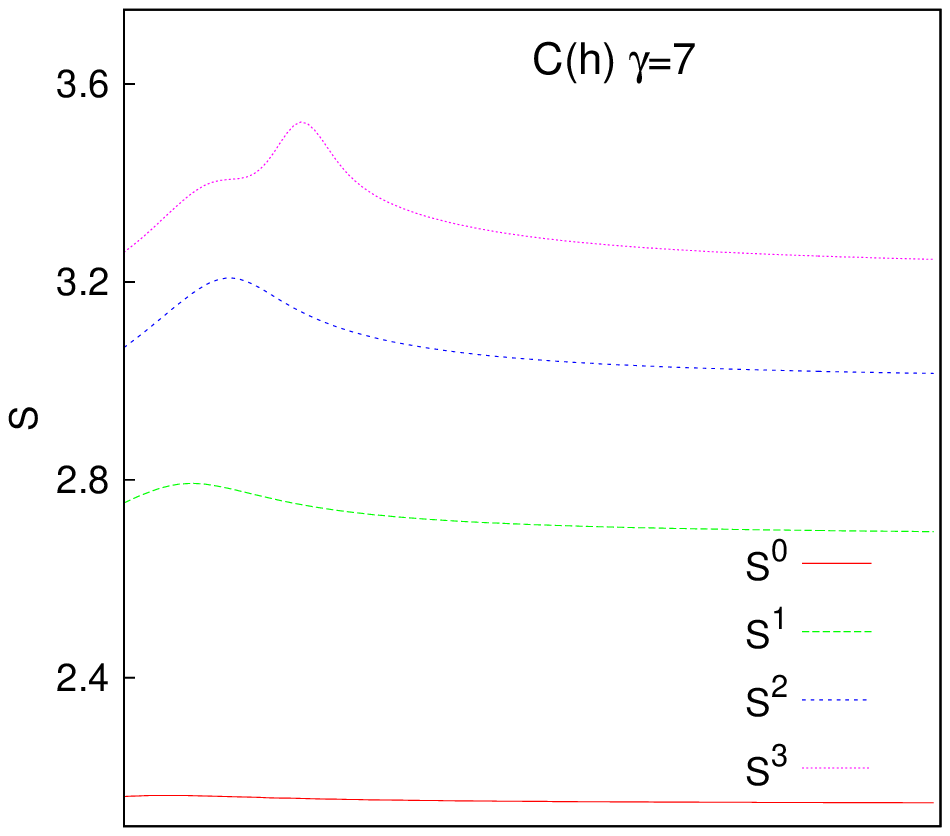}
\end{minipage}\hspace{0.10in}
\\[1pt]
\begin{minipage}[c]{0.18\textwidth}\centering
\includegraphics[scale=0.30]{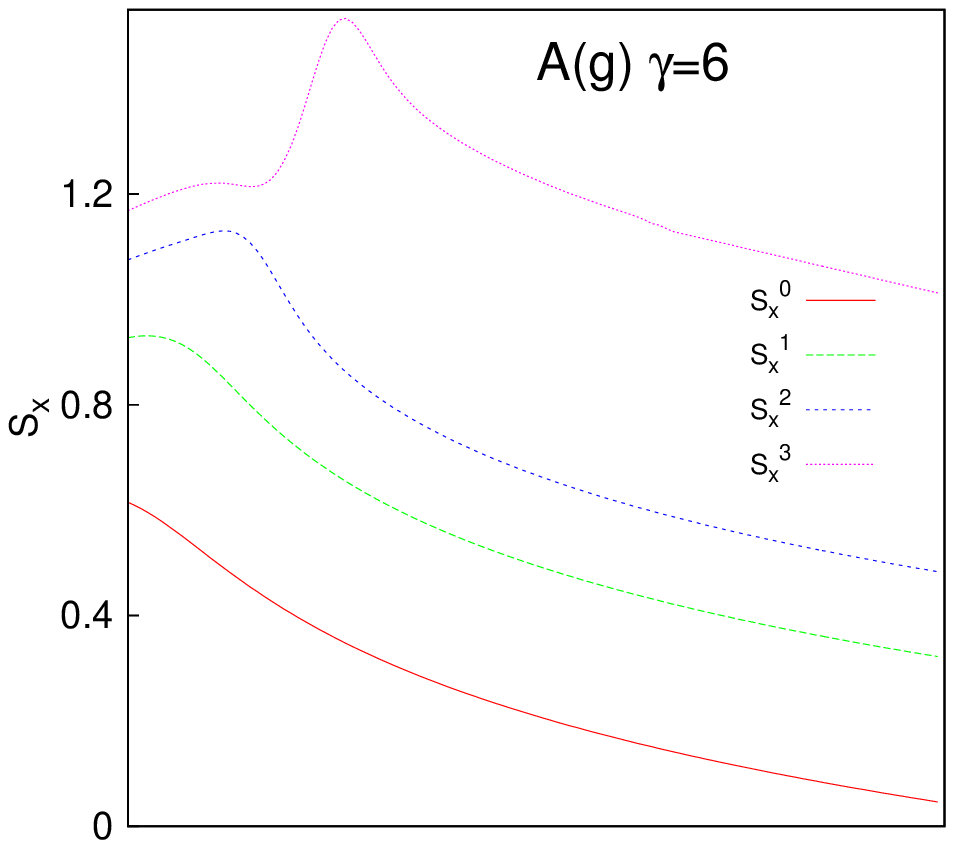}
\end{minipage}\hspace{0.10in}
\begin{minipage}[c]{0.18\textwidth}\centering
\includegraphics[scale=0.30]{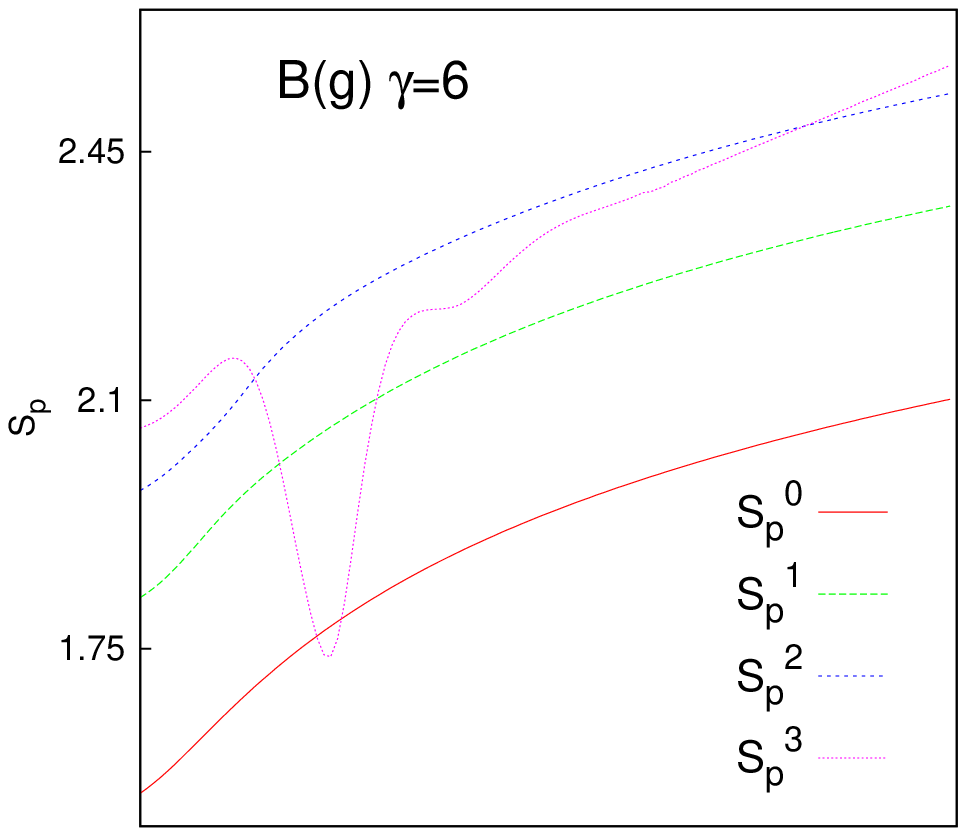}
\end{minipage}\hspace{0.10in}
\begin{minipage}[c]{0.18\textwidth}\centering
\includegraphics[scale=0.30]{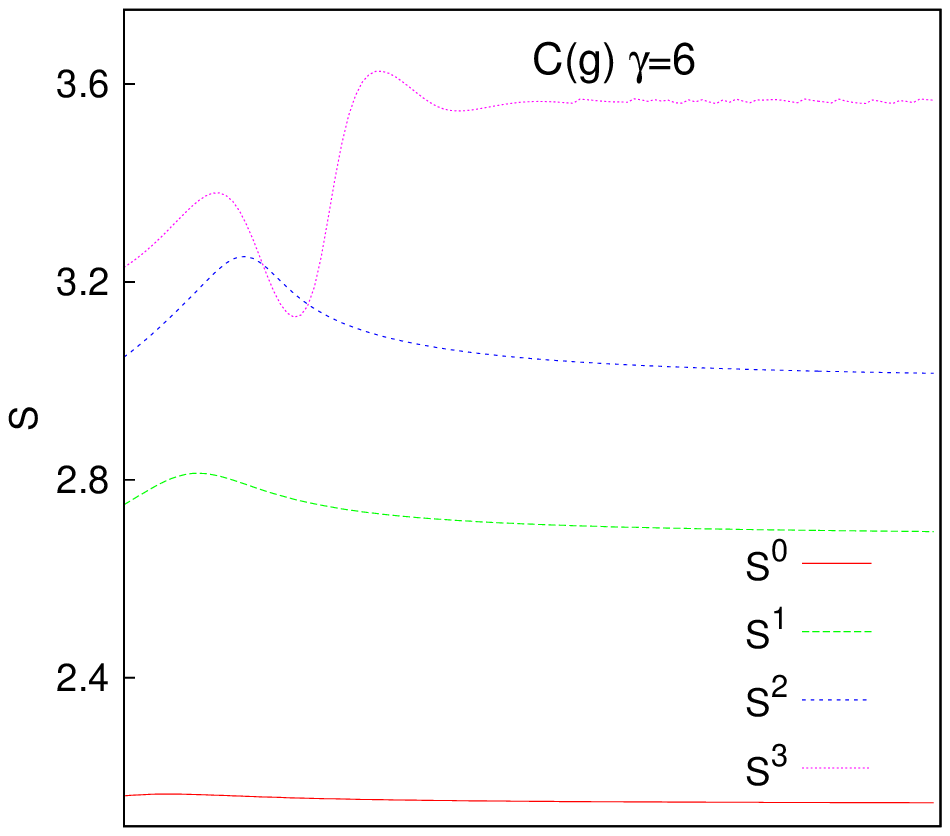}
\end{minipage}\hspace{0.10in}
\\[1pt]
\begin{minipage}[c]{0.18\textwidth}\centering
\includegraphics[scale=0.30]{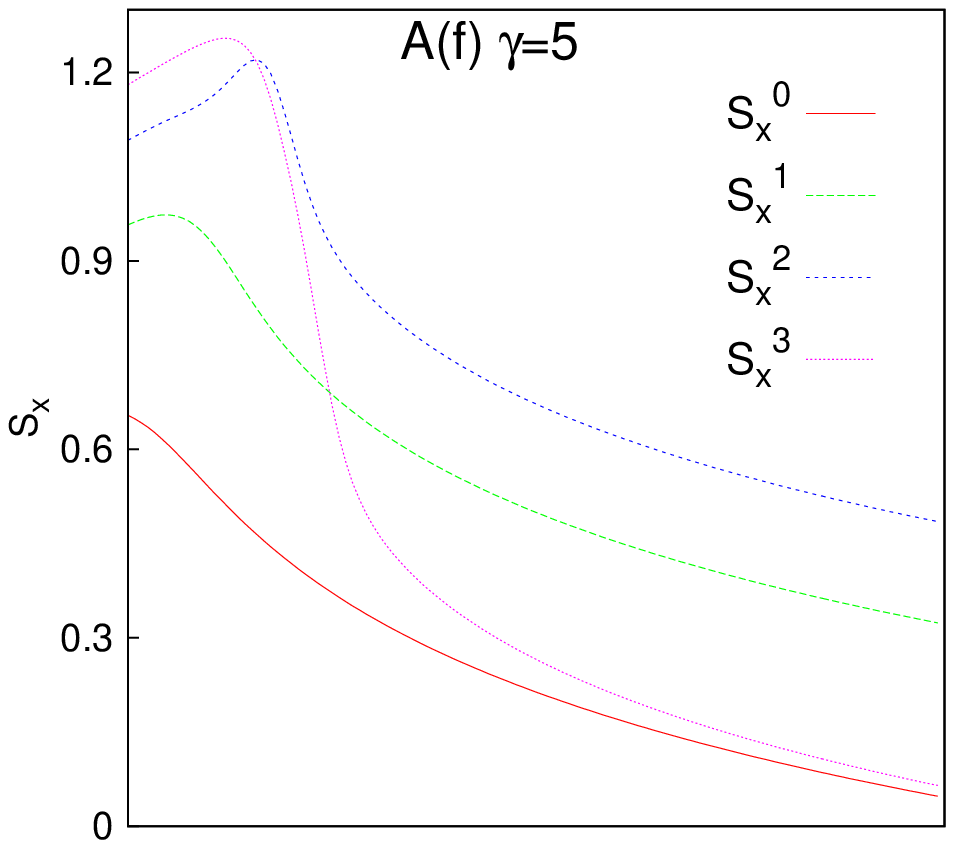}
\end{minipage}\hspace{0.10in}
\begin{minipage}[c]{0.18\textwidth}\centering
\includegraphics[scale=0.30]{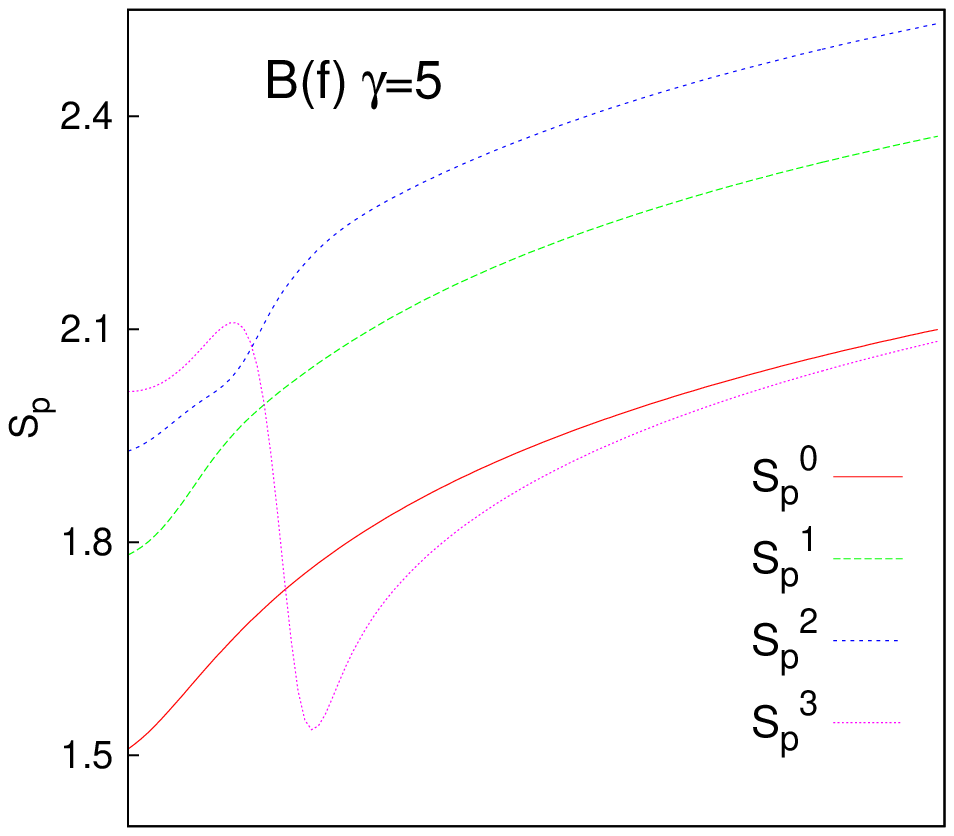}
\end{minipage}\hspace{0.10in}
\begin{minipage}[c]{0.18\textwidth}\centering
\includegraphics[scale=0.30]{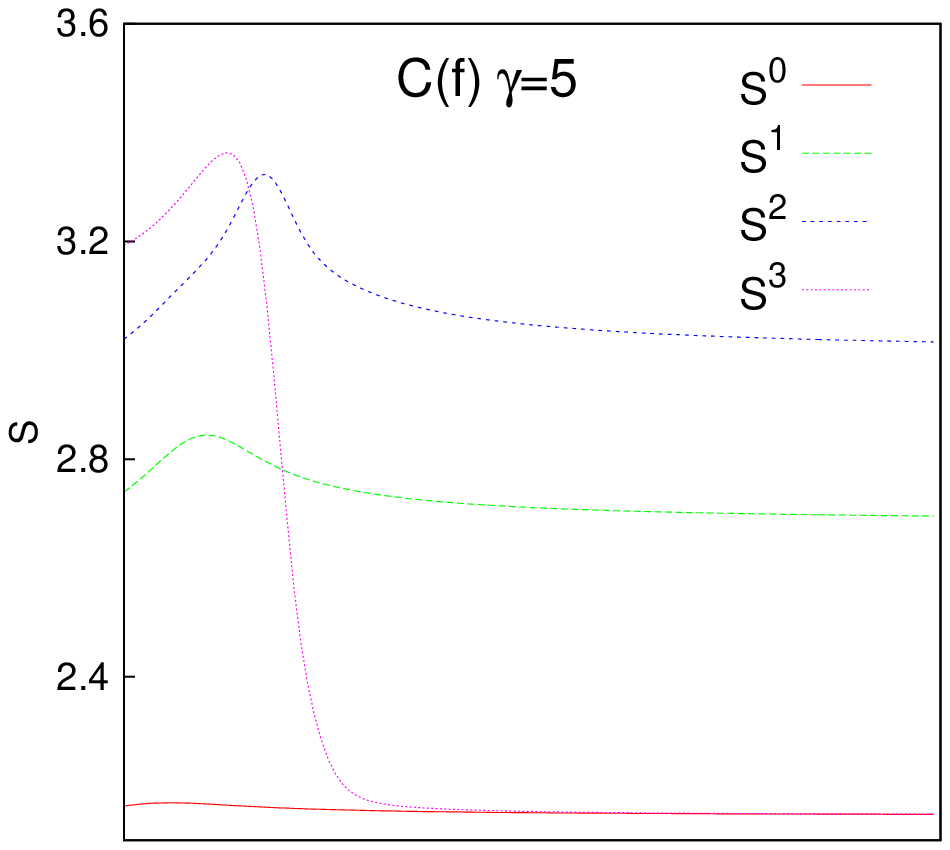}
\end{minipage}\hspace{0.10in}
\\[1pt]
\begin{minipage}[c]{0.18\textwidth}\centering
\includegraphics[scale=0.30]{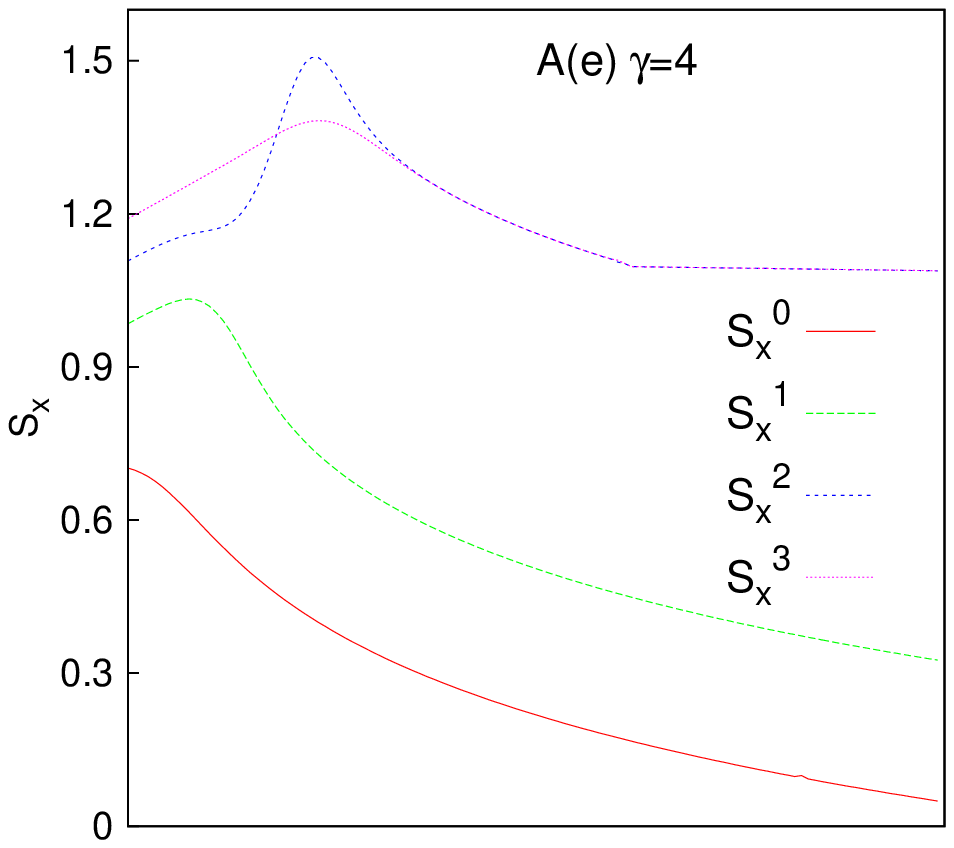}
\end{minipage}\hspace{0.10in}
\begin{minipage}[c]{0.18\textwidth}\centering
\includegraphics[scale=0.30]{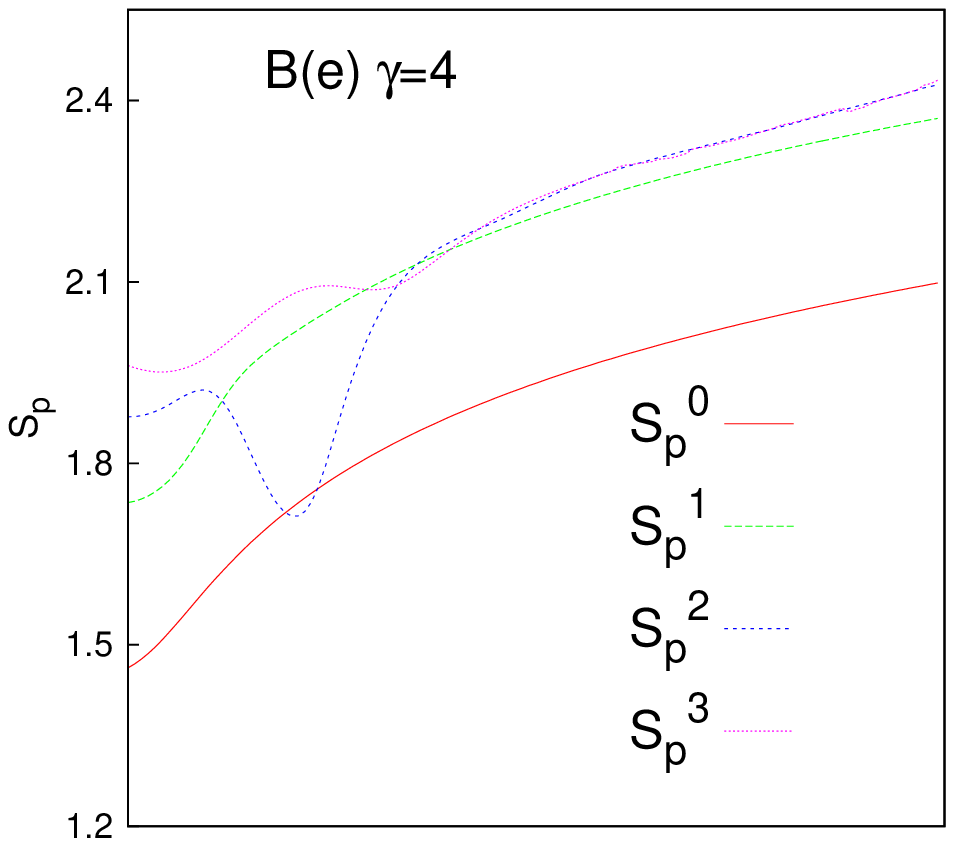}
\end{minipage}\hspace{0.10in}
\begin{minipage}[c]{0.18\textwidth}\centering
\includegraphics[scale=0.30]{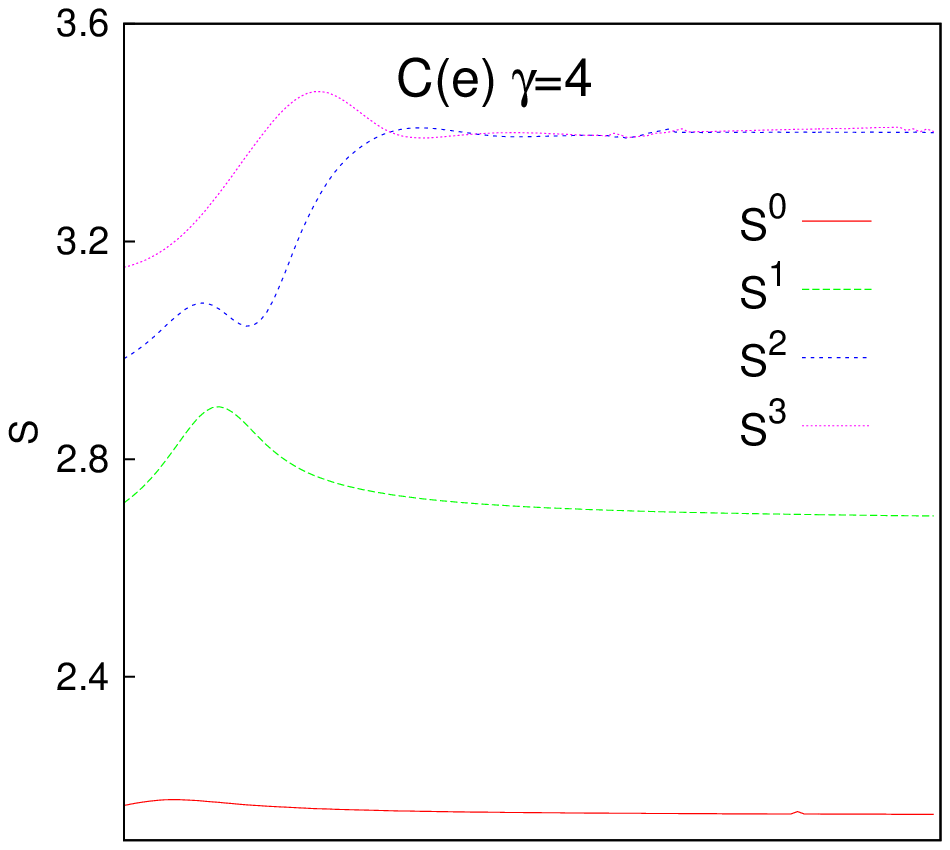}
\end{minipage}\hspace{0.10in}
\\[1pt]
\begin{minipage}[c]{0.18\textwidth}\centering
\includegraphics[scale=0.30]{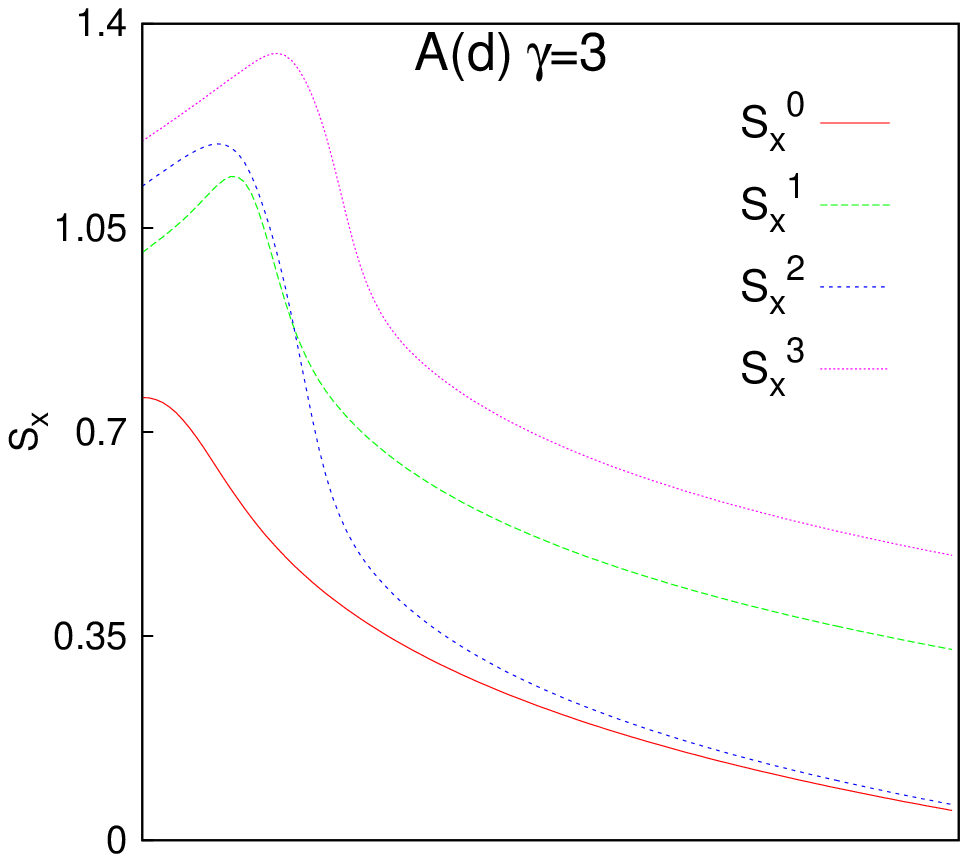}
\end{minipage}\hspace{0.10in}
\begin{minipage}[c]{0.18\textwidth}\centering
\includegraphics[scale=0.30]{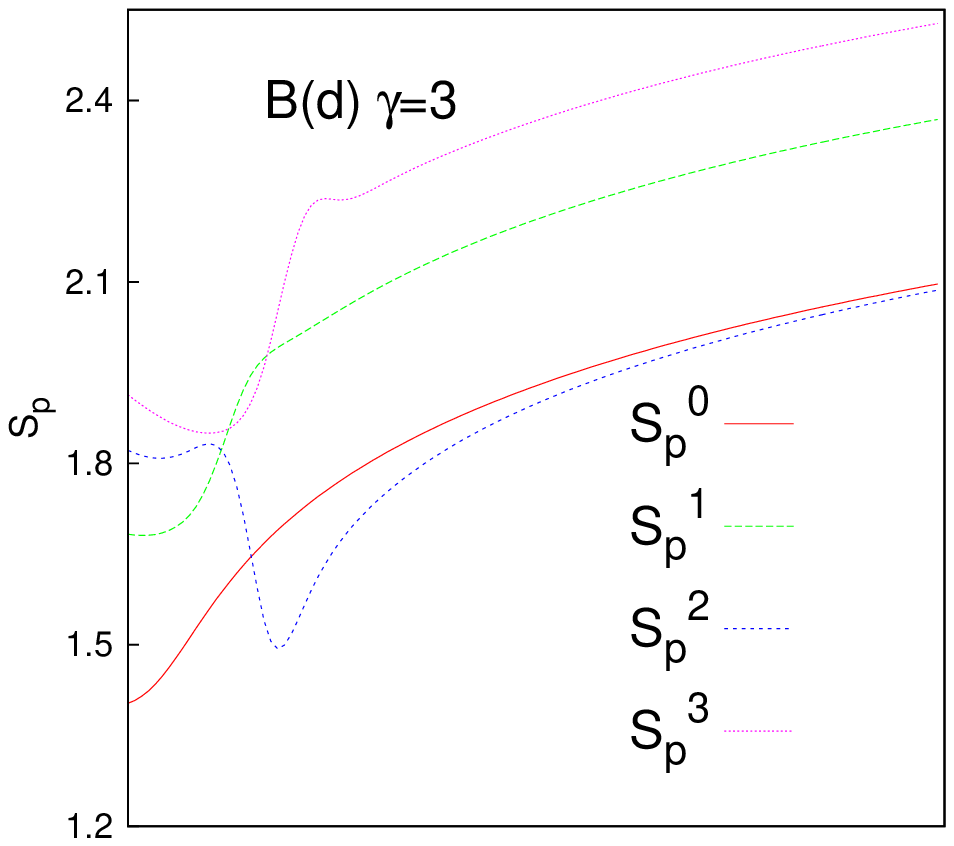}
\end{minipage}\hspace{0.10in}
\begin{minipage}[c]{0.18\textwidth}\centering
\includegraphics[scale=0.30]{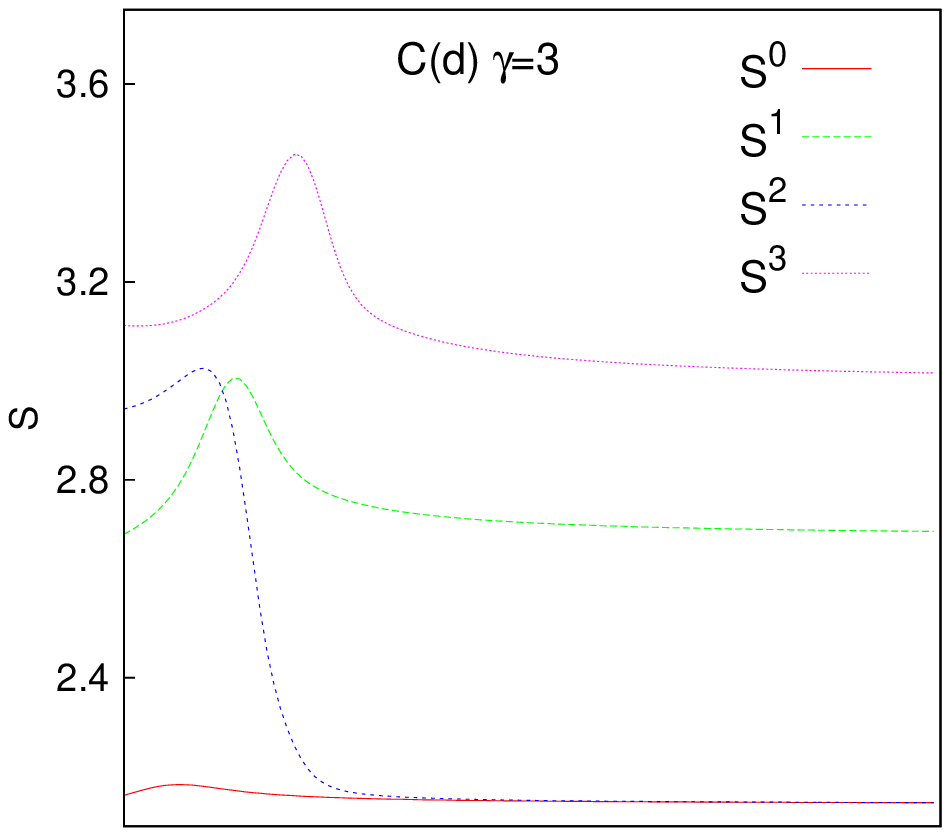}
\end{minipage}\hspace{0.10in}
\\[1pt]
\begin{minipage}[c]{0.18\textwidth}\centering
\includegraphics[scale=0.30]{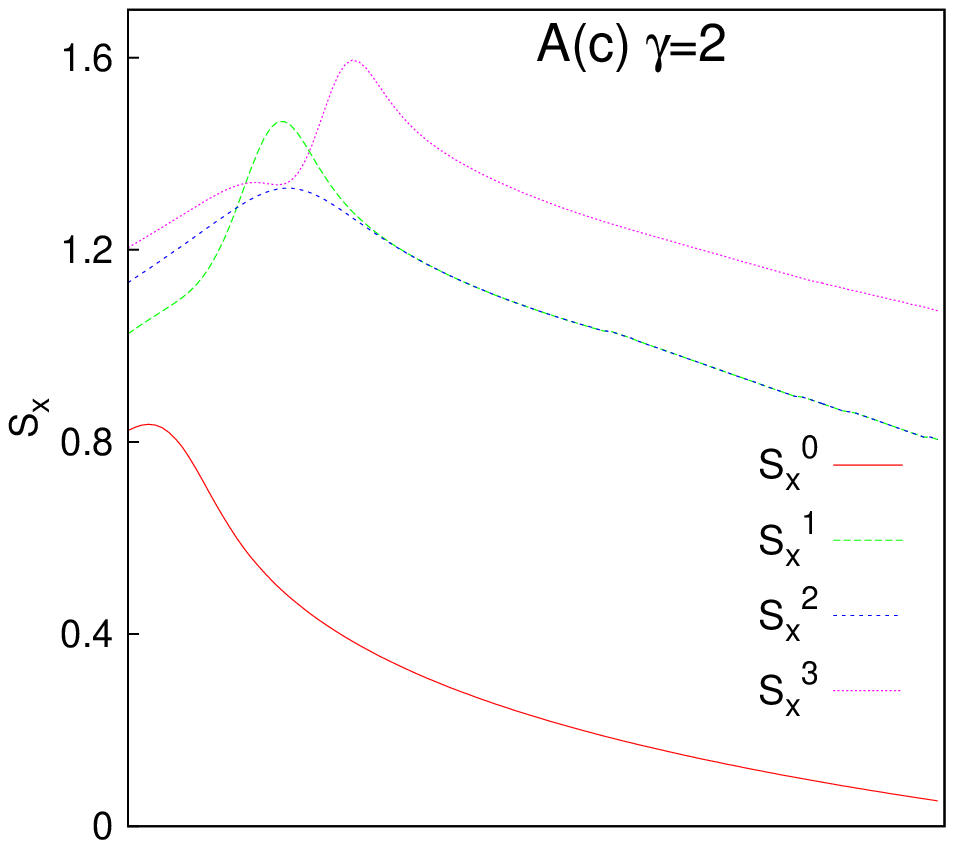}
\end{minipage}\hspace{0.10in}
\begin{minipage}[c]{0.18\textwidth}\centering
\includegraphics[scale=0.30]{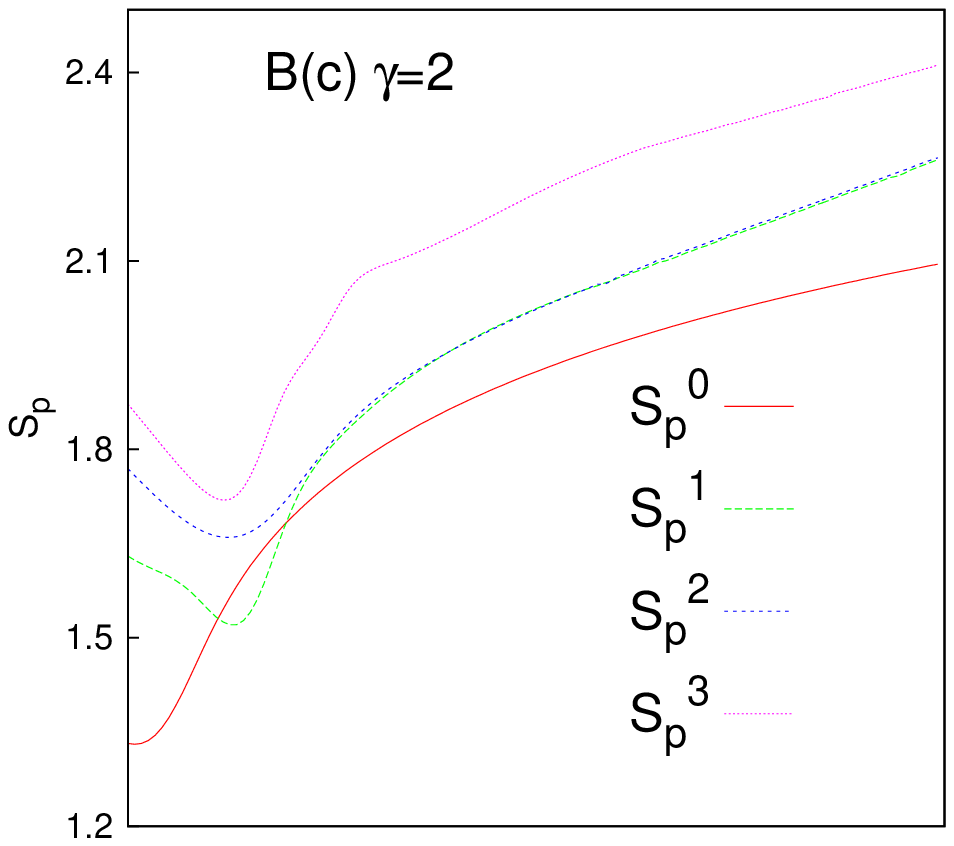}
\end{minipage}\hspace{0.10in}
\begin{minipage}[c]{0.18\textwidth}\centering
\includegraphics[scale=0.30]{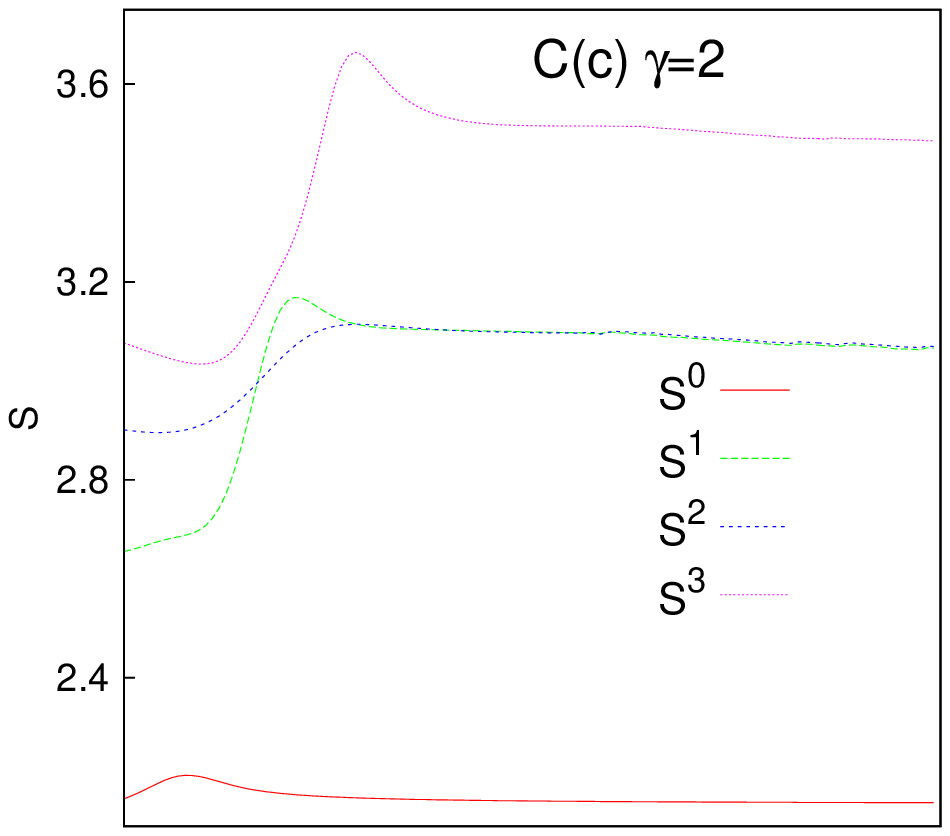}
\end{minipage}\hspace{0.10in}
\\[1pt]
\begin{minipage}[c]{0.18\textwidth}\centering
\includegraphics[scale=0.30]{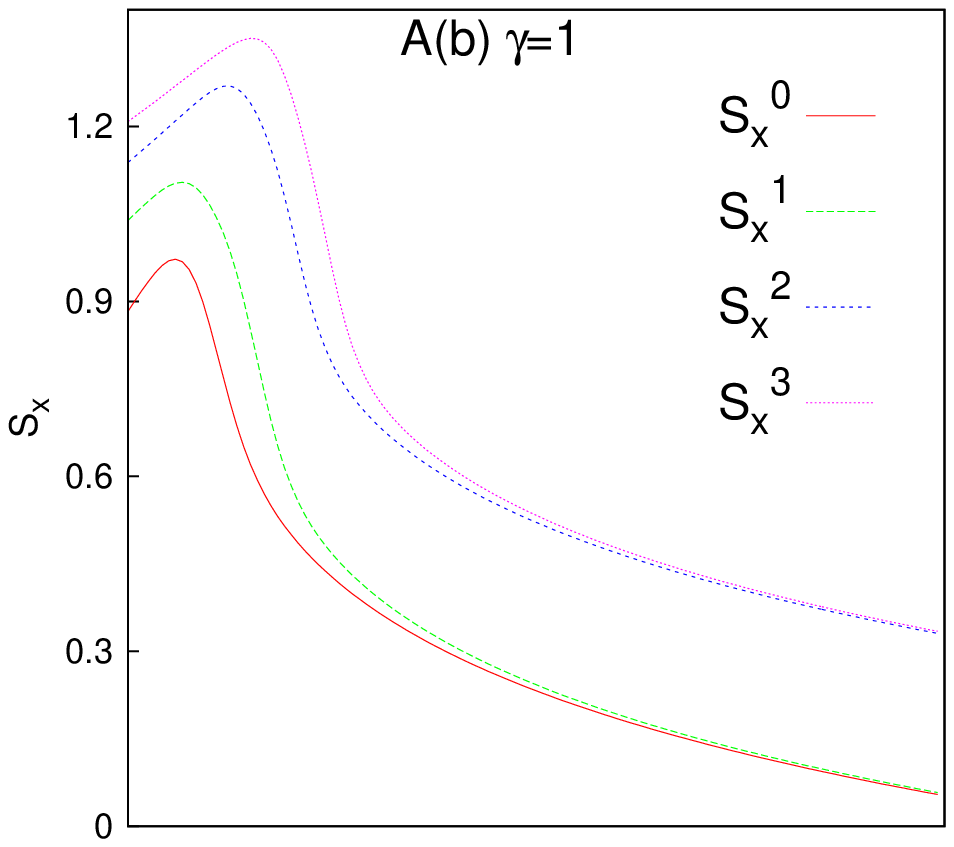}
\end{minipage}\hspace{0.10in}
\begin{minipage}[c]{0.18\textwidth}\centering
\includegraphics[scale=0.30]{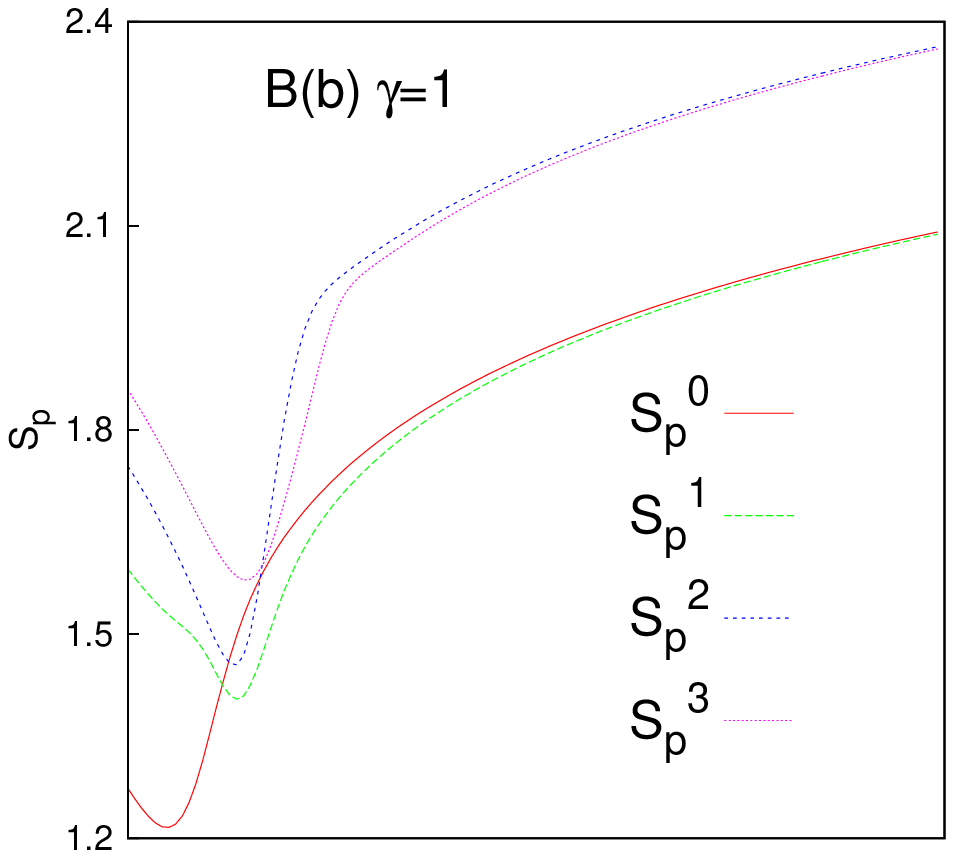}
\end{minipage}\hspace{0.10in}
\begin{minipage}[c]{0.18\textwidth}\centering
\includegraphics[scale=0.30]{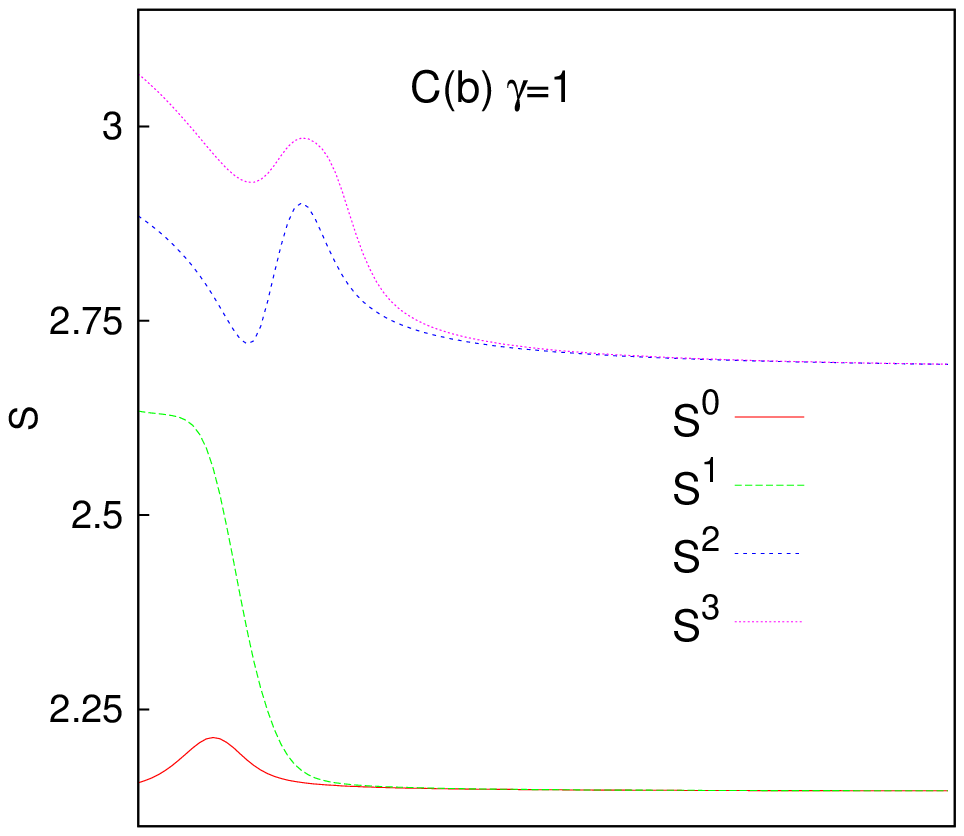}
\end{minipage}\hspace{0.10in}
\\[1pt]
\begin{minipage}[c]{0.18\textwidth}\centering
\includegraphics[scale=0.30]{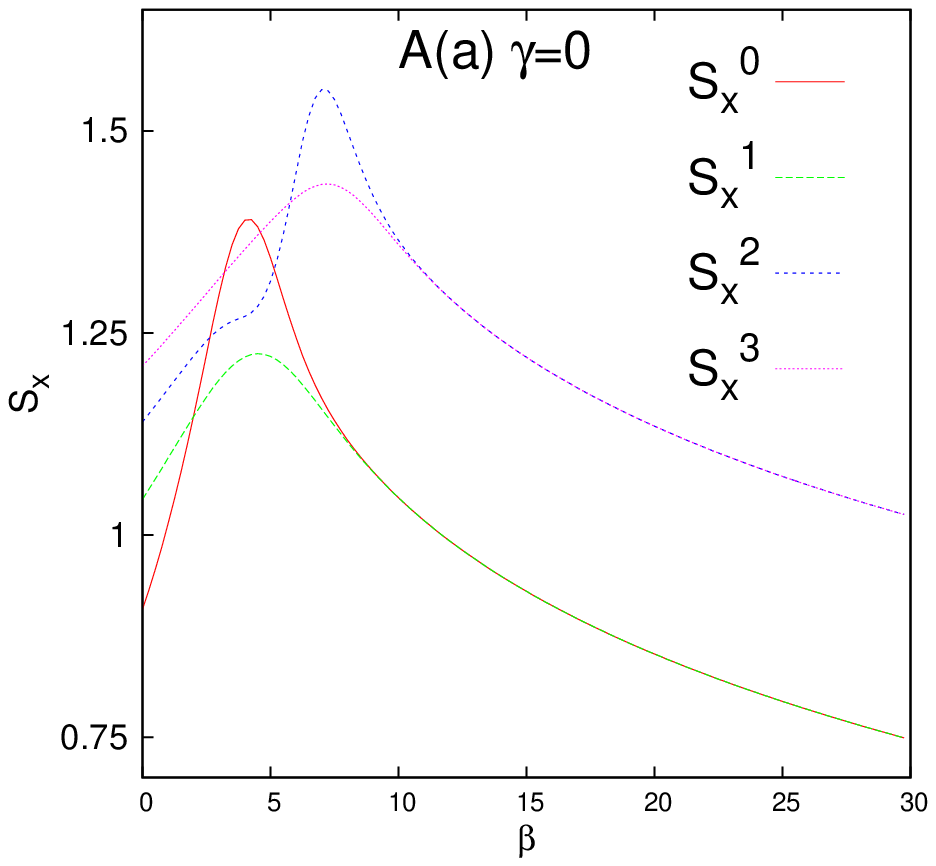}
\end{minipage}\hspace{0.10in}
\begin{minipage}[c]{0.18\textwidth}\centering
\includegraphics[scale=0.30]{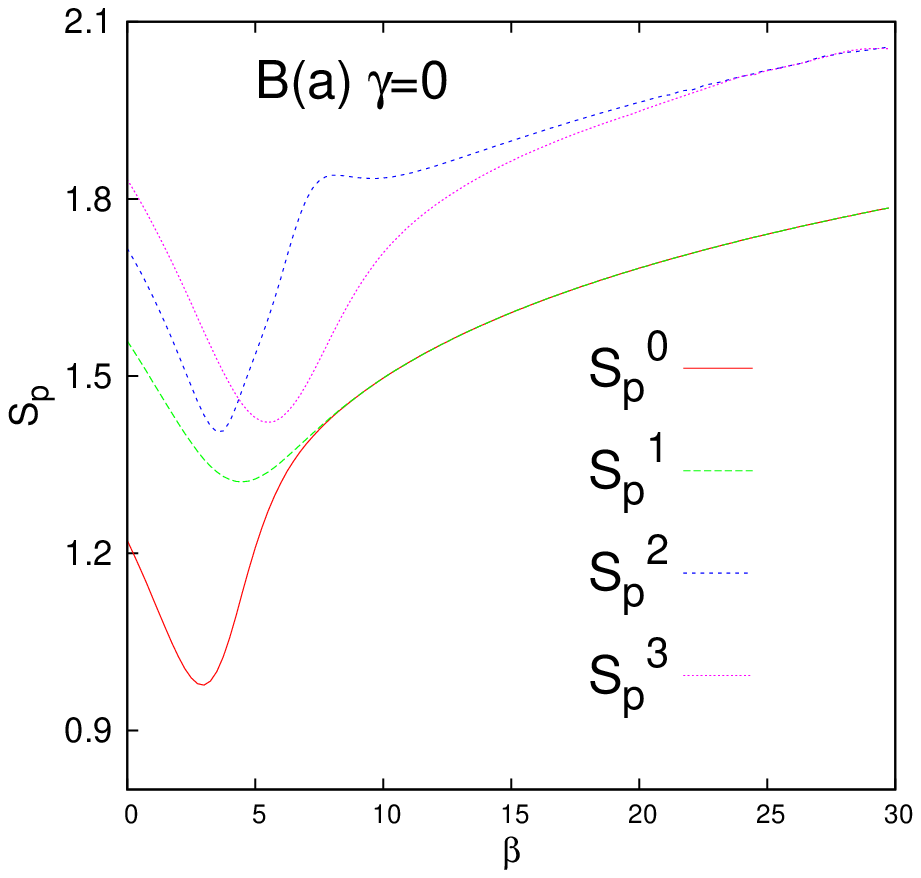}
\end{minipage}\hspace{0.10in}
\begin{minipage}[c]{0.18\textwidth}\centering
\includegraphics[scale=0.30]{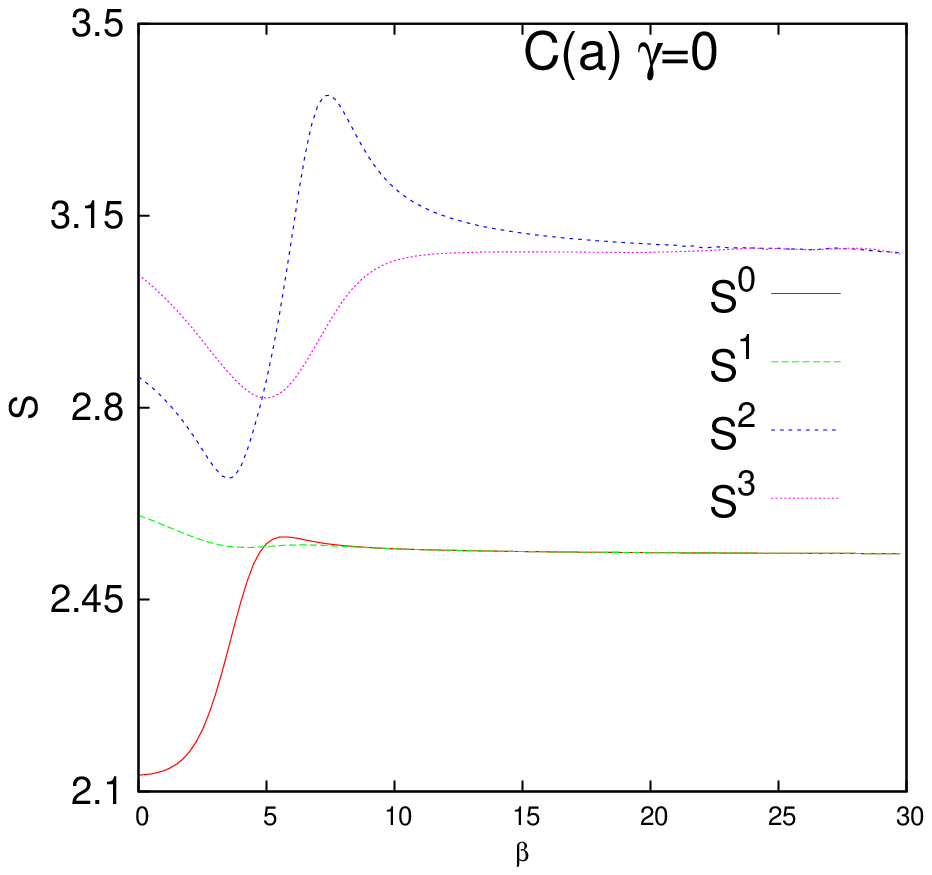}
\end{minipage}\hspace{0.10in}
\caption[optional]{$S_x$ (A), $S_p$ (B) and $S$ (C) for first four states, in left, middle and right
columns, plotted against $\beta$, for asymmetric DW potential, in Eq~(4) keeping $\alpha$ fixed at 1. Eight panels 
(a)-(h), in each column refer to eight $\gamma$, namely, 0,1,2,3,4,5,6,7 respectively. For more details, see text.}
\end{figure}

\subsection{Information-based uncertainty measures}
We now focus on information-based measures for first four energy states. Figure~(9) displays $S_{x}$ against $\gamma$ 
plots at six $\beta$ in top six rows, A(a)-A(f). A similar presentation strategy is followed as in traditional 
uncertainties in Fig.~(7). Increase in $\beta$ only makes the particle more confined within well~I 
or II. Generally, at certain $\gamma$, curve for $n$th state quickly jumps to a crest and then gradually decreases
by traveling through $n+1$ peaks, very similar to that encountered for $\Delta x$ in Fig.~(7). 
After that, particle eventually localizes in larger (I) well. Recall that these jumps at even $\gamma$ 
(integer $k$) signify transition points, which takes place only at fixed interval of $\Delta \gamma$, and observed upto a 
certain $\gamma$ (characteristics for a given state). At fractional $k$, change in $S_{x}$ is rather small because of 
confinement. Once, it finally rests in larger well ($k \! > \! n$), $S_{x}$ decreases with increase in $\gamma$ 
indicating gradual build-up of confinement. Next B(a)-B(f), demonstrate that slope of $S_{p}$ versus $\gamma $ curves 
changes noticeably at intervals of $\Delta \gamma$, in a fashion similar to that for $\Delta p$ in Fig.~(7).
This discontinuity implies transition of particle from one well to another. Again, an $n$th state leads to 
($n$+1) jumps in $S_{p}$ before localization happens in well~I. Analogous variations of $S$ as a function of $\gamma$ 
are depicted in bottom row. Like $S_{x}$, $S$ too shows pronounced peaks at intervals of $\Delta \gamma$ indicating transition 
points at integer $k$. Moreover, like $S_x$, an $n$th state gives rise to ($n$+1) peaks in $S_x$ before 
settling in well~I. When particle stays in a specific well, extent of confinement is not always same. 
Visibly, when $0 \! < \! k \! < \! 1$ ($ 0 \! < \! \gamma \! < \! 2 $), $S$ for $n \! = \! $0,1 remain very close. This is 
explained from the fact that, at this interval these two states act effectively as lowest state of wells~I, II. Likewise,
for $1 \! < \! k \! < \! 2$ ($ 2 \! < \! \gamma \! < \! 4$), $S$ of $n \! = \! 0$,2 are nearer to each other, as they
behave as lowest states of wells~I, II. As expected, in $2 \! < \! k \! < \! 3$, $S$ of $n \! = \! 0$,3 approach each
other quite closely. This observations could be interpreted from a consideration of number of effective nodes in 
Table~V in each of these states in respective interval of $\gamma$. Thus, at $n$th interval of $\Delta \gamma$, $n$th 
state behaves similar to ground state of a particular well. Note that there is a jump in $S_{x}^{0},S_{p}^{0}, S^{0}$ at 
$\gamma \! = \! 0$, indicative of a critical transition point, after which a particle remains in well~I. This also 
mirrors the fact that, symmetric DW is a special case of asymmetric DW potential.

\begin{figure}            
\centering
\hspace*{2mm}{(a)}
\begin{minipage}[c]{0.4\textwidth}\centering
\includegraphics[scale=1.1, width=5cm]{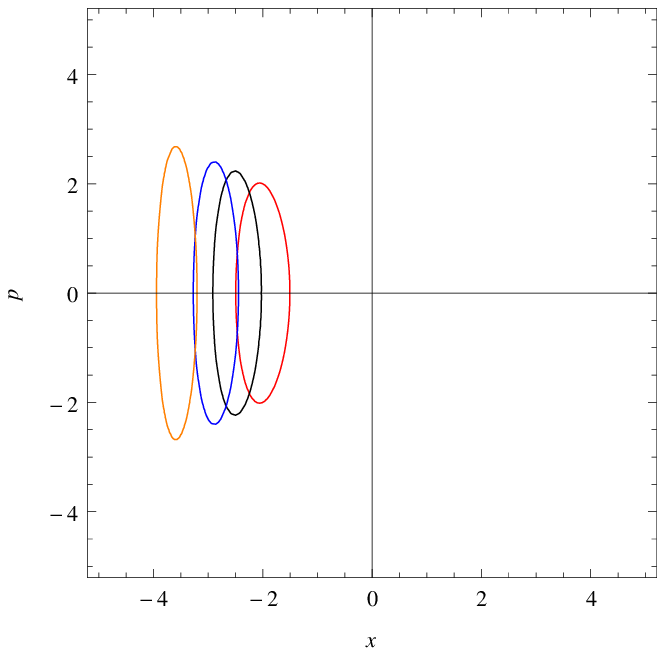}
\end{minipage}\hspace{0.05in}
\hspace*{2mm}{(b)}
\begin{minipage}[c]{0.4\textwidth}\centering
\includegraphics[scale=1.1,width=5cm]{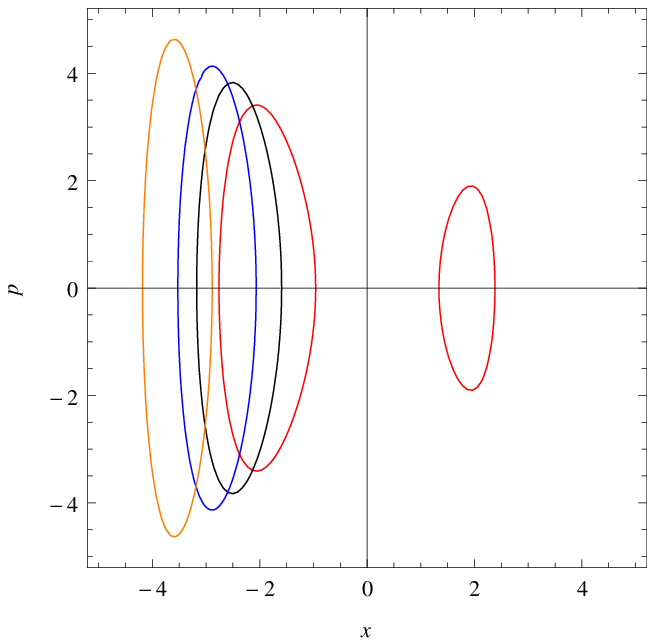}
\end{minipage}
\\[20pt]
\hspace*{2mm}{(c)}
\begin{minipage}[c]{0.4\textwidth}\centering
\includegraphics[scale=1.1,width=5cm]{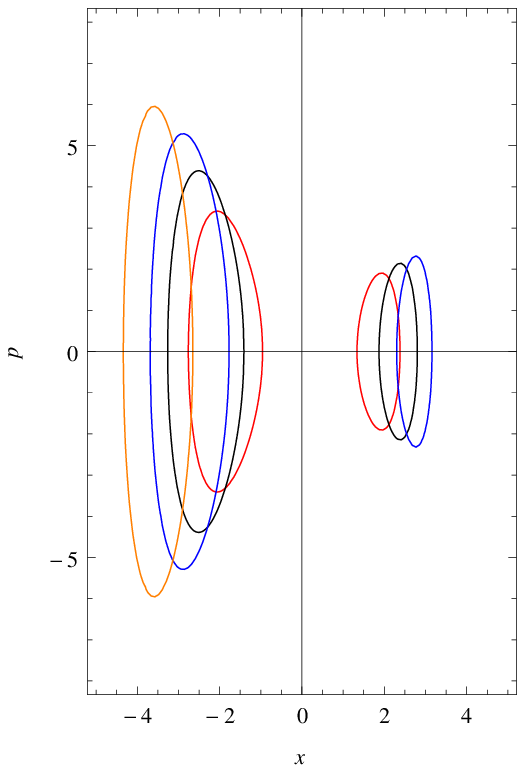}
\end{minipage}\hspace{0.05in}
\hspace*{2mm}{(d)}
\begin{minipage}[c]{0.4\textwidth}\centering
\includegraphics[scale=1.1,width=5cm]{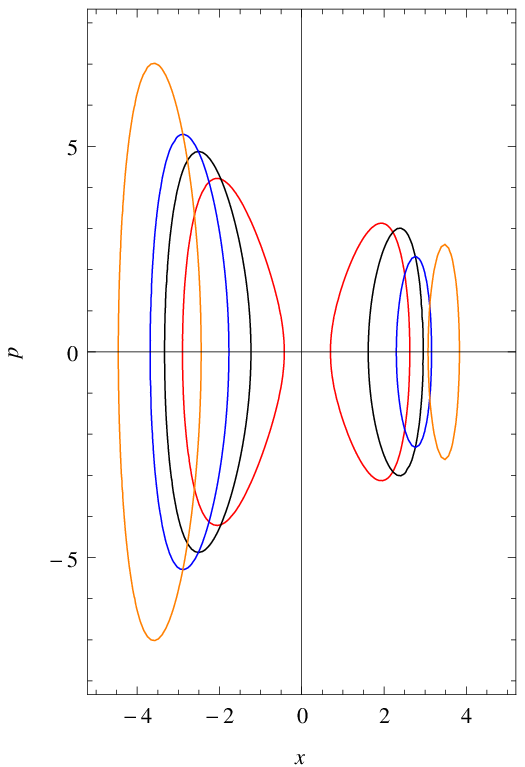}
\end{minipage}
\caption[optional]{Phase space for first four states (in panels a-d) of asymmetric DW potential, Eq.~(4), for four different sets of 
($\gamma$, $\beta$), namely (2,8) (in red), (3,12) (in black), (4,16) (in blue) and (6,25) (in orange), keeping $\alpha$ fixed at 1.}
\end{figure}

We now undertake a study of change of Shannon entropy with $\beta$ in Fig.~(10) through four lowest states 
eight selected $\gamma$. In all cases $S_{x}$ initially increases with $\beta$ before reaching a maximum and then decreases 
asymptotically. This arises due to the competing effect of $\beta$. Interestingly, positions of these maxima shift to left 
with increasing $\gamma$, for any given state; moreover they tend to disappear with $\gamma$, with higher state requiring 
larger $\gamma$. This is explained by considering the predominance of asymmetry over competing effect due to localization 
of particle in a definite well. Depending upon odd, even $\gamma$, two different trends in $S_{x}$ is observed beyond a 
certain threshold $\beta$. In A(a), for $k \! = \! 0$ ($\gamma \! = \! 0$), $S_{x}$ of $n \! = \! 0$,1 are very close; 
similarly at $k \! = \! 1$ ($\gamma \! = \! 2$), same happens for $n \! = \! 1$,2 in A(c); at $k \! = \! 2$ 
($\gamma \! = \! 4$), $n \! = \! 2$,3 are close in A(f). Continuing in this way, one can predict that, $S_{x}$ of 
$n \! = \! 3$,4 will approach each other at $k \! = \! 3$ or $\gamma \! = \! 6$ (not shown) and so on. 
This closeness of different sets of $S_{x}$ at certain $k$ is in agreement with quasi-degeneracy rule. On the other hand, 
when $\gamma$ is odd (1, 3, 5, etc.), then at $k \! = \! 0.5, 1.5, 2.5$, $S_{x}$ of $n \! = \! 0$ becomes closer to 
$n \! = \! 1$,2,3 states respectively in A(b), A(d), A(f). This, again is due to number of effective node of those states 
becoming same at those specific $k$.   

Next, B(a)-B(h), C(a)-C(h) illustrate changes of $S_{p}$ and $S$ with $\beta$. Generally, $S_{p}$ decreases with increase 
in $\beta$, reaches a minimum and then increases gradually. Again, positions of these minima move towards left with $\beta$. 
Beyond a certain $\gamma$, the extrema tend to die out. Also for higher $n$, there are some occasional humps observed. 
As in $S_x$, $S_p$ in certain states also converge depending on odd, even $\gamma$. Variation of $S$ with $\beta$, however, 
is not straightforward; complicated sequence of maxima, minmima is observed for different states for various $\beta$. 
All these occurs due to a balance of various effect present in such asymmetric DW. Again, $S$ merge for certain states at 
different $k$. For example, this takes place for $n \! = \! 0$,1 in C(a) (when $k \! = \! 0$); $n \! = \! 1$,2 in C(c)
(when $k \! = \! 1$); $n \! = \! 2$,3 in C(e) 
(when $k \! = \! 2$), etc. Following this trend we can expect this joining to occur $n \! = \! 3$,4 states when $k \! = \! 3$
(not shown in figure). These are well in accordance to quasi-degeneracy rules prescribed in IV.A. Also for odd $\gamma$, 
merging of $S$ occurs in agreement with rules of IV.B; this happens for $n \! = \! 0$,1 in C(b) when $\gamma \! = \! 1$ 
($k \! = \! 0.5$); $n \! = \! 0$,2 in C(d) when $\gamma \! = \! 3$ ($k \! = \! 1.5$); $n \! = \! 0$,3 in C(f) for 
$\gamma \! = \! 5$ ($k \! = \! 2.5$), etc. Thus, using above rule, one can expect that at $\gamma \! = \! 7$ ($k \! = \! 3.5$), 
$S$ of $n \! = \! 0$,4 states will coincide (not shown in figure). 

An analogous study was performed for variation of Fisher information, Onicescu energy and Onicescu-Shannon information 
measures with changes in $\gamma$ and $\beta$. In order to save space, these results
are provided in supporting document. Thus Figs.~(S1), (S3), (S5) give plots for $I$, $E$, $OS$ with respect to $\gamma$, 
while $S2$, $S4$, $S6$ refer to $\beta$ variations. A detailed analysis reveals that nature of $I_{x},E_{x}$, $OS_{x}$ 
versus $\gamma$ plots are qualitatively similar to those of $S_{p}$ against $\gamma$. Similarly, trends of $I_{p},E_{p}$,
$OS_{p}$ versus $\gamma$ plots resemble those of $S_{x}$ against $\gamma$. The conclusions from $\beta$ 
plots are also very similar to those from the $\beta$ plots of Shannon entropy. Overall all these measures lead to comparable 
conclusions as obtained through study of conventional uncertainty product and $S$ in position and momentum space.   

\begin{figure}            
\begin{minipage}[c]{0.20\textwidth}
\centering
\includegraphics[scale=0.30]{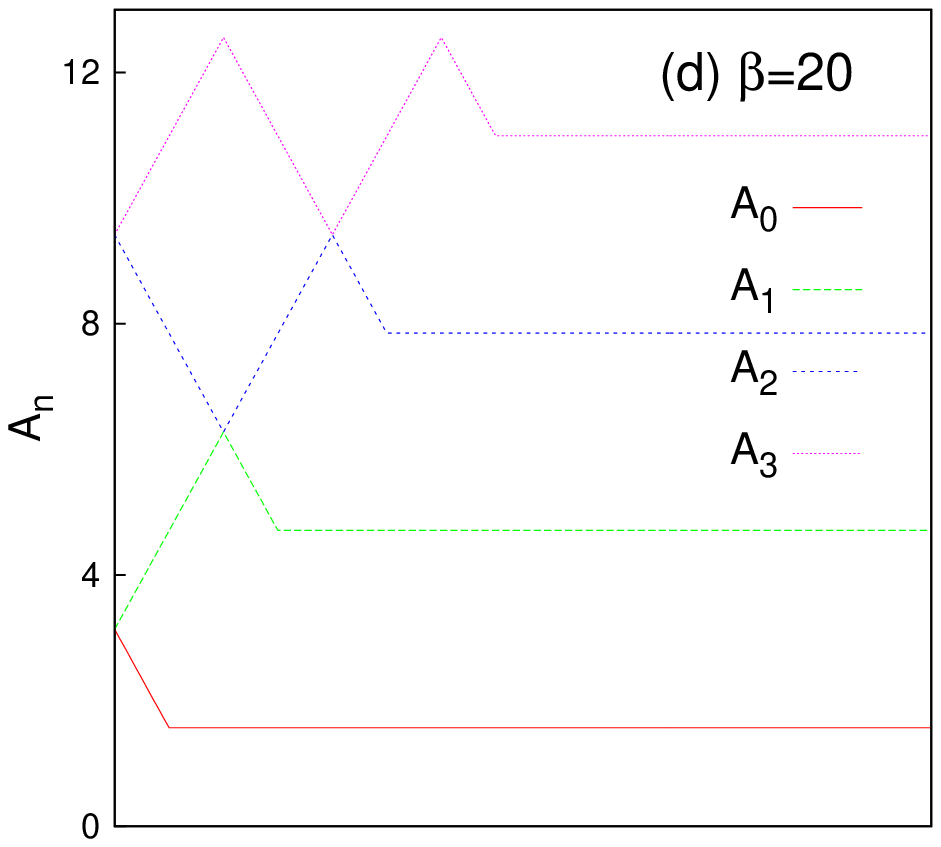}
\end{minipage}%
\hspace{0.1in}
\begin{minipage}[c]{0.20\textwidth}
\centering
\includegraphics[scale=0.30]{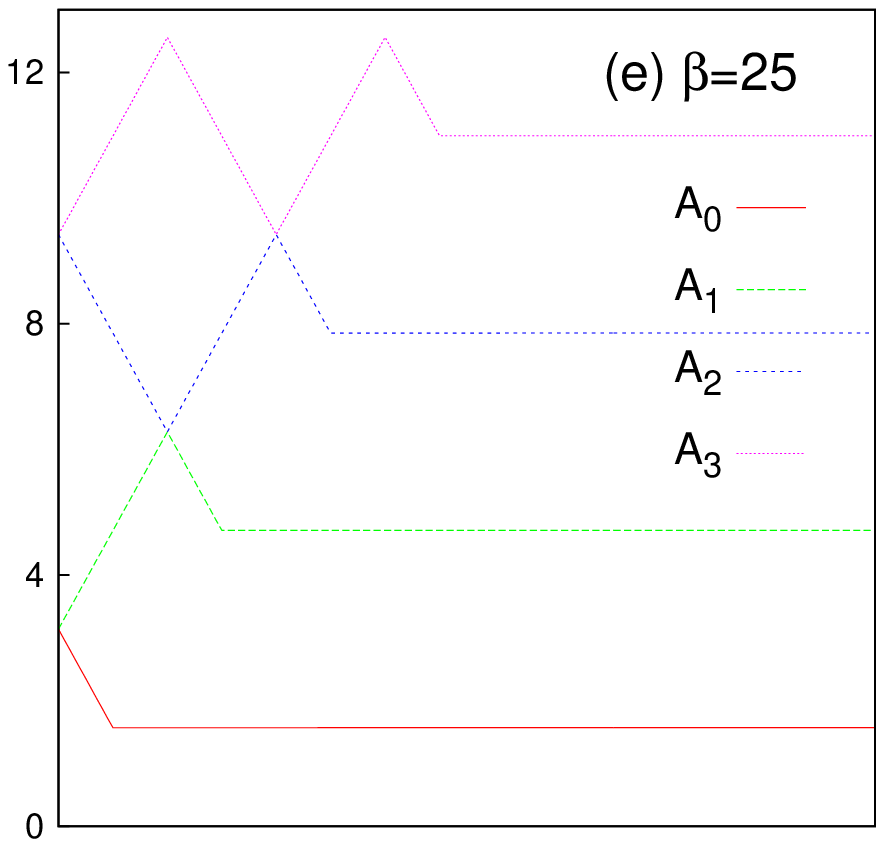}
\end{minipage}%
\hspace{0.1in}
\begin{minipage}[c]{0.20\textwidth}
\centering
\includegraphics[scale=0.30]{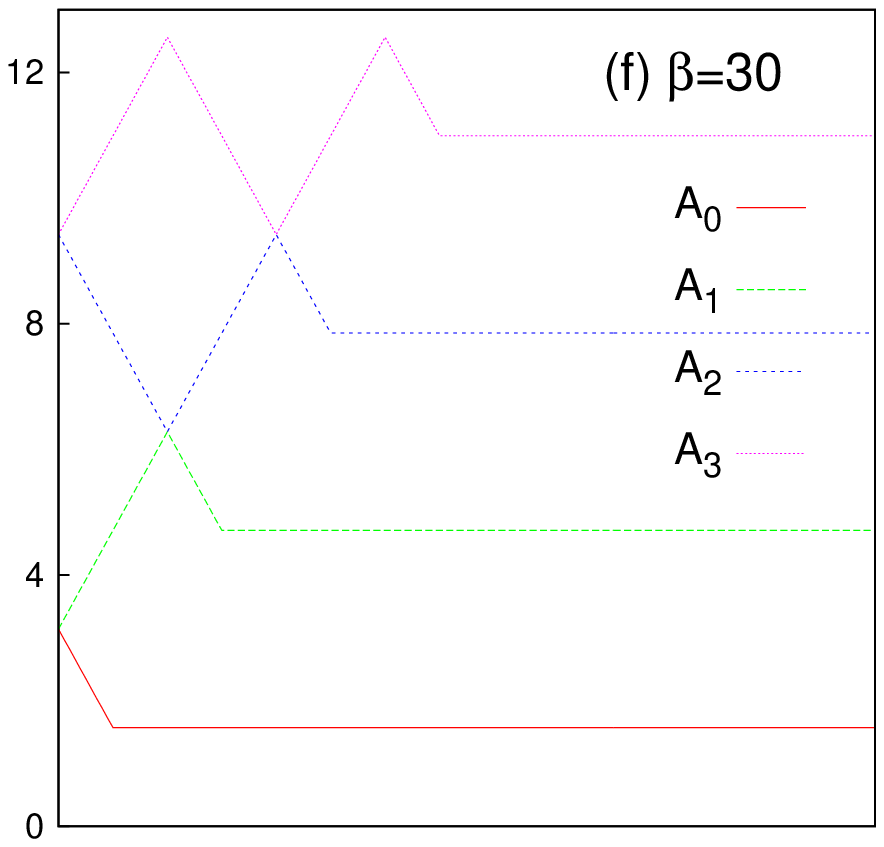}
\end{minipage}%
\\[10pt]
\begin{minipage}[c]{0.20\textwidth}
\centering
\includegraphics[scale=0.30]{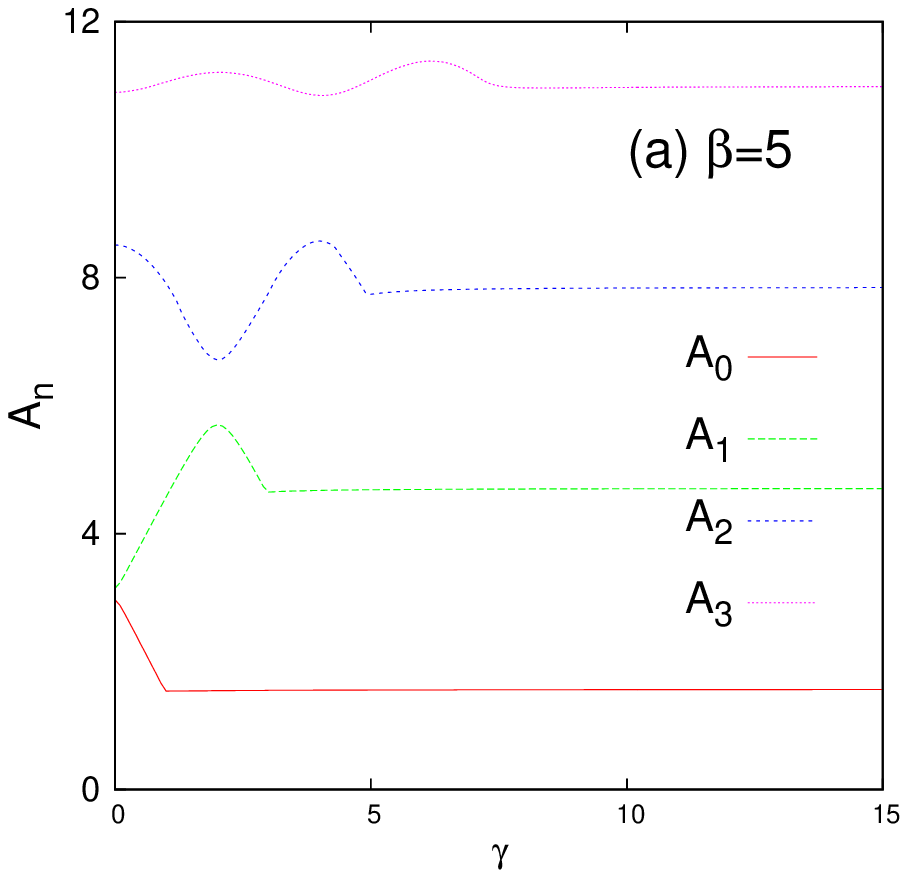}
\end{minipage}%
\hspace{0.1in}
\begin{minipage}[c]{0.20\textwidth}
\centering
\includegraphics[scale=0.30]{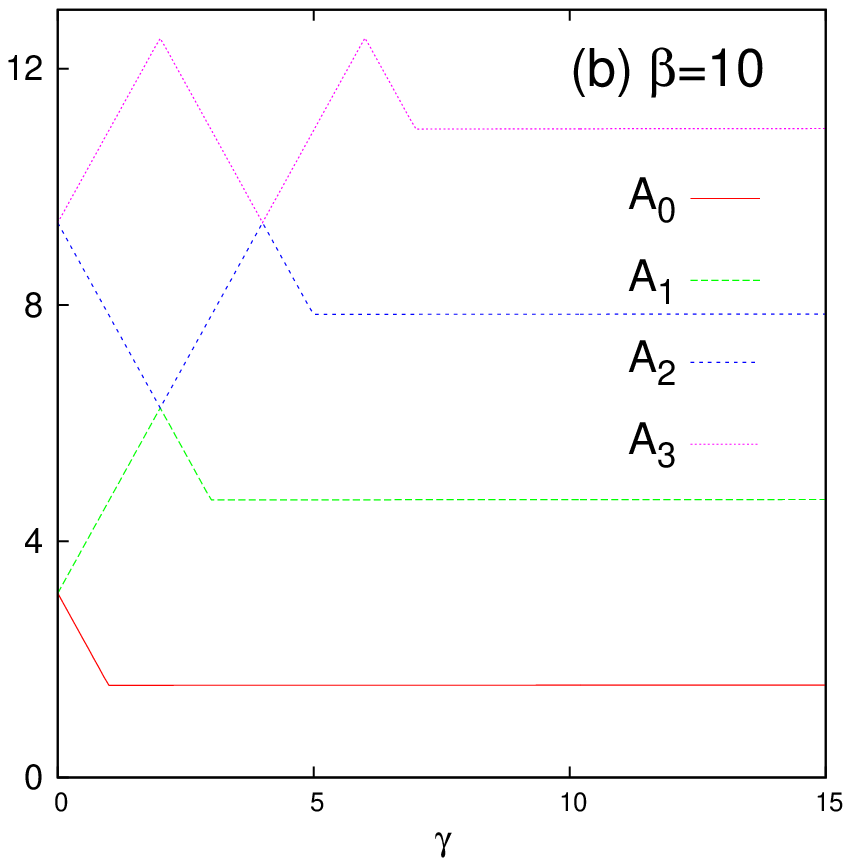}
\end{minipage}%
\hspace{0.1in}
\begin{minipage}[c]{0.20\textwidth}
\centering
\includegraphics[scale=0.30]{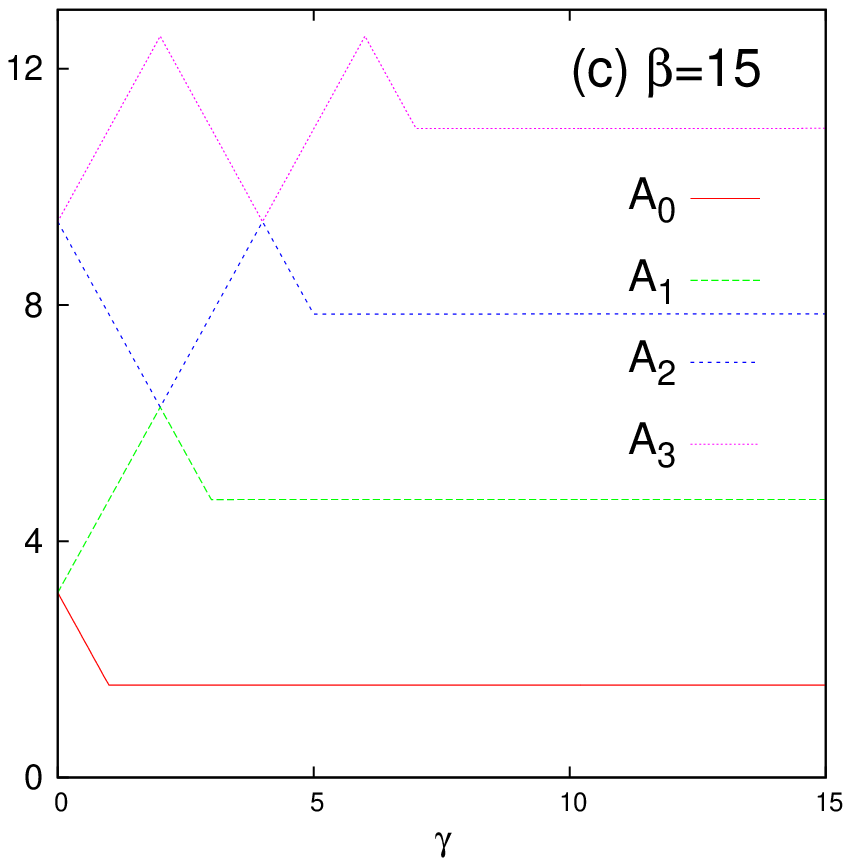}
\end{minipage}%
\caption[optional]{$A_n$ for first four states, plotted against $\gamma$, for DW potential in Eq.~(4) for $\alpha =1$.
Six plots (a)-(f), refer to six $\beta$, namely, 5,10,15,20,25,30 respectively. For more details, see text.}
\end{figure}

\subsection{Phase-space area}
Now we offer a semi-classical phase-space analysis for first four energy states of the asymmetric DW potential. Figure (11) 
shows this for four representative ($\gamma, \beta$) parameter sets, \emph{viz.,} (2,8) (red), (3,12) (black), (4,16) (blue)
and (6,25) (orange), covering the major interesting situations discussed in Sec.~(IV). Panels (a)-(d) correspond to four lowest 
states having $n=0,1,2,3$ respectively. When $n=0$, respective phase-spaces for all sets are characterized by a single closed 
lobe on left side of $p$-axis, indicating localisation of the particle in well~I, which seems to follow rule~(ii) proposed 
in (IV.B). For $n=1$, there appears two closed lobes in phase-space for $k=1$, resembling a situation where the particle 
can be distributed in both wells in a similar fashion as in rule (i.a). Once, $k>1$, it appears that all the phase-space plots 
are represented by single closed lobes. This again seems to follow from rule~(ii). Similarly, for $n=2$,
phase-space breaks into two closed lobes at $k=1,2$ corresponding to particle localisation in both wells (rule~(i.a)). 
At $k=3$, one finds a single closed lobe on left side (rule~(ii)). But, at fractional $k$ (1.5), there emerges two closed 
lobes. This observation is incompatible with particle distribution rule~(b) of (IV.B). 
Again, for $n=3$ there originates two closed lobes in $k$=1,1.5,2,3. Among these four cases, when $k$ is integer, it is 
evident from (i.a) that, there will appear two closed lobes. But, appearance of two closed lobes at $k$=1.5 is in contrast 
with rule~(b). Thus the quantum mechanical (in terms of energy and probability distribution) and semi-classical (in terms 
of phase-space) viewpoints do not necessarily complement each other always. 
 
Figure~(12) shows variation of phase-space area as a function of $\gamma$ at six different $\beta$ values. Here also we 
envisage that, the curve for an $n$th state undergoes $n$ number of jumps before localising in well~I. But the jumps are not 
as fast as in case of $\Delta x$ or $S_x$. Additionally, at $k=1,2$, curves for $n=1,2$ and $n=2,3$ respectively merge, indicating 
quasi-degeneracy of these states, as observed earlier in energy distribution rules; (i.a), (i.b) in (IV.A). 

Finally, Fig.~(13) portrays the change of phase-space area as a function $\beta$ at eight different $\gamma$. At 
fractional $k$, there appears a maximum followed by a minimum. Interestingly, at $k=0$, $n=0,1$ states, coalesce after 
certain $\beta$ value. Similarly, at $k=1$ and 2, $n=1,2$ and $n=2,3$ states converge after certain $\beta$ respectively. 
This again establishes the quasi-degeneracy in such systems, as a consequence of rules (i.a), (i.b) of (IV.A).

\begin{figure}            
\begin{minipage}[c]{0.20\textwidth}
\centering
\includegraphics[scale=0.30]{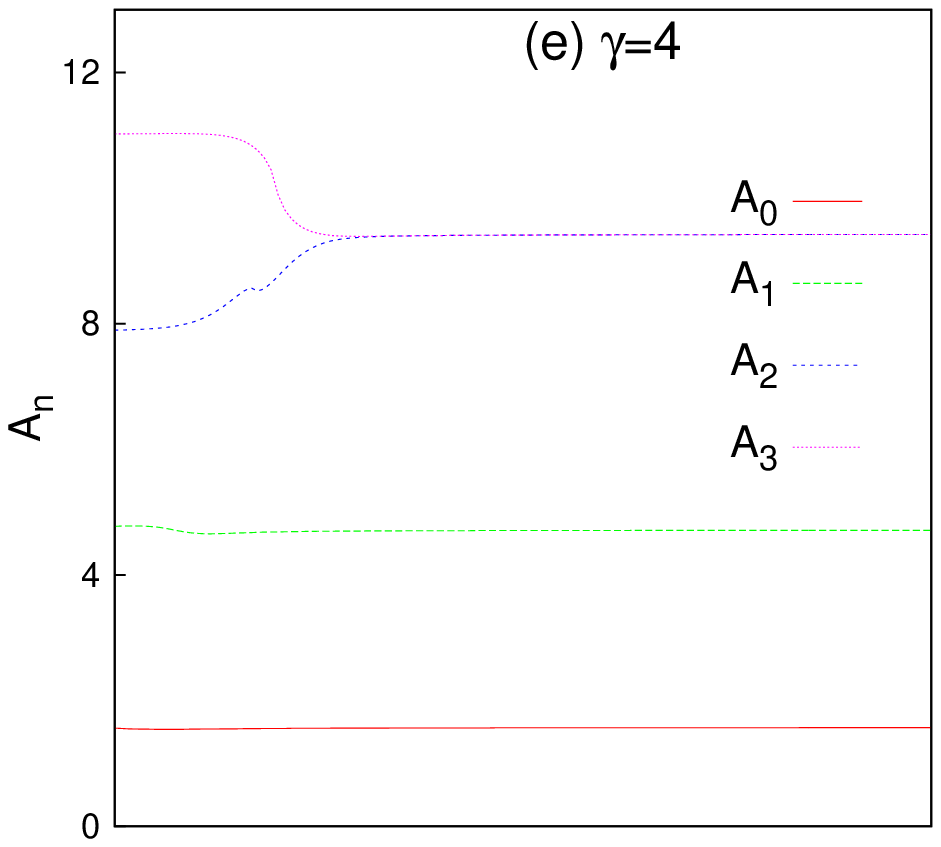}
\end{minipage}%
\hspace{0.1in}
\begin{minipage}[c]{0.20\textwidth}
\centering
\includegraphics[scale=0.30]{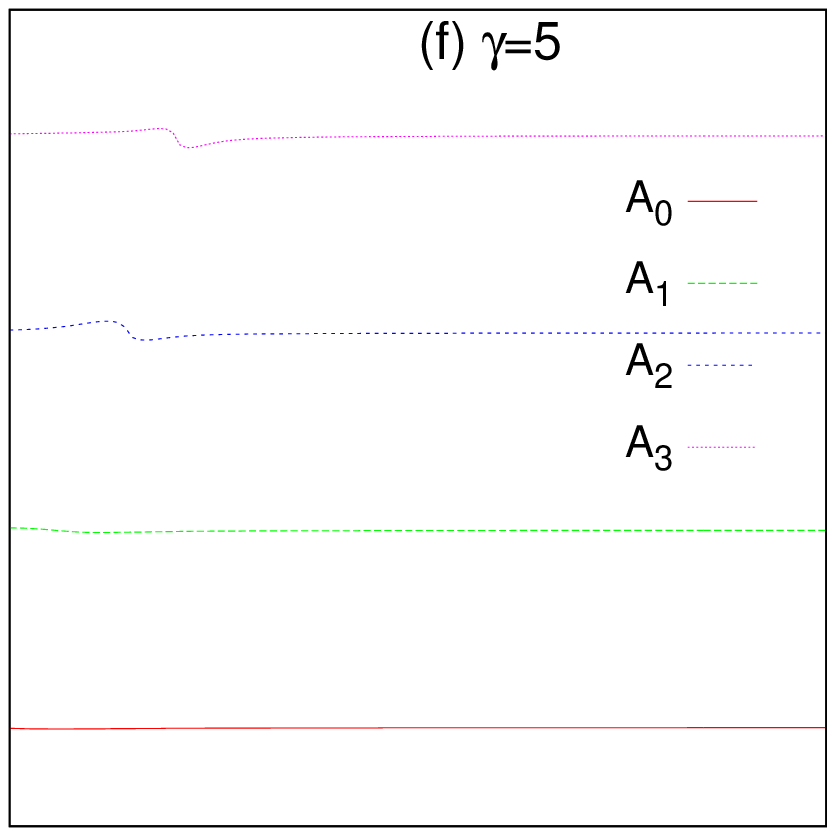}
\end{minipage}%
\hspace{0.1in}
\begin{minipage}[c]{0.20\textwidth}
\centering
\includegraphics[scale=0.30]{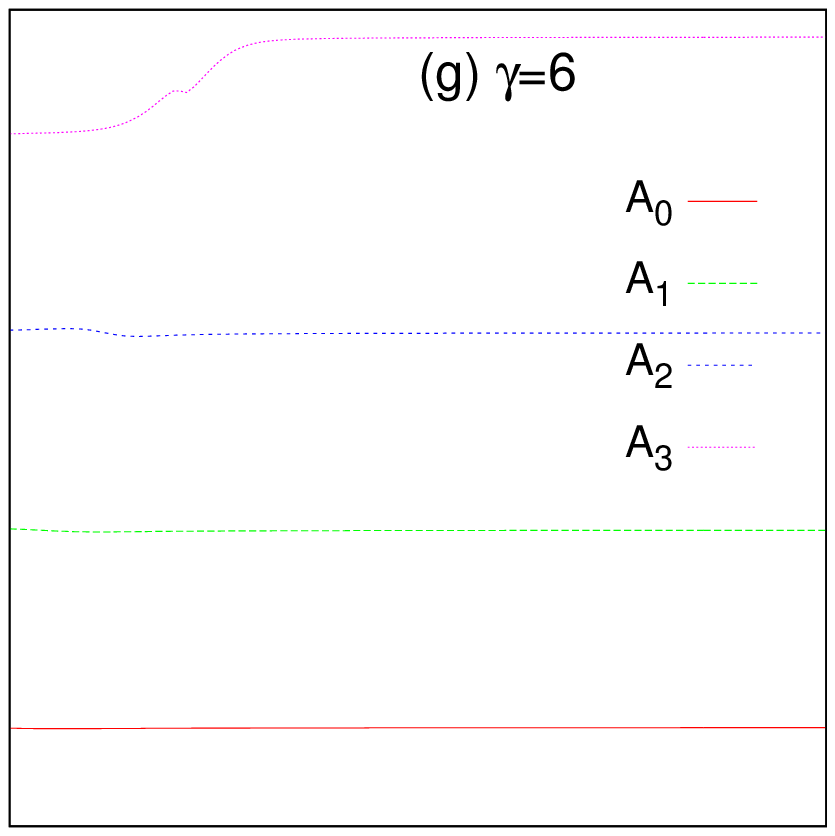}
\end{minipage}%
\hspace{0.1in}
\begin{minipage}[c]{0.20\textwidth}
\centering
\includegraphics[scale=0.30]{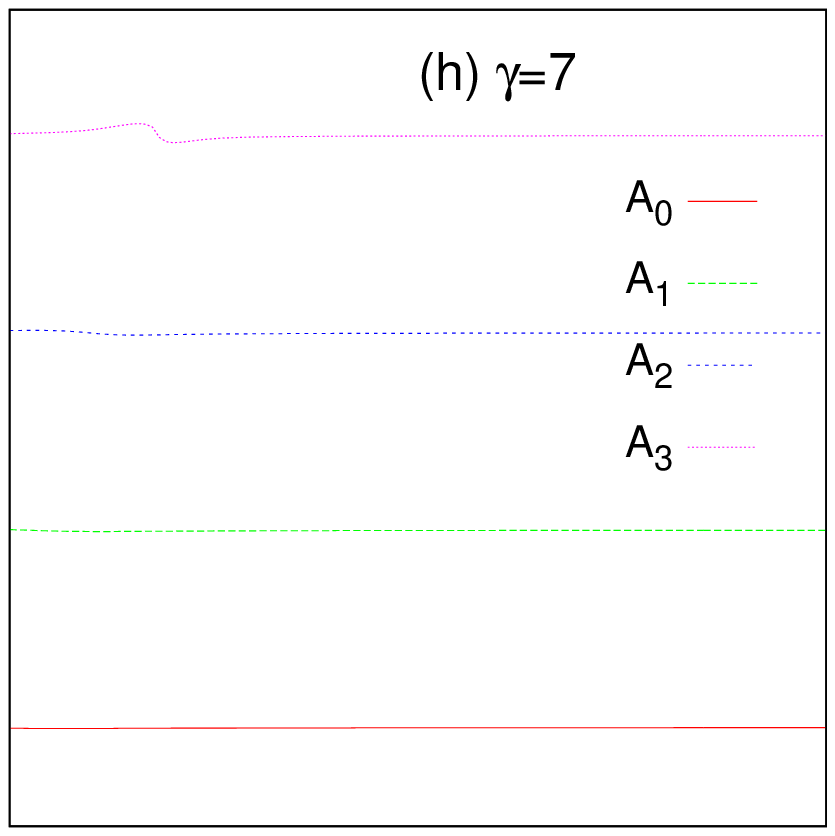}
\end{minipage}%
\\[10pt]
\begin{minipage}[c]{0.20\textwidth}
\centering
\includegraphics[scale=0.30]{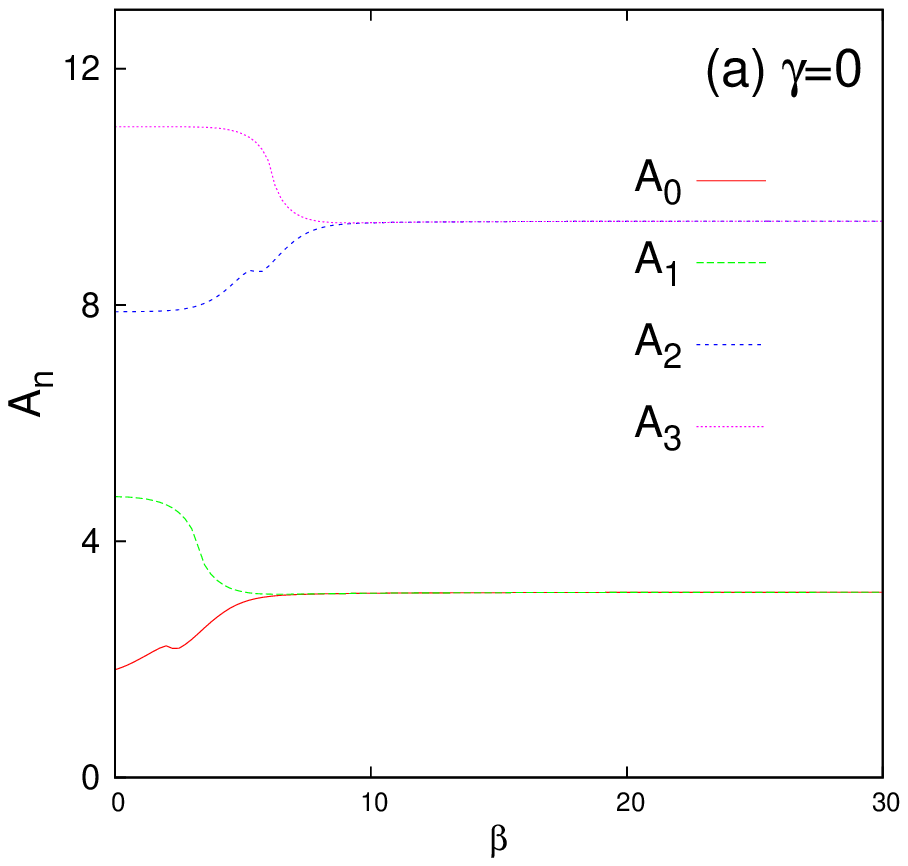}
\end{minipage}%
\hspace{0.1in}
\begin{minipage}[c]{0.20\textwidth}
\centering
\includegraphics[scale=0.30]{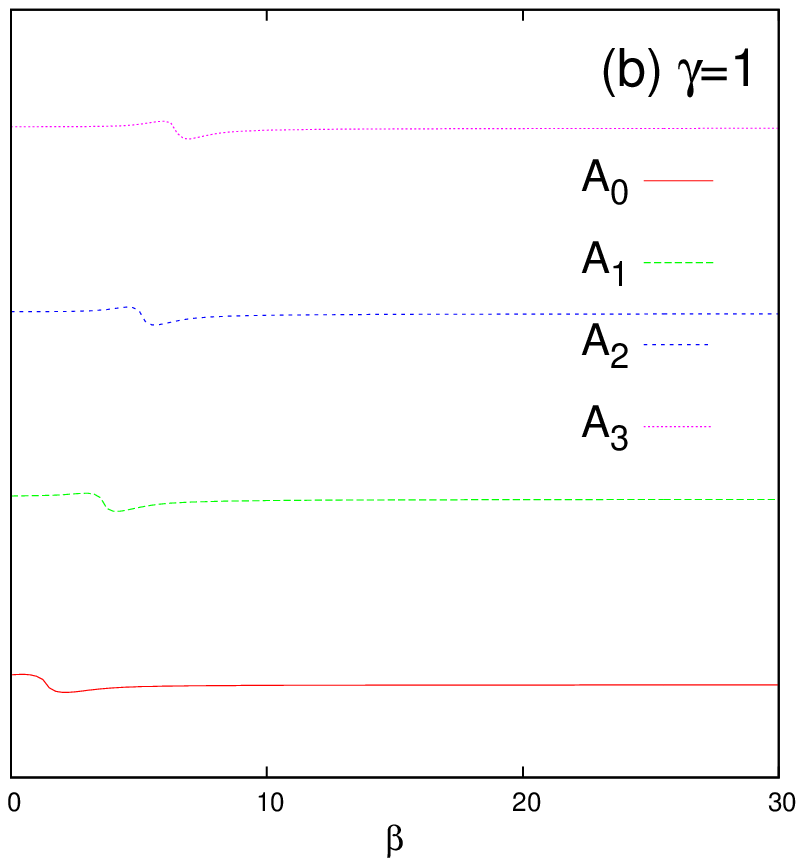}
\end{minipage}%
\hspace{0.1in}
\begin{minipage}[c]{0.20\textwidth}
\centering
\includegraphics[scale=0.30]{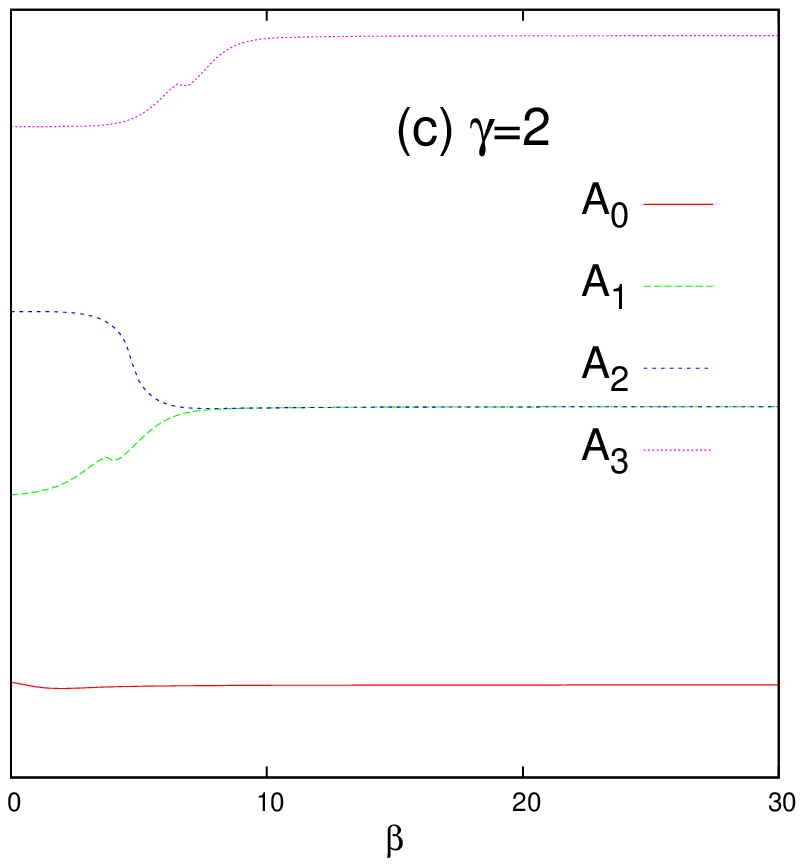}
\end{minipage}%
\hspace{0.1in}
\begin{minipage}[c]{0.20\textwidth}
\centering
\includegraphics[scale=0.30]{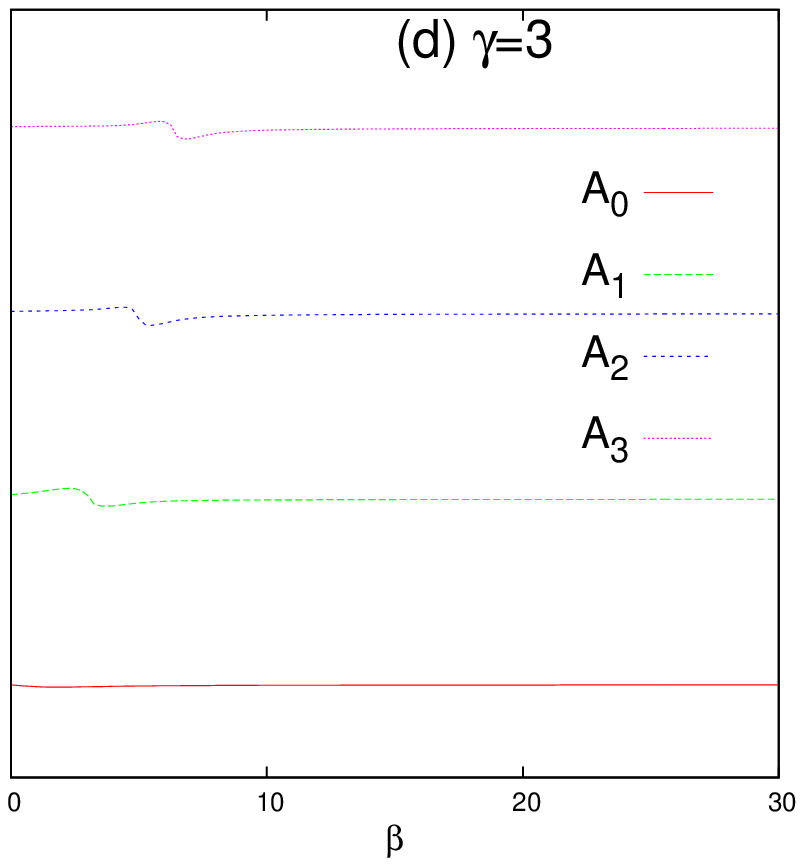}
\end{minipage}%
\caption[optional]{$A_n$ for first four states, plotted against $\beta$, for DW potential in Eq~(4) for $\alpha =1$.
Eight plots (a)-(h) refer to eight $\gamma$, namely, 0,1,2,3,4,5,6,7 respectively. For more details, see text.}
\end{figure}

\section{Conclusion}
Energy distribution and localisation of a particle in an asymmetric DW potential is studied. Conventional uncertainty 
relation and a host of entropy-based information measures such as, $S, I, E, OS$ have been invoked to correlate the 
results obtained through energy and probability-distribution analysis. Calculation in position and momentum space were 
carried out by means of a simple yet accurate variation-induced method. From a classical viewpoint, particle in such a 
potential will always reside in deeper well, but this work reveals that there may be a clearly defined rule which localizes 
it in either deeper or shallower well. A rule has also proposed for energy distribution as well. It is also established 
that at certain range of $\gamma$, two wells of DW effectively behave as two different potentials.

Study of wave function and probability completely explains distribution of particle in two wells. 
Uncertainty-based information measures can not fully explain the location of particle, but sufficient to describe the 
switching from one well to another. Asymmetry prevails over competing effect of localization/delocalization as 
particle usually resides in a definite well. Interestingly, there are certain transition points where, opposing effects 
predominate over asymmetry, causing the particle to reside in both wells. Convergence of such properties arises not 
only due to quasi-degeneracy, but also because of same number of effective nodes in those states.
Furthermore, a semi-classical analysis indicates that at integer $k$, phase-space can be split into two closed 
lobes, signifying quasi-degeneracy in such DW potentials. 
It is also shown that, symmetric DW may be treated as a special case of asymmetric DW. Further investigation of 
information-based quantities in 2D, 3D potentials or confinement situations may lead to interesting insight, some of which 
may be taken up later.  
\section{Acknowledgement} 
The authors sincerely thank the two anonymous referees for constructive comments and suggestions. NM acknowledges IISER 
Kolkata for a post-doctoral fellowship. It is a pleasure to thank Prof.~Raja Shunmugam for his kind support.

\end{document}